\documentclass[twoside,12pt]{article}
\usepackage{float}
\usepackage{epsfig}
\usepackage{axodraw}
\usepackage{amsmath}
\usepackage{amssymb}
\usepackage{eucal}
\usepackage{cite}
\usepackage[rflt]{floatflt}

\usepackage[T2A,OT1]{fontenc}
\usepackage[ot2enc]{inputenc}
\usepackage[russian,english]{babel}

\def\TCop{\textsuperscript{\textcopyright}}

\def\Journal#1#2#3#4{{#1} {#2} (#4) #3 }

\def\NPA{{\em Nucl. Phys.} A}

\def\PLB{{\em Phys. Lett.} B}

\def\PRL{\em Phys. Rev. Lett.}

\def\PREP{\em Phys. Rep.}

\def\PRD{{\em Phys. Rev.} D}
\def\PRC{{\em Phys. Rev.} C}

\newcommand\DD{D\hspace*{-3mm}/\hspace{1mm}}

 \newcommand{\MS}{\overline{\sf MS}}
\newcommand{\be}{\begin{equation}}
\newcommand{\ee}{\end{equation}}
\newcommand{\bea}{\begin{eqnarray}}
\newcommand{\eea}{\end{eqnarray}}
\newcommand{\nn}{\nonumber}
\newcommand{\ep}{\varepsilon}

\newcommand{\lsim}{\raisebox{-0.07cm   }
{$\, \stackrel{<}{{\scriptstyle\sim}}\, $}}
\newcommand{\gsim}{\raisebox{-0.07cm   }
{$\, \stackrel{>}{{\scriptstyle\sim}}\, $}}
\newcommand\GeV{\,\mbox{GeV}}
\newcommand\TeV{\,\mbox{TeV}}
\newcommand\GA{\,\mbox{\boldmath $\Gamma$}}
\newcommand{\ds}{\displaystyle}

\newcommand\xv{\mbox{\boldmath $\xi$}}

\newcommand\Mvec{\mbox{\boldmath ${\rm M}$}}

\newcommand\IV{\mbox{\boldmath $I$}} 
\newcommand\UV{\mbox{\boldmath $U$}} 
\newcommand\LV{\mbox{\boldmath $L$}} 
\newcommand\PV{\mbox{\boldmath $P$}} 
\newcommand\RV{\mbox{\boldmath $R$}}

\newcommand\FV{\,\mbox{\boldmath $F$}}

\newcommand\MV{\,\mbox{\boldmath $M$}}

\newcommand\qV{\,\mbox{\boldmath $q$}}
\newcommand\eV{\,\mbox{\boldmath $e$}}

\begin{document}
\sloppy
\thispagestyle{empty} 

\title{{\small DESY 12-096,~~DO-TH-12/19,~~SFB/CPP-12-38,~~LPN 12-056,~~{\tt arXiv:1208.6087[hep-ph]}
\hfill
\\ }
\vspace{1cm} 
The Theory of Deeply Inelastic Scattering}
\author{Johannes Bl\"umlein
\\
Deutsches Elektronen-Synchrotron, DESY, \\
Platanenallee 6, D--15738 Zeuthen, Germany
}
\maketitle
%-----------------------------------------------------------------------------------------------------------
\begin{abstract} 
\noindent
The nucleon structure functions probed in deep-inelastic scattering at large virtualities form an important
tool to test Quantum Chromdynamics (QCD) through precision measurements of the strong coupling constant 
$\alpha_s(M_Z^2)$ and the different parton distribution functions. The exact knowledge of these quantities is 
also of importance for all precision measurements at hadron colliders. During the last two decades very 
significant progress has been made in performing precision calculations. We review the theoretical 
status reached for both unpolarized and polarized lepton-hadron scattering based on perturbative QCD.
\end{abstract}
%-----------------------------------------------------------------------------------------------------------
%\eject
%\tableofcontents

\newcommand\Qvec{\mbox{\boldmath $Q$}}
\newcommand\qvec{\mbox{\boldmath $q$}}
\newcommand{\nll}{\nonumber\\}
\newcommand{\bq}{\begin{equation}}
\newcommand{\eq}{\end{equation}}
\newcommand{\gap}{\stackrel{>}{\sim}}
\newcommand{\lap}{\stackrel{<}{\sim}}
\newcommand{\pola}{\stackrel{\rightarrow}{\Rightarrow}}
\newcommand{\polb}{\stackrel{\rightarrow}{\Leftarrow}}
\newcommand{\polc}{\stackrel{\rightarrow}{\Uparrow}}
\newcommand{\pold}{\stackrel{\rightarrow}{\Downarrow}}
\newcommand{\fem}{$F_2^{em}$}
\newcommand{\lam}{$\Lambda$}
\newcommand{\qsq}{$Q^2$}
\newcommand{\alsq}{$\alpha_s(Q^2)$}
\newcommand{\dals}{$\delta \alpha_s$}
\newcommand{\dfeq}{\stackrel{\scriptsize =}{\scriptsize df}}
\newcommand{\als}{\alpha_s}
\newcommand{\aqu}{\langle Q^2 \rangle}
\newcommand\PB{\mbox{pb}}
\newcommand\MSbar{$\overline{\mbox{MS}}$}
\newcommand\order{{\cal O}}
\newcommand\mat{{\cal M}}
\newcommand\secu{\,\mbox{sec}}
\newcommand\CALL{{\mathcal L}}
\newcommand\KG{\kappa_G}
\newcommand\KA{\kappa_A}
\newcommand\LG{\lambda_G}
\newcommand\LA{\lambda_A}
\newcommand\PD{\not p}
\newcommand\MP{M^2_{\Phi}}
\newcommand\BB{\beta}
\newcommand\XL{\log \left| \frac{1 + \B}{1 - \B} \right |}
\newcommand\SM{\frac{\hat{s}}{M_{\Phi}^2}}
\newcommand\SMM{\frac{\hat{s}^2}{M_{\Phi}^4}}
\newcommand\SMMM{\frac{\hat{s}^3}{M_{\Phi}^6}}
\newcommand\SMMMM{\frac{\hat{s}^4}{M_{\Phi}^8}}
\newcommand\BC{\beta^2 \cos^2 \theta}
\newcommand\CB{\beta^2 \cos^2 \theta}
\newcommand\CBB{\beta^4 \cos^4 \theta}
\newcommand\CBBB{\beta^6 \cos^6 \theta}
\newcommand\CBBBB{\beta^8 \cos^8 \theta}
\newcommand\SH{\hat{s}}
\newcommand\uh{\hat{u}}
\newcommand\thh{\hat{t}}
\newcommand\eA{\varepsilon_1}
\newcommand\eB{\varepsilon_2}
\newcommand\Ea{\epsilon_1}
\newcommand\Eb{\epsilon_2}
\newcommand\shh{{\mathrm{sh}}}
\newcommand\thhh{{\mathrm{th}}}
\newcommand\st{\sin^2 \theta}
\newcommand\BR{\left(1 - \CB \right)}
\newcommand\kavec{\,\mbox{\boldmath $k$}}
\newcommand\kbvec{\,\mbox{\boldmath $k'$}}
\newcommand\svec{\,\mbox{\boldmath $s$}}
\newcommand{\ba}{\begin{eqnarray}}
\newcommand{\ea}{\end{eqnarray}}

%%%%%%%%%%%%%%%%%%%%%%%%%%%%%%%%%%%%%%%%%%%%%%%%%%%%%%%%%%%%%%%%%%%%%%%%%%%%%%%%%%%%%%%%%%
\section{Introduction}
\label{sec:1}
\renewcommand{\theequation}{\thesection.\arabic{equation}}
\setcounter{equation}{0}
%%%%%%%%%%%%%%%%%%%%%%%%%%%%%%%%%%%%%%%%%%%%%%%%%%%%%%%%%%%%%%%%%%%%%%%%%%%%%%%%%%%%%%%%%%

\noindent
Matter consists of regular structures at microscopic distances, which exhibit themselves at 
the crystalline, molecular, and atomic levels \cite{DALTON}. The discovery of $\alpha, \beta$ and 
$\gamma$ radioactivity \cite{Becquerel} provided new natural probes beyond the visible spectrum of 
light and $X$-rays to resolve even smaller structures of matter. In 1911 E. Rutherford discovered 
the atomic nucleus of a size much smaller than that of atoms through scattering of $\alpha$-particles at 
gold~\cite{RUTHERF}. Herewith the picture of matter at small distances changed dramatically rising 
the question for further sub-structures. The composite nature of nuclei could be explained after 
Chadwick's \cite{CHADWICK} discovery of the neutron and Yukawa's model for nuclear forces \cite{YUKAWA}. 
Another important discovery was made by Frisch and Stern in 1933 measuring the anomalous magnetic moment 
of the proton with a  different value from that of point-particles, like electrons \cite{ANOMp}. Later in 1939 
Alvarez and Bloch measured the anomalous magnetic moment of the neutron \cite{ANOMn}, both of 
which constituted 
first 
evidence on the compositeness of nucleons. The current values of the nucleon magnetic moments are 
\cite{Nakamura:2010zzi}
%------------------------------------------------------------------------------------------------
\begin{equation}
\mu_p = 2.792847356 \pm  0.000000023~\mu_N,~~~~~ \mu_n = -1.9130427 \pm 0.0000005~\mu_N,
\label{eq:anmom}
\end{equation}
%------------------------------------------------------------------------------------------------
with $\mu_N =  e \hslash/2 m_p$ the nuclear magneton. 

\begin{figure}[H]
\begin{center}
\mbox{\epsfig{file=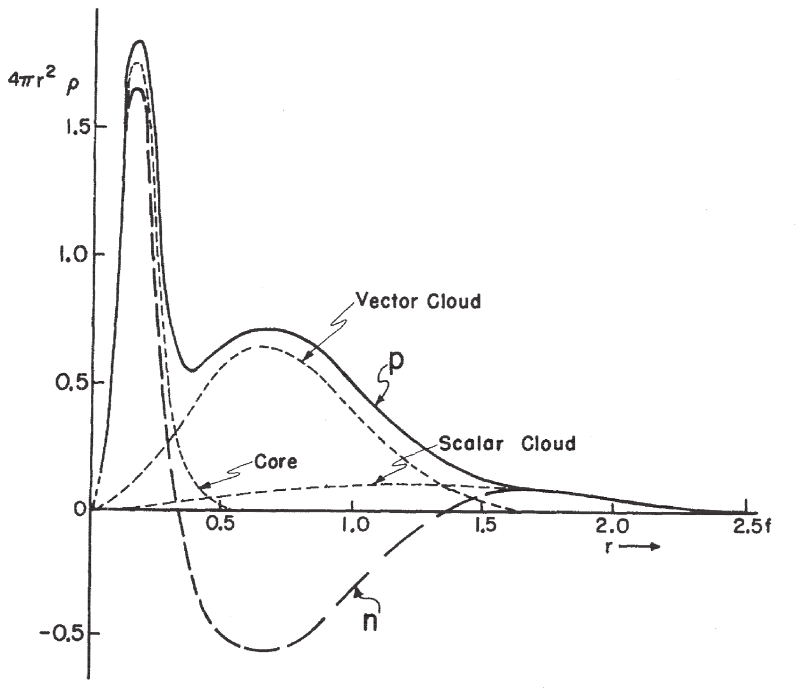,height=7cm,width=7cm}}
\end{center}
\caption[]{
\label{FIG:Schopper}
\small \sf
Charge distribution for the proton and the neutron implied by the form factors of Ref.~\cite{Olson:1961zz},
Figure 2(b); from \cite{Olson:1961zz}, \TCop (1961) by the American Physical Society.
}
\end{figure}

During the 1950ies the Hofstadter experiments \cite{Hofstadter:1963} operated at virtualities being
large enough to reveal the charge distribution inside nucleons, which is illustrated in 
Figure~\ref{FIG:Schopper}. 
A positive core distribution and 
tail are found both for the proton and neutron, with a positive vector cloud in case of the proton
and a negative one for the neutron, pointing to first details of the nucleon sub-structure. However,
the specific nature of these distributions remained yet unexplained.
%----------------------------------------------------------------------------------------------
\restylefloat{figure}
%----------------------------------------------------------------------------------------------

\begin{figure}[H]
\begin{center}
\includegraphics[scale=0.30,angle=-90]{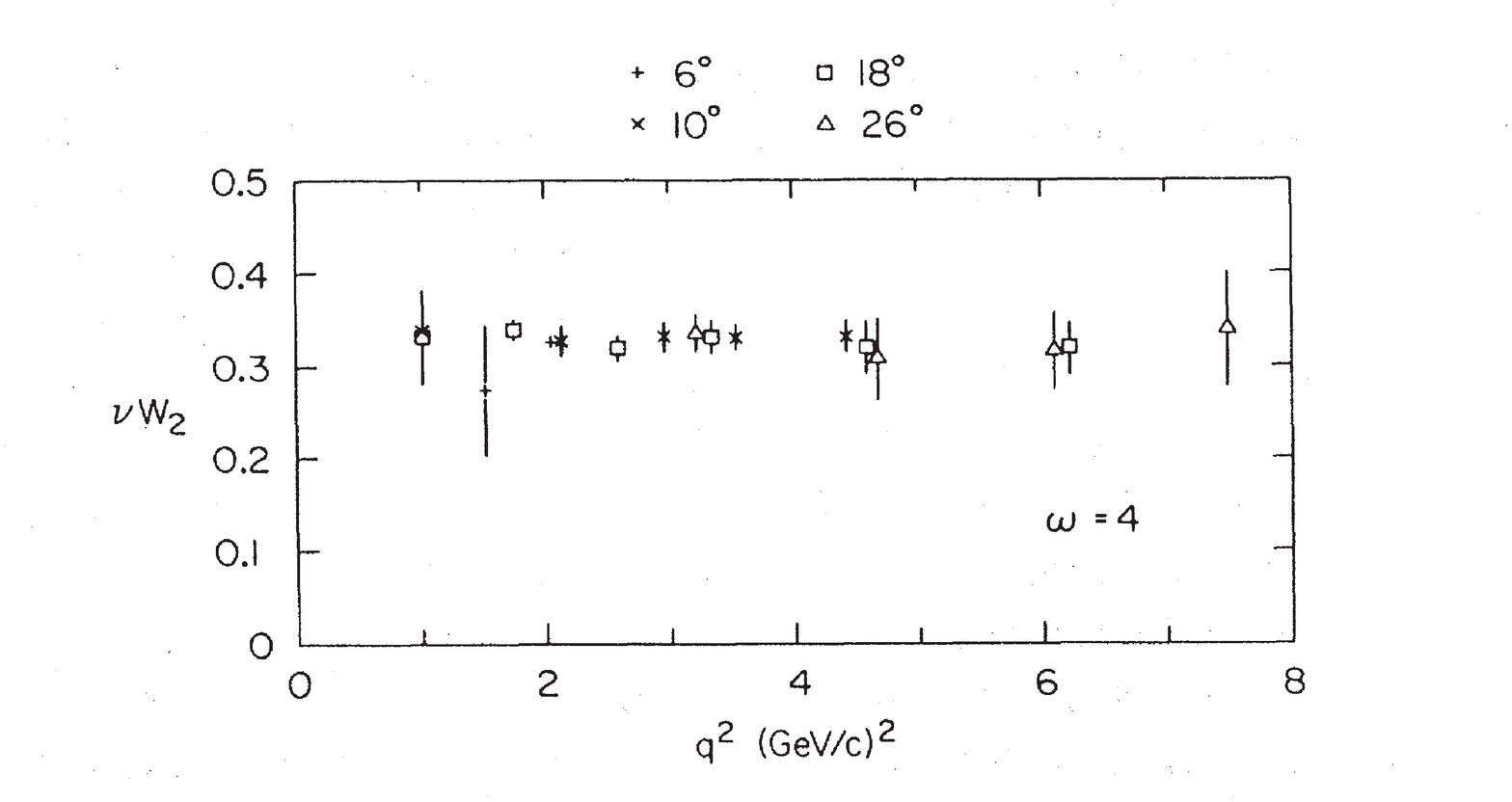}
\includegraphics[scale=0.30,angle=-90]{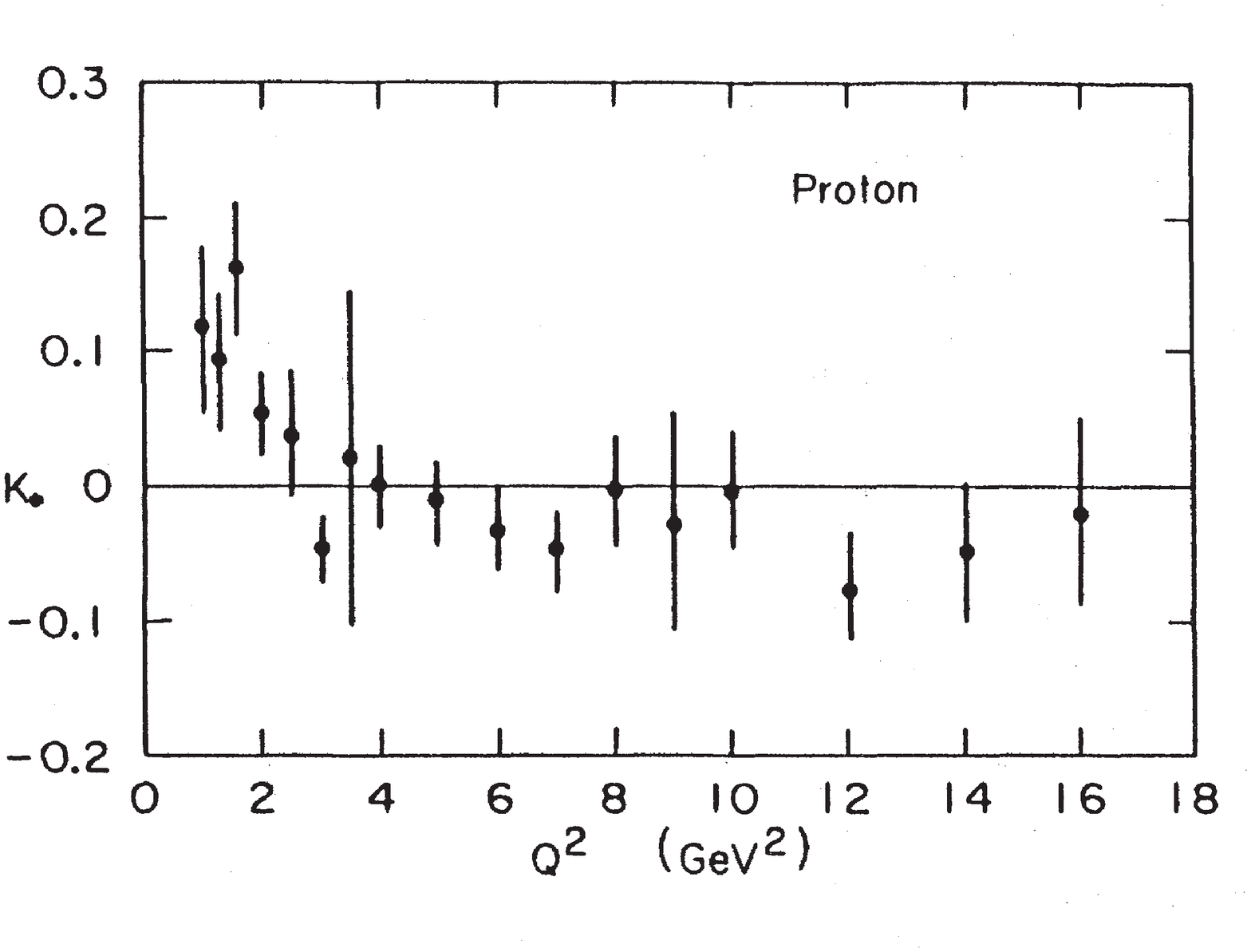}
\end{center}
\caption[]{
\label{FIG:scaling}
\small \sf
Left: An early observation of scaling: $\nu W_2$ for the proton
as a function of $-q^2$ for $W > 2 \GeV$, at $x = 1/\omega$ = 0.25; Right: 
The Callan-Gross relation: $K_0 = F_2/(2xF_1) - 1$ vs $-q^2$. 
These results established the spin of the partons
as $1/2$; from \cite{Kendall:1991np}, 
\TCop (1991) by the American Physical Society.
}
\end{figure}

In 1964 Gell-Mann \cite{GellMann:1964nj} and Zweig \cite{Zweig:1964jf} proposed the quarks~\footnote{
G.~Zweig named the hadron constituents {\it aces}.} as building blocks of hadrons to catalog the plethora of
observed mesons and baryons. During the late 1960ies the MIT-SLAC experiments 
\cite{Panofsky:1968pb,Taylor:1969xi,Bloom:1969kc,Breidenbach:1969kd,Kendall:1991np,Taylor:1991ew,
Friedman:1991nq} measured deep-inelastic
electron-nucleon scattering at the Stanford Linear Accelerator at much shorter distances and beyond the
resonance region. The important finding of these experiments were scaling and the observation
that the longitudinal structure function is small, Figure~\ref{FIG:scaling}, confirming a prediction
by Callan and Gross \cite{Callan:1969uq} for scattering off spin $1/2$ particles. The scaling behaviour 
of structure functions had been predicted by Bjorken using current algebra methods \cite{Bjorken:1968dy}. 
These new observations led Feynman to the parton model
\cite{FEYNMAN,Feynman:1973xc} of point-like fermionic 
constituents of the nucleons which react at high virtualities with the exchanged gauge bosons in the 
deep-inelastic process directly. 
%----------------------------------------------------------------------------------------------
\restylefloat{figure}
%----------------------------------------------------------------------------------------------

\begin{figure}[H]
\begin{center}
\mbox{\epsfig{file=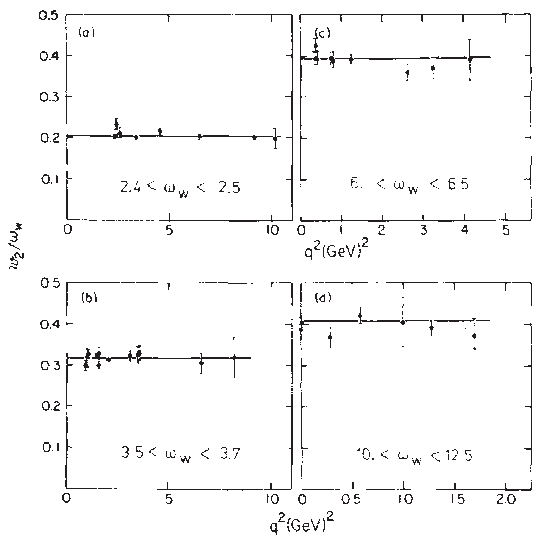,height=7cm,width=7cm}}
\end{center}
\caption[]{
\label{FIG:Brasse}
\small \sf
The function $\nu W_2$ plotted vs $|q^2|$ assuming $R = |q^2|/\nu^2$ for different fixed $\omega_W =
(1/x + M_N^2/|q^2|)/(1+ 0.2/|q^2|)$; from \cite{Brasse:1972wk}
\TCop (1972) by Elsevier Science.
}
\end{figure}

\noindent
Deep-inelastic scattering off constituent quarks has been discussed as early as 1967 \cite{BJ-HQ} in 
connection to data of that time \cite{Hand:1967}. After the discovery of scaling at SLAC also
data taken in other experiments were analyzed for this behaviour. One example concerns data taken at DESY 
at lower values of $|q^2|$ \cite{Brasse:1972wk}, cf. Figure~\ref{FIG:Brasse}, presented using the
Rittenberg-Rubinstein variable $\omega_W$. 
%----------------------------------------------------------------------------------------------
\restylefloat{figure}
%----------------------------------------------------------------------------------------------

The parton model introduced a new level of compositeness for fermions being confined inside hadrons 
and related to the strong interactions. The final quantum field theory of the strong interactions 
developed over a series of years. Already in 1965 Nambu \cite{Nambu:1966} proposed a Yang-Mills 
\cite{Yang:1954ek} $SU(3)$ gauge theory for the strong interactions, based on a three-valued charge 
degree of freedom \cite{Han:1965pf}. Before a symmetry was introduced using para-statistics 
\cite{Greenberg:1964pe} which later became color. At this time it was unknown whether Yang-Mills theories 
could be renormalized. The formalism by Faddeev and Popov \cite{Faddeev:1967fc} needed for their quantization in 
covariant gauges has been found two years later only. The renormalization of massless Yang-Mills theories
was proven by 't Hooft in 1971 \cite{'tHooft:1971fh} and Quantum Chromodynamics (QCD) as the theory of
strong interactions was proposed by Fritzsch and Gell-Mann in 1972 \cite{Fritzsch:1972jv} and 
Fritzsch, Gell-Mann and Leutwyler \cite{Fritzsch:1973pi}. In 1973 Gross, Wilczek \cite{Gross:1973id} and 
Politzer \cite{Politzer:1973fx} studied the running of the strong coupling constant of color octet
Yang-Mills theory with color triplet quarks and found asymptotic freedom, see 
also~\cite{Khriplovich:1969aa,tHooft:unpub}. The Lagrangian of QCD, referring to the covariant $R_\xi$-gauges,
is given by \cite{YND}
%-------------------------------------------------------------------------------------------------------------
\begin{eqnarray}
{{\CALL}}_{\rm QCD} = \sum_q \bar{\psi}_{q,j}(x) \left[i\DD^{jk} - m_q\right] \psi_{q,k}(x) 
- \frac{1}{4} F_a^{\mu\nu}(x) F^a_{\mu\nu}(x)
- \frac{1}{2\xi} (\partial_\mu A_a^\mu(x))^2
+ \partial_\mu \chi^a_\mu(x) D^{ab,\mu} \chi_b(x)~,
\end{eqnarray}
%-------------------------------------------------------------------------------------------------------------
where $\psi_q(x)$ denotes the quark fields, $A_a^\mu(x)$ the gluon fields, 
$F_a^{\mu\nu} = \partial^\mu A_a^\nu - \partial^\nu A_a^\mu + g f_{abc} A^{b,\mu} A^{c,\nu}$ the field 
strength tensor, $f_{abc}$ the structure constants of $SU(3)_c$, the gauge group of QCD, $\xi \in {\mathbb R}$
the gauge parameter, $\chi_a(x)$ the ghost field, and the covariant derivatives
$D_\mu^{ab} = \delta^{ab} \partial_\mu - g f^{abc} A_{c,\mu}(x), \DD^{jk} = \gamma_\mu \left[\delta_{jk} 
\partial^\mu - ig A_a^\mu t_{jk}^a\right]$, with $t^a$ the generators of $SU(3)_c$. 
Based on this, perturbative calculations in Quantum 
Chromodynamics can be performed at large virtualities. Due to their high complexity these calculations
are usually being performed using computer algebra programs, a first dedicated of which was 
{\tt SCHOONSCHIP} by M.~Veltman \cite{Veltman:1963}.

At short distances, resp. large scales of the momentum transfer, the nucleon structure functions 
$F_i(x,Q^2)$ obey the light--cone expansion \cite{Wilson:1969zs,Zimmermann:1970,Brandt:1970kg,
Frishman:1971qn} through which they are represented in terms of Wilson coefficients $C_i^k$
and non-perturbative operator matrix elements (OMEs) $f_k$ of local operators, characterized by their 
twist \cite{Gross:1971wn}. At leading twist the functions $f_k$ are the parton densities and at 
higher twist they correspond to parton correlation functions. In this separation an arbitrary scale 
$\mu_F^2$, the factorization scale, emerges. Applying the renormalization group equations 
\cite{Stueckelberg:1953dz,Symanzik:1970rt,Callan:1970yg,Petermann:1979} to the structure functions one obtains 
separate evolution equations 
for the Wilson coefficients and operator matrix elements which are usually of matrix-type due to the
mixing of different operators. The solution of these evolution equations exhibits scaling 
violations~\footnote{Scaling violations emerge naturally through radiative 
corrections in all renormalizable field theories. Drell and collaborators at the end of the 1960ies 
were seeking for scaling describing structure functions by fermion-meson interactions with loop
corrections and found scaling violations in 
general~\cite{DRELL60,Yan:1969gn}, see also \cite{DISS:WVN}.}, 
implied by the renormalization group equations, and the running of the strong coupling constant 
and finite quark masses.

During the last 40 years the deep--inelastic structure functions both in neutrino and charged 
lepton--nucleon scattering were measured at a steadily increasing precision and allow at present for 
important QCD precision tests at the per cent level. The strong coupling constant $\alpha_s(M_Z^2)$ can 
be measured at an accuracy of $\sim 1\%$ from the scaling violations of the
deep--inelastic structure functions. Very accurate extractions of the twist-2 parton densities are
possible, which are an essential ingredient for the physics at hadron colliders as Tevatron and the LHC,
and are thus instrumental in the search for new elementary particles. 

At the theoretical side, the deep--inelastic structure functions are best described at the level 
of the twist-2 contributions. The QCD coupling constant is known to 4-loop order 
\cite{vanRitbergen:1997va,Czakon:2004bu}. The anomalous dimensions and Wilson coefficients for the 
unpolarized massless case were calculated to 3-loop order 
\cite{Larin:1993vu,Larin:1996wd,Retey:2000nq,Blumlein:2004xt,Moch:2004pa,Vogt:2004mw,Vermaseren:2005qc}.
There are first results at 4-loop order \cite{Baikov:2006ai,CHET_priv1,Velizhanin:2011es}. Heavy quark 
corrections are known to 2-loop order \cite{HEAV2,Riemersma:1994hv} and for larger scales a series of
Mellin moments has been calculated to 3-loop order \cite{Bierenbaum:2009mv}. In the polarized case the 
level of 2-loop corrections has been reached for massless and massive corrections 
\cite{Zijlstra:1993sh,Mertig:1995ny,VOGELS:POL,Buza:1996xr,Bierenbaum:2007pn}, with first results at 
3-loop order \cite{Vogt:2008yw}. The present massless and massive 3-- and 4--loop calculations require
powerful computer algebra systems and packages like {\tt FORM} \cite{Vermaseren:2000nd}, {\tt MATAD} 
\cite{Steinhauser:2000ry}, and {\tt Sigma} \cite{SIGMA}, cf. also \cite{Harlander:1998dq}, to perform these 
voluminous calculations which amount even in parallel form to several CPU years and request 
$O(100~{\rm Gbyte})$ RAM at multi-processor systems. QED corrections to the deep-inelastic scattering cross 
sections are large in certain kinematic regions and have to be known at the accuracy defined by the 
measured data. The region of small values of $x$ or large hadronic masses at a given
virtuality requires special attention because of potentially large perturbative corrections. In the large 
$x$ region target mass effects and higher twist corrections contribute. 
For large values of $x$ higher order resummations are important. In the polarized case 
the structure function $g_2(x,Q^2)$ provides an 
inclusive observable at which twist--3 contributions can be studied. Structure functions and parton 
distributions obey a series of sum rules, to which also QCD corrections are available and one may perform 
QCD tests using these relations. The nucleon spin problem \cite{EMCp1} rouse the question of also angular 
momentum contributions. These can be accessed in deeply-virtual Compton
scattering and similar reactions \cite{NONF} as has been shown in \cite{Ji:1996ek}.
  
In this article we review the present theoretical understanding of the deep--inelastic process, mainly  
in the region of large scales. In Section~\ref{sec:2} the scattering cross sections are described and a 
brief survey on the present deep inelastic data is given. Section~3 deals with the light-cone expansion
and in Section~4 the parton model is described. The renormalization of deep-inelastic structure functions
in the twist-2 approximation is discussed in Section~5. Related to this, a survey on the perturbative 
expansion of the strong coupling constant $\alpha_s(Q^2)$, the anomalous dimensions, resp. splitting 
functions, and the massless Wilson coefficients are given in Section~6--8. In Section~9 we discuss QED 
and electro-weak corrections to deep inelastic scattering. The heavy quark corrections are described in 
Section~10. Section~11 deals with the target mass corrections. The solution of the evolution 
equations is outlined in Section~12. Here we also discuss the results of different NNLO QCD analyses and 
the status of the determination of $\alpha_s(M_Z^2)$ from deep-inelastic data. In Section~13 a series of 
aspects of small-$x$ resummations are considered. Resummations in the region of large values of $x$
are discussed in Section~\ref{sec:largX}. Major sum rules and integral relations for deep-inelastic 
structure functions are summarized in Section~15. Higher twist corrections and aspects of 
nuclear parton distributions are considered in Sections~16 and 17.
In this review we will fully concentrate on firmly established results on the basis of perturbative
Quantum Chromodynamics and will not discuss other approaches of a more phenomenological character, 
including other assumptions.

%%%%%%%%%%%%%%%%%%%%%%%%%%%%%%%%%%%%%%%%%%%%%%%%%%%%%%%%%%%%%%%%%%%%%%%%%%%%%%%%%%%%%%%%%%
\section{The deep-inelastic process}
\label{sec:2}
\renewcommand{\theequation}{\thesection.\arabic{equation}}
\setcounter{equation}{0}
%%%%%%%%%%%%%%%%%%%%%%%%%%%%%%%%%%%%%%%%%%%%%%%%%%%%%%%%%%%%%%%%%%%%%%%%%%%%%%%%%%%%%%%%%%

\noindent
The deep-inelastic process, at Born level, can be illustrated by the diagram shown in Figure~4. 
A lepton $l = e^\pm \mu^\pm, \nu_i (\overline{\nu}_i)$ with momentum $k_1$ scatters off a nucleon $N$ 
exchanging an electro-weak gauge boson $V = \gamma, Z^0, W^\pm$ to a lepton  $l'$ with momentum $k_2 = k_1 
- q$ and an ensemble of hadrons $X$ in an effective $2 \rightarrow 2$ process $k_1 + p_1 \rightarrow k_2 
+ p_2$. Here $p_1$ is the momentum of the incoming nucleon and $p_2$ the momentum of the outgoing 
hadrons. 

In many deep-inelastic scattering experiments the kinematic variables can be measured 
only from a sub-set of the momenta of the external particles. This applies in particular to the 
momentum of outgoing neutrinos but also for the momentum $p_2$ in fixed target experiments.
As will be shown later in Section~\ref{sec:QED} the size of the radiative corrections strongly 
depends on the choice of the kinematic variables. For the measurement at HERA a wide variety 
of sets of kinematic variables was designed to allow for cross checks and to limit the size of 
QED radiative corrections. 

%-----------------------------------------------------------------------------------------
\begin{center}
\setlength{\unitlength}{1mm}
\begin{picture}(60,40)(0,0)
\ArrowLine(0,100)(60,80)
\ArrowLine(60,80)(120,100)
\Photon(60,80)(60,40) 5 5
\ArrowLine(0,20)(60,40)
\ArrowLine(60,40)(120,40)
\put(55,0){\makebox(0,0)[b]{$\left \} \right.$
\sf spectators}}
\put(-7,34){\makebox(0,0)[b]{$l$}}
\put(57,13){\makebox(0,0)[b]{\sf current jet}}
\put(27,20){\makebox(0,0)[b]{\Large\sf V}}
\put(33,35){\makebox(0,0)[b]{$k_2$}}
\put(43,17){\makebox(0,0)[b]{$p_{q_f}$}}
\put( 9,35){\makebox(0,0)[b]{$k_1$}}
\put( 7,12){\makebox(0,0)[b]{$p_{q_i} =
x p_1$}}
\put(15,20){\makebox(0,0)[c]{$q$}}
\ArrowLine(0,10)(120,10)
\ArrowLine(0,0)(120,0)
\makebox(-15,12)[b]{\Large\sf N {\LARGE $\left \{ \right.$}}
\end{picture}
\end{center}

\vspace{5mm}
\noindent
\begin{center}
{\small
Figure~4: \sf 
\sf Diagram describing deep inelastic $lN$ scattering. 
}
\end{center}
\setcounter{figure}{4}
%-----------------------------------------------------------------------------------------
At Born level the unpolarized scattering cross section depends on the virtuality of the exchanged
gauge boson $V$, $q^2 = - Q^2$ and the inelasticity $y$. At a given cms energy $s = (k_1 + p_1)^2$ 
these quantities define the Bjorken variable $x$,
%-------------------------------------------------------------------------------------------------
\begin{eqnarray}
\label{eq:y}
y &=& \frac{p_1.(k_1 - k_2)}{p_1.k_1} \\  
\label{eq:x}
x &=& \frac{Q^2}{s y}~.
\end{eqnarray}
%-------------------------------------------------------------------------------------------------
The different sets of variables are~: 
%-------------------------------------------------------------------------------------------------
\begin{center}
\begin{tabular}{rll}
{\it i)}  & {\it Leptonic variables}  & $q \equiv q_l = k_2 - k_1,~~y_l = p_1.(k_1-k_2)/p_1.k_1$\\
%---------------
{\it ii)}  & {\it Hadronic variables} \cite{JBMK} & $q \equiv q_h = p_2 - 
p_1,~~y_l = p_1.(p_2-p_1)/p_1.k_1$\\
%---------------
{\it iii)} & {\it Jacquet-Blondel variables} \cite{Amaldi:1979yh}  & $Q^2_{JB} = 
(\vec{p}_{2,\perp})^2/(1-y_{JB}),~~y_{JB} = \Sigma/(2 E(k_1))$\\
&  & $\Sigma = \sum_h (E_h - p_{h,z})$ \\
%---------------
{\it iv)} & {\it Mixed variables} \cite{JBMK}    & $q = q_l, y_m = 
y_{JB}$\\
%---------------
{\it v)}  & {\it Double angle method} \cite{BEK}  
& $Q^2_{DA} = \frac{\ds 4 E(k_2)^2 \cos^2(\theta(k_2)/2)}{\ds \sin^2(\theta(k_2)/2) 
+ 
\sin(\theta(k_2)/2)
\cos(\theta(k_2)/2)
\tan(\theta(p_2)/2)}$,
\\
& & $y_{DA} = 1 - \frac{\ds \sin(\theta(k_2)/2)}{\ds \sin(\theta(k_2)/2) 
+ 
\cos(\theta(k_2)/2)
\tan(\theta(p_2)/2)}$,\\
%---------------
{\it vi)} & {\it $\theta y$ method} \cite{Blumlein:1994ii}   & $ Q^2_{\theta y} = 4 E(k_2)^2 
(1-y_{JB}) 
\frac{\ds 1 + \cos(\theta(k_2))}{\ds 1 - \cos(\theta(k_2))},~~y_{\theta y} = y_{JB}$\\
%---------------
{\it vii)}  & {\it $\Sigma$ method}  \cite{Bassler:1994uq}    &  
$Q^2_\Sigma = \frac{\ds (\vec{k}_{2,\perp})^2}{\ds 1 - y_\Sigma},~~y_\Sigma = 
\frac{\ds \Sigma}{\ds \Sigma + E(k_2) [1-\cos(\theta(k_2))]}$\\
%---------------
{\it viii)} & {\it $e\Sigma$ method} \cite{Bassler:1994uq}    & $Q_{e\Sigma}^2 = Q_l^2,~~y_{e\Sigma} 
= \frac{\ds Q_l^2}{\ds s x_\Sigma} $
\end{tabular}
\end{center}
%-------------------------------------------------------------------------------------------------
Here $E(l_i)$ and $\theta(l_i)$ are measured in the detector's rest frame. The invariant mass $W$ of
the final state hadrons is given by
%-------------------------------------------------------------------------------------------------
\begin{eqnarray}
W^2 = p_2^2 = (q + p_1)^2 = M^2 + 2q.p_1 - Q^2 = M^2 + Q^2 (1-x)/x~,
\end{eqnarray}
%-------------------------------------------------------------------------------------------------
and $M$ is the nucleon mass. The inclusive lepton-nucleon process is denoted as deep inelastic if 
$W > 2 \GeV$ and the values of $Q^2$ are sufficiently large, normally $Q^2 > 4 \GeV^2$. If $W \approx M$ 
the process is called (quasi-)elastic, allowing also for proton--neutron transitions. Here the scattering 
cross section is described by the (quasi-)elastic nucleon form-factors 
\cite{FORMFACT,Schmitz}.
The resonance region is characterized by $M < W \lsim 2\GeV$. Corresponding models for the scattering 
cross sections were given in \cite{RESON}.

The double differential scattering Born cross section off unpolarized nucleons read 
\cite{SCX,Arbuzov:1995id,Schmitz}
%------------------------------------------------------------------------------------------
\begin{eqnarray}
\label{eq:ncB}
\frac{d^2 \sigma_{\rm NC}^{l^\pm N}}{dx dy} &=& \frac{2 \pi \alpha^2 s}{Q^4} \left\{ \left[ 2(1-y) - 2xy 
\frac{M^2}{S} \right] \hat{F}_2(x,Q^2) + Y_- x\hat{F}_3(x,Q^2) + y^2\left(1 - \frac{2m_l^2}{Q^2}\right) 
2x\hat{F}_1(x,Q^2)
\right\} 
\nonumber\\
\\
\label{eq:ncNU}
\frac{d^2 \sigma_{\rm NC}^{\nu(\bar{\nu}N}}{dx dy} &=& \frac{G_F^2 s}{16 \pi} \left[ \frac{M_Z^2}{Q^2 + 
M_Z^2}\right]^2
\left\{Y_+ W_2^{\rm NC}(x,Q^2) \pm Y_- xW_3^{\rm NC}(x,Q^2) - y^2 W_L^{\rm NC}(x,Q^2)\right\}
\\
\label{eq:ccB}
\frac{d^2 \sigma_{\rm CC}}{dx dy} &=& \frac{G_F^2 s}{4 \pi} \left[ \frac{M_W^2}{Q^2 + M_W^2}\right]^2
\left\{Y_+ W_2^{\rm CC}(x,Q^2) \pm Y_- xW_3^{\rm CC}(x,Q^2) - y^2 W_L^{\rm CC}(x,Q^2)\right\},
\end{eqnarray}
%------------------------------------------------------------------------------------------
where $\alpha$ and $G_F$ denote the fine-structure and Fermi constants, $M_{W,Z}$ and $m_l$ are the 
$W,Z$-boson and lepton masses and
%------------------------------------------------------------------------------------------
\begin{eqnarray}
Y_\pm = 1 \pm (1-y)^2~.
\end{eqnarray}
%------------------------------------------------------------------------------------------
Three of the electro-weak parameters in (\ref{eq:ncB}--\ref{eq:ccB}) are independent. One choice is
given by the set $(\alpha, M_Z, G_F)$. The others are expressed by corresponding relations, 
cf. \cite{Arbuzov:1995id}. 
In case of neutral current scattering the functions $\hat{F}_i$ are propagator-weighted structure 
functions, \cite{Arbuzov:1995id}~:
%------------------------------------------------------------------------------------------
\begin{eqnarray}
\hat{F}_{1,2}(x,Q^2) &=& F_{1,2}(x,Q^2) + 2|Q_e|(v_e + \lambda a_e) \chi(Q^2) G_{1,2}(x,Q^2)
                        + 4 (v_e^2 + a_e^2 + 2 \lambda v_e a_e) \chi^2(Q^2) H_{1,2}(x,Q^2) 
\nonumber\\ 
\\
x\hat{F}_{3}(x,Q^2) &=& - 2 {\rm sign}(Q_e) \left\{|Q_e|(a_e + \lambda v_e) \chi(Q^2) xG_{3}(x,Q^2)
                        + [2 v_e a_e + \lambda(v_e^2 + a_e^2)] \chi^2(Q^2) 
x H_{3}(x,Q^2)  \right\},
\nonumber\\ 
\end{eqnarray}
%------------------------------------------------------------------------------------------
with $Q_e = -1, \lambda = {\rm sign} \xi,  v_f = 1 - 4 |Q_f|\sin^2 \theta_w^{\rm eff}, a_f = 1,
\chi(Q^2) = G_F M_Z^2 Q^2/(\sqrt{2}~8 \pi \alpha (Q^2 + M_Z^2))$. $\theta_w^{\rm eff}$ denotes the
effective weak mixing angle, $M_Z$ the $Z$-boson mass, and $\xi$ the lepton polarization.
The structure function $2xF_1$  can be expressed using 
%------------------------------------------------------------------------------------------
\begin{eqnarray}
2x{F}_1(x,Q^2) = {F}_2(x,Q^2) - {F}_L(x,Q^2)
\end{eqnarray}
%------------------------------------------------------------------------------------------
and similar for $G_1, H_1, W^{\pm}_1$. At this level the Callan-Gross relation \cite{Callan:1969uq} 
implies vanishing longitudinal structure functions, which take finite values due to QCD corrections and 
target mass effects, see Sections~\ref{sec:WILSC},\ref{sec:TM}.

The scattering cross sections (\ref{eq:ncB}--\ref{eq:ccB}) can be represented in the 
general form
%------------------------------------------------------------------------------------------
\begin{eqnarray}
\label{eq:scat}
\sum_i \frac{d^3 \sigma_i}{dx dy d\phi} &=& \sum_i P_i(s,Q^2) L^{\mu\nu}_i W_{\mu\nu,i}~,
\end{eqnarray}
%------------------------------------------------------------------------------------------
where $P_i(s,Q^2)$ denotes a propagator term  and $L_{\mu\nu,i}$ and $W^{\mu\nu}_i$ are
the respective contributions to the leptonic and hadronic tensor. Here we specified also the azimuthal 
angle of the final state lepton which emerges in the scattering cross section of polarized leptons off 
transversely polarized nucleons. The leptonic tensor can be calculated perturbatively. The hadronic tensor 
is given by
%----------------------------------------------------------------------
\begin{eqnarray}
W_{\mu\nu,i} = \frac{1}{4\pi}\int d^4x e^{iqx} 
\langle PS\mid[{J_\mu^{i_1}(x)}^\dagger ,J_\nu^{i_2}(0)]\mid PS\rangle~.
\label{eq:HADTEN}
\end{eqnarray}
%----------------------------------------------------------------------
Here $S$ denotes the four--vector of the nucleon spin with $S\cdot P=0$ and the normalization 
$S^2 = -M^2$. 
In the  framework of the quark--parton model the currents $J_{\mu}^j$ are
given by
%----------------------------------------------------------------------
\ba
\label{eqCUR1}
J_\mu^j(x)=\sum_{f,f'}\bar q'(x)\gamma_\mu(v_q^{j}+a_q^{j}
\gamma_5)q(x)U_{ff'}~,
\ea
%----------------------------------------------------------------------
with $v_q^{j}$ and $a_q^{j}$ the vector and axial-vector couplings of the quarks.
For charged current interactions $U_{ff^\prime}$ denotes the Cabibbo-Kobayashi-Maskawa matrix, 
whereas for neutral current interactions  $U_{ff^\prime} = \delta_{ff^\prime}$. The hadronic tensor
obeys the following properties~\cite{Geyer:1977gv,Tangerman:1994eh}
%------------------------------------------------------------------------------------------
\begin{eqnarray}
{\rm Covariance:}        && W_{\mu'\nu'}(q',P') = \Lambda_{\mu'}^\mu \Lambda_{\nu'}^\nu 
W_{\mu\nu}(q,P),~~\Lambda \in L_+^{\uparrow} \\
{\rm Hermiticity:}       && W_{\mu\nu}(q,P) = W^*_{\nu\mu}(q,P) \\
{\rm Spectrality:}       && W_{\mu\nu}(q,P) = 0,~~~-\frac{q^2}{2 P.q} > 1\\
{\rm Causality:}         && \tilde{W}_{\mu\nu}(x,P) = \int d^4 q e^{-iqx} W_{\mu\nu}(q,p)\\
{\rm T}-{\rm invariance:}     && W_{\mu\nu}(\bar{q},\bar{P},\bar{S}) = \left[W^{\mu\nu}(q,P,S)\right]^*, 
\end{eqnarray}
%------------------------------------------------------------------------------------------
where $L_+^{\uparrow}$ is the orthochronous Lorentz group and $\bar{a}_\mu = a^\mu$. In general,
%------------------------------------------------------------------------------------------
\begin{eqnarray}
\hspace*{-3.cm} 
{\rm Parity:}                 && W_{\mu\nu}(\bar{q},\bar{P},-\bar{S}) = 
W^{\mu\nu}(q,P,S) \\
\hspace*{-3.5cm} 
{\rm Symmetry:}               && W_{\mu\nu}(q,P) = W_{\mu\nu}(-q,P)\\
\hspace*{-3.5cm} 
{\rm Current~conservation:}   && q_\mu W^{\mu\nu} = W^{\mu\nu} q_\nu = 0~.
\end{eqnarray}
%------------------------------------------------------------------------------------------
are not obeyed. We construct the following hadronic tensor including the case of polarized
targets, \cite{Blumlein:1998nv}: 
%----------------------------------------------------------------------
\begin{eqnarray}
W_{\mu\nu} =&& \left(-g_{\mu\nu}+\frac{q_\mu q_\nu}{q^2}\right)F_1(x,Q^2)
  + \frac{\hat P_\mu\hat P_\nu}{P\cdot q} F_2(x,Q^2)
-i\varepsilon_{\mu\nu\lambda\sigma}\frac{q^\lambda P^\sigma}{2P\cdot q}
F_3(x,Q^2)
\nonumber \\
&&+\frac{q_\mu q_\nu}{P\cdot q} F_4(x,Q^2)
+\frac{(p_\mu q_\nu+p_\nu q_\mu)}{2 P\cdot q} F_5(x,Q^2) \nonumber \\
&& +i\varepsilon_{\mu\nu\lambda\sigma}\frac{q^\lambda S^\sigma}{P\cdot q}
g_1(x,Q^2)
+i\varepsilon_{\mu\nu\lambda\sigma}
\frac{q^\lambda (P\cdot q S^\sigma-S\cdot q P^\sigma)}{(P\cdot q)^2}
g_2(x,Q^2) \nonumber \\ 
&&+\left[\frac{\hat P_\mu\hat S_\nu+\hat S_\mu\hat P_\mu}{2} -
S\cdot q\frac{\hat P_\mu\hat P_\nu}{P\cdot q}\right] \frac{g_3(x,Q^2)} 
{P\cdot q}\nonumber \\
&&+S\cdot q\frac{\hat P_\mu\hat P_\nu}{(P\cdot q)^2} g_4(x,Q^2) +
(-g_{\mu\nu}+\frac{q_\mu q_\nu}{q^2})\frac{(S\cdot q)}{P\cdot q} g_5(x,Q^2),
\nonumber\\
&& + i\varepsilon_{\mu\nu\lambda\sigma} \frac{P_\sigma S_\lambda}{P\cdot q}
g_6(x,Q^2) +
S\cdot q\frac{q_\mu q_\nu}{(P\cdot q)^2} g_7(x,Q^2) \nonumber \\
&&+\frac{(p_\mu q_\nu+q_\mu p_\nu) S\cdot q}{2 (P\cdot q)^2} g_8(x,Q^2)
+\frac{S_\mu q_\nu+S_\nu q_\mu}{2 P\cdot q} g_9(x,Q^2)~,
\label{eq:HADT}
\end{eqnarray}
%--------------------------------------------------------------------------
with
%--------------------------------------------------------------------------
\begin{eqnarray}
\hat P_\mu = P_\mu-\frac{P\cdot q}{q^2}q_\mu~,~~~~~~~~~~~~~~~~~
\hat S_\mu = S_\mu-\frac{S\cdot q}{q^2}q_\mu~.
\nonumber
\end{eqnarray}
%--------------------------------------------------------------------------
The choice of the polarized structure functions beyond $g_{1,2}$ is not unique in
the literature, cf.~\cite{Blumlein:1996vs}. For the functions $g_{3,4,5}$ we used
the convention of \cite{Blumlein:1996vs}, and for the further structure functions
that of \cite{Ji:1993ey,Maul:1996dx}. The spin-vector in the longitudinal and transversal 
case is defined by
%----------------------------------------------------------------------
\begin{eqnarray}
S_L &=& (0, 0, 0, M)~, \nonumber\\
S_T &=& M (0, \cos\alpha, \sin\alpha,0)~.
\end{eqnarray}
%----------------------------------------------------------------------
For the purely polarized nucleon contributions one obtains the following 
differential scattering cross sections, \cite{Blumlein:1998nv}~:
%----------------------------------------------------------------------
\begin{eqnarray}
\label{eq:Plong}
\frac{d^2\sigma(\lambda, \pm S_L)}{dxdy}
&=& \pm
2 \pi s  \frac{\alpha^2}{Q^4}
\sum_i C_i \eta_i(Q^2)
\nn\\
&\times& \left [
-2\lambda y \left( 2-y-\frac{2 x y M^2}{S} \right) xg_1^i +
8 \lambda \frac{y x^2 M^2}{S} g_2^i
 + \frac{4 x M^2}{s} \left ( 1 - y - \frac{x y M^2}{s} \right)
 g_3^i \right.
\nn\\
&-& 2
\left (1 + \frac{2 x M^2}{s} \right)
\left (1 - y - \frac{x y M^2}{s} \right ) g_4^i
- 2 x y^2 \left (1 + \frac{2 x M^2}{s} \right) g_5^i \nn\\
&+&   \left.
4\lambda\frac{x y M^2}{s} g_6^i -2\left(1-y-\frac{x y M^2}{s}\right)
g_9^i \right]~,
\\
%------------
\label{eq:Ptran}
\frac{d^3\sigma(\lambda, \pm S_T)} {dx dy d\phi} &=&
\pm
s \frac{\alpha^2}{Q^4} \sum_i
C_i \eta_i(Q^2)
\nn\\
&\times&\!\!\!
 2 \sqrt{\frac{M^2}{s}} \sqrt{x y \left [1 - y - \frac{x y M^2}{s}
\right]} \cos(\alpha-\phi) \left [
-2\lambda y x g_1^i -
4\lambda x  g_2^i \right.
\\     
&-& \left.\!\!\!\frac{1}{y} \left (2 - y
- \frac{2 x y M^2}{s} \right) g_3^i
+ \frac{2}{y}  \left (1 - y - \frac{x y M^2}{s} \right) g_4^i
+  2 x y g_5^i -2\lambda g_6^i -g_9^i  \right]~.
\nn
\end{eqnarray}
%----------------------------------------------------------------------

\noindent
Here $\lambda$ denotes the lepton polarization and $C^\gamma = 1, C^{\gamma Z} = g_V + \lambda g_A,
g_V^\gamma = 1, g_A^\gamma = 0, g_V^Z = (1 - 4 Q_l \sin^2\theta_W^{\rm eff})/2, g_A^Z = -1/2, 
g_V^{W^-} = 1, g_A^{W^-} = -1, C^{Z} = (g_V + \lambda g_A)^2, 
C^{W^\pm} = (1 \pm \lambda), \eta^{|\gamma|^2} = 1, 
\eta^{\gamma Z} = G_F Q^2/(2 \sqrt{2} \pi \alpha) (M_Z^2/(Q^2 +M_Z^2),
\eta^{|Z|^2} = (\eta^{\gamma Z})^2, \eta^{|W^\pm|^2} = [G_F Q^2/(4 \pi \alpha) (M_W^2/(Q^2 
+M_W^2)]^2$.~\footnote{A factor of 1/2 has to be corrected for $\eta^{|W^\pm|^2}$ in 
\cite{Blumlein:1996vs,Blumlein:1998nv}, 
cf. \cite{Forte:2001ph}.} The structure functions $g_7$ to $g_9$ do not contribute in case of 
vanishing lepton masses and the structure functions $F_{4,5}$ are related to $F_{1,2,3}$.

The individual structure functions can be measured form the scattering cross sections 
(\ref{eq:ncB}--\ref{eq:ccB},\ref{eq:Plong}, \ref{eq:Ptran}) applying the QED and electro-weak 
radiative corrections to the data, cf. Section~\ref{sec:QED}.  
Current data analyses in the unpolarized case refer to the structure function data on $F_2$ and 
charged current scattering cross sections for proton and deuteron targets taken at SLAC 
%\cite{Atwood:1976ys, 
%Bodek:1979rx,Mestayer:1982ba,Whitlow:1990dr, Whitlow:1991uw, Gomez:1993ri, Dasu:1993vk}
\cite{SLAC,Dasu:1993vk}, from BCDMS \cite{Benvenuti:1989rh,Benvenuti:1989fm},
NMC \cite{NMC,Arneodo:1996qe}
%\cite{ Arneodo:1993kz, Arneodo:1995cq, Arneodo:1996qe,Arneodo:1996kd} 
at CERN, and HERA \cite{herapdf:2009wt}. Surveys on earlier neutrino and charged 
lepton deep-inelastic data were given in \cite{SURV1}.
%%Eisele:1986uz,Diemoz:1986kt,Sloan:1988qj,Mishra:1989jc,WINTER}. 

Moreover, there are data on the longitudinal structure function $F_L(x,Q^2)$ 
\cite{
Benvenuti:1989rh,
Whitlow:1990gk,
Dasu:1993vk,
Arneodo:1996qe,
Adloff:1996yz,
Abe:1998ym,
Aaron:2010ry}.
To separate the contributions due to the different sea-quark flavors Drell-Yan data 
\cite{DY:EXP}
%\cite{Moreno:1990sf,Towell:2001nh,Webb:2003ps} 
and di-muon data \cite{DIMUON}
%\cite{Bazarko:1994tt, Goncharov:2001qe,Tzanov:2005kr, Mason:2006qa} 
are used. In some analyses \cite{Martin:2009iq,Ball:2011uy} also other sets 
of inclusive and semi-inclusive deep-inelastic scattering data, including data on $F_2^{Q\bar{Q}}(x,Q^2),~Q 
= c,b$, data from E665, CHORUS, and Tevatron data on 
weak boson production
 \cite{EXP1a} are used.
%\cite{Adloff:2000qk,Chekanov:2001qu,
%Chekanov:2008aa,Chekanov:2009gm,Adloff:2003uh,Adloff:2000qj,Breitweg:1998dz,Chekanov:2002ej,Chekanov:2003yv,
%Chekanov:2003vw} 
%\cite{Breitweg:1999ad,Chekanov:2003rb,Chekanov:2008yd,Chekanov:2009kj,Adloff:1996xq,
%Adloff:2001zj,Aktas:2005iw,Aktas:2004az}, $ep$ jet-production 
%\cite{H1:2007pb,Chekanov:2002be,Chekanov:2006xr}, 
%\cite{Adams:1996gu,Onengut:2005kv}, 
%\cite{Abazov:2007pm,Acosta:2005ud,CDF_Z,Aaltonen:2009ta,Aaltonen:2010zza,Abazov:2007jy}.
Various PDF-fitting groups use also the Tevatron jet data 
\cite{TEVJET}
%\cite{Abulencia:2007ez,Aaltonen:2008eq,D0:2008hua,Abazov:2010fr,Affolder:2001hn}
in their analysis.
%----------------------------------------------------------------------------------------------
\restylefloat{figure}
\begin{figure}[H]
\begin{center}
\includegraphics[scale=0.40,angle=0]{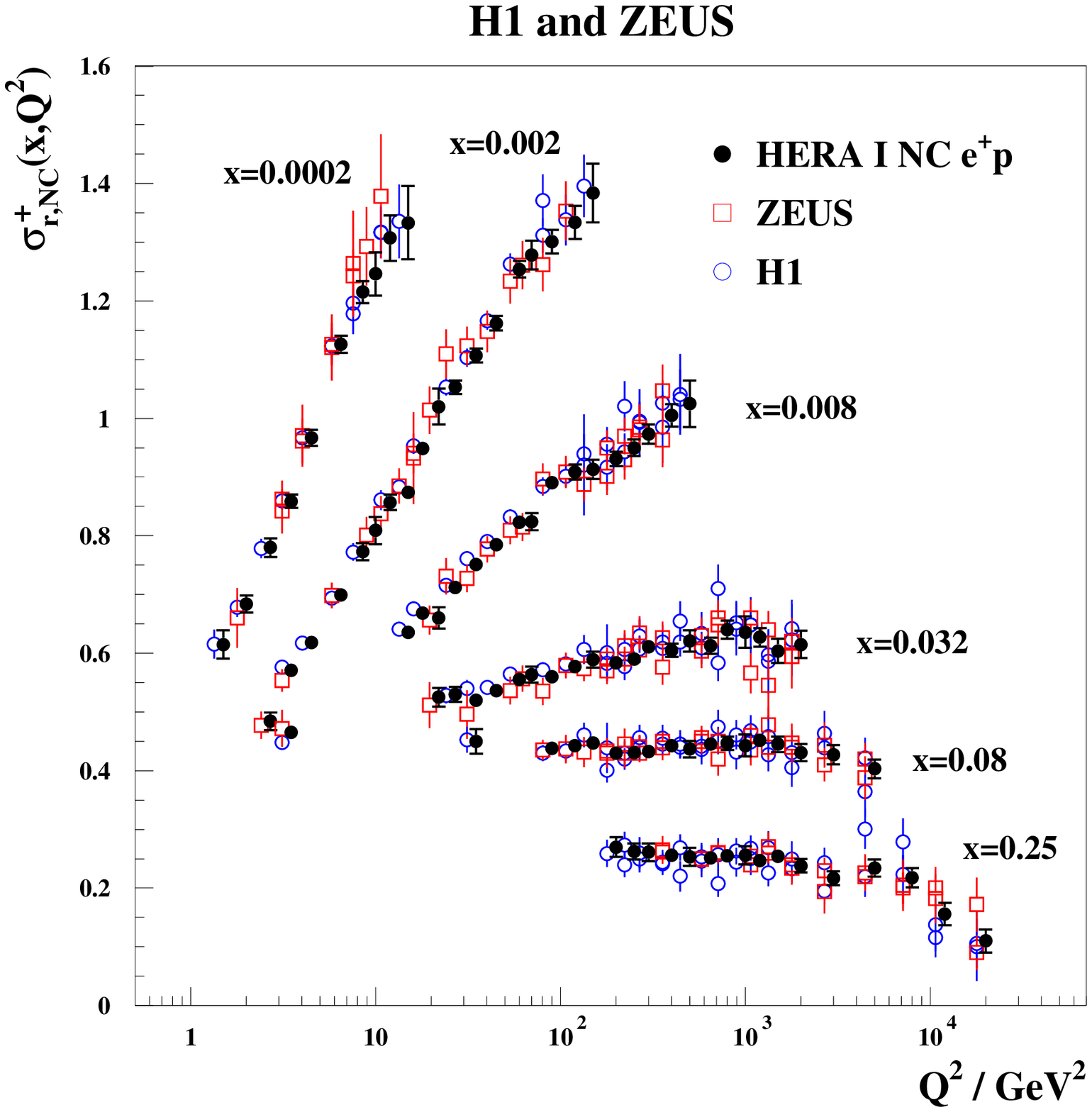}
\includegraphics[scale=0.40,angle=0]{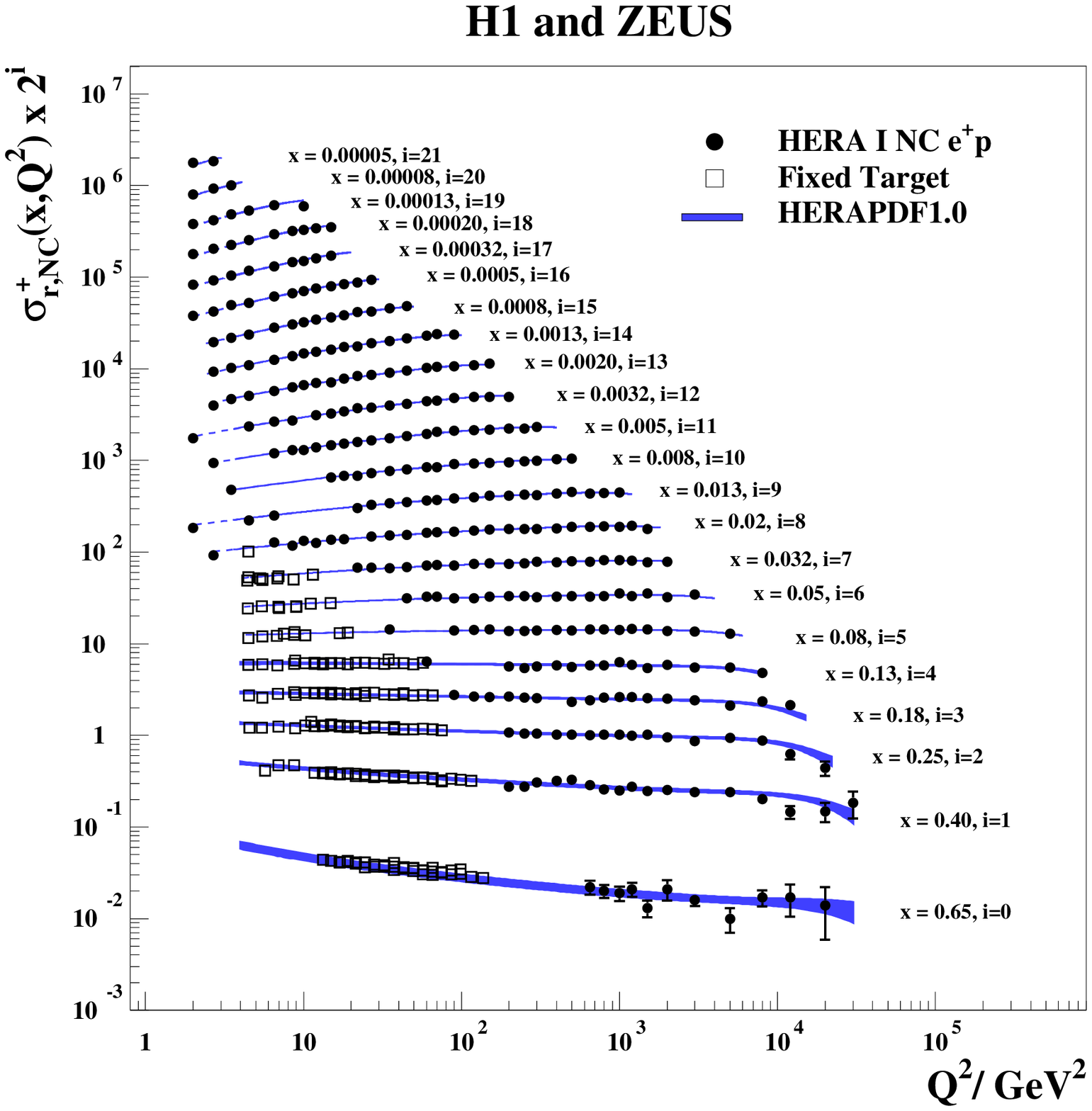}
\end{center}          
\caption[]{
\label{DIS1:FIG}
\small \sf
Left:~
HERA combined NC $e^+p$ reduced cross section as a function of $Q^2$ for six $x$-bins
compared to the separate H1 and ZEUS data input to the averaging procedure. The error bars
indicate the total experimental uncertainty. The individual measurements are displaced 
horizontally for better visibility; Right~: The combined data with the HERAPDF1.0 fit 
\cite{HERAPDF1.0} is superimposed. The bands represent the total uncertainty of the fit. 
Dashed lines are shown for $Q^2$ values not included in the QCD analysis;
from \cite{herapdf:2009wt} \TCop (2009) Springer Verlag.
}
\end{figure}
%----------------------------------------------------------------------------------------------

%----------------------------------------------------------------------------------------------
\restylefloat{figure}
\begin{figure}[H]
\begin{center}
\includegraphics[scale=0.40,angle=0]{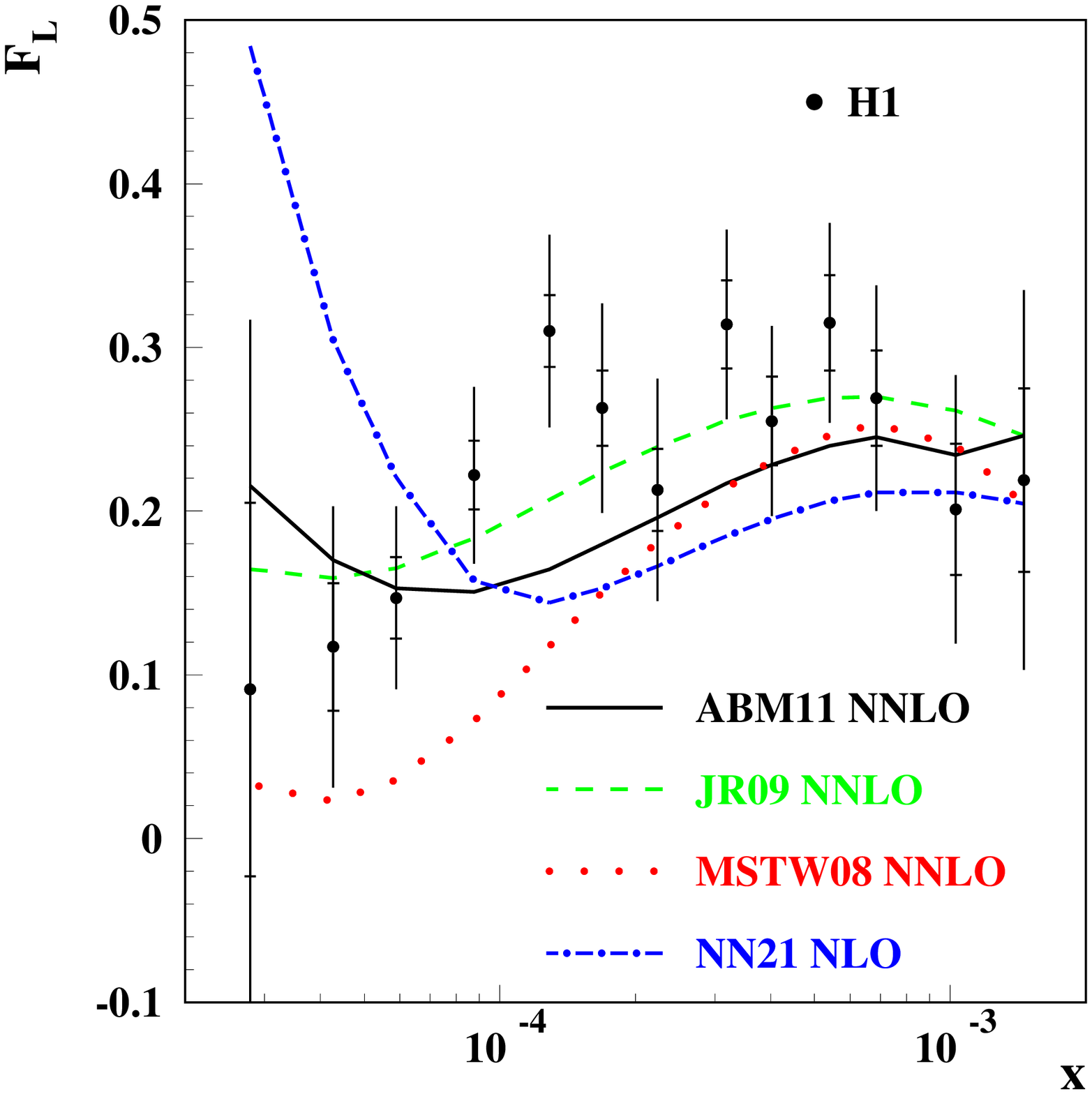}
\end{center}          
  \caption{\small   
    \label{fig:flh1} 
\sf      
     The data on $F_L$ versus $x$ obtained by the H1 collaboration~\cite{Aaron:2010ry}
     confronted with the 3-flavor scheme NNLO predictions based on the
     different parton distributions functions (PDFs). Solid line: 
     ABM11~\cite{Alekhin:2012ig}, dashes:
     JR09~\cite{JimenezDelgado:2008hf}, dots: MSTW~\cite{Martin:2009iq}).
     The NLO predictions based on the 3-flavor NN21
     PDFs~\cite{Ball:2011mu} are given for comparison (dashed dots).
     The value of $Q^2$ for the data points and the curves in the plot
     rises with $x$ in the range of $1.5 \div 45~{\rm GeV}^2$; from \cite{Alekhin:2012ig}.
  }
\end{figure}
%----------------------------------------------------------------------------------------------
\noindent
In the polarized case the data stem from SLAC \cite{E142n,E154n,E143pd,E155d,E155p}
EMC, SMC, COMPASS \cite{EMCp1,EMCp2,SMCpd,COMPd,COMP1} at CERN, CLAS \cite{JLABn,CLA1pd,CLA2pd},
and HERMES \cite{HERMn,HERMpd} at HERA, taken at $p$, $d$, and $n$ ($^3$He) targets on the 
polarization asymmetry $A_1$, the ratio $g_1/F_1$ or for the structure function $g_1$. 
Data on the structure function $g_2(x,Q^2)$ were taken by the SLAC experiments, SMC and HERMES
\cite{EXP:G2,E143pd}.
%%\cite{Adams:1997tq,E143pd,Anthony:2002hy,Airapetian:2011wu}. 

%----------------------------------------------------------------------------------------------
\restylefloat{figure}
\begin{figure}[H]
\begin{center}
\includegraphics[scale=0.25,angle=0]{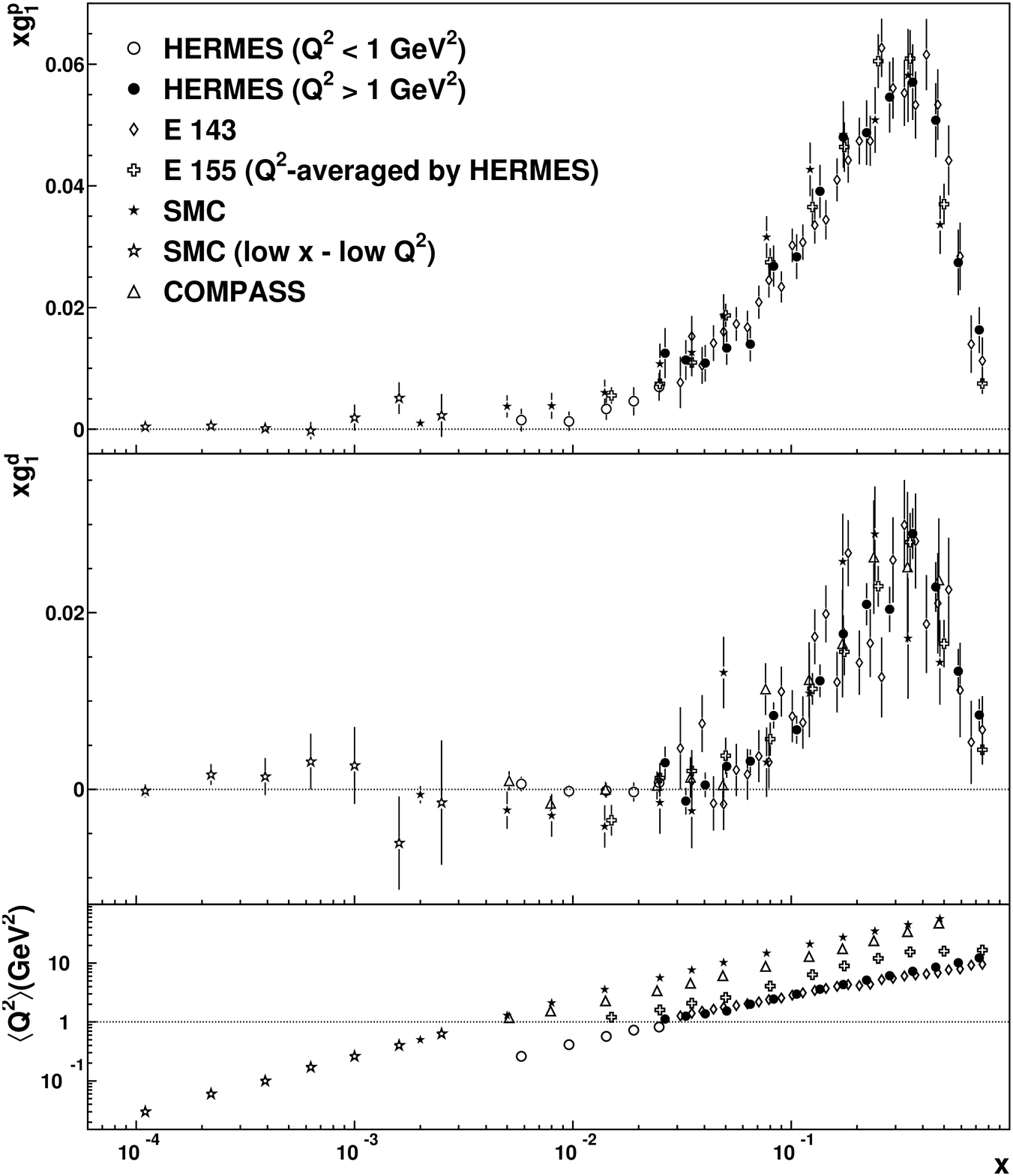}
\end{center}          
\caption[]{
\label{DIS3:FIG}
\small \sf
HERMES results on $xg_1^p$ and $xg_1^d$ vs $x$, shown on separate panels, compared to data from SMC 
\cite{SMCpd}, E143 \cite{E143pd}, E155 \cite{E155d,E155p} and COMPASS \cite{Ageev:2005gh}.
The error bars represent the sum in quadrature of statistical and systematic uncertainties.
The HERMES data points shown are statistically correlated by unfolding QED radiative 
and detector smearing effects; the statistical uncertainties shown are obtained from only 
the diagonal elements of the covariance matrix. The E143 and E155 data points are correlated 
due to the method for correcting for QED radiation. For the HERMES data the closed
(open) symbols represent values derived by selecting events with $Q^2 > 1 \GeV^2 (Q^2 < 1 \GeV^2 )$;
from \cite{HERMpd}
\TCop (2007) by the American Physical Society.
}
\end{figure}
%----------------------------------------------------------------------------------------------

The current precision in the measurement of the unpolarized structure function $F_2(x,Q^2)$ is illustrated 
in Figure~\ref{DIS1:FIG} showing the combined HERA data together with the world fixed target data. Also the 
individual measurements of H1 and ZEUS are compared. The present experimental errors reach the 1\% level
and the scaling violations of $F_2(x,Q^2)$ are clearly visible over a wide range in $Q^2$. 

In Figure~\ref{fig:flh1} recent measurements of the structure function $F_L(x,Q^2)$ at low values of $x$ 
by the H1 experiment are shown together with predictions by different PDF-fits. The structure function
$F_L(x,Q^2)$ can only be measured changing the beam energy in given bins of $(x,Q^2)$ to separate it
from $F_2(x,Q^2)$ via the $y$-dependence of the scattering cross section. A precise measurement
requires high luminosities and can be expected at the EIC, and at even smaller values of $x$ at a future
high energy $ep$ collider.

Finally, we illustrate the current precision of the measurement of the polarized structure function 
$g_1(x,Q^2)$ at $p$ and $d$ targets in Figure~\ref{DIS3:FIG}. If compared to the structure function
$F_2(x,Q^2)$ the experimental errors are larger since $g_1(x,Q^2)$ is measured from a polarization 
asymmetry. The range in $Q^2$ of these measurements is smaller than in case of $F_2(x,Q^2)$  since all
measurements were performed in fixed target experiments.

\begin{figure}[H]
\begin{center}
\includegraphics[scale=0.35,angle=0]{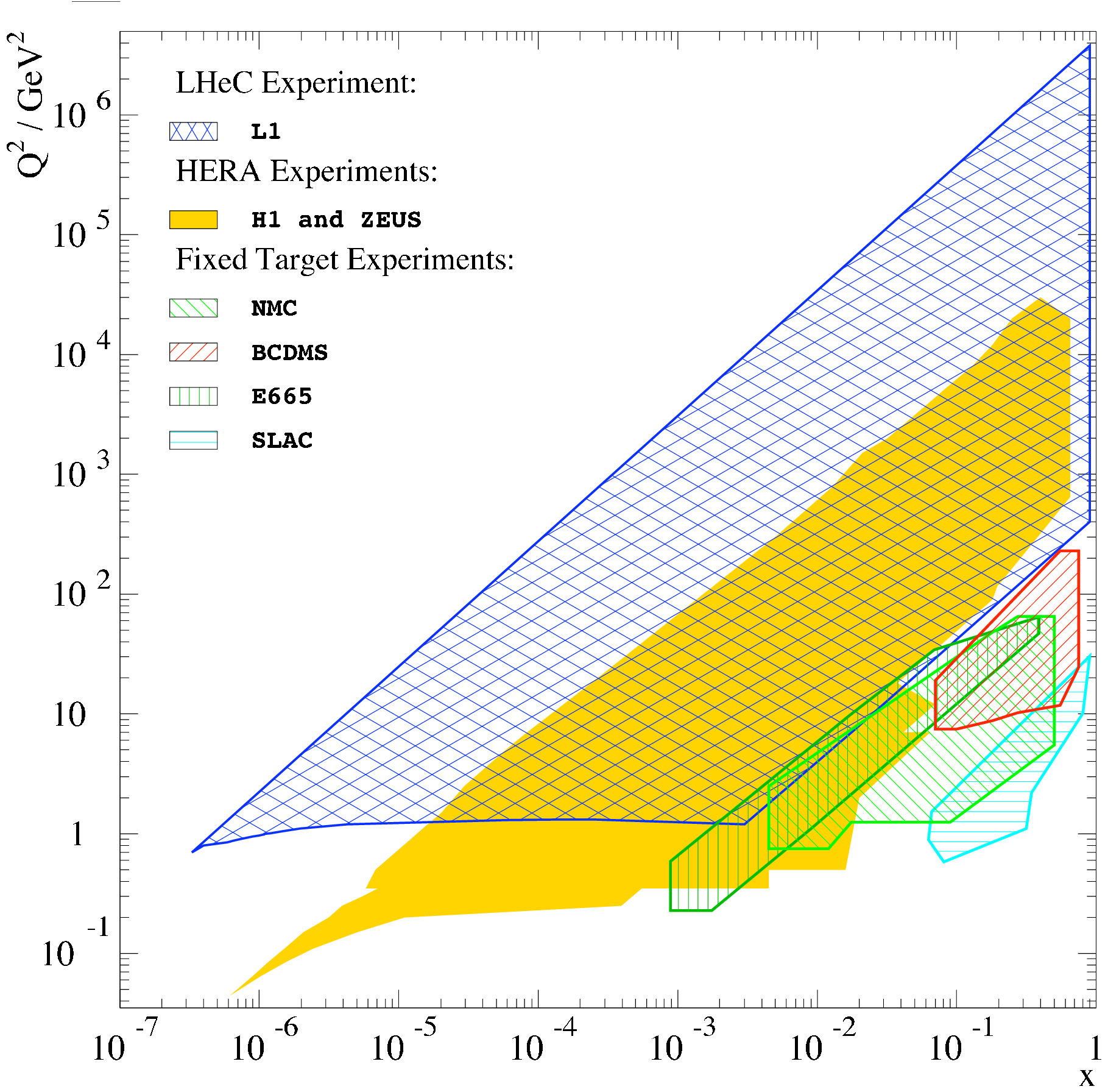}
\end{center}          
\caption[]{
\label{FIG:KIN}
\small \sf
The kinematic range in $x$ and $Q^2$ probed by the deep-inelastic scattering experiments 
at HERA, CERN, and SLAC. The region accessible to a possible future $e^{\pm}p$ experiment
with $E_p = 7 \TeV$ and $E_e = 140 \GeV$ is also shown; by courtesy of M.~Klein and
E.~Lobodzinska.
}
\end{figure}

In Figure~\ref{FIG:KIN} the kinematic region which has been probed by the different fixed target 
experiments at CERN, SLAC and the HERA experiments is illustrated. For virtualties $Q^2 \geq 4 \GeV^2$
values of $x \simeq 4 \cdot 10^{-4}$ are reached. The highest values of $Q^2$ reached are $\sim 20.000 
\GeV^2$. There is a proposal for an $ep$ experiment in the LHC ring operating at $E_p = 7 \TeV$ and $E_e = 
140 \GeV$, which would extend the present $x$ and $Q^2$ ranges by about 1.5 orders of magnitude
\cite{LHEC}.

The kinematic region for polarized data ranges for $0.005 < x < 0.75$ and $1 \lsim Q^2 \lsim 70~\GeV^2$ 
with a kinematic correlation between these two variables. At a future facility like the Electron-Ion 
Collider EIC \cite{EIC,Boer:2011fh} this region will be extended significantly. Moreover, the high 
luminosity
available will allow precision measurements also for polarized scattering. 
%----------------------------------------------------------------------------------------------
\restylefloat{figure}
%----------------------------------------------------------------------------------------------
%%%%%%%%%%%%%%%%%%%%%%%%%%%%%%%%%%%%%%%%%%%%%%%%%%%%%%%%%%%%%%%%%%%%%%%
\section{The Light-Cone Expansion}
\renewcommand{\theequation}{\thesection.\arabic{equation}}
\setcounter{equation}{0}
\label{sec:3}
%%%%%%%%%%%%%%%%%%%%%%%%%%%%%%%%%%%%%%%%%%%%%%%%%%%%%%%%%%%%%%%%%%%%%%%

\vspace{1mm}
\noindent
We consider the hadronic tensor (\ref{eq:HADTEN}) and limit the discussion for brevity to the case of pure 
photon exchange. It is given by the absorptive part of the forward Compton-amplitude $T_{\mu\nu}$
%------------------------------------------------------------------------------------------
\begin{eqnarray}
W_{\mu\nu}(P,q) = \frac{1}{\pi} {\sf Im} T_{\mu\nu}(P,q)
\end{eqnarray}
%------------------------------------------------------------------------------------------
with
%------------------------------------------------------------------------------------------
\begin{eqnarray}
\label{eq:Tmn}
T_{\mu\nu}(P,q) &=& i \int d^4\xi e^{iq\xi} \langle P| {\sf T} J_\mu(\xi)J_\nu(0)|P\rangle \\ 
&=& \frac{1}{2x} \left(g_{\mu\nu} + \frac{q_\mu q_\nu}{Q^2}\right) T_1(x,Q^2)
%\nonumber\\ & &
+ \frac{2x}{Q^2} \left(P_\mu P_\nu + \frac{P_\mu q_\nu + P_\nu q_\mu}{2x} -\frac{Q^2}{4x^2} g_{\mu\nu}\right)
T_2(x,Q^2)~.
\end{eqnarray}
%------------------------------------------------------------------------------------------
One may integrate (\ref{eq:Tmn}), \cite{Frishman:1971qn}, which yields
%------------------------------------------------------------------------------------------
\begin{eqnarray}
\label{eq:Tmn1}
T_{\mu\nu}(P,q) 
= T_{\mu\nu}(q^2,\nu) 
&=& 4 \pi \int_{-\infty}^{+\infty} d\xi_0 \int_0^\infty d|\xv|~|\xv|\left[
\frac{e^{i\nu \xi_0}
    \sin(\sqrt{\nu^2-q^2}|\xv|)}{\sqrt{\nu^2-q^2}}\right] 
\nonumber\\ && \hspace*{5cm}
\times
\langle P|{\sf T}[J_\mu(x) J_\nu(0)]|P\rangle~.
\end{eqnarray}
%-----------------------------------------------------------------------------------------
For $\nu \rightarrow \infty$ the exponentials behave like
%------------------------------------------------------------------------------------------
\begin{eqnarray}
e^{i \nu \xi_0} e^{\pm i \sqrt{\nu^2 - q^2}|\xv|} \approx 
e^{i \nu(\xi_0 \pm |\xv|)} e^{\pm i M |\xv| x}~.
\end{eqnarray}
%-----------------------------------------------------------------------------------------
Due to the Riemann-Lebesgue theorem \cite{RL} the dominant contributions to (\ref{eq:Tmn1}) 
come from
%------------------------------------------------------------------------------------------
\begin{eqnarray}
|\xi_0 \pm |\xv|| \lsim \frac{1}{\nu},~~~~~~|\xv|           \lsim \frac{1}{M x}~,
\end{eqnarray}
%------------------------------------------------------------------------------------------
with $x$ the Bjorken variable. Thus 
%------------------------------------------------------------------------------------------
\begin{eqnarray}
\xi^2 = \xi_0^2 - \xv^2 \lsim \frac{1}{Q^2}
\end{eqnarray}
%------------------------------------------------------------------------------------------
and in the Bjorken limit $\nu, Q^2 \rightarrow \infty$ the contributions very near to the light 
cone dominate.

At very short distances $\xi^2 \approx 0$ the light-cone expansion 
of the time ordered product of currents has the following representation, 
\cite{Wilson:1969zs,Zimmermann:1970,Brandt:1970kg,Frishman:1971qn}~:
%------------------------------------------------------------------------------------------
\begin{eqnarray}
\label{eq:COMP1}
\lim_{\xi^2 \rightarrow 0} {\sf T} J(x)J(0) \sim \sum_{i,N,\tau} \bar{C}_{i,\tau}^N(\xi^2,\mu^2) \xi_{\mu_1} 
\ldots \xi_{\mu_N} O_{i,\tau}^{\mu_1, \ldots, \mu_N}(0,\mu^2)
\end{eqnarray}
%------------------------------------------------------------------------------------------
Here we consider general currents and
$O_{i,\tau}^{\mu_1, \ldots, \mu_N}$ denote local operators, which are finite as $\xi^2 
\rightarrow 0$, and $\bar{C}_{i,\tau}^N(\xi^2,\mu^2)$ are the corresponding Wilson coefficients. Let $D_O$ 
and $D_J$ be the canonical dimensions of the operators and currents and $N$ their global spin. The 
twist $\tau$ \cite{Gross:1971wn} of the operator is given by
%------------------------------------------------------------------------------------------
\begin{eqnarray}
\tau = D_O - N~.
\end{eqnarray}
%------------------------------------------------------------------------------------------
The Wilson coefficients behave then like 
%------------------------------------------------------------------------------------------
\begin{eqnarray}
\bar{C}_{i,\tau}^N(\xi^2,\mu^2) \approx \left(\frac{1}{\xi^2}\right)^{-\tau/2 + d_J}~.
\end{eqnarray}
%------------------------------------------------------------------------------------------
The local operators of lowest twist are given by
%------------------------------------------------------------------------------------------
\begin{eqnarray}
\label{eq:op1}
O^{\sf NS}_{q;r;\mu_1, \ldots, \mu_n}(0) &=& i^{N-1}{\bf S} \left[\bar{\psi} \gamma_P \gamma_{\mu_1} 
D_{\mu_2} \ldots D_{\mu_N} \frac{\lambda_r}{2}\psi\right]
\\ 
\label{eq:op2}
O^{\sf S}_{q;r;\mu_1, \ldots, \mu_n}(0) &=& i^{N-1}{\bf S} \left[\bar{\psi} \gamma_P \gamma_{\mu_1} 
D_{\mu_2} \ldots D_{\mu_N} \psi\right]
\\ 
\label{eq:op3}
O^{\sf S}_{g;r;\mu_1, \ldots, \mu_n}(0) &=& 2 i^{N-2}
{\bf S} {\bf Sp} \left[E_{\mu_1 \alpha}^{\beta \gamma} F^a_{\beta \gamma} D_{\mu_2} \ldots D_{\mu_{N-1}} 
F^{\alpha,a}_{\mu_N}\right]~.
\end{eqnarray}
%------------------------------------------------------------------------------------------
Here the indices $q$ and $g$ refer to quark and gluon field operators, respectively, and $\lambda_r$ denotes 
the Gell-Mann matrix of the corresponding light flavor representation; $\psi$
is the quark field, $F_{\mu\nu}^a$ the gluonic field strength tensor in QCD, $D_\mu$ the covariant derivative,
{\bf S} the symmetry operator for all Lorentz indices and {\bf Sp} the color-trace, where the index $a$ is
the color index in the adjoint representation. In the unpolarized case $\gamma_P = {\bf 1}$ and 
$E_{\mu_1 \alpha}^{\beta \gamma} = \delta_{\mu_1}^\beta \delta_{\alpha}^\gamma$,while in the polarized case 
$\gamma_P = \gamma_5$ and $E_{\mu_1 \alpha}^{\beta \gamma} = \frac{1}{2} \varepsilon_{\mu_1 \alpha}
^{\beta \gamma}$, where $\varepsilon_{\alpha_1 \alpha_2 \alpha_3 \alpha_4}$ is the Levi-Civita symbol.
The operators (\ref{eq:op1}--\ref{eq:op3}) still contain trace terms, which have to be subtracted. 
More generally than just by applying the symmetry operation a rigorous twist decomposition can be obtained
applying the method of \cite{GEYLA}.  

Let us consider the case of twist-2 operators in the following. We form the expectation value of the 
operator (\ref{eq:COMP1}) between nucleon states and perform the 
Fourier transform of the Wilson coefficients through which
%------------------------------------------------------------------------------------------
\begin{eqnarray}
\label{eq:COMP2}
T(x,Q^2) = \sum_N C^N\left(\frac{Q^2}{\mu^2}\right) A_N\left(\frac{\mu^2}{P^2}\right) \frac{1}{x^N}
\end{eqnarray}
%------------------------------------------------------------------------------------------
is obtained, where
%------------------------------------------------------------------------------------------
\begin{eqnarray}
\int d^4\xi e^{i\xi q} \xi_{\mu_1} \ldots \xi_{\mu_N} \bar{C}^N(\xi^2, \mu^2) 
&=& \frac{1}{i} \left(\frac{2}{Q^2}\right)^N q_{\mu_1} \ldots q_{\mu_N} C^N\left(\frac{Q^2}{\mu^2}\right) 
\\
\langle P|O_{\mu_1, \ldots \mu_N}(0,\mu^2)|P\rangle &=& P_{\mu_1} \ldots P_{\mu_N} 
A_N\left(\frac{\mu^2}{P^2}\right)~. 
\end{eqnarray}
%------------------------------------------------------------------------------------------
The sum in (\ref{eq:COMP2}) runs over even or odd moments starting with a value $N_0 \geq 0$ depending
on the crossing relation of the currents, cf. e.g.~\cite{Blumlein:1996vs}. 
Here $\mu^2$ denotes the factorization scale. This scale is arbitrary and cancels between the Wilson 
coefficients and the operator matrix elements since the structure functions do not depend on it.
Performing a contour integral in $x$ around the singularities of (\ref{eq:COMP2}) one obtains the 
expressions for the structure functions in Mellin space~: 
%------------------------------------------------------------------------------------------
\begin{eqnarray}
\label{eq:COMP3}
F_{2,L}(N,Q^2) = \Mvec[F_{2,L}(x,Q^2)](N) = \sum_i C_i^N\left(\frac{Q^2}{\mu^2}\right) 
A_{i,N}\left(\frac{\mu^2}{P^2}\right)~.
\end{eqnarray}
%------------------------------------------------------------------------------------------
Eq.~(\ref{eq:COMP3}) shows the factorized form of the deep-inelastic structure function. 
The factorization theorems \cite{FACT} state this form is remaining under higher order corrections.

Here the Mellin transformation \cite{MELLIN} is given by
%------------------------------------------------------------------------------------------
\begin{eqnarray}
\label{eq:MELLIN1}
\Mvec[A(x)](N) = \int_0^1 dx x^{N-1} A(x)~,
\end{eqnarray}
%------------------------------------------------------------------------------------------
with the Mellin-convolution
%------------------------------------------------------------------------------------------
\begin{eqnarray}
\label{eq:MELLIN2}
A(x) \otimes B(x) &=& \int_0^1 dx_1 \int_0^1 dx_2 \delta(x - x_1 x_2) A(x_1) B(x_2)\\
\label{eq:MELLIN3}
\Mvec[A(x) \otimes B(x)](N) &=& \Mvec[A(x)](N) \Mvec[B(x)](N)~.  
\end{eqnarray}
%------------------------------------------------------------------------------------------
The Wilson coefficients are perturbatively calculable and account for both the massless and massive quark 
contributions. The operator matrix elements $A_{i,N}$ are of non-perturbative nature. In case of  twist-2 
they also denote the moments of the parton distributions for massless quarks and 
gluons, labeled by the index $i$,
%------------------------------------------------------------------------------------------
\begin{eqnarray}
\label{eq:par1}
A_{i,N}\left(\frac{\mu^2}{P^2}\right) \equiv f_i(N, \mu^2)
\end{eqnarray}
%------------------------------------------------------------------------------------------
being discussed in the following Section.
%%%%%%%%%%%%%%%%%%%%%%%%%%%%%%%%%%%%%%%%%%%%%%%%%%%%%%%%%%%%%%%%%%%%%%%%%%%%%%%%%%%%%%%%%%
\section{Parton Models}
\label{sec:parton}
\renewcommand{\theequation}{\thesection.\arabic{equation}}
\setcounter{equation}{0}
%%%%%%%%%%%%%%%%%%%%%%%%%%%%%%%%%%%%%%%%%%%%%%%%%%%%%%%%%%%%%%%%%%%%%%%%%%%%%%%%%%%%%%%%%%

\noindent
The SLAC-MIT experiments \cite{Bloom:1969kc,Breidenbach:1969kd} found the strict correlation 
between the variables $\nu \equiv p.q/M \nu$ and $Q^2$ for large enough values and 
at the fixed ratio
%--------------------------------------------------------------------------------------
\begin{eqnarray}
\label{eq:COR}
                    \omega = \frac{2 M \nu}{Q^2} = \frac{1}{x}~.  
\end{eqnarray}
%--------------------------------------------------------------------------------------
In the Bjorken limit \cite{Bjorken:1968dy} the structure functions  are given by 
%--------------------------------------------------------------------------------------
\begin{eqnarray}
\label{eq:PAR3}
\lim_{Q^2, \nu \rightarrow \infty} M W_1(Q^2,\nu)   &\rightarrow& F_1(x) 
\\
\lim_{Q^2, \nu \rightarrow \infty} \nu W_2(Q^2,\nu) &\rightarrow& F_2(x)~. 
\end{eqnarray}
%--------------------------------------------------------------------------------------

In the (naive) parton model by Feynman \cite{FEYNMAN,Feynman:1973xc} one assumes that the 
correlation (\ref{eq:COR}) is {\it exact}. A nucleon at short distances is assumed to constitute of individual 
partons which are charged and move collinear to the nucleon momentum with momentum fractions $x_i,~\sum_i 
x_i = 1$. Let the parton which interacts with the virtual photon carry charge $e_1$ and momentum 
fraction $x_1$. The hadronic tensor is then given by
%--------------------------------------------------------------------------------------
\begin{eqnarray}
\label{eq:PARmain}
W_{\mu\nu} &=& \frac{4 \pi}{M} \left[p_\mu p_\nu W_2 - g_{\mu\nu} M^2 W_1\right] = 
\int_0^1 dx_1 f(x_1) \frac{2E}{2E_1} |M|^2 2 \pi \delta((x_1 p + q)^2 - m^2)
\\
|M|^2      &=& \frac{1}{2}  4 e_1^2 \left[p_{1 \mu} p'_{1 \nu} + p_{1 \nu} p'_{1 \mu} - g_{\mu\nu} p_1.q
\right]~,
\end{eqnarray}
%--------------------------------------------------------------------------------------
with $p_1' = p_1 + q$ and $m$ the mass of the parton. $f(x_1)$ denotes the number density of the 
struck parton. The $\delta$-distribution, which incorporates the correlation (\ref{eq:COR}) in 
(\ref{eq:PARmain}), can be written as
%--------------------------------------------------------------------------------------
\begin{eqnarray}
\label{eq:PAR1}
\delta((x_1 p + q)^2 -m^2) = \delta(Q^2 - 2 x_1 p.q) \equiv \frac{1}{2 M \nu} \delta(x_1 - x)~.
\end{eqnarray}
%--------------------------------------------------------------------------------------
After performing the integral in (\ref{eq:PARmain}) one obtains
%--------------------------------------------------------------------------------------
\begin{eqnarray}
\label{eq:PAR2}
\nu W_2(Q^2,\nu) = \sum_i e_i^2 x f_i(x), ~~~~~~~~ 2M W_1(Q^2,\nu) = \sum_i e_i^2 f_i(x)~.
\end{eqnarray}
%--------------------------------------------------------------------------------------
Here, we summed over all charged partons and anti-partons contained in the nucleon.
In the Bjorken limit the structure functions are thus described by
%--------------------------------------------------------------------------------------
\begin{eqnarray}
\label{eq:PAR3a}
F_1(x) = \frac{1}{2} \sum_i e_i^2 f_i(x),~~~~~~~F_2(x) = \sum_i e_i^2 x f_i(x).
\end{eqnarray}
%--------------------------------------------------------------------------------------
The parton distribution function $f_i(x)$ agrees with the Mellin inversion of Eq.~(\ref{eq:par1}).
Furthermore, the Callan-Gross relation \cite{Callan:1969uq} 
%--------------------------------------------------------------------------------------
\begin{eqnarray}
\label{eq:CG1}
F_2(x) = 2x F_1(x)
\end{eqnarray}
%--------------------------------------------------------------------------------------
holds. The above argument could have been also given assuming a micro-canonical ensemble
in (\ref{eq:PARmain}), \cite{MUENST}.

In the parton model the unpolarized structure functions at the Born level in 
(\ref{eq:ncB}--\ref{eq:ccB}) 
are given by~: 
%------------------------------------------------------------------------
\begin{equation}
\begin{tabular}{rcll}
$F_2(x,Q^2)$&=& $x {\displaystyle
\sum_q }
Q_q^2[q(x,Q^2) + \overline{q}(x,Q^2)]$
 &~~~~~$|\gamma|^2$ \\
$G_2(x,Q^2)$  &=& $x {\displaystyle \sum_q}
|Q_q| v_q [q(x,Q^2) + \overline{q}(x,Q^2)]$
 &~~~~~$|\gamma Z|$  \\
$H_2(x,Q^2)$&=& $x 
{\displaystyle \sum_q}
\frac{1}{4}(v_q^2 + a_q^2) [q(x,Q^2)
+ \overline{q}(x,Q^2)]$  &~~~~~$|Z|^2$ \\
$xG_3(x,Q^2)$  &=& $x {\displaystyle \sum_q}
|Q_q| a_q [q(x,Q^2) - \overline{q}(x,Q^2)]$
 &~~~~~$|\gamma Z|$  \\
$xH_3(x,Q^2)$&=& $x 
{\displaystyle \sum_q }
\frac{1}{2}
v_q a_q [q(x,Q^2)
- \overline{q}(x,Q^2)]$  &~~~~~$|Z|^2$
\\
$W_2^{{\rm NC}}(x,Q^2)$&=& $ x{\displaystyle \sum_i } (v_i^2 + a_i^2)
[q_i(x,Q^2) 
+ \overline{q}_i(x,Q^2)]$  &~~~~~$|Z|^2$
\\
$xW_3^{{\rm NC}}(x,Q^2)$&=& $2x {\displaystyle \sum_i } v_i a_i
[q_i(x,Q^2)
- \overline{q}_i(x,Q^2)]$  &~~~~~$|Z|^2$
\\
$W_2^{{\rm CC},+}(x,Q^2)$&=& $2x {\displaystyle \sum_i }
[d_i(x,Q^2)
+ \overline{u}_i(x,Q^2)]$  &~~~~~$|W^+|^2$
\\
$W_2^{{\rm CC},-}(x,Q^2)$&=& $2x {\displaystyle \sum_i }
[u_i(x,Q^2)
+ \overline{d}_i(x,Q^2)]$  &~~~~~$|W^-|^2$
\\
$xW_3^{{\rm CC},+}(x,Q^2)$&=& $2x {\displaystyle \sum_i }
[d_i(x,Q^2)
- \overline{u}_i(x,Q^2)]$  &~~~~~$|W^+|^2$
\\
$xW_3^{{\rm CC},-}(x,Q^2)$&=& $2x {\displaystyle \sum_i }
[u_i(x,Q^2)
- \overline{d}_i(x,Q^2)]$  &~~~~~$|W^-|^2$ 
\end{tabular}
\end{equation}
%------------------------------------------------------------------------
Examples for the polarized structure functions (\ref{eq:Plong},\ref{eq:Ptran})
are~ \cite{Blumlein:1996vs,Blumlein:1996tp,Stratmann:1995fn}:
%------------------------------------------------------------------------
\begin{equation}
\begin{tabular}{rcll}
\\
$g_1^{|\gamma|^2}(x,Q^2)$&=& ${\displaystyle
\sum_q }
Q_q^2[\Delta q(x,Q^2) + \Delta \overline{q}(x,Q^2)]$ & \\
%---
$g_1^{|Z|^2}(x,Q^2)$&=& $
{\displaystyle
\sum_q }\frac{1}{4} 
(v_q^2 + a_q^2) [\Delta q(x,Q^2) + \Delta \overline{q}(x,Q^2)]$ &
\\
%---
$g_5^{|Z|^2}(x,Q^2)$&=& ${\displaystyle\sum_q }  \frac{1}{2} v_q a_q \left[
\Delta q(x,Q^2) - \Delta \bar{q}(x,Q^2)\right],~{\rm etc.}$
\end{tabular}
\end{equation}
%------------------------------------------------------------------------
Here $q_i(x,Q^2)$ and $\bar{q}_i(x,Q^2)$ denote the quark and antiquark distributions of the up ($u$) and
down ($d$) type for four active flavors in the massless limit. $\Delta q_i(x,Q^2)$ and 
$\Delta \bar{q}_i(x,Q^2)$ are the corresponding polarized parton densities. $Q_q$ denotes the 
charge of the quark. Furthermore,
%------------------------------------------------------------------------
\begin{equation}
g_4(x,Q^2) = 2x g_5(x,Q^2)
\end{equation}
%------------------------------------------------------------------------
holds. The remaining structure functions contain also twist-3 contributions 
\cite{Blumlein:1996vs,Blumlein:1996tp}, cf.~Sections~\ref{sec:polt3},16.

The applicability of the parton picture rests on the comparison of two times  \cite{Drell:1970yt}:
$\tau_{\rm int}$ - the interaction time of the virtual gauge boson with the hadron and
$\tau_{\rm life}$ - the life-time of individual partons. Let these times be measured in 
an infinite momentum frame, with $p$ the large longitudinal momentum.
Applying old fashioned perturbation theory, cf. \cite{Weinberg:1966jm}, they are given by
%--------------------------------------------------------------------------------------
\begin{eqnarray}
\tau_{\rm int}  &\sim& \frac{1}{q_0} = \frac{4 p x}{Q^2(1-x)} \\
\tau_{\rm live} &\sim& \frac{1}{\sum_i (E_i - E)} = \frac{2 p}{\sum_i(k^2_{\perp,i} +m_i^2)/x_i -M^2}~. 
\end{eqnarray}
%--------------------------------------------------------------------------------------
These are non-covariant quantities. Note that in this approach energy is not conserved across a vertex, 
while momentum is conserved. Here, $q_0$ is the energy component of the exchanged gauge 
boson and $E_i$ denote the energies of the individual quantum fluctuations over the hadronic 
background and $E$ is the average energy. The individual momentum fractions are $x_i$ with  
$\sum_i x_i = 1$. $m_i$ denote the partonic masses and $M$ is the nucleon mass.
The parton model is applicable if
%--------------------------------------------------------------------------------------
\begin{eqnarray}
\label{eq:Pmain}
R_\tau = \frac{\tau_{\rm live}}{\tau_{\rm int}} \gg 1~.
\end{eqnarray}
%--------------------------------------------------------------------------------------
This ratio is a covariant quantity. Yet $p x$ has to be a large momentum. If (\ref{eq:Pmain}) is not
fulfilled the exchanged gauge boson is unable to resolve an individual parton.

In the massless case
$m_i, M = 0$ and assuming that all partons having the same transverse momentum one obtains
%--------------------------------------------------------------------------------------
\begin{eqnarray}
\label{eq:Rt1}
R_\tau =  \frac{Q^2 (1-x)}{\displaystyle 2 k_\perp^2 x \sum_i \frac{1}{x_i}}
\approx \frac{Q^2 (1-x)^2}{\displaystyle 2 k_\perp^2}
\end{eqnarray}
%--------------------------------------------------------------------------------------
The last expression in (\ref{eq:Rt1}) is obtained assuming two `essential' parton with $x_1 = x,
x_2 = 1 - x$. Eq.~(\ref{eq:Rt1}) is valid for $Q^2 \gg k_\perp^2$ and if $x$ neither becomes 
large or very small, sufficiently away from the elastic and the high energy region. These arguments 
are qualitative of course. However, close to the excluded regions multi-parton and higher twist 
effects are expected to contribute essentially.

The momentum fraction $z$ of the nucleon momentum $p$ carried by the single parton depends
on the mass of the parton before ($m_I$) and and after ($m_F$) the interaction with the intermediate 
gauge boson as well as the nucleon mass $M$ \cite{ZKIN}
%--------------------------------------------------------------------------------------
\begin{eqnarray}
\label{eq:z1}
z = \frac{Q^2 + m_F^2 - m_I^2 + \sqrt{(Q^2 + m_F^2 - m_I^2)^2 + 4 m_I^2 Q^2}}
         {\displaystyle 2(\nu + \sqrt{\nu^2 + M^2 Q^2})}~.
\end{eqnarray}
%--------------------------------------------------------------------------------------
Important special cases are $M = 0, m_I = 0$ and $M = 0, m_I = m_F$,
%--------------------------------------------------------------------------------------
\begin{eqnarray}
\label{eq:z2}
z(M = 0, m_I = 0)   &=& x \left[1 + \frac{m_F^2}{Q^2}\right] \\
z(M = 0, m_I = m_F) &=& \frac{x}{2} \left[1 + \sqrt{1 + \frac{2 m_F^2}{Q^2}} \right]~.
\end{eqnarray}
%--------------------------------------------------------------------------------------
Since $z \leq 1$ in both cases $x$ is constrained to values smaller than one.
For $m_I = m_F = 0, M \neq 0$, $z$ coincides with the Nachtmann variable $\xi$, 
(\ref{eq:NACHT}). In the fully massless case the momentum fraction is given by the Bjorken variable
%--------------------------------------------------------------------------------------
\begin{eqnarray}
\label{eq:z3}
z = x~. 
\end{eqnarray}
%--------------------------------------------------------------------------------------

In many calculations the so-called {\it collinear parton model} is used. Here the partons
carry the momentum
%--------------------------------------------------------------------------------------
\begin{eqnarray}
\label{eq:colp}
p' = z p~,~~z \in [0,1]~.
\end{eqnarray}
%--------------------------------------------------------------------------------------
This Ansatz is sufficient in cases where mass effects or transverse degrees of freedom can be safely 
neglected.

The {\it covariant parton model} generalizes (\ref{eq:colp}) accounting for transverse momentum 
effects and the finite nucleon mass \cite{COVP,Jackson:1989ph,Blumlein:1996tp}. In Sudakov variables 
the parton momentum $k$ in terms of the momenta $p$ and $q$, with $p^2 = M^2,~p.k_\perp = q.k_\perp = 
0, k_\perp.k_\perp  = - k_\perp^2$ is given by
%--------------------------------------------------------------------------------------
\begin{eqnarray}
\label{eq:covp}
k  = x p + \frac{k^2 + k_\perp^2 - x^2 M^2}
                {\displaystyle \frac{Q^2}{2} \left[1 + \sqrt{1 + \frac{4 x^2 M^2}{Q^2}}\right]
                 +2 x^2 M^2} \left[\frac{1}{2} \left(1 + \sqrt{1 + \frac{4 x^2 M^2}{Q^2}}\right) q
                 +x p \right] + k_\perp~.
\end{eqnarray}
%--------------------------------------------------------------------------------------
Here $k^2$ denotes the off-shellness of the parton. This representation is used to derive the 
correct integral relations among polarized structure functions which are sensitive to transverse degrees of 
freedom, like $g_{2}(x,Q^2)$, cf.~Sections~15,16.
%%%%%%%%%%%%%%%%%%%%%%%%%%%%%%%%%%%%%%%%%%%%%%%%%%%%%%%%%%%%%%%%%%%%%%%
\section{Renormalization and Factorization and Deep-Inelastic \newline Structure Functions}
\label{sec:RGE}
\renewcommand{\theequation}{\thesection.\arabic{equation}}
\setcounter{equation}{0}
%%%%%%%%%%%%%%%%%%%%%%%%%%%%%%%%%%%%%%%%%%%%%%%%%%%%%%%%%%%%%%%%%%%%%%%

\noindent
The quark masses and the strong coupling are scale dependent quantities due to renormalization, while
observables like the structure functions are independent of these arbitrary scales. The scale dependence
can be expressed by the operator ${\cal D}(\mu^2)$ \cite{Symanzik:1970rt,Callan:1970yg}
%------------------------------------------------------------------------------------------
\begin{eqnarray}
\label{eq:ren1}
{\cal D}(\mu^2) := \mu^2 \frac{\partial}{\partial \mu^2} 
                 + \beta(a_s(\mu^2)) \frac{\partial}{\partial a_s(\mu^2)} 
                 - \gamma_m(a_s(\mu^2)) m(\mu^2) \frac{\partial}{\partial m(\mu^2)} 
\end{eqnarray}
%------------------------------------------------------------------------------------------
with
%------------------------------------------------------------------------------------------
\begin{eqnarray}
\label{eq:ren1a}
\beta(a_s(\mu^2))    &=&   \mu^2 \frac{\partial a_s(\mu^2)}{\partial \mu^2}\\ 
\gamma_m(a_s(\mu^2)) &=& - \frac{\mu^2}{m(\mu^2)} \frac{\partial m(\mu^2)}{\partial \mu^2}~, 
\end{eqnarray}
%------------------------------------------------------------------------------------------
such that 
%------------------------------------------------------------------------------------------
\begin{eqnarray}
\label{eq:ren2}
\left[{\cal D}(\mu^2) + \gamma_{J_1} + \gamma_{J_2} - n_\psi \gamma_\psi - n_A \gamma_A\right] 
F_i(x,Q^2) = 0~.
\end{eqnarray}
%------------------------------------------------------------------------------------------
Referring to (\ref{eq:COMP3}) one obtains the following two renormalization group equations for the
operator matrix elements and the Wilson coefficients 
%------------------------------------------------------------------------------------------
\begin{eqnarray}
\label{eq:ren3}
\sum_j \left[{\cal D}(\mu^2) \delta_{ij} + \gamma_{ij}^{\rm S,NS}
- n_\psi \gamma_\psi - n_A \gamma_A
\right] A_j(N,\mu^2) &=& 0 \\
\label{eq:ren4}
\sum_j \left[{\cal D}(\mu^2) \delta_{ij} + \gamma_{J_1} + \gamma_{J_2} - \gamma_{ij}^{\rm S,NS}\right] 
C_i\left(N,\frac{Q^2}{\mu^2}\right) 
&=& 0~.
\end{eqnarray}
%------------------------------------------------------------------------------------------
Here $\gamma_\psi$, $\gamma_A$ and $\gamma_{J_{1,2}}$ denote the anomalous dimension of external quarks, 
gluons, and the currents, which can be non-zero if the currents are not conserved, cf. e.g. \cite{CURNC}. 
Here the scale $\mu^2 = \mu^2_F$ refers to the 
factorization scale. Furthermore, one may introduce different scales for the renormalization of the coupling 
and/or the masses. Often these more general scale choices are used for rough estimates of the remaining 
uncertainties due to higher order corrections. 

In the twist-2 approximation the
deep-inelastic structure functions obey the following generic representation~: 
%-----------------------------------------------------------------------
\begin{eqnarray}
F_i(x,Q^2) = \sum_l  C_{i,l}\left(a_s,\frac{Q^2}{\mu_F^2},m_c,m_b,x\right)
                     \otimes f^l\left(a_s,z,\frac{\mu_F^2}{P^2}\right)~,
\label{eq:FACT}
\end{eqnarray}
%-----------------------------------------------------------------------
with $a_s = a_s(\mu^2_R),  m_c = m_c(\mu^2_R), m_b = m_b(\mu^2_R)$, 
and $\mu_{F,R}$ the factorization and renormalization scales and the sum runs over the corresponding
partonic combinations. Furthermore, one has to specify the scheme in which $a_s$, resp. $m_{c,b}$ are defined.
Performing the perturbative calculations to a given order one may match the $\mu^2$-dependence and 
formulate so-called scheme independent evolution equations for observables, 
cf.~\cite{Furmanski:1981cw,Grunberg:1982fw,Catani:1996sc,Blumlein:2000wh}.

Let us now consider some aspects of the renormalization of the parton distribution functions
and the Wilson coefficients. In the calculation we will refer to the $\overline{\rm MS}$ scheme in $D=4+\ep$ 
dimensions for the regularization of all singularities, expanding in the dimensional 
parameter $\ep$. In the calculation of operator matrix elements and Wilson coefficients a {\it universal} 
factor
%-----------------------------------------------------------------------
\begin{eqnarray}
S_\ep = \exp \left[\frac{\ep}{2} \left(\gamma_E - \ln(4\pi)\right)\right] 
\end{eqnarray}
%-----------------------------------------------------------------------
occurs in each loop order, which is set to one, unlike in the ${\rm MS}$-scheme. Here $\gamma_E$ 
denotes the Euler-Mascheroni constant. To  account for mass effects usually different schemes will be 
used in intermediary steps. The renormalized parton densities $f^l$
correspond to strictly massless partons and the coupling constant is that of the $\overline{\rm MS}$ 
scheme. So-called heavy quark parton densities will be introduced at a later stage. All heavy quark 
effects are contained in the Wilson coefficients $C_{i,l}$. Let us now define the massless flavor part of an 
(inclusive) structure function $F_i(x,Q^2)$, $F_i^{\rm massless}(x,Q^2)$. It is given by (\ref{eq:FACT}) in 
case the Wilson coefficients do not contain {\it any} direct or indirect heavy quark mass effects. Note 
that this definition is different from one sometimes used in experiment, where one requests all final state 
fermions being massless. Correspondingly, the massive part of $F_i$ is given by
%-----------------------------------------------------------------------
\begin{eqnarray}
F_i^{\rm massive}(x,Q^2) = F_i(x,Q^2) - F_i^{\rm massless}(x,Q^2)~. 
\end{eqnarray}
%-----------------------------------------------------------------------

The local twist-2 operators (\ref{eq:op1}--\ref{eq:op3}), resp. the parton distribution functions,
are renormalized by
%-----------------------------------------------------------------------
\begin{eqnarray}
O_q^{\rm NS} = Z^{\rm NS}(\mu^2)    \hat{O}^{\rm NS}_q,~~~
O_i^{\rm S}  = Z^{\rm S}_{ij}(\mu^2) \hat{O}^{\rm S}_j
\end{eqnarray}
%-----------------------------------------------------------------------
in the non-singlet and singlet case respectively. The $Z$-factors account for the ultraviolet
singularities 
%-----------------------------------------------------------------------
\begin{eqnarray}
Z^{\alpha}_{ij}(a_s,\ep) = \delta_{ij} + \sum_{k=0}^\infty a_s^l \sum_{n=1}^l \frac{z^{l,n}_\alpha}{\ep^n}
\end{eqnarray}
%-----------------------------------------------------------------------
and the anomalous dimensions given by
%-----------------------------------------------------------------------
\begin{eqnarray}
\gamma_{qq}^{\rm NS} = Z^{-1,\rm NS}(\mu^2) \mu^2 \frac{\partial}{\partial \mu^2} Z^{\rm 
NS}(\mu^2),~~~
\gamma_{ij}^{\rm S} = {Z^{-1,\rm S}_{il}(\mu^2)} \mu^2 \frac{\partial}{\partial \mu^2} 
Z^{\rm S}_{lj}(\mu^2)~.
\end{eqnarray}
%-----------------------------------------------------------------------

The unrenormalized coupling $\hat{a}_s =g_s^2/(16 \pi^2)$ is related to the renormalized one by
%-----------------------------------------------------------------------
\begin{eqnarray}
\hat{a}_s = Z_g^{2}(a_s,\ep) {a}_s~. 
\end{eqnarray}
%-----------------------------------------------------------------------

The heavy quark masses in the Wilson coefficients are renormalized either in the
on-shell scheme or the $\overline{\rm MS}$ scheme, 
%-----------------------------------------------------------------------
\begin{eqnarray}
\hat{m}  = Z_m m = m\left[ 1 + \sum_{k=1}^\infty \hat{a}_s^k \left(\frac{m^2}{\mu^2}\right)^{k \ep/2}
\delta m_k(\ep)\right]~.
\end{eqnarray}
%-----------------------------------------------------------------------
The functions $\delta m_{1,2}(\ep)$ have been calculated in \cite{MASSREN}.

At large enough scales $Q^2 \gg m_q^2$ the heavy flavor Wilson coefficients factorize 
into massive operator matrix elements and the massless Wilson coefficients \cite{Buza:1996xr}.
In case of the structure function $F_2(x,Q^2)$ this is the case for $Q^2 \gsim 10~m_Q^2$, with $m_Q$ the
heavy quark mass.
In this kinematic regime the renormalization of the heavy flavor Wilson coefficients can be traced
back to the massless flavor ones and that of the scheme-independent massive operator matrix elements.
As has been lined out in \cite{Bierenbaum:2009mv} charge renormalization is somewhat more complicated
and it is useful to first perform it in a {MOM}-scheme using the background field method \cite{BGF}
before one transforms to the $\overline{\rm MS}$-scheme. The massive OMEs contain collinear singularities
due to massless sub-graphs, which have to be factorized by the matrices $\Gamma^{-1}_{ij}(\mu^2)$.
Unlike the massless case, where

\vspace*{-2mm}
%-----------------------------------------------------------------------
\begin{eqnarray}
\Gamma^{-1}_{ij}(\mu^2) = Z_{ij}(\mu^2),
\end{eqnarray}
%-----------------------------------------------------------------------
this is not the case for the massive OMEs. 
The renormalization of the massless Wilson coefficients is performed by $\Gamma^{-1}_{ij}(\mu^2)$ taken in the 
massless case, cf.~\cite{Vermaseren:2005qc} to 3-loop order. 
It is evident that the fragile framework
of renormalization scarcely allows for any manipulations on `phenomenological' grounds ad hoc. 
In any case, in this way a {\sf new} scheme is chosen.  
%%%%%%%%%%%%%%%%%%%%%%%%%%%%%%%%%%%%%%%%%%%%%%%%%%%%%%%%%%%%%%%%%%%%%%%
\section{The strong coupling constant}
\label{sec:alpha}
\renewcommand{\theequation}{\thesection.\arabic{equation}}
\setcounter{equation}{0}
%%%%%%%%%%%%%%%%%%%%%%%%%%%%%%%%%%%%%%%%%%%%%%%%%%%%%%%%%%%%%%%%%%%%%%%

\noindent
The strong coupling $a_s(\mu^2) = g_s^2/(16 \pi^2) = \alpha_s(\mu^2)/4\pi$ is a central parameter in QCD. 
It is not an observable itself but the various hard scattering processes are parameterized by it in perturbation 
theory. Different renormalization schemes may be chosen, cf.~\cite{Duke:1984ge}.
Most commonly $\alpha_s(\mu^2)$ is expressed in the $\overline{\rm MS}$--scheme~\cite{Bardeen:1978yd}.
All quark flavors are treated as effectively massless.

The running coupling is obtained as the solution of the equation
%------------------------------------------------------------------------
\begin{equation}
\label{eAL1}
\frac{d a_s(\mu^2)}{d \ln \mu^2} = 
- \beta_0 a_s^2
- \beta_1 a_s^3
- \beta_2 a_s^4
- \beta_3 a_s^5
+ \ldots
\end{equation}
%------------------------------------------------------------------------
So far the contributions to the $\beta$-function have been calculated up to 4--loop order in the 
$\overline{\rm MS}$--scheme, where
the LO~\cite{Gross:1973id,Politzer:1973fx,Khriplovich:1969aa,tHooft:unpub},
NLO~\cite{Caswell:1974gg,Jones:1974mm}, NNLO~\cite{Tarasov:1980au,Larin:1993tp}, 
and N$^3$LO \cite{vanRitbergen:1997va,Czakon:2004bu} 
terms are given by~\footnote{For Quantum Electrodynamics also the coefficient 
$\beta_4 = 195067/486 + (800/3) \zeta_3 + (416/3) \zeta_4 - (6800/3) \zeta_5$ has been 
calculated in \cite{CHET11}.}~:
%------------------------------------------------------------------------
\begin{eqnarray}
\label{eAL2}
\beta_0 &=& 11 - \frac{2}{3} N_f,
\\
\beta_1 &=& 102 - \frac{38}{3} N_f,  \\
\beta_2 &=& \frac{2857}{2} -\frac{5033}{18} N_f + \frac{325}{54} N_f^2, \\
\beta_3 &=& 
   \left(\frac{149753}{6} + 3564 \zeta_3 \right)
 - \left(\frac{1078361}{162} + \frac{6508}{27} \zeta_3 \right) N_f %%\nonumber\\ &&
 + \left(\frac{50065}{162} + \frac{6472}{81} \zeta_3\right) 
N_f^2  + \frac{1093}{729} N_f^3.
\nonumber\\
\end{eqnarray}
%------------------------------------------------------------------------
Here we refer to the color coefficients in $SU(3)_c$ and $N_f$ denotes the number of active flavors.
The solution of (\ref{eAL1}) reads \cite{Blumlein:1994kw}
%------------------------------------------------------------------------
\begin{equation}
\label{eAL3}
\frac{1}{a_s(Q^2)} = \frac{1}{a_s(Q^2_0)} +
\beta_0 \ln  \left( \frac{Q^2}{Q^2_0} \right )
+ \Phi^{(n)}(a_s(Q^2);\beta_i)
- \Phi^{(n)}(a_s(Q^2_0);\beta_i).
\end{equation}
%------------------------------------------------------------------------
The superscript $n$ denotes the term at which the expansion
of the $\beta$-function in (\ref{eAL1}) was truncated. In NNLO one
obtains
%------------------------------------------------------------------------
\begin{eqnarray}
\label{eAL4}
\Phi^{(2)}(x;\beta_i) &=&
 - \frac{\beta_1}{2 \beta_0} \ln \left |
\frac{x^2}{\beta_0 + \beta_1 x + \beta_2 x^2}
\right | %%\nonumber \\
%& &~~
+ \frac{\beta_1^2 - 2 \beta_0 \beta_2}{\beta_0 \sqrt{
4 \beta_2 \beta_0 - \beta_1^2}} \arctan \left ( \frac{\beta_1 +
2 \beta_2  x} {\sqrt{4 \beta_0 \beta_2 - \beta_1^2} } \right ).
\nonumber\\
\end{eqnarray}
%------------------------------------------------------------------------
Note that
%------------------------------------------------------------------------
\begin{eqnarray}
\label{eAL5}
N_f \leq 5: & &  4\beta_0 \beta_2 - \beta_1^2 > 0 \nonumber \\
N_f = 6: & &  4\beta_0 \beta_2 - \beta_1^2 < 0.
\end{eqnarray}
%------------------------------------------------------------------------
This explains the smaller change in $a_s(\mu^2)$ from NLO to NNLO
at scales below the top threshold, if compared to the logarithmic corrections
characteristic for NLO.

Often (\ref{eAL1}) is solved expanding the approximate polynomial in $a_s$ after
the separation of variables. Let us define $L = \ln(\mu^2/\Lambda^2)$. Then, according to 
the convention in \cite{Bardeen:1978yd}, one obtains \cite{Chetyrkin:1997sg}~:
%------------------------------------------------------------------------
\begin{eqnarray}
a_s(\mu^2) &=& \frac{1}{\beta_0 L} - \frac{\beta_1}{\beta_0^3 L^2} \ln(L)
               + \frac{1}{\beta_0^3 L^3}\left[\frac{\beta_1^2}{\beta_0^2} \left(\ln^2(L) -\ln(L) - 1\right)
+\frac{\beta_2}{\beta_0}\right]
\nonumber\\
& & + \frac{1}{\beta_0^4 L^4} \left[\frac{\beta_1^3}{\beta_0^3} \left( - \ln^3(L) + \frac{5}{2} \ln^2(L) +2 
\ln(L) - \frac{1}{2}\right) - 3 \frac{\beta_1 \beta_2}{\beta_0^2} \ln(L) + \frac{\beta_3}{2 \beta_0}\right]~. 
\end{eqnarray}
%------------------------------------------------------------------------
Here $\Lambda \equiv \Lambda(N_f)$ denotes the QCD scale. In applications the scale $\mu^2$ varies over 
a wider range and may pass flavor thresholds. As a convention, one identifies the scale $\mu$ at which
a new quark flavor becomes active by $m_q,~~q = c,b,t$. This definition implies matching conditions in 
$a_s(\mu^2)$. At LO and NLO they are given by $a_s(N_f-1) = a_s(N_f)$, while in higher orders more specific
conditions apply \cite{Bernreuther:1981sg,Chetyrkin:1997sg,Marciano:1983pj}.
For 4-loop running the 3-loop matching conditions are
%------------------------------------------------------------------------
\begin{eqnarray}
\frac{a_s(N_f-1)}{a_s(N_f)} =  1 + C_2 a_s^2(N_f) + C_3(N_f) a_s^3(N_f)~,
\end{eqnarray}
%------------------------------------------------------------------------
with $C_2 = -14/3, C_3 = -340.7289736 + 16.79813197 (N_f-1)$ for $\mu = m_q$.
In Section~\ref{sec:SOL} we will summarize the current status of the determination of $\alpha_s(M_Z^2)$
from deep-inelastic data.
%%%%%%%%%%%%%%%%%%%%%%%%%%%%%%%%%%%%%%%%%%%%%%%%%%%%%%%%%%%%%%%%%%%%%%%
\section{Anomalous Dimensions and Splitting Functions}
\label{sec:SPLIT}
\renewcommand{\theequation}{\thesection.\arabic{equation}}
\setcounter{equation}{0}
%%%%%%%%%%%%%%%%%%%%%%%%%%%%%%%%%%%%%%%%%%%%%%%%%%%%%%%%%%%%%%%%%%%%%%%

\noindent
The anomalous dimensions of local twist-2 operators $\gamma_{ij}(a_s,N)$
have the representation
%-----------------------------------------------------------------------
\begin{eqnarray}
\gamma_{ij}^{l}(a_s, N) = \sum_{k=1}^\infty a_s(\mu^2) \gamma_{ij}^{l,(k-1)}(N)~,
\end{eqnarray}
%-----------------------------------------------------------------------
where $N$ denotes the Mellin variable. They are related to the splitting functions
in $z$-space by
%-----------------------------------------------------------------------
\begin{eqnarray}
\gamma_{ij}^{l,(k-1)}(N) = - \int_0^1 dz z^{N-1} P_{ij}^{l,(k-1)}(z)
\end{eqnarray}
%-----------------------------------------------------------------------
uniquely \cite{CARLS}, cf.~also~\cite{MEL:UN}.
%\cite{Parisi:1973nx,Gross:1974fm,Parisi:1974sq,Zee:1974du}.
$z$ denotes the collinear momentum fraction of the parton compared to its emitting particle.

Here $l$ labels the three flavor non-singlet (NS$^\pm$, NS$^{\rm v}$) and singlet (S) cases, 
with partonic transitions $j \rightarrow i$. The three flavor non-singlet splitting functions are
given by, cf.~\cite{Moch:2004pa},
%-----------------------------------------------------------------------
\begin{eqnarray}
\gamma_{qq}^{\rm NS, \pm} (N) &=& \gamma_{qq}^{\rm v}(N) \pm \gamma_{q\bar{q}}^{\rm v}(N) \\ 
\gamma_{qq}^{\rm NS, v}(N)    &=& \gamma_{qq}^{\rm v}(N) - \gamma_{q\bar{q}}^{\rm v}(N) + 
N_f(\gamma_{qq}^{\rm s}(N)  
-\gamma_{q\bar{q}}^{\rm s}(N))
\end{eqnarray}
%-----------------------------------------------------------------------
with
%-----------------------------------------------------------------------
\begin{eqnarray}
\label{eq:3s1}
\gamma_{q_{i} q_j}(N) &=& \gamma_{\bar{q}_{j} \bar{q}_i}(N) = \delta_{ij} \gamma_{qq}^{\rm v}(N) + 
\gamma_{qq}^{\rm s}(N) \\ 
\label{eq:3s2}
\gamma_{q_{i} \bar{q}_j}(N) &=& \gamma_{\bar{q}_{i} {q}_j}(N) =
\delta_{ij} \gamma_{q\bar{q}}^{\rm v}(N) + \gamma_{q\bar{q}}^{\rm s}(N)~. 
\end{eqnarray}
%-----------------------------------------------------------------------
$\gamma_{q\bar{q}}^{\rm v}(N)$ contributes for the first time at NLO and the non-zero difference 
$\gamma_{qq}^{\rm s}(N) -\gamma_{q\bar{q}}^{s}(N)$ emerges with NNLO. Up to exceptions the anomalous
dimensions are different in the unpolarized and polarized cases. Note that there are
sometimes different conventions being used in the literature parameterizing the
anomalous dimensions, splitting functions and Wilson coefficients, due to a different
normalization of the strong coupling constant or using the operator $\mu \partial/\partial \mu$ 
instead of $\mu^2 \partial/\partial \mu^2$ in the renormalization group equation. Fermion-number 
conservation implies
%-----------------------------------------------------------------------
\begin{eqnarray}
\gamma_{qq}^{\rm NS,-}(1) &=& 0 \\
\gamma_{qq}^{\rm NS,v}(1) &=& 0~.
\end{eqnarray}
%-----------------------------------------------------------------------
%%%%%%%%%%%%%%%%%%%%%%%%%%%%%%%%%%%%%%%%%%%%%%%%%%%%%%%%%%%%%%%%%%%%%%%%
\subsection{Leading Order}
%%%%%%%%%%%%%%%%%%%%%%%%%%%%%%%%%%%%%%%%%%%%%%%%%%%%%%%%%%%%%%%%%%%%%%%%

\vspace*{1mm}
\noindent
The leading order QCD anomalous dimensions in the unpolarized case have been calculated
in \cite{Gross:1973ju,Gross:1974cs,Georgi:1951sr}. They are given by 
%-----------------------------------------------------------------------
\begin{eqnarray}
\gamma_{qq}^{(0)}(N) &=&  C_F \left[ 4 S_1(N) - \frac{3 N^2 + 3 N +2}{N(N+1)}\right] = 
\gamma_{qq}^{(0), \rm NS}(N) \\
\gamma_{qg}^{(0)}(N) &=& -4 T_F N_f \frac{2 +N +N^2}{N (N+1) (N+2)}\\
\gamma_{gq}^{(0)}(N) &=& -2 C_F \frac{2 +N +N^2}{(N-1) N (N+1)}\\
\gamma_{gg}^{(0)}(N) &=& 
 C_A \left[
- \frac{24 +2 N + 13 N^2 +22 N^3 + 11 N^4)}{3 (N-1) N (1 + N) (2 + N)} + 4 S_1(N) \right]
+ \frac{8}{3} T_F N_f,
\end{eqnarray}
%-----------------------------------------------------------------------
resp. in $z$-space 
%------------------------------------------------------------------------
\begin{eqnarray}
\label{eSP1}
P_{qq}^{(0)}(z) &=& 2 C_F \left ( \frac{1 + z^2}{1-z}\right)_+ \\
P_{qg}^{(0)}(z) &=& 4 T_f N_f
 \left [ z^2 + (1 -z)^2  \right] 
\\
P_{gq}^{(0)}(z) &=& 2 C_F \frac{1 + (1 -z)^2}{z} \\
P_{gg}^{(0)}(z) &=& 4 C_A \left [ \frac{1}{(1-z)_+} +\frac{1}{z} - 2 + z - z^2\right] +  2 \beta_0  \delta(1 - 
z).
\end{eqnarray}
%------------------------------------------------------------------------
Here, $S_1(N)$ denotes the simplest of the nested harmonic sums \cite{HSUM}.
%%\cite{Vermaseren:1998uu,Blumlein:1998if},
%------------------------------------------------------------------------
\begin{eqnarray}
S_{b,a_1, ..., a_n}(N) = \sum_{k=1}^N \frac{({\rm sign}(b))^k}{k^{|b|}} S_{a_1,...,a_n}(k),~~~S_{\emptyset} = 1~.
\end{eqnarray}
%------------------------------------------------------------------------
The structure of the splitting functions $P_{qq}^{(0)}, P_{qg}^{(0)},P_{gq}^{(0)}$
has been known from QED, \cite{QED1},
%\cite{Fermi:1924tc,Williams,vonWeizsacker:1934sx,KESSLER,LANDAU4} 
see also~\cite{Gribov:1972rt,QED2}.
%%Baier:1973ms,Chen:1975sh}. 
Following  \cite{DRELL60,Yan:1969gn}
splitting functions for fermion-pseudoscalar and fermion-abelian vector theories 
%-----------------------------------------------------------------------
\begin{eqnarray}
{\cal L} = \bar{\psi} \gamma_5 \psi \Phi,~~~~~{\cal L} = \bar{\psi} \gamma_\mu \psi A^\mu,
\end{eqnarray}
%-----------------------------------------------------------------------
were calculated in \cite{Gribov:1972ri,CHRIST72,Lipatov:1974qm,Bukhvostov:1974uu}. For Quantum 
Chromodynamics they were computed in Refs.~\cite{Parisi:1976qj,Altarelli:1977zs,Kim:1977hp} in the 
space-like resp. in \cite{Dokshitzer:1977sg} in the time-like case. In \cite{Altarelli:1977zs,Bukhvostov:1985rn} 
also the individual helicity contributions were given, \cite{DKMT}. The calculation of the 
QCD splitting functions \cite{Altarelli:1977zs} was of importance to extend the naive parton model
to the QCD improved parton model. The notion of splitting functions, furthermore, forms a more 
intuitive picture, if compared to the more formal description obtained from QCD corrections to
the amplitudes in the light cone expansion, and contributed a lot to the detailed  understanding of the 
respective processes.

The following relations apply for $z < 1$~:
%-----------------------------------------------------------------------
\begin{eqnarray}
P_{qq}(z) &=& P_{gq}(1-z)\\
P_{qg}(z) &=& P_{qg}(1-z)\\
P_{gg}(z) &=& P_{gg}(1-z)~.
\end{eqnarray}
%-----------------------------------------------------------------------
Furthermore, the integral relations
%-----------------------------------------------------------------------
\begin{eqnarray}
\label{eq:FNC}
\int_0^1 dz z\left[P_{qq}(z) + P_{gq}(z)\right] &=& 0 \\
\int_0^1 dz z\left[2 N_f P_{qg}(z) + P_{gg}(z)\right] &=& 0 
\end{eqnarray}
%-----------------------------------------------------------------------
hold. 

In the polarized case the leading order singlet anomalous dimensions were calculated in
\cite{Sasaki:1975hk,Ahmed:1975tj}~\footnote{The foregoing paper \cite{Ito:1975pf} was not fully 
correct.} using the operator 
approach and in Ref.~\cite{Altarelli:1977zs} for the splitting functions~:
%------------------------------------------------------------------------
\begin{eqnarray}
\label{ePO1}
\Delta \gamma_{qq}^{(0)}(N) &=& \gamma_{qq}^{(0), \rm NS}(N) =
\Delta \gamma_{qq}^{(0), \rm NS}(N) \\
 \Delta \gamma_{qg}^{(0)}(z) &=& - 4 T_f N_f \frac{N+2}{N(N+1)}\\
 \Delta \gamma_{gq}^{(0)}(z) &=& - 2 C_F \frac{N-1}{N(N+1)} \\
 \Delta \gamma_{gg}^{(0)}(z) &=& 4 C_A \left[ S_1(N) - \frac{2}{N(N+1)} \right] - 2 \beta_0~.
\end{eqnarray}
%------------------------------------------------------------------------
%%%%%%%%%%%%%%%%%%%%%%%%%%%%%%%%%%%%%%%%%%%%%%%%%%%%%%%%%%%%%%%%%%%%%%%%
\subsection{Next-to-Leading Order}
%%%%%%%%%%%%%%%%%%%%%%%%%%%%%%%%%%%%%%%%%%%%%%%%%%%%%%%%%%%%%%%%%%%%%%%%

\noindent
The  splitting functions at NLO were calculated in \cite{SPLIT:NLO,Moch:1999eb}.
%\cite{Floratos:1977au,Floratos:1978ny,GonzalezArroyo:1979df,GonzalezArroyo:1979ng,
%Curci:1980uw,Furmanski:1980cm,Floratos:1980hk,Floratos:1981hs,GonzalezArroyo:1979he,
%Floratos:1980hm,Hamberg:1991qt,Ellis:1996nn}
For space-like virtualities they read~:  
%---------------------------------------------------------------------------------
\begin{eqnarray}
\gamma_{qq}^{\rm NS,-,(1)}(N) &=&
C_A C_F \Biggl[
- 16 S_{-2,1}
- \frac{51 N^5+102 N^4+655 N^3+484 N^2+12 N+144}{18 N^3 (N+1)^2}
+ 8 S_{-3}
+ \frac{268}{9} S_1
\nonumber\\ &&
+ \left(16 S_1 - \frac{8}{N (N+1)}\right) S_2
-\frac{44}{3} S_2
+8 S_3
-\frac{8 (-1)^N}{(N+1)^3}\Biggr]
%---
+ C_F^2
\Biggl[
32 S_{-2,1}
+ \Biggl(\frac{8 (2 N+1)}{N^2 (N+1)^2}
\nonumber\\&&
-16 S_2 \Biggr) S_1
+\frac{4 \left(3 N^2+3 N+2\right)}{N(N+1)} S_2
-\frac{3 N^6+9 N^5+9 N^4-5 N^3-24 N^2-32 N-24}{2 N^3(N+1)^3}
\nonumber\\ &&
-16 S_{-3}
 + \left(\frac{16}{N (N+1)} - 32 S_1\right) S_2
-16 S_3(N) +\frac{16(-1)^N}{(N+1)^3}
\Biggr]
\nonumber\\ &&
%---
+  C_F N_f \left(\frac{3 N^4+6 N^3+47 N^2+20 N-12}{9 N^2 (N+1)^2}
  -\frac{40}{9} S_1
  +\frac{8}{3} S_2 \right)
\\
%%---------------------------
\gamma_{qq}^{{\rm PS},(1)}(N) &=& 
-4 C_F N_f
\frac{(2 + 5 N + N^2) (4 + 4 N + 7 N^2 + 5 N^3)}{(-1 + N) N^3 (1 + N)^3 (2 + N)^2}
\\
%%---------------------------
\gamma_{qg}^{(1)}(N) &=& 
%%---------------------------
C_A N_f
\Biggl\{
\frac{16 (-1)^N (5 + 7 N + 4 N^2 + N^3)}{(1 + N)^3 (2 + N)^3} 
- \frac{4 P_1(N)}
       {(-1 + N) N^3 (1 + N)^2 (2 + N)^3} 
\nonumber\\ &&
- \frac{16 (3 + 2 N)}{(1 + N)^2 (2 + N)^2} S_1 
+ \frac{4 (2 + N + N^2)}{N (1 + N) (2 + N)} \left(2 S_{-2} + S_1^2 + S_2\right)
\Biggr\}
%%-----------------------------------------------------------------------------------------
%Pqg1cfNf
\nonumber\\ &&
+ C_F N_f 
\Biggl\{
\frac{-2 (4 + 8 N + 25 N^2 + 51 N^3 + 36 N^4 + 15 N^5 + 5 N^6)}
     {N^3 (1 + N)^3 (2 + N)} 
\nonumber\\ & & 
+ \frac{8}{N^2} S_1
+ \frac{4 (2 + N + N^2)}{N (1 + N) (2 + N)} \left(- S_1^2 + S_2\right) 
\Biggr\}
\\
%%---------------------------
\gamma_{gq}^{(1)}(N) &=&
C_F^2
\Biggl\{
\frac{2 (-4 - 12 N - N^2 + 28 N^3 + 43 N^4 + 30 N^5 + 12 N^6)}{(-1 + N) N^3 (1 + N)^3} 
\nonumber\\ &&
- \frac{4 (10 + 17 N + 8 N^2 + 5 N^3)}{(-1 + N) N (1 + N)^2} S_1 
+ \frac{4 (2 + N + N^2)}{(-1 + N) N (1 + N)} \left(S_1^2 + S_2\right) \Biggr\}
%%---------------------------
\nonumber\\ &&
%Pgq1cacf
+ C_A C_F
\Biggl\{
\frac{-8 (-1)^N (2 + 6 N + 5 N^2 + 3 N^3)}{(-1 + N) N^2 (1 + N)^3} 
- \frac{4 P_2(N)}{9 (-1 + N)^2 N^3 (1 + N)^2 (2 + N)^2} 
\nonumber\\ &&
+ \frac{4 (-12 - 22 N + 41 N^2 + 17 N^4)}{3 (-1 + N)^2 N^2 (1 + N)} S_1 
+ \frac{4 (2 + N + N^2)}{(-1 + N) N (1 + N)} \left( S_{-2} - S_1^2 +S_2\right) 
%%---------------------------
\nonumber\\ &&
%Pgq1cfNf
+ C_F N_f
\Biggl\{
\frac{8 (16 + 27 N + 13 N^2 + 8 N^3)}{9 (-1 + N) N (1 + N)^2} 
- \frac{8 (2 + N + N^2)}{3 (-1 + N) N (1 + N)} S_1\Biggr\}
\\
%------------------------
\gamma_{gg}^{(1)}(N) &=& 
C_A^2
\Biggl\{
\frac{-16 (-1)^N (8 + 36 N + 61 N^2 + 61 N^3 + 36 N^4 + 12 N^5 + 2 N^6)}
{(-1 + N) N^2 (1 + N)^3 (2 + N)^3} 
\nonumber\\ &&
- \frac{2 P_3(N)
}{9 (-1 + N)^2 N^3 (1 + N)^2 (2 + N)^3} 
- 8 S_{-3}  
+ \left[\frac{ 32 (1 + N + N^2)}{(-1 + N) N (1 + N) (2 + N)} - 16 S_1\right] S_{-2}
\nonumber
\end{eqnarray}
\begin{eqnarray}
&&
+ \left[
   \frac{4 P_4(N)}
        {9 (-1 + N)^2 N^2 (1 + N)^2 (2 + N)^2} - 16 S_2\right] S_1
+ \frac{32 (1 + N + N^2)}{(-1 + N) N (1 + N) (2 + N)} S_2 
\nonumber\\
&&
- 8 S_3 
+ 16 S_{-2,1}
\Biggr\}
\nonumber\\
%%---------------------------
%Pgg1caNf
&&
+ C_A N_f
\left[
\frac{8 (6 + 28 N + 41 N^2 + 29 N^3 + 22 N^4 + 9 N^5 + 3 N^6)}{9 (-1 + N) N^2 (1 + N)^2 (2 + N)}
- \frac{40}{9} S_1
\right]
\nonumber\\
%%---------------------------
%Pgg1cfNf
& & + 2 C_F N_f
\frac{P_5(N)}
{(-1 + N) N^3 (1 + N)^3 (2 + N)}
\end{eqnarray}
%-----------------------------------------------------------------------------------------
with
%-----------------------------------------------------------------------------------------
\begin{eqnarray}
%%---------------------------
P_1(N) &=&
16 + 48 N + 56 N^2 + 52 N^3 + 25 N^4 + 23 N^5 + 14 N^6 + 5 N^7 + N^8
\\
P_2(N) &=& 144 + 432 N - 152 N^2 - 900 N^3 - 275 N^4 + 592 N^5 + 834 N^6 + 512 N^7 + 109 N^8
\\
P_3(N) &=&
576 + 1488 N + 1088 N^2 - 536 N^3 - 848 N^4 + 711 N^5 + 2075 N^6 + 1949 N^7 + 
937 N^8 \nonumber\\ &&
+ 288 N^9 + 48 N^{10}
\\
P_4(N) &=& 
-144 - 144 N + 772 N^2 + 844 N^3 - 109 N^4 - 392 N^5 + 134 N^6 + 268 N^7 + 67 N^8
\\
P_5(N) &=& -8 - 8 N - 10 N^2 - 22 N^3 - 3 N^4 + 6 N^5 + 8 N^6 + 4 N^7 + N^8~.
\end{eqnarray}
%-----------------------------------------------------------------------------------------
%----------------------------------------------------------------------------------------
In the polarized case the anomalous dimensions were obtained in \cite{Mertig:1995ny,VOGELS:POL}~:
%----------------------------------------------------------------------------------------
\begin{eqnarray}
\Delta \gamma_{qq}^{{\rm NS,-},(1)}(N) &=& \gamma_{qq}^{{\rm NS,-},(1)}(N) \\
\Delta \gamma_{qq}^{{\rm PS},(1)}(N) &=&
16 C_F T_F N_f \frac{(N+2) \left(N^3+2 N+1\right)}{N^3 (N+1)^3}
\\
%-------------
\Delta \gamma_{qg}^{(1)}(N) &=&
16 C_A T_F N_f\Biggl\{\frac{N-1}{N(N+1)} \left[S_1^2 + S_2 +2 S_{-2}\right]
-\frac{4}{N (N+1)^2} S_1
\nonumber\\ &&
-\frac{N^5+N^4-4 N^3+3 N^2-7 N-2}{N^3 (N+1)^3}\Biggr\}
+8 C_F T_F N_f \Biggl\{4\frac{(N-1)}{N^2 (N+1)} S_1 
\nonumber\\ &&
-\frac{(N-1)(5N^4+10N^3-N+2)}{N^3 (N+1)^3}
+2 \frac{(N-1)}{N (N+1)}
   \left[S_2-S_1^2\right]\Biggr\}
\\
%-------------
\Delta \gamma_{gq}^{(1)}(N) &=&
8 C_A C_F \Biggl\{\frac{N+2}{N (N+1)}\left[-S_1^2 + S_2 + 2 S_{-2}\right]
+\frac{11 N^2+22 N+12}{3 N^2 (N+1)} S_1 
\nonumber\\ &&
- \frac{76 N^5+271
   N^4+254 N^3+41 N^2+72 N+36}{9 N^3 (N+1)^3}\Biggr\}
+4 C_F^2 (N+2) \Biggl\{2 \frac{1}{N(N+1)} 
\nonumber\\ &&
\times[S_1^2+S_2]
- 2 \frac{3N+1}{N(N+1)^2} S_1 
   +\frac{(3N+1)(3N^3+3N^2-N-2)}{N^3(N+1)^3}\Biggr\}
\nonumber\\ &&
+32 C_F N_f T_F (N+2)
   \left[\frac{5N+2}{9 N (N+1)^2}-\frac{1}{3 N(N+1)}S_1\right]
\nonumber\\
%-------------
\Delta \gamma_{gg}^{(1)} &=&
4 C_A^2 \Biggl\{\frac{2 \left(67 N^4+134 N^3+67 N^2+144
   N+72\right)}{9 N^2 (N+1)^2} S_1
- 8 S_1 [S_2+S_{-2}]
\nonumber\\ & &
   -\frac{48 N^6+144 N^5+469 N^4+698 N^3+7 N^2+258 N+144}{9 N^3 (N+1)^3}
+16 \frac{S_2 + S_{-2}}{N (N+1)}
\nonumber
\end{eqnarray}
\begin{eqnarray}
& &
+ 8 S_{-2,1}-4 S_3 -4 S_{-3}\Biggr\}
\nonumber\\ & &
+ 32 C_A 
   T_F N_f \left[\frac{3 N^4+6 N^3+16 N^2+13 N-3}{9 N^2
   (N+1)^2}-\frac{5}{9} S_1\right]
\nonumber\\ & &
+ 8 C_F T_F N_f
\frac{N^6+3 N^5+5 N^4+N^3-8 N^2+2 N+4}{N^3 (N+1)^3}
\end{eqnarray}
%----------------------------------------------------------------------------------------
In the above expressions one may identify the leading behaviour at small $x$
given by poles at $N = 1$ resp. $N = 0$, cf. Section~13. The leading large $x$ behaviour
manifests itself by powers of $S_1(N) \propto \ln(N)$ in the limit $N \rightarrow \infty$,
cf.~Section~14. 
%%%%%%%%%%%%%%%%%%%%%%%%%%%%%%%%%%%%%%%%%%%%%%%%%%%%%%%%%%%%%%%%%%%%%%%%
\subsection{3-Loop Order}
%%%%%%%%%%%%%%%%%%%%%%%%%%%%%%%%%%%%%%%%%%%%%%%%%%%%%%%%%%%%%%%%%%%%%%%%

\begin{figure}[H]
\begin{center}
\mbox{\epsfig{file=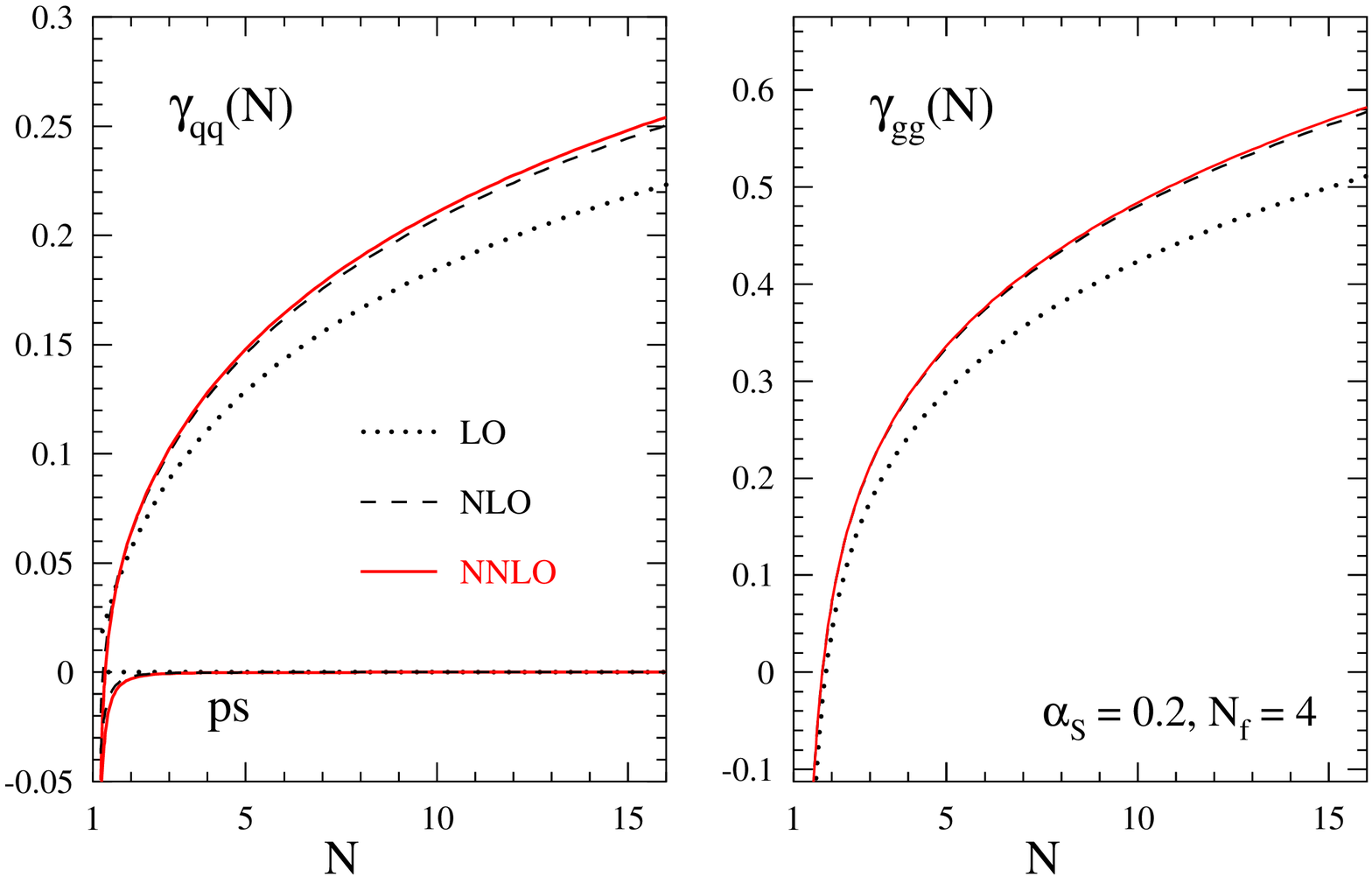,,width=13cm}}
\end{center}
\caption[]{
\label{FIG:SPLIT3}
\small \sf
The perturbative expansion of the diagonal anomalous dimensions $\gamma_{qq}(N)$ and $\gamma_{gg}(N)$
for four flavors at $\alpha_s = 0.2$. The pure-singlet (ps) contribution to $\gamma_{qq}(N)$ is shown 
separately. From \cite{Vogt:2004mw},  \TCop (2004) by Elsevier Science.}
\end{figure}

\noindent
The unpolarized 3-loop anomalous dimensions have first been calculated for a series 
of fixed moments in  \cite{Larin:1993vu,Larin:1996wd,Retey:2000nq,Blumlein:2004xt} and then in complete form
in \cite{Moch:2004pa,Vogt:2004mw}. Independent checks for the moments were obtained as a by-product
of the calculation of the massive operator matrix elements \cite{Bierenbaum:2009mv} for all color factors
$\propto T_F$ and for the leading $N_f$ dependence \cite{Gracey:1994nn,Ablinger:2010ty,BHS12a}. In 
Mellin-space
the anomalous dimensions may be represented by basic harmonic sums up to weight {\sf w = 5}
\cite{Blumlein:2009ta}. Analytic continuations for these functions to complex values of $N$ were given in 
\cite{ANCONT}. In $z$-space they are given by harmonic polylogarithms \cite{Remiddi:1999ew}. The corresponding 
expressions are very lengthly and will not be presented here. At a given value of $\alpha_s$ their size 
can be compared to the corresponding values at LO and NLO, cf. Figure~\ref{FIG:SPLIT3}. As present QCD-anlyses
show, the seemingly small difference between the curves at NLO and NNLO account for a difference
of $\Delta \alpha_s(M_Z^2) \approx 0.005$, which is essential given the experimental accuracy of $\sim 1\%$
being reached at present, see Section~\ref{sec:SOL}.
%----------------------------------------------------------------------------------------------
\restylefloat{figure}
%----------------------------------------------------------------------------------------------
\noindent
First numeric results on the quarkonic contributions to the 3-loop polarized anomalous dimensions were 
given in \cite{Vogt:2008yw}.
%%%%%%%%%%%%%%%%%%%%%%%%%%%%%%%%%%%%%%%%%%%%%%%%%%%%%%%%%%%%%%%%%%%%%%%%
\subsection{4-Loop Order}
%%%%%%%%%%%%%%%%%%%%%%%%%%%%%%%%%%%%%%%%%%%%%%%%%%%%%%%%%%%%%%%%%%%%%%%%

\noindent
At the 4--loop level at present the 2nd and 3rd moment of the non-singlet anomalous dimension
have been calculated \cite{Baikov:2006ai,CHET_priv1,Velizhanin:2011es}. Recently the general color
structure of $\gamma^{(3),\rm NS}(N=2)$ has been given in \cite{Velizhanin:2011es}. 
%---------------------------------------------------------------------------------------------
\begin{eqnarray}
\gamma^{(3),\rm NS}_{N_f=3}(N=2) &=&
\frac{1680283336}{177147} - \frac{23873952}{6561} \zeta_3 + \frac{5120}{3} \zeta_4
- \frac{56960}{243} \zeta_5, \text{\cite{Baikov:2006ai}},
\\
\gamma^{(3),\rm NS}_{SU(3)_c}(N=2) &=&
\left[\frac{3100369144}{177147}
+ \frac{26060864}{6561} \zeta_3
- \frac{7040}{27} \zeta_4
- \frac{1249280}{243} \zeta_5
\right]
\nonumber\\ &&
+
\left[
- \frac{167219672}{59049}
- \frac{6322976}{2187} \zeta_3
+ \frac{64640}{81} \zeta_4
+ \frac{14720}{9} \zeta_5 
\right] N_f
\nonumber\\ &&
+ \left[ \frac{1084904}{19683}
     + \frac{2560}{27} \zeta_3
     - \frac{1280}{27} \zeta_4
\right] N_f^2
+
\left[-\frac{4096}{6561} + \frac{512}{243} \zeta_3 \right] N_f^3,
\text{\cite{Velizhanin:2011es}}~.
\end{eqnarray}
%---------------------------------------------------------------------------------------------
In numerical studies the 4-loop \cite{Baikov:2006ai,CHET_priv1}  
moments were compared to the result of a Pad\'{e} approximation \cite{Blumlein:2006be}
%---------------------------------------------------------------------------------------------
\begin{eqnarray}
\label{eq:NS_PAD}
\gamma_{\rm NS}^{(3), \text{{\rm Pad\'e}}}(N) = 
\frac{{\gamma_{\rm NS}^{(2)}(N)}^2}{\gamma_{\rm NS}^{(1)}(N)}
\end{eqnarray}
%---------------------------------------------------------------------------------------------
for $N_f = 3$, as genuine number of massless flavors. For the second moment the Pad\'{e} 
approximation and the exact result deviate by 20.9 \% and for the third moment by 14.9 \%
only. In Ref.~\cite{Blumlein:2006be} an uncertainty of $\pm 100\%$ has been assumed for 
(\ref{eq:NS_PAD}).
%%%%%%%%%%%%%%%%%%%%%%%%%%%%%%%%%%%%%%%%%%%%%%%%%%%%%%%%%%%%%%%%%%%%%%%%%
\subsection{Anomalous dimensions in the large $N_f$ limit}
%%%%%%%%%%%%%%%%%%%%%%%%%%%%%%%%%%%%%%%%%%%%%%%%%%%%%%%%%%%%%%%%%%%%%%%%%

\noindent
The leading and sub-leading coefficients in $1/N_f$ can be calculated for the 
anomalous dimensions and the $\beta$-function using the method of Refs.~\cite{VPH} 
to all orders in the coupling constant in the $\MS$--scheme. The result of these computations 
give very useful predictions for the complete diagrammatic calculation, which is more 
difficult. Results were obtained for the fermion mass anomalous dimensions 
\cite{Gracey:1993sn,Ciuchini:1999wy,Ali:2001ng}, the QCD $\beta$-function 
\cite{Gracey:1996up}, the unpolarized anomalous dimensions of composite twist-2 operators 
\cite{Gracey:1994nn,Bennett:1997ch}, the flavor non-singlet Wilson coefficient of $F_L(x,Q^2)$ 
\cite{Gracey:1996tb}, the polarized anomalous dimensions \cite{GRAC2}, and 
the anomalous dimensions for transversity \cite{GRACTR}. Here, the electron and quark mass 
anomalous dimension \cite{Gracey:1993sn,Ciuchini:1999wy} have been calculated to $O(1/{N_f^2})$. 
As one example we show the prediction for the 4--loop flavor non-singlet anomalous 
dimension
%-----------------------------------------------------------------------------------------------
\begin{eqnarray}
\gamma^{(3),\rm NS}(N) &=&
C_F (T_R N_f)^3 \Biggl\{
\frac{2}{27} S_4(N) - \frac{10}{81} S_3(N) - \frac{2}{81} S_2(N) - \frac{2}{81} S_1(N)
+ \frac{131}{1296} 
\nonumber\\ &&
+ \left[\frac{4}{27} S_1(N) - \frac{2}{27 N(N+1)} - \frac{1}{9} \right] \zeta_3 
+ \frac{(2N-1)(2 N^5 - 5 N^4 - 10 N^3 + 7 N + 3)}{81 N^4 (N+1)^4}\Biggr\}~,
\nonumber\\
\end{eqnarray}
%-----------------------------------------------------------------------------------------------
which has been confirmed by an explicit calculation for $N=2$ \cite{Baikov:2006ai,Velizhanin:2011es}
and $N=3$ \cite{CHET_priv1}.
Another example concerns a combination of gluonic anomalous dimensions. At 3-loop order one obtains
%-----------------------------------------------------------------------------------------------
\begin{eqnarray}
\tilde{\gamma}_{gg}^{(2)} + \tilde{\gamma}_{gq}^{(2)} 
\frac{\gamma_{qg}^{(0)}}{\tilde{\gamma}_{gg}^{(0)}} &=&
C_F T_R^2 N_f^2 \Biggl\{\frac{64}{3} \frac{(N^2 + N + 2)^2}{(N+2)(N+1)^2(N-1)N^2} S_1^2
                       -\frac{4}{27} \frac{p_1(N)}{(N+2)(N+1)^4(N-1)N^4} 
\nonumber \end{eqnarray} \begin{eqnarray}
&&
                       -\frac{64}{9} \frac{10 N^6 + 30 N^5 +109 N^4 +168 N^3 +155 N^2 +76 N + 12}
                                          {(N+2)(N+1)^3(N-1)N^3} S_1
\Biggr\} \nonumber\\
&& +C_A T_R^2 N_f^2 \Biggl\{\frac{8}{27} \frac{8 N^6 + 24 N^5 - 19 N^4 -78 N^3 - 253 N^2 -210 N 
-96}{(N+2)(N+1)^2(N-1)N^2} S_1 
\nonumber\\ &&
-\frac{2}{27} \frac{p_2(N)}{(N+2)(N+1)^3(N-1)N^3}\Biggr\} 
\\
p_1(N) &=& 33 N^{10} +165 N^9 - 32 N^8 - 1118 N^7 -5807 N^6 -12815 N^5 - 16762 N^4 -13800 N^3
\nonumber\\ &&
        - 7112 N^2 - 2112 N - 288
\\
p_2(N) &=& 87 N^8 +348 N^7 +848 N^6 +1326 N^5 +2609 N^4 + 3414 N^3 + 2632 N^2 
        + 1088 N + 192~.
\nonumber\\
\end{eqnarray}
%-----------------------------------------------------------------------------------------------
Here $\tilde{\gamma}_{ij}^{(l)}$ denotes the respective part of highest power in $N_f$. The 
result
was confirmed in \cite{Vogt:2004mw}, and more recently also in \cite{BHS12a}.
%%%%%%%%%%%%%%%%%%%%%%%%%%%%%%%%%%%%%%%%%%%%%%%%%%%%%%%%%%%%%%%%%%%%%%%
\section{Coefficient functions}
\label{sec:WILSC}
\renewcommand{\theequation}{\thesection.\arabic{equation}}
\setcounter{equation}{0}
%%%%%%%%%%%%%%%%%%%%%%%%%%%%%%%%%%%%%%%%%%%%%%%%%%%%%%%%%%%%%%%%%%%%%%%

\noindent
The Wilson coefficients in Mellin space obey the following perturbative expansion
%-----------------------------------------------------------------------------------------
\begin{eqnarray}
C_i(N,a_s) = \delta_{iq} + \sum_{k=1}^\infty a_s(\mu^2) C_i^{(k)}(N)~.
\end{eqnarray}
%-----------------------------------------------------------------------------------------
Similar expressions are also obtained for other hard scattering inclusive cross sections
depending on a single momentum-fraction scale like for the Drell-Yan and hadronic
Higgs-boson production cross section in the heavy top-quark approximation \cite{DY,HIGGS}.
%%%%%%%%%%%%%%%%%%%%%%%%%%%%%%%%%%%%%%%%%%%%%%%%%%%%%%%%%%%%%%%%%%%%%%%
\subsection{First Order}
%%%%%%%%%%%%%%%%%%%%%%%%%%%%%%%%%%%%%%%%%%%%%%%%%%%%%%%%%%%%%%%%%%%%%%%

\noindent
The first order unpolarized coefficient functions in the massless case in the $\overline{\rm MS}$-scheme 
are given by, cf.\cite{Furmanski:1981cw}~:
%-----------------------------------------------------------------------------------------
\begin{eqnarray}
%%---------------------------
C_{2,q}^{(1)}(N) &=& C_F \Biggl\{
- \frac{9 N^3 + 2 N^2 - 5 N -2}{N^2 (1 + N)} 
+ \frac{3 N^2 + 3 N -2}{N (1 + N)} S_1 + 2 S_1^2 - 2 S_2
\Biggr\}
\\
%%---------------------------
C_{2,g}^{(1)}(N) &=& N_f \Biggl\{ 
\frac{-2 (-2 - N - 4 N^2 + N^3)}{N^2 (1 + N)  (2 + N)} 
- \frac{2 (2 + N + N^2)}{N (1 + N) (2 + N)} S_1
\Biggr\}\\
%%---------------------------
C_{L,q}^{(1)}(N) &=& \frac{4 C_F}{1 + N}
;~~~~~~~
C_{L,g}^{(1)}(N) = \frac{8 N_f}{(1 + N) (2 + N)} 
\\
%%---------------------------
C_{3,q}^{(1)}(N) &=& C_{2,q}^{(1)}(N) - 2 C_F \frac{2 N+1}{N(N+1)}~.
\end{eqnarray}
%-----------------------------------------------------------------------------------------
Likewise the LO massless polarized Wilson coefficients for the twist-2 structure function $g_1(x,Q^2)$ 
read, \cite{Altarelli:1979ub,Humpert:1980uv,Bodwin:1989nz,Vogelsang:1990ug}, see also \cite{Gluck:1995yr}~:
%-----------------------------------------------------------------------------------------
\begin{eqnarray}
\Delta C_{1,q}^{(1)}(N) &=& C_{3,q}^{(1)}(N)\\
\Delta C_{1,g}^{(1)}(N) &=& 2 N_f \frac{N-1}{N(N+1)}
\left[\frac{1}{N} - 1 - S_1\right]~.
\end{eqnarray}
%-----------------------------------------------------------------------------------------
The first moment of $\Delta C_{1,g}^{(1)}(N)$ vanishes, which is also observed for $\Delta C_{1,g}^{(2)}(N)$,
\cite{Zijlstra:1993sh}.
%%%%%%%%%%%%%%%%%%%%%%%%%%%%%%%%%%%%%%%%%%%%%%%%%%%%%%%%%%%%%%%%%%%%%%%
\subsection{Higher Orders}
%%%%%%%%%%%%%%%%%%%%%%%%%%%%%%%%%%%%%%%%%%%%%%%%%%%%%%%%%%%%%%%%%%%%%%%

\noindent
The massless Wilson coefficients for the unpolarized structure functions at $O(a_s^2)$ were
calculated in \cite{F2FL,Moch:1999eb} for $F_2(x,Q^2)$, in  \cite{F2FL,FL_I,Moch:1999eb} for 
$F_L(x,Q^2)$, and for $xF_3(x,Q^2)$ in \cite{Zijlstra:1992kj}. A series of moments was calculated in 
\cite{Larin:1991fv}. The polarized $O(a_s^2)$ Wilson coefficients for the structure function $g_1(x,Q^2)$ 
were computed in \cite{Zijlstra:1993sh}. All Wilson coefficients can be expressed in terms of harmonic 
sums up to weight {\sf w = 4}. As an example we show the coefficient functions in case of $F_L(x,Q^2)$, 
with the quarkonic, the pure-singlet and the gluonic contribution~: 
%-----------------------------------------------------------------------------------------
\begin{eqnarray}
C_{L,q}^{(2)}(N) &=&
C_A C_F
\Biggl\{
\left(\frac{32}{N+1} S_1
     -\frac{32 \left(N^4+2 N^3-N^2-2 N-6\right)}{(N-2) N (N+1)^2 (N+3)}\right) S_{-2}
+\frac{16}{N+1} [S_{-3} + S_3]
\nonumber\\ &&
+\frac{92}{3 (N+1)} S_1
-\frac{32}{N+1} S_{-2,1}
-\frac{16 (-1)^N P_6(N)}{5 (N-2) (N-1)^2 N^2 (N+1)^4 (N+2)^2 (N+3)^3}
\nonumber\\ &&
+\frac{2 P_7(N)}{45 (N-1)^2 N^2 (N+1)^4 (N+2)^2 (N+3)^3}
+ \frac{48}{N+1} \zeta_3
\Biggr\}
%---
+ C_F^2 \Biggl\{
-\frac{4 \left(9 N^2+13 N+2\right)}{N (N+1)^2} S_1
\nonumber\\ &&
+        \left(\frac{64 \left(N^4+2 N^3-N^2-2 N-6\right)}{(N-2) N (N+1)^2 (N+3)} 
-\frac{64}{N+1} S_1 \right) S_{-2} 
+\frac{8}{N+1} S_1^2
-\frac{32}{N+1} [S_{-3} + S_3]
\nonumber\\ &&
-\frac{8}{N+1} S_2
+\frac{64}{N+1} S_{-2,1}
+\frac{32 (-1)^N P_6(N)}
      {5 (N-2)(N-1)^2 N^2 (N+1)^4 (N+2)^2 (N+3)^3}
\nonumber\\ &&
-\frac{2 P_8(N)
}{5 (N-1)^2 N^2 (N+1)^4 (N+2)^2 (N+3)^3} 
+ \frac{96}{N+1} \zeta_3
\Biggr\}
%---
\nonumber\\ && 
+ C_F N_F \left[
  \frac{-4(19 N^2 + 7 N -6)}{9 N (N+1)^2} 
- \frac{8}{3 (N+1)} S_1 
  \right]
\\
%-----------------------------------------------------------------------------------------
C_{L,q}^{(2),\rm PS}(N) &=& C_F N_F 
  \left[
  \frac{-16 ( N^5 + 2 N^4 + 2 N^3 - 5 N^2 - 12 N -4 )}
       {(N-1) N^2 (N+1)^3 (N+2)^2} 
- \frac{16 (N^2 + N +2 )}{(N-1) N (N+1)^2 (N+2)} S_1\right]
\nonumber\\
\\
%-----------------------------------------------------------------------------------------
C_{L,g}^{(2)}(N) &=&
C_A C_F \Biggl\{
 \frac{32 \left(2 N^3-2 N^2-N-1\right)}{(N-1) N (N+1)^2 (N+2)} S_1
+\frac{16}{(N+1)(N+2)} S_1^2
-\frac{32}{(N+1) (N+2)} S_{-2}
\nonumber\\ &&
-\frac{16}{(N+1)(N+2)} S_2
+\frac{32 (-1)^N \left(N^3+4 N^2+7 N+5\right)}{(N+1)^3(N+2)^3}
\nonumber\\ &&
-\frac{16 \left(2 N^5+9 N^4+5 N^3-12 N^2-20 N-8\right)}{(N-1) N^2 (N+1)^2 (N+2)^3}
\Biggr\}
+ C_F N_f
\Biggl\{
-\frac{8 \left(3 N^2+3 N+2\right)}{N(N+1)^2(N+2)} S_1
\nonumber\\
&&
+\frac{32 (N-1)}{(N-2) (N+1)(N+3)} S_{-2}
+\frac{8 P_9(N)}{15 (N-1)^2 N^2 (N+1)^3 (N+2)^2 (N+3)^3}
%---
\nonumber\\
&&
+\frac{32 (-1)^N P_{10}(N)}
      {15 (N-2) (N-1)^2 N^2 (N+1)^3 (N+2)^2 (N+3)^3}\Biggr\}
\end{eqnarray}
%-----------------------------------------------------------------------------------------
%-----------------------------------------------------------------------------------------
\begin{eqnarray}
P_6(N) &=& 2 N^{11}+41 N^{10}+226 N^9+556 N^8+963 N^7+2733 N^6+7160 N^5+8610 N^4+1969 N^3
\nonumber\\ &&
-2748 N^2-864 N-216 \\
P_7(N) &=& 1075 N^{12}+14390 N^{11}+73464 N^{10}+160740 N^9+35682 N^8-516984 N^7-979012 N^6
\nonumber\\ &&
-627068 N^5+84099 N^4+300258 N^3+119124 N^2-648 N+7776 \\
%
%\left(2 N^{11}+41 N^{10}+226 N^9+556 N^8+963 N^7+2733 N^6+7160
%       N^5+8610 N^4+1969 N^3-2748 N^2-864 N-216\right)}
%
\\
P_8(N) &=& 85 N^{12}+1130 N^{11}+5472 N^{10}+9300 N^9-13574 N^8-85432
       N^7-149336 N^6
\nonumber
\end{eqnarray}
\begin{eqnarray}
&&
-114524 N^5
-11383 N^4+44894 N^3+22992 N^2-4104
       N-432
\\
P_9(N) &=& 26 N^9+539 N^8+3244 N^7+8465 N^6+9342 N^5+841 N^4-5720 N^3-2193 N^2
\nonumber\\
&&
+2484 N+1404\\
P_{10}(N) &=& N^{10}-13 N^9-39 N^8+222 N^7+1132 N^6+1787 N^5+913 N^4+392 N^3+645
       N^2
\nonumber\\ &&
-324 N-108
\end{eqnarray}
%-----------------------------------------------------------------------------------------
The other Wilson coefficients have a similar structure, cf.~\cite{Blumlein:2009tj,JBVR}.

\begin{figure}[H]
\begin{center}
\mbox{\epsfig{file=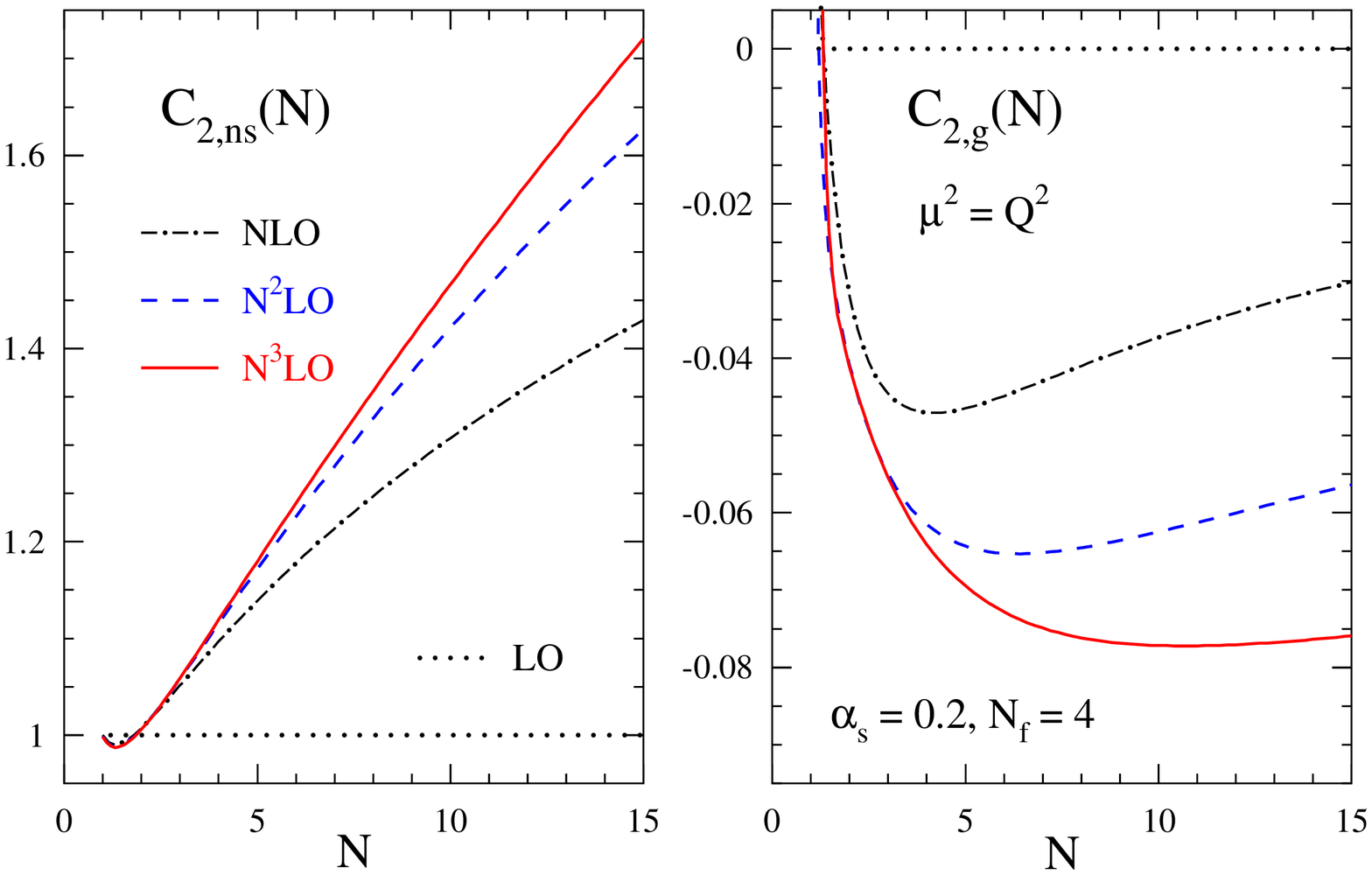,,width=13cm}}
\end{center}
\caption[]{
\label{FIG:COEFF3}
\small \sf
The perturbative expansion of the non-singlet (left) and gluon (right) $N$-space coefficient
functions for $F_2/x$ at $\alpha_s = 0.2, N_f = 4$. $O(\alpha_s)$ dash-dotted lines, $O(\alpha_s^2)$ 
dashed lines, $O(\alpha_s^3)$ full lines. In the non-singlet case LO corresponds to
$\alpha_s^0$, From \cite{Vermaseren:2005qc}, \TCop (2005) by Elsevier Science.}
\end{figure}

At $O(a_s^3)$ a series of moments for the Wilson coefficients of the structure functions was calculated
in \cite{Larin:1993vu,Larin:1996wd,Retey:2000nq,Blumlein:2004xt}. The complete expressions were computed
in \cite{Vermaseren:2005qc} and combinations of charged current structure functions in \cite{XF3}.
In Figure~\ref{FIG:COEFF3} the relative size of the QCD corrections to two Wilson coefficients is 
illustrated comparing the corrections  $O(a_s)$ to $O(a_s^3)$. In the flavor non-singlet case the 
corrections
for the Wilson coefficients at NNLO are larger than in case of the anomalous dimension for large values
of $N$, resp. in the large $x$ region.
%----------------------------------------------------------------------------------------------
\restylefloat{figure}
%----------------------------------------------------------------------------------------------
The massless Wilson coefficient at $O(a_s^3)$ can be represented in terms of harmonic sums up to weight 
{\sf w = 6}, resp. harmonic polylogarithms in $z$-space. The corresponding expressions are very large. 
Structural simplifications can be obtained applying algebraic \cite{Blumlein:2003gb} and structural 
relations \cite{Blumlein:2009ta} of these quantities. The Wilson coefficients obey difference equations 
of order $\sim 35$ and degree $\sim 1000$, which can be found in principle determining somewhat more than 
5000 moments, \cite{Blumlein:2009tj}. The corresponding difference equations can be solved with the package 
{\tt Sigma} \cite{SIGMA}. 

%%%%%%%%%%%%%%%%%%%%%%%%%%%%%%%%%%%%%%%%%%%%%%%%%%%%%%%%%%%%%%%%%%%%%%%%
\section{QED and Electro-weak Radiative Corrections to Deep- \newline
Inelastic Scattering}
\label{sec:QED}
\renewcommand{\theequation}{\thesection.\arabic{equation}}
\setcounter{equation}{0}
%%%%%%%%%%%%%%%%%%%%%%%%%%%%%%%%%%%%%%%%%%%%%%%%%%%%%%%%%%%%%%%%%%%%%%%%

\noindent
The QED radiative corrections to the deep--inelastic scattering cross
sections become rather large in part of the kinematic region due to 
logarithmic terms of $O(\alpha \ln(Q^2/m_l^2))$, with $m_l$ the charged lepton mass. The corrections in case 
of $e^\pm N$ are larger than for $\mu^\pm N$ scattering. They have to be 
known precisely since their unfolding is usually performed for the scattering 
cross section itself prior to the QCD analyses of the  deep--inelastic structure 
functions.

A first dedicated calculation of the QED radiative corrections to
deep inelastic $e^\pm N$~scattering was performed by Mo and Tsai~\cite{RC1} 
and applied in the analysis of the MIT-SLAC experiments, \cite{Bloom:1969kc,Breidenbach:1969kd}. 
Later calculations were carried out in \cite{RC2,RC3} for $l^{\pm}N$ 
scattering. 

The detailed knowledge of QED and electroweak radiative corrections was
of special importance also for the measurements of the electroweak parameters in 
deep inelastic $\nu(\overline{\nu})N$ scattering. Early calculations were 
carried out in \cite{RC55}.
%\cite{RC4,RC5,RC6,RC7,RC8,RC9,RC10,RC11}. 
More recent calculations
have been performed in \cite{RC66}
%\cite{Diener:2003ss,Arbuzov:2004zr,Diener:2005me,Park:2009ft}
in relation to the NuTeV anomaly \cite{Zeller:2001hh}. The QED bremsstrahlung corrections apply to the 
lepton lines and the incoming quark line, since the corrections to the inclusive hadronic final state vanish
according to the Kinoshita-Lee-Nauenberg theorem \cite{KLN}. In \cite{RC4} the leading logarithmic
corrections in $O(\alpha)$ were calculated, suggesting first to absorb the quarkonic QED corrections into 
the scaling violations of the quark distributions.

With the advent of HERA the radiative corrections were partly recalculated and dedicated calculations for 
deep inelastic neutral and charged current $e^{\pm}p$ scattering were carried out by different groups, also 
using a variety techniques,~\cite{RC3,
Bardin:1987rz,
Bardin:1988by,
Bardin:1989vz,
Kuraev:1988xn,
BEBEN,
Blumlein:1989gk,
Blumlein:1990wz,
Blumlein:1991ag,
Montagna:1991ku,
Kripfganz:1987yu,
Kripfganz:1990vm,
Blumlein:1993ef,
Akhundov:1995na,
Blumlein:1994ii,
Akhundov:1994my,
Bardin:1995mf,
Arbuzov:1995id,
Bohm:1986na,
Bohm:1987cg,
Spiesberger:1990fa,
Kwiatkowski:1990es}.
These approaches include both
semi-analytic 
calculations~\cite{
RC3,
Bardin:1987rz,
Bardin:1988by,
Bardin:1989vz,
Kuraev:1988xn,
BEBEN,
Blumlein:1989gk,
Blumlein:1990wz,
Blumlein:1991ag,
Montagna:1991ku,
Kripfganz:1987yu,
Kripfganz:1990vm,
Blumlein:1993ef,
Akhundov:1995na,
Blumlein:1994ii,
Akhundov:1994my,
Bardin:1995mf,
Arbuzov:1995id}
and calculations
based on Monte Carlo methods~\cite{
Bohm:1986na,
Bohm:1987cg,
Spiesberger:1990fa,
Kwiatkowski:1990es}.
The virtual electro-weak 1-loop corrections to neutral and charged current deep-inelastic
scattering were calculated in Refs.~\cite{
Bohm:1986na,
Bohm:1987cg,
Bardin:1988by,
Bardin:1989vz}.
The inclusive bremsstrahlung corrections are often presented integrating over the phase space 
of the emitted photons or lepton pairs, which can be performed analytically 
\cite{
BEBEN,
Bardin:1987rz,
Bardin:1988by,
Bardin:1989vz,
Kuraev:1988xn,
Blumlein:1989gk,
Blumlein:1990wz,
Blumlein:1991ag,
Montagna:1991ku,
Kripfganz:1987yu}.

Dominant contributions to the QED radiative corrections may be obtained
using leading log (LLA)
techniques~\cite{RC4,RC3,Blumlein:1990wz,BEBEN,Blumlein:1991ag,Montagna:1991ku,
Blumlein:1989gk,Kripfganz:1990vm,Blumlein:1994ii}. This approach,
which is
based on the factorization of (collinear) fermion mass
singularities, allows to
determine the terms $\propto \alpha \ln(Q^2/m_f^2)$ in a
straightforward way for different settings of the measured kinematic
variables. Also higher order terms were calculated within this
approach~\cite{Kripfganz:1990vm,Blumlein:1994ii}.

The first order terms are described by:
%------------------------------------------------------------------------
\begin{equation}
\frac{d^2 \sigma^{ini(fin),1loop}}{dxdy}
=
\frac{\alpha}{2\pi} L_e
\int\limits_{0}^{1}dz
P_{ee}^{(0)}(z)
\left\{
\theta(z-z_0)
{\cal J}(x,y,Q^2)
\left.
\frac{d^2 \sigma^{0}}{dxdy}\right|_{x={\hat x}, y={\hat y}, S={\hat S}}
-
\frac{d^2 \sigma^{0}}{dxdy}
\right\},
\label{eRC2}
\nll
\end{equation}
%------------------------------------------------------------------------
where $P_{ij}^{(0)}(z)$ denote the leading order QED splitting functions with $i,j \in 
\{e,\gamma\}$. They are obtained from the QCD splitting functions given in Sect.~\ref{sec:SPLIT} setting 
$C_A=0, C_F=1, T_F=1$. The scale of the correction is set by the logarithm
%------------------------------------------------------------------------
\begin{eqnarray}
L_e
= {\ln}  \frac{Q^2}{m_e^2} -1.
\label{eRC4}
\end{eqnarray}
%------------------------------------------------------------------------
\noindent
The second order corrections  ${\cal O}((\alpha L_e)^2)$
are:
%---------------------------------------------------------------------
\begin{eqnarray}
\frac{d^2 \sigma^{l,2loop}}{dx dy} &=&
 \left[ \frac{\alpha}{2 \pi}
L_e\right]^2
\int_0^1 dz
 P_{ee}^{(2,1)}(z)
\left \{
\theta(z - z_0) {\cal J}(x,y,z)
\left.\frac{d^2 \sigma^{0}}{dx dy} \right|_{x=\hat{x},
y=\hat{y},S=\hat{S}}  -  \frac{d^2 \sigma^{0}}{dx dy} \right \}
\nonumber
\end{eqnarray}

%-------------------------------------
\begin{equation}
\label{eRC7}
+
\left( \frac{\alpha}{2 \pi} \right )^2
\int_{z_0}^1 dz
 \left \{
L_e^2
P_{ee}^{(2,2)}(z)
+  L_e
\sum_{f=l,q} \ln\frac{Q^2}{m^2_f} \,
P_{ee,f}^{(2,3)}(z) \right \}
{\cal J}(x,y,z)
\left.\frac{d^2 \sigma^{0}}{dx dy} \right|_{x=\hat{x},
y=\hat{y},S=\hat{S}}.
\end{equation}
%--------------------------------------------------------------------

%----------------------------------------------------------------------------------------------
\noindent
\restylefloat{table}
\begin{table}[H]
\renewcommand{\arraystretch}{1}
\centering
\scalebox{0.81}{%
\begin{tabular}{|l|c|c|c|c|}
\hline
\multicolumn{5}{|c|}{Initial State Radiation $\hat{S} = z S$} \\
\hline
\multicolumn{1}{|c}{Kinematics} &
\multicolumn{1}{|c}{$\hat{Q}^2$} &
\multicolumn{1}{|c}{$\hat{y}$} &
\multicolumn{1}{|c}{$\hat{z_0}$} &
\multicolumn{1}{|c|}{${\cal J}(x,y,z)$} \\
\hline
leptonic variables  
& $z Q_l^2$ 
& $\frac{\ds z + y_l -1}{\ds z}$ 
& $\frac{\ds 1- y_l }{\ds 1 - x_l y_l}$ 
& $\frac{\ds y_l}{\ds z + y_l -1}$
\\
%---
mixed    variables  
& $z Q_l^2$ 
& $\frac{\ds y_{JB}}{\ds z}$ 
& $y_{JB}$ 
& 1 
\\
%---
hadronic variables  
& $Q_h^2$ 
& $\frac{\ds y_h}{\ds z}$ 
& $y_h$ 
& $\frac{1}{z}$ 
\\
%---
JB       variables  
& $\frac{\ds Q_{JB}^2 (1-y_{JB})z}{\ds z-y_{JB}}$
& $\frac{\ds y_{JB}}{\ds z}$ 
& $\frac{\ds y_{JB}}{\ds 1 -x_{JB}(1-y_{JB})}$ 
& $\frac{\ds 1-y_{JB}}{\ds z - y_{JB}}$ 
\\
%---
double angle method 
& $z^2 Q^2_{DA}$ 
& $y_{DA}$ 
& $0$ 
& $z$
\\
%---
  $\theta_l, y_{JB}$  
& $Q_{\theta y}^2 \frac{\ds z(z-y_{JB})}{\ds 1-y_{JB}}$ 
& $\frac{\ds y_{JB}}{\ds z}$ 
& $y_{JB}$ 
& $\frac{\ds z-y_{JB}}{\ds 1- y_{JB}}$  
\\
%---
$\Sigma$ method     
& $Q_\Sigma^2$ 
& $y_\Sigma$ 
& $x_\Sigma$ 
& $\frac{1}{z}$
\\
%---
$e\Sigma$ method    
& $z Q_l^2$
& $z y_{e\Sigma}$ 
& $x_\Sigma$ 
& $1$ 
\\
\hline
\multicolumn{5}{|c|}{Final State Radiation $\hat{S} = S$} \\
\hline
leptonic variables  
& $\frac{\ds Q_l^2}{\ds z}$  
& $\frac{\ds z+y_l-1}{\ds s}$ 
& $1 - y_l(1-x_l)$ 
& $\frac{\ds y_l}{\ds z(z+y_l-1)}$ 
\\  
%---
mixed    variables  
& $\frac{\ds Q_l^2}{\ds z}$  
& $y_{JB}$ 
& $x_m$
& $\frac{1}{z}$
\\
%---
$\Sigma$ method     
& $Q_\Sigma^2 \frac{\ds (1-y_\Sigma(1-z))}{\ds z^2}$ 
& $\frac{\ds y_\Sigma z}{\ds 1- y_\Sigma(1-z)}$
& $z_0^{\Sigma, f}$
& $\frac{1}{z^2}$
\\ 
%---
$e\Sigma$ method    
& $\frac{\ds Q_l^2}{\ds z}$  
& $\frac{\ds y_{e\Sigma} z^2}{\ds (1- y_\Sigma(1-z))^2}$ 
& $z_0^{\Sigma, f}$
& $\frac{\ds 1 + y_{e\Sigma}(1-z)}{\ds (1-y_{e\Sigma}(1-z))z}$ 
\\
%---
\hline
\end{tabular}
}
\renewcommand{\arraystretch}{1}
\caption[]{\label{TAB:QED}
\sf \small
Scaling properties of various sets of kinematic variables for leptonic initial and final state 
radiation, cf. \cite{Arbuzov:1995id}.}
\end{table}
%----------------------------------------------------------------------------------------------
This notion reproduces the soft photon terms of complete
calculations in leptonic variables (cf. e.g.~\cite{Arbuzov:1995id}).
The shifted variables $\hat{x}, \hat{y}$, the threshold $z_0$ and the Jacobian  ${\cal J}$
depend on the choice of the external kinematic variables, see Table~\ref{TAB:QED}. 
$S$ denotes the cms energy and the hats refer to variables in the sub--system.

Here the different second order splitting kernels are given by
%--------------------------------------------------------------------
\begin{eqnarray}
\label{eRC8}
P_{ee}^{(2,1)}(z) &=& \frac{1}{2}
\left [P_{ee}^{(0)} \otimes P_{ee}^{(0)} \right ](z)
= \frac{1 + z^2}{1 - z} \left [ 2\ln(1-z) - \ln z + \frac{3}{2} \right ]
+ \frac{1}{2}(1 + z) \ln z - (1 -z),
\\
%-----------------------------------------
\label{eRC9}
P_{ee}^{(2,2)}(z) &=& \frac{1}{2}
\left [
P_{e \gamma}^{(0)} \otimes P_{\gamma e}^{(0)}
\right ](z) \equiv
(1 + z) \ln z +
\frac{1}{2}(1 - z) + \frac{2}{3} \frac{1}{z} (1 - z^3),
\\
%---------------------------------
\label{eRC10}
P_{ee,f}^{(2,3)}(z) &=&
N_c(f) Q_f^2
\frac{1}{3} P_{ee}^{(0)}(z)
                        \theta \left( 1 - z - \frac{2 m_f}{E_e}
                        \right),
\end{eqnarray}
%-----------------------------------------------
denoting double-photon radiation, scattering of a fermion  into
a fermion by a collinear photon, and collinear fermion pair production.
Here, $m_f$ is the mass of the produced fermion, $Q_f$ its charge, $N_c(f) = 3$ 
for quarks, $N_c(f) = 1$ for leptons, respectively, and
$\otimes$ denotes the Mellin-convolution (\ref{eq:MELLIN2}).
The soft-photon exponentiation
is performed solving the non-singlet
evolution equation in the range $z \rightarrow 1$ analytically,
cf.~e.g.~\cite{Gribov:1972rt}. Since the terms up to ${\cal O}(\alpha^2)$ were
taken into account in Eq.~(\ref{eRC7})
already
the corresponding
contributions have to be subtracted. One obtains~\cite{Blumlein:1994ii}:
%--------------------------------------------------------------------
\begin{equation}
\label{eRC12}
\frac{d^2 \sigma^{(>2,soft)}}{dx dy} =
 \int_0^1  dz
P_{ee}^{(>2)}(z,Q^2) \left \{
\theta(z - z_0) {\cal J}(x,y,z)
\left. \frac{d^2 \sigma^{(0)}}{dx dy} \right|_{x=\hat{x},
y=\hat{y},S=\hat{S}}  -  \frac{d^2 \sigma^{(0)}}{dx dy}
\right\} ,
\end{equation}
%--------------------------------------------------------------------
with
%--------------------------------------------------------------------
\ba
\label{eRC13}
P_{ee}^{>2}(z,Q^2) &=& 
\zeta (1 - z)^{\zeta - 1}
\frac{\exp\left [ \frac{1}{2} \zeta \left (
\frac{3}{2} - 2 \gamma_E \right ) \right ]} {\Gamma(1 + \zeta)},
-
\frac{\alpha}{2 \pi} L_e \,
\frac{2}{1 - z} \left \{ 1 +
\frac{\alpha}{2\pi} L_e \,
\left [ \frac{11}{6} + 2 \ln(1 - z) \right ] \right \},
\nll
\ea
%--------------------------------------------------------------------
with $\zeta = -3 \ln \left [ 1 - (\alpha/3\pi) L_e \right ]$. 
%----------------------------------------------------------------------------------------------

\vspace*{-5mm}
\restylefloat{figure}
\begin{figure}[H]
\begin{center}
\includegraphics[scale=0.40,angle=0]{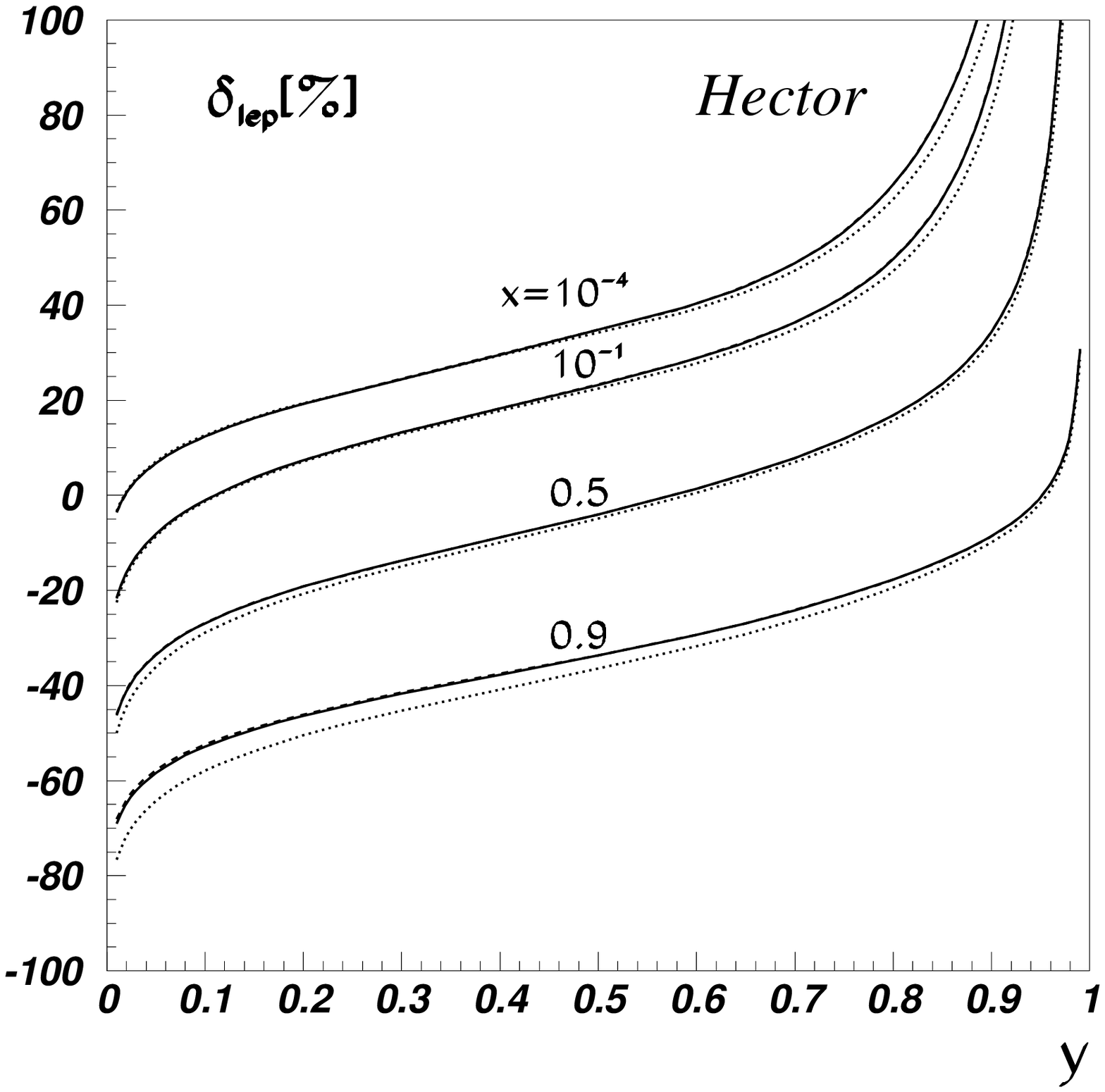}
\includegraphics[scale=0.45,angle=0]{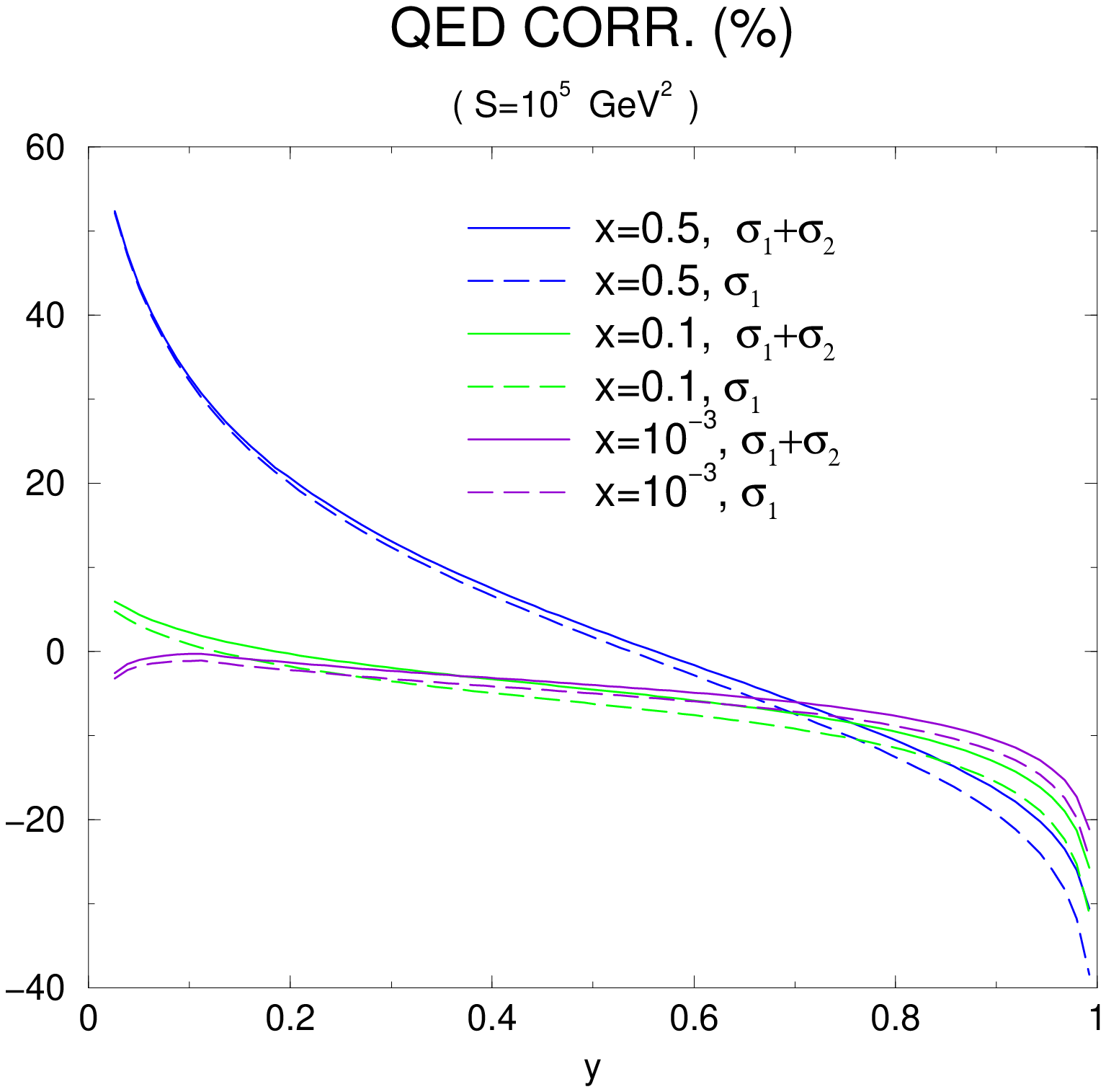}
\end{center}
\caption[]{
\label{QED:FIG}
\small \sf
a)~Radiative corrections in leptonic variables in per cent for $E_e = 26.8$~GeV, $E_p = 
820$~GeV. Dotted lines: $O(\alpha)$, dashed lines: $O(\alpha^2)$, solid lines: in addition soft 
photon exponentiation, from \cite{Arbuzov:1995id}, \TCop (1995) by Elsevier Science;
~~b) Radiative corrections in mixed variables in per cent for $E_e = 26.8$~GeV, $E_p =
820$~GeV. The lines left from above to below correspond to $x = 0.5, 0.1, 0.001$ in consecutive
order. Full lines $O(\alpha) + O(\alpha^2 L^2) +  O(\alpha^2 L)$; dashed lines: $O(\alpha)$; 
Figure by courtesy of H. Kawamura.}
\end{figure}
%----------------------------------------------------------------------------------------------
Finally, there is also a fermion conversion term in 
${\cal O}(\alpha^2 L_e^2)$, $P(z,Q^2; e^- \rightarrow e^+) = \left(\frac{\alpha}{2 \pi} \right )^2
L_e^2 P_{ee}^{(2,2)}(z)$, \cite{Blumlein:1994ii}.

In Figure~\ref{QED:FIG} we illustrate the size of QED radiative corrections for two
types of the measurement of the kinematic variables in the case
of neutral current deep inelastic scattering. In the first case the variables $x$ and $Q^2$ 
are measured from the scattered lepton only, while in the second case mixed variables, cf. Table~\ref{TAB:QED},
are applied, resulting into corrections of different type. The radiative
corrections for leptonic variables become very large at small $x$
and high $y$. Choosing the double-angle method, one obtains corrections which behave rather flat in $y$,
cf. \cite{Blumlein:1994ii}.
The different methods to calculate the QED radiative corrections
have been compared in ${\cal O}(\alpha)$ for a  variety of kinematic
measurements and are well understood. In 2nd order so far only LLA
results are available~\cite{Kripfganz:1990vm,Blumlein:1994ii} for the full set of external
kinematic variables studied by the HERA experiments. Only in the case of mixed variables
also the $O(\alpha^2 L)$ corrections were calculated \cite{Blumlein:2002fy}.

In the region in which the virtuality $Q^2$ of the exchanged photon becomes very small the observed final state
consists of a $p_\perp$-balanced electron-photon pair, with little hadronic activity near the beam-pipe.
This contribution is called Compton peak and has been studied first in \cite{RC1}. It is given by
%------------------------------------------------------------------------
\ba
\frac{d^2 \sigma^{C}}{d x_l d y_l}
&=&
\frac{\alpha^3}{x_l S}
\left[1+(1-y_l)^2\right]
\ln \left( \frac{Q^2_l}{M^2_N}  \right)
\int \limits_{x_l}^{1} \frac{dz}{z^2}
\frac{z^2+(x_l - z)^2}{ x_l(1 - y_l)}
\sum_f \left[q_f(z,Q^2_l) + {\bar q}_f(z, Q^2_l)\right]
\label{eRC5}
\ea
%------------------------------------------------------------------------
cf.~\cite{Blumlein:1989gk} for leptonic variables. The scale setting is the same as being used
in \cite{Bardin:1987rz} referring to the kinematic approach in which the hadronic structure is dealt 
with inclusively. A more refined expression than (\ref{eRC5}) was derived in Ref.~\cite{Blumlein:1993ef}.
LLA higher order corrections are easily obtained applying the corresponding radiators. The process was 
described using the parton model in \cite{BEBEN} regulating the collinear singularity by finite light 
quark masses.

In the initial calculations of the QED corrections to deep-inelastic scattering based on 
the parton model quark lines were treated as fermion lines with mass $m_q$ in the on-shell scheme
\cite{Bohm:1986na,Bohm:1987cg,Bardin:1988by,Bardin:1989vz}. This led to logarithmic corrections
of $O(\alpha e_q^2 \ln(Q^2/m_q^2))$ behaving flat in $y$ and growing with $x$ to $\sim 7-12\%$ for $x = 0.9 ... 
0.99$. Here $e_q$ denotes the light quark's charge. The correct way
in treating these contributions is to absorb them into the scaling violations of the partons
along with the QCD corrections.
At leading order one obtains the radiator 
%----------------------------------------------------------------------------------------------------
\begin{eqnarray}
\alpha_s(\mu^2) \left[1 + e_q^2 \frac{\alpha(\mu^2)}{C_F\alpha_s(\mu^2)}\right] P_{qq}^0(z) \ln\left(
\frac{\mu^2}{m_q^2}\right)~. 
\end{eqnarray}
%----------------------------------------------------------------------------------------------------
In this way the correction becomes independent of the values of the light quark masses, 
cf.~\cite{RC4,Kripfganz:1987yu,Blumlein:1989gk}. 
The overall effect is bounded by $< 1.5\%$ at HERA
and hard to be resolved \cite{Kripfganz:1988bd}.
Later numerical studies of the resummation of the leading order QED effect were performed in 
\cite{QUARK:lat}. %Spiesberger:1994dm,Roth:2004ti}. 

The integral leptonic QED corrections to polarized deep-inelastic $l^\pm N$ scattering, 
using the approach of form factors, have been calculated in \cite{Bardin:1996ch}, also presenting
the leading log results. Here also the effect of the exchange of electro-weak gauge bosons was 
included, which is of relevance at high energy lepton-nucleon colliders with polarized targets.
Leading log results were given in \cite{Fadin:1989dw}. Monte Carlo codes for the QED corrections 
in case of polarized deep-inelastic were designed in \cite{MC:POLQED}.
%Akushevich:1998dz,Akushevich:2011zy}.

The well-known resummation of the soft corrections \cite{Gribov:1972rt} to all orders in $O((\alpha L)^k)$ 
has been extended including the hard corrections to $O((\alpha L)^3)$ in \cite{Jezabek:1991bx,Skrzypek:1992vk}, 
to $O((\alpha L)^5)$ in the flavor non-singlet case in 
\cite{QED5th,Blumlein:2002bg}, %\cite{Przybycien:1992qe,Arbuzov:1999cq,
in the unpolarized flavor singlet case
in \cite{Blumlein:2007kx}, and in the polarized singlet case in \cite{Blumlein:2002bg}.
These corrections are universal and can be applied as well for other scattering cross sections.
For cms energies up to the TeV range these corrections are sufficient. However, depending on the process
corrections to next-to-leading log may be needed.

Another higher order correction is due to small-$x$ resummation in case of QED. 
These are the resummed non--singlet corrections of $O((\alpha \log^{2}(x))^l)$, 
cf.~\cite{Blumlein:1996yz}.

%%%%%%%%%%%%%%%%%%%%%%%%%%%%%%%%%%%%%%%%%%%%%%%%%%%%%%%%%%%%%%%%%%%%%%%%%%%%%%%%%%%%%%%%%%
\section{Heavy Flavor Wilson Coefficients}
\label{sec:HEAVY}
\renewcommand{\theequation}{\thesection.\arabic{equation}}
\setcounter{equation}{0}
%%%%%%%%%%%%%%%%%%%%%%%%%%%%%%%%%%%%%%%%%%%%%%%%%%%%%%%%%%%%%%%%%%%%%%%%%%%%%%%%%%%%%%%%%%

\noindent
The heavy flavor contributions to the deep-inelastic structure functions are rather large in the 
unpolarized case for smaller values of $x$. Due to the different scaling violations of the massless and
massive contributions to the structure functions the exact knowledge of the heavy flavor Wilson coefficients
is important. At leading order they were calculated in the unpolarized case in \cite{HQ:LO} and in 
the polarized case in \cite{Watson:1981ce,Vogelsang:1990ug,Blumlein:2003wk}. They
are given by~:
%--------------------------------------------------------------------------------
\begin{eqnarray}
H_{g,F_2}^{(1)}\left(z,\frac{m^2}{Q^2}\right) &=& 
8T_F \Biggl\{v\left[-\frac{1}{2} + \left(4
+  \frac{2m^2}{Q^2}\right) z(1-z)\right] 
+ \left[-\frac{1}{2} +z(1-z) +2\frac{m^2}{Q^2}z(3z-1) + 4\frac{m^4}{Q^4} z^2 \right]
\nonumber\\
&& \times
\ln\left(\frac{1-v}{1+v}\right)\Biggr\}
\\
%----
H_{g,F_L}^{(1)}\left(z,\frac{m^2}{Q^2}\right) &=& 16 T_F \Biggl\{v z(1-z) + 2 \frac{m^2}{Q^2} z^2
\ln\left(\frac{1-v}{1+v}\right)\Biggr\}
%\\
\\
%----
H_{g,g_1}^{(1)}\left(z,\frac{m^2}{Q^2}\right) &=& 4T_F \Biggl\{v (3-4z) +(1-2z) 
\ln\left(\frac{1-v}{1+v}\right)\Biggr\}~,
\end{eqnarray}
%--------------------------------------------------------------------------------
with $v = \sqrt{1 - 4 m^2 z/(Q^2(1-z))}$ and $m$ the heavy quark mass. The heavy flavor contribution to the 
structure functions at LO is given by
%--------------------------------------------------------------------------------
\begin{eqnarray}
F_{2,L}^{Q\bar{Q}}(x,Q^2,m^2) = e_Q^2 a_s \int_{ax}^1 \frac{dz}{z} 
H_{g,F_{2,L}}^{(1)}\left(\frac{x}{z},\frac{m^2}{Q^2}\right) G(z,Q^2),~~a = 1 + \frac{4m^2}{Q^2}~,
\end{eqnarray}
%--------------------------------------------------------------------------------
and analogously for $g_{1}^{Q\bar{Q}}(x,Q^2,m^2)$. In the polarized case the heavy-flavor contributions 
to $g_1(x,Q^2)$ and $g_2(x,Q^2)$ show an oscillatory behavior \cite{Blumlein:2003wk} since the first moment 
of $H_{g,g_1}^{(1)}$ vanishes and $H_{g,g_2}^{(1)}$ obeys a Wandzura-Wilczek relation 
\cite{Wandzura:1977qf}. 
In the unpolarized case the heavy-flavor Wilson coefficients have been calculated at NLO in semi-analytic 
form in \cite{HEAV2,Riemersma:1994hv}.\footnote{A fast numerical implementation in Mellin-space has been 
given in \cite{Alekhin:2003ev}.} A detailed proof of hard-scattering factorization including heavy
quarks was given in \cite{Collins:1998rz}. 

In the region $Q^2 \gg m^2$ the heavy-flavor Wilson coefficients factorize into massive operator matrix 
elements with massless external lines\footnote{Massive OMEs with massive external lines have been dealt 
with in \cite{Blumlein:2011mi}.} and the corresponding massless Wilson coefficients \cite{Buza:1996xr}. For 
the structure function
$F_2(x,Q^2)$ this representation holds at the 1\% level for $Q^2 \gsim 10~m^2$ \cite{Buza:1996xr}, while
for $F_L(x,Q^2)$ this is valid only at much higher scales $Q^2 \gsim 800~m^2$. The results of 
\cite{Buza:1996xr} were confirmed in \cite{Bierenbaum:2007qe} and the $O(a_s^2 \ep)$ corrections were 
derived in \cite{Bierenbaum:2008yu}. In the polarized case the NLO heavy flavor Wilson coefficients were 
computed for the asymptotic region only \cite{Buza:1996xr,Bierenbaum:2007pn}. At LO and NLO also 
heavy flavor corrections for the charged current case have been calculated 
\cite{CCHQ}.

Because of the high accuracy to which the structure function $F_2(x,Q^2)$ is measured, the heavy flavor 
Wilson coefficients have to be computed to NNLO, like the massless contributions. This is analytically 
possible using the method of massive operator matrix elements in the region $Q^2 \gsim 10~m^2$.
At 3-loop order the five Wilson coefficients $L_{q,i}^{\rm NS}, L_{q,i}^{\rm PS}, L_{g,i}^{\rm S}, 
H_{q,i}^{\rm PS}, H_{g,i}^{\rm S}$ contribute, cf.~\cite{Bierenbaum:2009mv}.
In \cite{Bierenbaum:2009mv} a series of Mellin moments has been calculated for the corresponding 
massive OMEs ranging up to $N = 10, ..., 14$, depending on the process, applying the code {\tt MATAD}
\cite{Steinhauser:2000ry}.  
First 3-loop results for general values of $N$ were derived in \cite{Ablinger:2010ty,BHS12a}. These
calculations require completely different technologies. In \cite{Ablinger:2010ty} the Wilson coefficients 
$L_{q,i}^{\rm PS}$ and  $L_{g,i}^{\rm S}$ were calculated completely. All logarithmic contributions 
$\propto \ln^k(Q^2/m^2), k = 1,2,3$ in the 3-loop case have been computed for general values of $N$ 
in \cite{Bierenbaum:2010jp}. The technology to calculate the ladder topologies is already available 
\cite{ABHKSW12} and first results have been obtained for other more involved 3-loop topologies.
Here codes like {\tt Sigma} and {\tt HarmonicSums} \cite{SIGMA,ABS12} are essential, which were developed 
further
to cope with problems of this kind. In the asymptotic region also the 3-loop contributions for 
the heavy flavor Wilson coefficients to $F_L(x,Q^2)$ were calculated \cite{Blumlein:2006mh}.
Furthermore, a series of moments of the massive OMEs contributing to transversity at $O(a_s^3)$ were 
computed in \cite{Blumlein:2009rg}. 
The heavy flavor Wilson coefficients at higher order can be expressed for the pole or running mass, which is
implemented through the renormalization procedure, cf. Sect.~\ref{sec:RGE}, up to 3-loop order in 
\cite{Bierenbaum:2009mv}; for a phenomenological application at NLO see \cite{Alekhin:2010sv}.
In the threshold region of heavy-quark pair production one may estimate higher order corrections
using soft resummations. They have been studied in the unpolarized case in \cite{THR:UNP} 
and the polarized case in \cite{Eynck:2000gz}.

The above calculations were all performed in the fixed-flavor number scheme. As has been shown in
\cite{Gluck:1993dpa,Alekhin:2009ni} this description is sufficient throughout the kinematic range at HERA
for charm and bottom. Still one may consider to introduce a so-called variable flavor number scheme, in 
particular to match the universal contributions on the way a single massive quark is becoming 
light.
Here one assumes that the transition always occurs for one heavy quark at the time. One problem in this
context is that $m_c^2/m_b^2 \approx 1/10$ and one cannot consider charm quarks as massless at $\mu^2 = 
m_b^2$. 
The matching conditions for the flavor non-singlet and singlet parton distributions have been given at NLO in
\cite{Buza:1996wv} and NNLO in \cite{Bierenbaum:2009mv} and the transition functions in 
\cite{Buza:1996xr,Bierenbaum:2007pn,Buza:1996wv,Bierenbaum:2009zt,Bierenbaum:2009mv}, with first general $N$ results 
at NNLO in \cite{Ablinger:2010ty,BHS12a}. Different descriptions to interpolate 
between the heavy-quark threshold region and the asymptotic region are considered in the literature. As an 
example, in the BMSN-scheme \cite{Buza:1996wv,Alekhin:2009ni} the interpolation in $F_2(x,Q^2)$ is given by
%--------------------------------------------------------------------------------
{\begin{eqnarray}
F_{2}^{Q\bar{Q},\rm BMSN}(x,Q^2,N_f+1) = 
F_{2}^{Q\bar{Q}}(x,Q^2,N_f)
+ F_{2}^{Q\bar{Q},\rm asymp}(x,Q^2,N_f+1)
- F_{2}^{Q\bar{Q},\rm asymp}(x,Q^2,N_f)~.
\end{eqnarray}}
%--------------------------------------------------------------------------------
\noindent
Here $F_{2}^{Q\bar{Q},\rm asymp}$ denotes the heavy flavor structure function in the limit of vanishing 
power corrections.
Other schemes were proposed in \cite{ACOT,TR,Forte:2010ta}. The schemes differ w.r.t. the way how
fast a massive quark becomes light. One may infer the correct behaviour from precision data.
The matching scales are usually chosen by $\mu = m_q$. However, this scale is usually process dependent and 
may turn out to be much larger as has been pointed out in \cite{Blumlein:1998sh} comparing exact 
calculations with the interpolative description. In any case, the reordering of terms in the interpolation 
schemes, cf.  \cite{Buza:1996wv,Alekhin:2009ni,ACOT,TR,Forte:2010ta}, have to be compatible with the 
$\overline{\rm MS}$ scheme. Otherwise an according scheme-transformation has to be provided 
since in the PDF-fits usually $\overline{\rm MS}$ parton distribution functions are determined and nearly 
all partonic scattering cross sections in physical applications are computed in this scheme.

Finally we would like to mention that at 3-loop order also diagrams with two different fermion lines 
contribute \cite{MBMC}. Here the moments $N = 2, 4, 6$ were calculated based on the code {\tt qexp} 
\cite{QEXP}. These contributions are even universal. However, due to the fact the $m_b$ is not very much 
larger than $m_c$ one obtains here double-logarithmic contributions, which may not simply be absorbed into
either the charm or the bottom distribution.
%%%%%%%%%%%%%%%%%%%%%%%%%%%%%%%%%%%%%%%%%%%%%%%%%%%%%%%%%%%%%%%%%%%%%%%%%
\section{Target Mass Corrections}
\label{sec:TM}
\renewcommand{\theequation}{\thesection.\arabic{equation}}
\setcounter{equation}{0}
%%%%%%%%%%%%%%%%%%%%%%%%%%%%%%%%%%%%%%%%%%%%%%%%%%%%%%%%%%%%%%%%%%%%%%%%%

\vspace*{1mm} 
\noindent
At low 4-momentum transfer $Q^2$ and large values of $x$ the nucleon mass $M$ 
modifies the scattering kinematics and therefore the nucleon structure functions.
The corresponding relations have first been worked out by Nachtmann \cite{Nachtmann:1973mr}
in the unpolarized case.\footnote{The same method has been applied to the polarized case in 
\cite{Wandzura:1977ce}.} Besides the Bjorken variable $x$ the new variable 
%--------------------------------------------------------------------------------
\begin{eqnarray}
\label{eq:NACHT}
\xi = \frac{2 x}{1+\sqrt{1+4 x^2 M^2/Q^2}} \equiv \frac{2x}{1+r}
\end{eqnarray}
%--------------------------------------------------------------------------------
emerges. Here the mass of the quark in the scattering process is assumed to be the same, $m_i = m_f$. 
For $x \ll 1$ or $Q^2 \rightarrow \infty$ both variables approach each other. The corrections for the 
different structure functions can be derived using the local operators as in the massless case, 
cf.~Sect.~\ref{sec:3}, \cite{Georgi:1976ve}. In the following we discuss the corrections for deep 
inelastic scattering off unpolarized and polarized targets.

Let us define the integrals \cite{Georgi:1976ve,Blumlein:1998nv} 
%--------------------------------------------------------------------------------
\begin{eqnarray}
\label{eq:TM2}
&&G_1(\xi) = \int_\xi^1 \frac{d\xi'}{\xi'} F(\xi'),~~~~
G_2(\xi) = \int_\xi^1 \frac{d\xi'}{\xi'} \int_{\xi'}^1 \frac{d\xi''}{\xi''} F(\xi''),~~~
G_3(\xi) = \int_\xi^1 d\xi' \int_{\xi'}^1 \frac{d\xi''}{\xi''} \int_{\xi'''}^1 
\frac{d\xi'''}{\xi'''}F(\xi'''),
\nonumber\\
&& 
H_1(\xi) = \int_\xi^1 d\xi' F(\xi'),~~~~
H_2(\xi) = \int_\xi^1 d\xi' \int_{\xi'}^1 d\xi'' F(\xi'')
\end{eqnarray}
%--------------------------------------------------------------------------------
and the operators
%--------------------------------------------------------------------------------
\begin{eqnarray}
\label{eq:TMop}
&&O_{F_1}^{(1)} = \frac{x}{2 r},~~O_{F_1}^{(2)} = -\frac{M^2}{Q^2} x^2 \frac{d}{dx} \frac{x}{\xi 
r},~~
O_{F_2} = x^2 \frac{d^2}{dx^2} \frac{x^2}{\xi^2 r},~~ O_{F_3} = -x \frac{d}{dx} \frac{x}{\xi r}
\nonumber\\ 
&&O_{g_1} =  x\frac{d}{dx} x\frac{d}{dx} \frac{x}{r\xi},~~
O_{g_2} = -x\frac{d^2}{dx^2}\frac{x^2}{r\xi},~~
O_{g_3} = 2x^2\frac{d^2}{dx^2}\frac{x^2}{r\xi^2},~~
O_{g_4} = -x^2\frac{d}{dx}x\frac{d^2}{dx^2}\frac{x^2}{r\xi^2}, \nonumber
\end{eqnarray}
\begin{eqnarray}
%\\
&&O_{g_5}^{(1)} = -x\frac{d}{dx} \frac{x}{r\xi},~~
O_{g_5}^{(2)} = \frac{M^2}{Q^2} x^2\frac{d^2}{dx^2} \frac{x^2}{r\xi}~.
\end{eqnarray}
%--------------------------------------------------------------------------------
Here a generic function $F(\xi)$ has been introduced, which reproduces the corresponding structure 
function in the massless case containing the respective electro-weak couplings and parton content.
As an example, the function $F$ in case of the unpolarized structure functions $F_1$ and $F_2$ is the
same. The target mass corrections can now be written in the following compact form for the unpolarized structure 
functions $F_{1,2,3}$ and the polarized structure functions $g_{1,...,5}$ in the twist-2 approximation~:
%--------------------------------------------------------------------------------
\begin{eqnarray}
\label{eq:TMF1}
F_1(x,Q^2)   &=& O_{F_1}^{(1)}  F(\xi) + O_{F_1}^{(2)} H_2(\xi) \\
\label{eq:TMF2}
F_2(x,Q^2)   &=& O_{F_2}^{(1)}  H_2(\xi)\\ 
\label{eq:TMF3}
F_3(x,Q^2)   &=& O_{F_3}^{(1)}  H_1(\xi) \\
\label{eq:TMg12}
g_{i}(x,Q^2) &=& O_{g_i} G_2(\xi),~~~~~~~~~~~~~~~~~~~~~~~i = 1,2\\
\label{eq:TMg34}
g_{i}(x,Q^2) &=& O_{g_i} G_3(\xi),~~~~~~~~~~~~~~~~~~~~~~~i = 3,4\\
\label{eq:TMg5}
g_{5}(x,Q^2) &=& O_{g_5}^{(1)}  G_1(\xi) + O_{g_5}^{(2)} G_3(\xi)~.
\end{eqnarray}
%--------------------------------------------------------------------------------
These relations easily transform into the corresponding integral representations.\footnote{We
agree with the results of  \cite{Georgi:1976ve} up to the obvious typos in (4.19) and (4.22), cf.
also \cite{Schienbein:2007gr}.}
For the implementation of the target mass corrections in Mellin-space one may refer
to contour integral representations directly, see \cite{Georgi:1976ve,Blumlein:1998nv}. 
In the unpolarized case they are given by 
%--------------------------------------------------------------------------------
\begin{eqnarray}
\label{eq:MOM1TM}
F_1(x,Q^2) &=& \frac{1}{2\pi i} \int_{-i \infty}^{i\infty} dN~x^{-N} 
\sum_{j=0}^\infty \left(\frac{M^2}{Q^2}\right)^j \binom{N+j}{j} 
\left[\frac{1}{2} a_{N+2j}^{(2)} 
+x\frac{M^2}{Q^2} \frac{N a_{N+2j+1}^{(2)} 
}{(N+2j)(N+2j+1)}\right],\nonumber\\
\\
%-------------------------
F_2(x,Q^2) &=& \frac{1}{2\pi i} \int_{-i \infty}^{i\infty} dN~x^{-N+1} 
\sum_{j=0}^\infty \left(\frac{M^2}{Q^2}\right)^j \binom{N+j}{j} \frac{N(N-1)}{(N+2j)(N+2j-1)}
a_{N+2j}^{(2)}, \text{\cite{Georgi:1976ve}}, \nonumber\\
\\
%-------------------------
\label{eq:MOM2TM}
F_3(x,Q^2) &=& \frac{1}{2\pi i} \int_{-i \infty}^{i\infty} dN~x^{-N}
\sum_{j=0}^\infty \left(\frac{M^2}{Q^2}\right)^j \binom{N+j}{j} \frac{N}{N+2j}
a_{N+2j}^{(3)}~,
\end{eqnarray}
%--------------------------------------------------------------------------------
with $a_k = \int_0^1 dz z^{k} F(z)$. In the limit $M^2 \rightarrow 0$ the structure functions
$F_{1,3}$ contain one power in $x$ less than $F_2$. Note that the definition of the OMEs $a_k$ differ 
in the literature. The target mass corrections for the longitudinal structure function are obtained by
\cite{Georgi:1976ve}
%--------------------------------------------------------------------------------
\begin{eqnarray}
2x F_L =  r^2  F_2 - 2x F_1~.
\end{eqnarray}
%--------------------------------------------------------------------------------
The corresponding relations
in the polarized case were given in \cite{Blumlein:1998nv}. For lower values of $x < 0.7$ and not too 
small values of $Q^2 \gsim 1 \GeV^2$ the infinite series in (\ref{eq:MOM1TM}--\ref{eq:MOM2TM})
may be approximated using the first five terms to reach sufficient precision. In the large-$x$ 
domain convergence is reached using up to $\sim 250$ terms. This requires economic implementations
of the OMEs $a_k$, changing during fits to data, for the respective complex values of $k$ along the 
integration contour, which has been given in \cite{Blumlein:2012se} recently.

Target mass corrections lead to a violation of the Callan-Gross relation \cite{Callan:1969uq}.
Furthermore, one may derive integral relations in the polarized case, cf. Sect.~\ref{sec:polt3}. 
In the limit $x \rightarrow 1$ the structure functions (\ref{eq:TMF1}--\ref{eq:TMg5}) do not vanish 
unlike the partonic functions $F(x)$. This has led to numerous discussions in the literature 
\cite{Bitar:1978cj,Miramontes:1988fz}. Effects in this region cannot be considered without of
a careful study of higher twist effects. In \cite{Accardi:2008ne} besides the usual target mass 
corrections contributions of the jet-function at large values of $x$ were considered to improve the
description in this region. However, this approach has to be viewed in comparison with others
w.r.t. a consequent twist expansion and the fact, that the {\it whole} hadronic final state Fock-space 
has to be summed over.
  
Let us mention that in \cite{Blumlein:1998nv} also the corresponding 
relations for the twist-3 contributions to the polarized structure functions were derived. 
The target mass corrections in case of twist-2 and 3, 
(\ref{eq:TMg12}), were also derived in \cite{Piccione:1997zh}.
For non-forward and diffractive scattering they have been calculated in \cite{Belitsky:2001hz,Geyer:2004bx} 
and \cite{Blumlein:2008di}.
%%%%%%%%%%%%%%%%%%%%%%%%%%%%%%%%%%%%%%%%%%%%%%%%%%%%%%%%%%%%%%%%%%%%%%%
\section{Solution of the Evolution Equations and Parton Distribution Functions}
\label{sec:SOL}
\renewcommand{\theequation}{\thesection.\arabic{equation}}
\setcounter{equation}{0}
%%%%%%%%%%%%%%%%%%%%%%%%%%%%%%%%%%%%%%%%%%%%%%%%%%%%%%%%%%%%%%%%%%%%%%%

\noindent
In Mellin-space the singlet evolution equations can be solved analytically. 
We follow~\cite{Blumlein:1997em} and express the evolution equation (\ref{eq:ren3}) 
in terms of $a_s = \alpha_s(Q^2)/4\pi$. The r.h.s. is expanded in the
coupling constant. The singlet evolution equation reads~:
%-----------------------------------------------------------------------
{\small
\begin{eqnarray}
 \frac{\partial \qV (a_s,N)}{\partial a_s}
 &=& \frac{a_s \PV_0(N) + a_s^2 \PV_1(N) + a_s^3 \PV_2(N)+ \ldots}
  {-a_s^2 \,\beta_0 - a_s^3 \,\beta_1 - a_s^4 \,\beta_2 - \ldots}
  \: \qV (a_s,N) %\nonumber\\
 = -\frac{1}{\beta_0 a_s}
  \bigg[ \PV_0 (N) + a_s \bigg( \PV_1 (N)- \frac{\beta_1}{\beta_0}
  \PV_0 (N)\bigg) \nonumber \\ & & %\mbox{} \hspace*{10mm}
  + a^2_s \bigg( \PV_2 (N) - \frac{\beta_1}{\beta_0} \PV_1 (N)
  + \bigg\{ \bigg( \frac{\beta_1}{\beta_0} \bigg)^2 - \frac{\beta_2}
  {\beta_0} \bigg\} \PV_0 (N) \bigg) + \ldots \bigg]
  \, \qV (a_s,N) \nonumber \\
  &=&  -\frac{1}{a_s}
  \bigg[ \RV_0 (N) + \sum_{k=1}^{\infty} a_s^k \RV_k (N) \bigg]
  \, \qV (a_s,N) ,
\label{eq:EVO1}
\end{eqnarray}}
%-----------------------------------------------------------------------

\noindent
with $\RV_0 \equiv  \PV_0/\beta_0, \RV_k \equiv  [\PV_k - \sum_{i=1}^{k}
                  \beta_{i} \RV_{k-i}]/\beta_0$.\footnote{The solution up to $k=3$ was given in \cite{EKL}.}
Here $\beta_k, k \geq 0$ denote the expansion coefficients of the $\beta$-function and
$\PV_k, k \geq 0$ are the singlet matrices of the splitting functions and
%-----------------------------------------------------------------------
\begin{eqnarray}
\qV(a_s,N) = \left(\begin{array}{c} \Sigma(a_s,N) \\ G(a_s,N)\end{array} \right),
\end{eqnarray}
%-----------------------------------------------------------------------
with $\Sigma(a_s,N) = \sum_{i=1}^{N_f} \left[q(a_s,N) + \bar{q}(a_s,N)\right],G(a_s,N)$ the flavor 
singlet and gluon distributions.

One may obtain the evolution equations for the three flavor non-singlet cases by replacing
the matrices in (\ref{eq:EVO1}) by scalars. In the singlet case the matrices $\RV_k$ do not commute in 
general. Firstly, the leading order solution is found by
%-----------------------------------------------------------------------
\begin{eqnarray}
\label{sol0}
 \qV^{\,\rm LO} (a_s,N) =
 \left( {a_s}/{a_0} \right)^{{\footnotesize -\RV}_0 (N)}
 \qV (a_0,N) \equiv \LV (a_s,a_0,N) \, \qV (a_0,N),
\end{eqnarray}
%-----------------------------------------------------------------------
with the starting distribution $\qV (a_0,N)$ and 
%-----------------------------------------------------------------------
\begin{eqnarray}
 \LV (a_s,a_0,N) = \eV_{-}(N) \left({a_s}/{a_0}\right)^{-r_{-}(N)}
   + \eV_{+}(N) \left({a_s}/{a_0}\right)^{-r_{+}(N)}~.
\end{eqnarray}
%-----------------------------------------------------------------------
The projectors $\eV_\pm$ are 
%-----------------------------------------------------------------------
\begin{eqnarray}
 \eV_{\pm} = \frac{1}{r_{\pm}-r_{\mp}} \Big[ \RV_0-r_{\mp}\IV \Big];~~  r_{\pm } = \frac{1}{2 \beta_0} \bigg[ P_{qq}^{(0)} + P_{gg}^{(0)}
  \pm \sqrt{ \Big( P_{qq}^{(0)} - P_{gg}^{(0)}\Big) ^2 +
  4 P_{qg}^{(0)} P_{gq}^{(0)} } \, \bigg]~.
\end{eqnarray}
%-----------------------------------------------------------------------
Here $r_{\pm }$ denote the leading order eigenvalues.
The general solution is given by
%-----------------------------------------------------------------------
\begin{eqnarray}
\label{sol1}
 \qV (a_s,N) &=& \UV (a_s,N) \LV (a_s,a_0,N) \UV^{-1}(a_0, N) \:
 \qV (a_0,N) \\
 &=& \Big[ 1 + \sum_{k=1}^{\infty} a^k_s\UV_k (N) \Big] \LV (a_s,a_0,N)
 \Big[ 1 + \sum_{k=1}^{\infty} a_0^k \UV_k (N) \Big]^{-1} \qV (a_0,N)
  . \nonumber
\end{eqnarray}
%-----------------------------------------------------------------------
The matrices $\UV_k$ obey
%-----------------------------------------------------------------------
\begin{eqnarray}
\big[ \UV_k, \RV_0 \big] & = & \RV_k + \sum_{i=1}^{k-1}
            \RV_{k-i} \UV_i + k \UV_k
  \equiv  \widetilde{\RV}_k + k \UV_k,~k \geq 1  \nonumber\\
  \UV_k &=&
  - \frac{1}{k} \Big[ \eV_{-} \widetilde{\RV}_k \eV_{-} +
    \eV_{+} \widetilde{\RV}_k \eV_{+} \Big]
  + \frac{\eV_{+} \widetilde{\RV}_k \eV_{-}}{r_{-} - r_{+} - k}
  + \frac{\eV_{-} \widetilde{\RV}_k \eV_{+}}{r_{+} - r_{-} - k}~.
\end{eqnarray}
%-----------------------------------------------------------------------
Potential poles in $\UV_k(N)$  are canceled by those in $\UV^{-1}$ in (\ref{sol1}).
The perturbative solution in N$^k$LO is obtained expanding (\ref{sol1}) up to the $k$th common power in 
$a_s$ and $a_0$ keeping $\LV(a_s,a_0,N)$.

The different structure functions in $N$-space are represented by
%-----------------------------------------------------------------------
\begin{eqnarray}
F_l(a_s,a_0,N) = \sum_{n=1}^3 C_{n,l}^{\rm NS}(a_s,N) q_n^{\rm NS}(a_s,a_0,N)
+ C_{l}^{\rm S}(a_s,N) \Sigma(a_s,a_0,N)
+ C_{l}^{\rm g}(a_s,N) G(a_s,a_0,N),
\label{eq:EVO2}
\end{eqnarray}
%-----------------------------------------------------------------------
where $C_{n,l}^{\rm NS, S,g}$ denote the corresponding Wilson coefficients and $q_n^{\rm NS}$ are the 
flavor non-singlet distributions. In a consistent representation one also expands  (\ref{eq:EVO2})
in powers of $a_s, a_0$, similar to (\ref{sol1}), matching the factorization scales. All these operations
can be performed analytically. 

To transform to $x$-space a single precise contour integral is carried out numerically
around the singularities of the problem. Usually these are poles on the real axis for ${\sf 
Re(z)} < c$, with a 
given constant $c$. If small-$x$ resummations are included the singularities may be located in the 
complex plane, cf.~Section~\ref{sec:smx}. The contour integral is given by
%-----------------------------------------------------------------------
\begin{eqnarray}
\label{minv}
 xf(x) = \frac{1}{\pi} \int^{\infty}_0 \!\! dz \, {\rm Im}
 [ e^{i \phi} x^{-C} f(N\! =\! C) ] \: ,
\end{eqnarray}
%-----------------------------------------------------------------------
since usually $f^*(N) = f(N^*)$ is obeyed, \cite{Gluck:1989ze}, and $C = c + z e^{i \phi}$.
The solution of the evolution equations in Mellin space allows for fast and very precise numerical 
implementations. In other approaches the evolution equations are solved in momentum-fraction space.
Public codes for the solution of the evolution equations are e.g. {\tt QCD-Pegasus} \cite{Vogt:2004ns}, {\tt 
Hoppet} \cite{Salam:2008qg}, 
{\tt QCDNUM} \cite{Botje:2010ay}, and {\tt Openqcdrad} \cite{OPENQCDRAD}.

The distributions $ q_n^{\rm NS}(a_0,N), \Sigma(a_0,N)$ and $G(a_0,N)$ are non-perturbative quantities, 
which have to be fitted to the world precision data on deep-inelastic scattering and suitable other 
hard scattering processes. Their shape is a prioiri unknown and the corresponding parameterizations, valid  
in the kinematic region to be analyzed, have to 
be found. One way to determine these shapes is to express the distribution functions in terms of orthogonal 
polynomials, which has been studied in detail in the past \cite{ORTHPOL,FP}. The Laguerre polynomials 
${\rm L}_n[\ln(1/x)]$ converge fastest \cite{FP}. In general the number of orthogonal polynomials needed is 
too large compared to the possible amount of parameters which can be fitted from the data. One way out 
consists in designing shapes in terms of polynomials or related functions, which are suggested by the 
orthogonal polynomial analysis and extend these forms gradually. The NNPDF collaboration uses neural 
network techniques \cite{Forte:2002fg} to find the correct shape of the parton densities and to estimate
their errors. Due to the finite experimental accuracy and the limited amount of data points available it 
seems that the present number of parameters cannot exceed 30--40 in the unpolarized case. In the polarized 
case this number is even lower. The errors of the statistical and systematic errors of the parton 
distribution functions can be determined in $\chi^2$-analyses.
%---------------------------------------------------------------------------------------------- 
\restylefloat{figure} 
\begin{figure}[H] 
\begin{center} 
\mbox{\epsfig{file=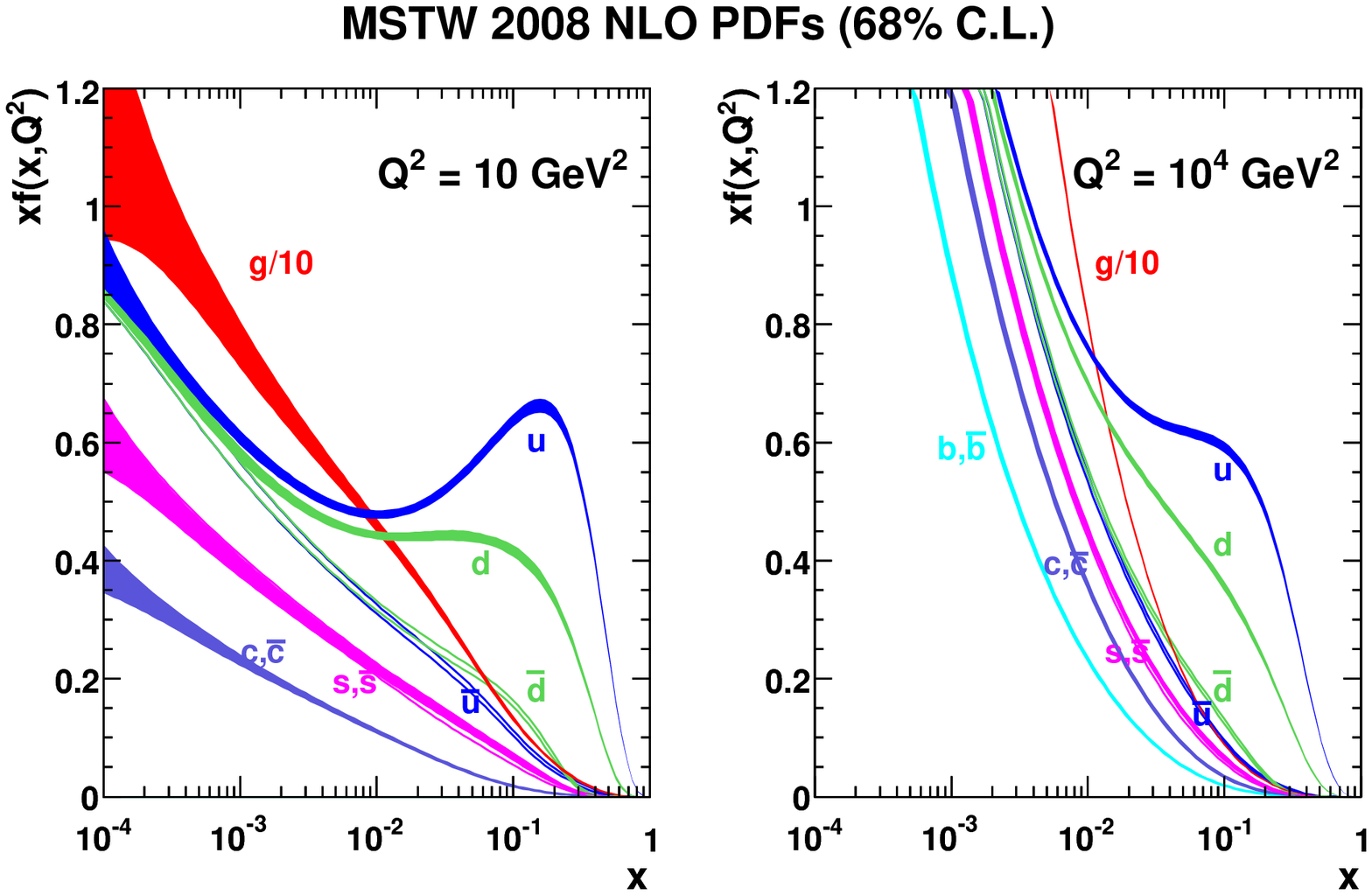,width=9cm}} 
\end{center} 
\caption[]{\label{FIG:mstw} 
\small \sf
The MSTW 2008 NLO PDFs at $Q^2 = 10 \GeV^2$ and $10^4 \GeV^2$; from \cite{Martin:2009iq}, \TCop 
(2009) by Springer Verlag.} 
\end{figure} 
%---------------------------------------------------------------------------------------------- 

\noindent
The valence quark distributions $u_v = u - \bar{u}$ and $d_v = d - \bar{d}$ are constrained by the following 
sum rules in case of the proton~:
%-----------------------------------------------------------------------
\begin{eqnarray}
\int_0^1 dx u_v(x,Q^2) = 2,~~~\int_0^1 dx d_v(x,Q^2) = 1~.
\end{eqnarray}
%-----------------------------------------------------------------------
Furthermore, assuming twist-2 dominance, the momentum sum rule reads~: 
%-----------------------------------------------------------------------
\begin{eqnarray}
\int_0^1 dx x \left[\Sigma(x,Q^2) + G(x,Q^2)\right] = 1~.
\end{eqnarray}
%-----------------------------------------------------------------------

In the following we give a brief summary on the status reached for unpolarized\footnote{For  
extensive recent reviews see \cite{DeRoeck:2011na,Perez:2012um}.}
and polarized twist-2 parton 
distribution 
functions (PDFs). They are obtained by the QCD analysis of the deep inelastic structure functions, 
supplemented by other hard scattering processes. The latter data sets mainly serve the purpose to 
resolve the flavor structure of the quarkonic sea and also might give better constraints on the gluon. 
%---------------------------------------------------------------------------------------------- 
\restylefloat{figure} 
\begin{figure}[H] 
\begin{center} 
\includegraphics[scale=0.35,angle=0]{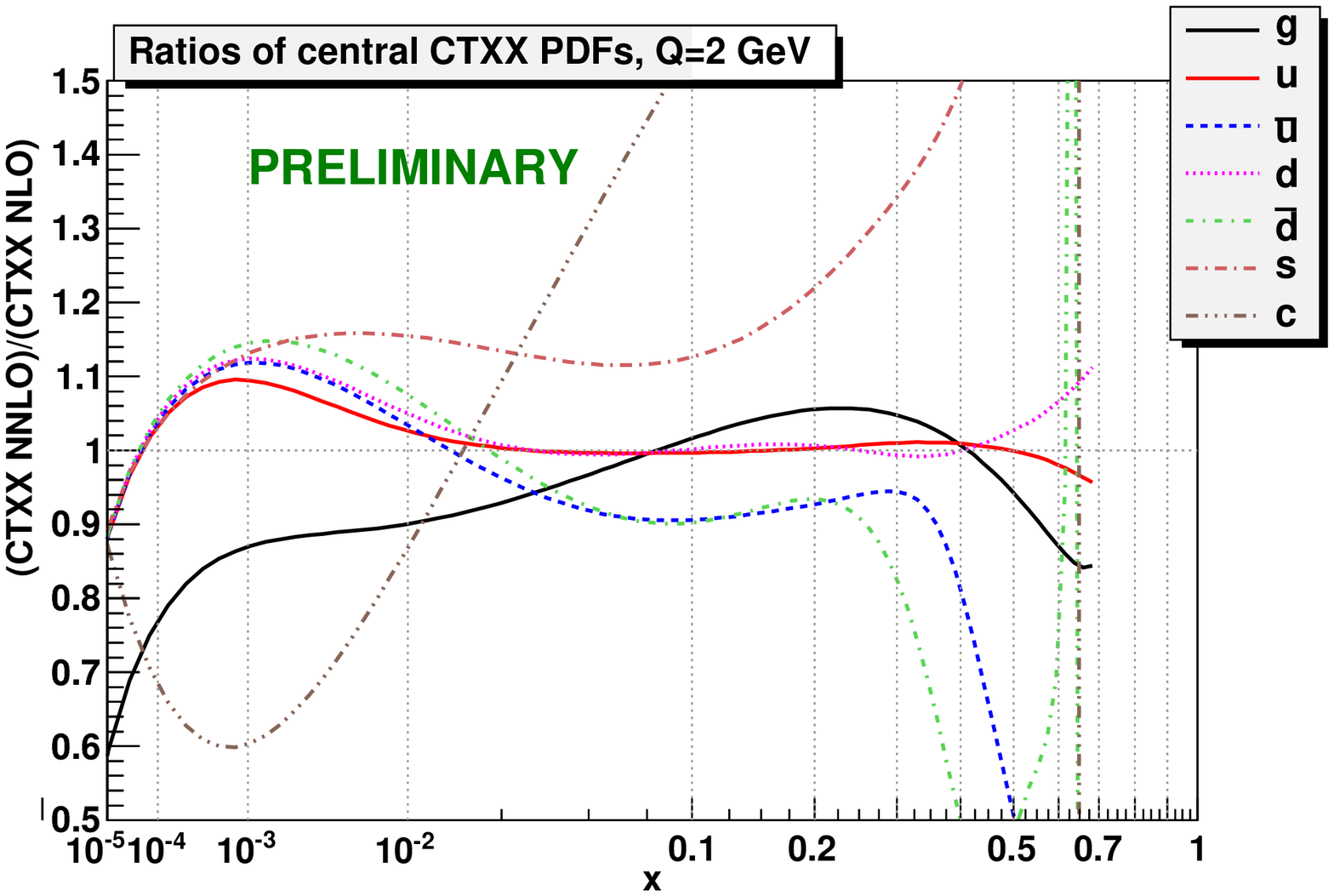} \hspace*{10mm}
\includegraphics[scale=0.35,angle=0]{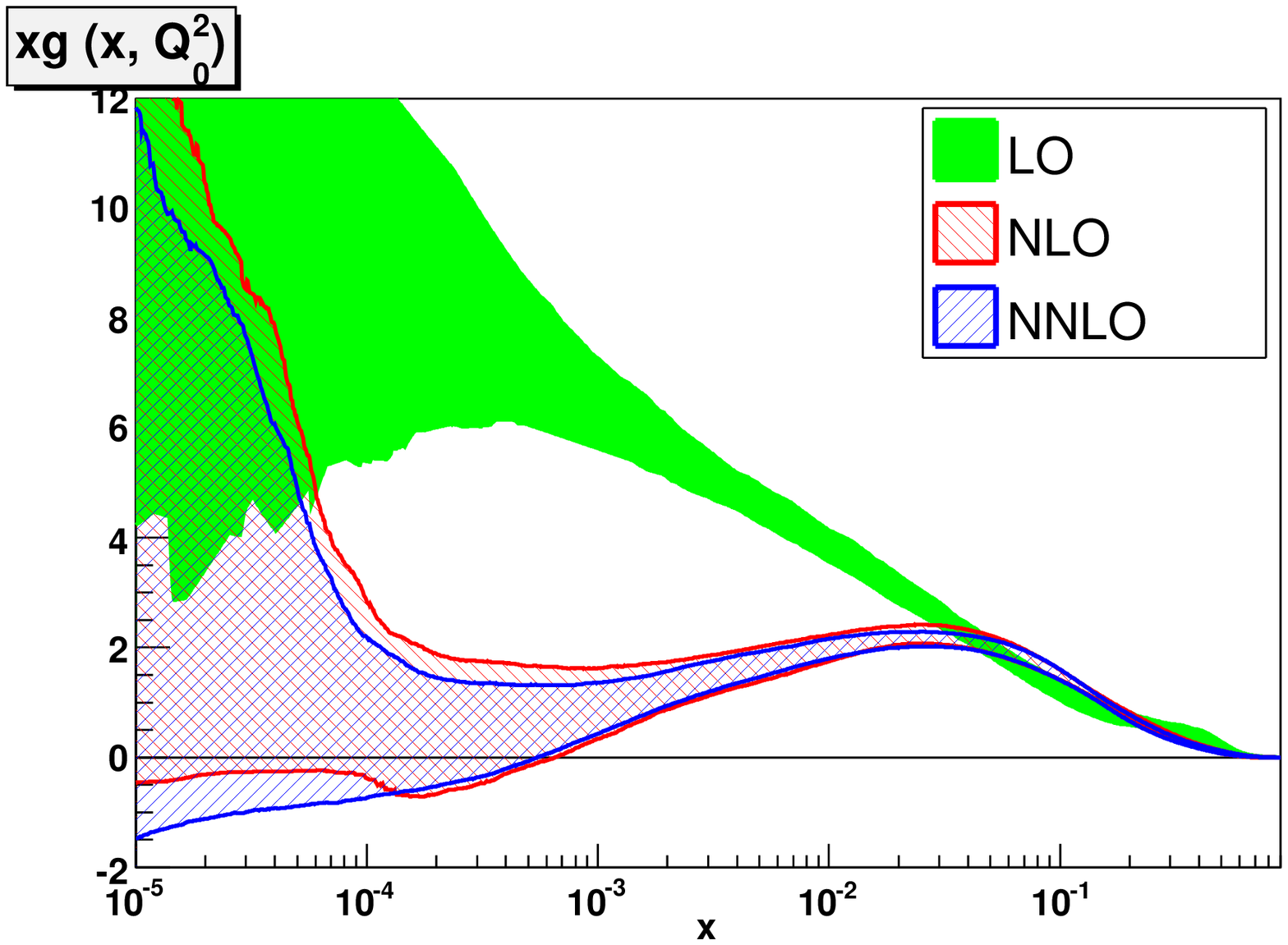} 
\end{center} 
\caption[]{ \label{FIG:cteq_nnpdf} 
\small \sf 
Left~: Ratios of the PDFs at NNLO and NLO by CTEQ for $Q^2 = 4 \GeV^2$; courtesy P. Nadolsky. 
Right~: Comparison of the gluon distribution by NNPDF2.1 at LO, NLO and NNLO at $Q^2 = 2 \GeV^2$; from 
\cite{Ball:2011uy}, \TCop (2012) by Elsevier Science.
} 
\end{figure} 
%---------------------------------------------------------------------------------------------- 
%%%%%%%%%%%%%%%%%%%%%%%%%%%%%%%%%%%%%%%%%%%%%%%%%%%%%%%%%%%%%%%%%%%%%%%
\subsection{Unpolarized Nucleons}
%%%%%%%%%%%%%%%%%%%%%%%%%%%%%%%%%%%%%%%%%%%%%%%%%%%%%%%%%%%%%%%%%%%%%%%

\noindent
Early QCD analyses of deep-inelastic structure functions have been performed starting in the late 1970ies 
in LO, and later in NLO, to derive first principal shapes of the parton densities, cf.~\cite{Buras:1977yj,
Gluck:1980cp,Duke:1983gd,Eichten:1984eu}. Analyses of the Dortmund group \cite{Gluck:1980cp,GRV}, the MRS 
group \cite{Martin:1987vw}, and the CTEQ (then Morfin-Tung) group \cite{MT} followed. In the 
first analyses the charm quark contributions were either treated as massless or at LO, since higher order 
corrections were not yet know.

At present the differential scattering cross sections of unpolarized deep-inelastic scattering in the massless
case are known to NNLO and the heavy flavor corrections to NLO. The present highest order parton fits
are based on this approximation. We note that some of the data contain significant higher twist contributions
which have to be quantified, or, if possible for the corresponding reaction, cut away. Furthermore, target 
mass corrections have to be applied, since part of the scaling violations in the large $x$ and lower $Q^2$ 
region are due to these, cf.~\cite{Blumlein:2006be}. 

At present the six collaborations  
ABM     \cite{Alekhin:2012ig},
CTEQ    \cite{Lai:2010vv},
HERAPDF \cite{HERAPDF1.5},
JR      \cite{JimenezDelgado:2008hf},
MSTW    \cite{Martin:2009iq}, and
NNPDF   \cite{Ball:2011mu} perform NNLO analyses, partly using different data sets. In 
Figure~\ref{FIG:mstw} an overview on the principal behaviour of the parton distributions  
as a function of $x$ are given for two typical scales of $Q^2$ by MSTW at NLO. The distributions
rise towards small values of $x$ and become steeper with growing values of $Q^2$. 

%---------------------------------------------------------------------------------------------- 
\restylefloat{figure} 
\begin{figure}[H] 
\begin{center} 
\mbox{\epsfig{file=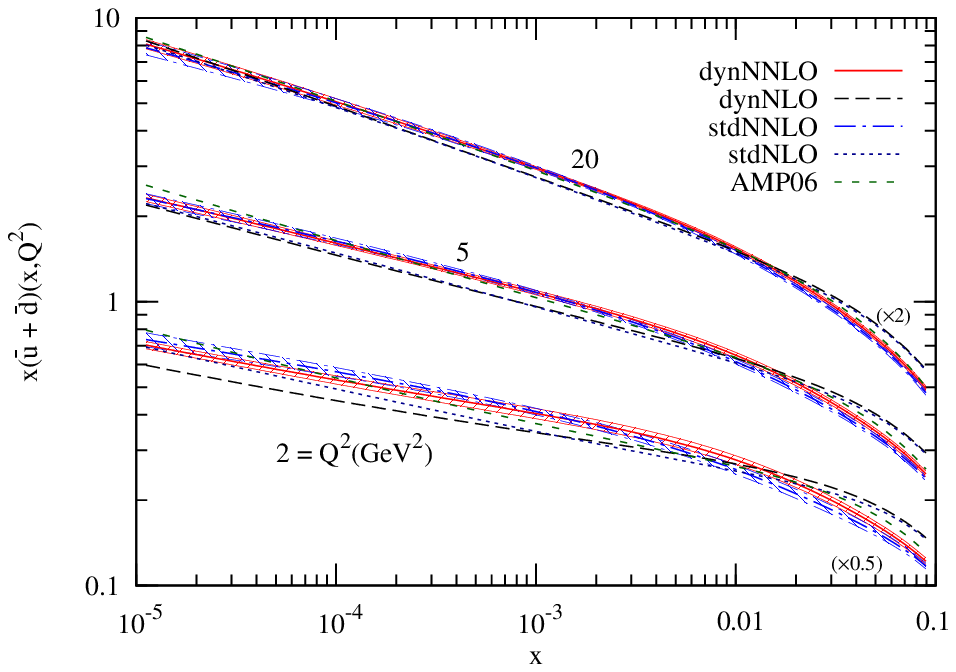,width=8cm}}
\mbox{\epsfig{file=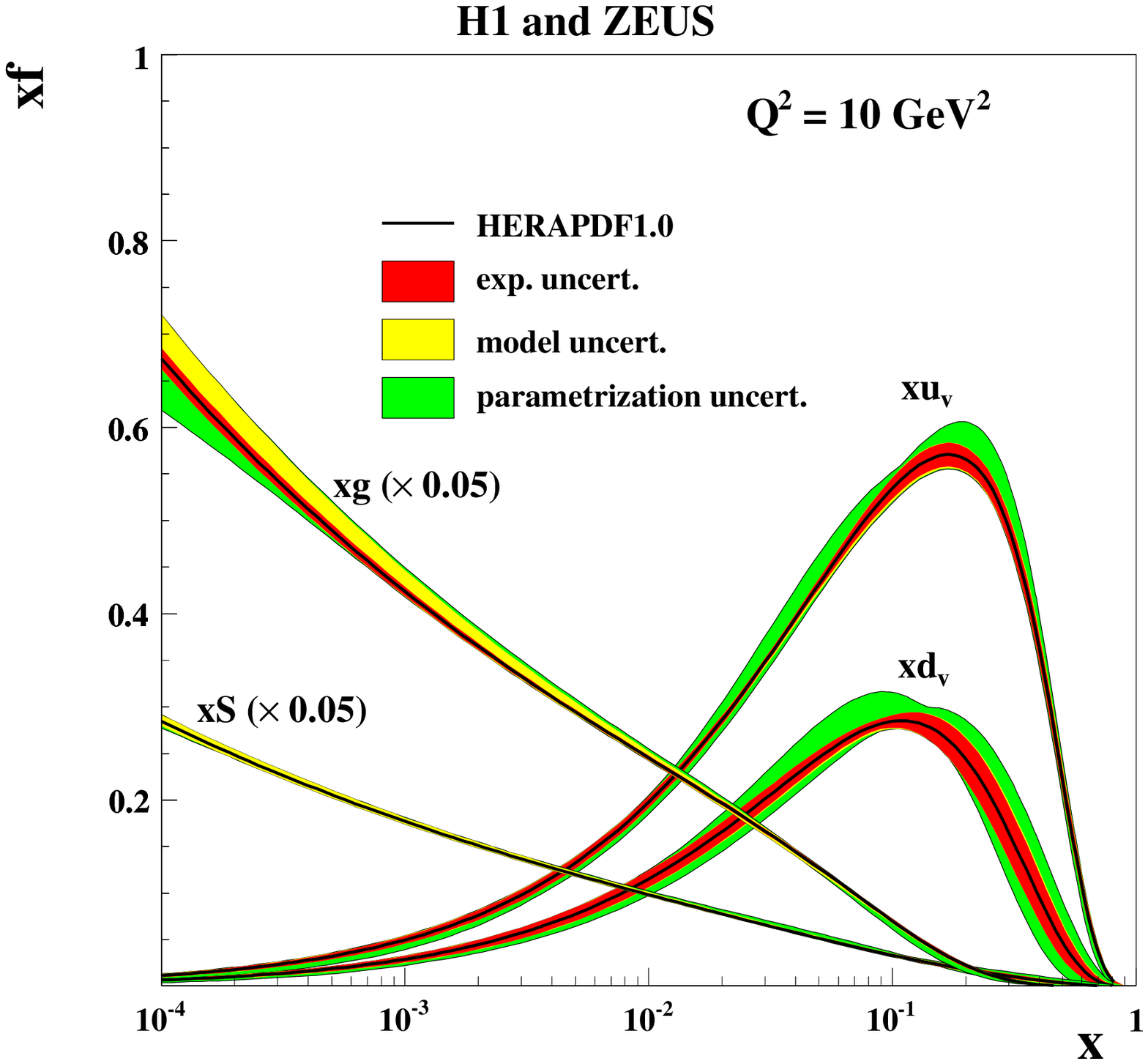,width=7cm}}
\end{center} 
\caption[]{ 
\label{FIG:JR_herapdf} 
\small \sf 
Left~: Comparing the $1\sigma$ error bands of the dynamical (dyn) and standard (std)
NNLO gluon JR distributions \cite{JimenezDelgado:2008hf} at small $x$ for various fixed values of $Q^2$. Note 
that $Q^2 = 2 \GeV^2$ is the input scale of the standard fit. The central NLO results 
are taken from \cite{Gluck:2007ck} with uncertainties comparable to the ones shown for 
NNLO for $Q^2$ above $2 \GeV^2$. For comparison the `standard' NNLO results of AMP06 \cite{Alekhin:2006zm}  
are shown as well. The results at $Q^2 = 2$ and $20 \GeV^2$ have been multiplied by 0.5 and 2, respectively, 
as indicated in the figure; from \cite{JimenezDelgado:2008hf}, \TCop (2009) by the American Physical 
Society.
Right~: The parton distribution functions from HERAPDF1.0, $xu_v, xd_v, xS = 2x(\bar{U} + \bar{D}), xg$,
at $Q^2 = 10 \GeV^2$. The gluon and sea distributions are scaled down by a factor 20. The experimental, 
model and parameterization uncertainties are shown separately; from \cite{herapdf:2009wt}, \TCop (2009) by 
Elsevier Science.} 
\end{figure} 
%----------------------------------------------------------------------------------------------
Furthermore, the 
heavier flavors, charm and bottom, contribute above the corresponding thresholds. At large values of $x$
the up- and down-quark distributions show a different behaviour due to the valence quark contributions
and the various sea quark contributions are of different size.
In Figure~\ref{FIG:cteq_nnpdf} recent results of the CTEQ and NNPDF analyses are shown. The size of the 
ratios of the PDFs by CTEQ obtained in NNLO and NLO are compared at a low scale for the different partons 
showing a particular sensitivity in the region of small and large values of $x$. The gluon density is an
important distribution for many processes at hadron colliders. A comparison is shown for the 
shapes obtained for $G(x,Q_0^2)$ in the NNPDF analysis from LO to NNLO. Here the gluon distribution is
becoming lower at higher orders in the small $x$ region, with still a wide error band.
In Figure~\ref{FIG:JR_herapdf} the evolution of the light sea-quark distributions of the JR NNLO analysis 
are shown in the small-$x$ region. Both the dynamical approach and the standard fit do agree rather well.
The results of the HERAPDF 1.0 NLO analysis, which are based on data taken at HERA only, are shown also
quantifying the parameterization uncertainty. It turns out, that the latter is of the size of the present
experimental uncertainty. 
Figure~\ref{FIG:abmPDF} shows the $1\sigma$ error bands obtained in the ABM11 analysis and compares the 
different massless parton distribution to the results found by the latest JR, MSTW, NNPDF analyses
at the scale $\mu  = 2 \GeV$. While the valence distributions and the sum of the up and down sea-quarks do 
agree rather well in all the four fits, there are still important differences in the gluon distribution
at lower values of $x$. In particular the gluon distribution of MSTW is taking low values in the small $x$
region, and eventually becomes negative unlike in all the other analyses. There are also differences in the 
$d_s - u_s = \bar{d} - \bar{u}$
distribution and for the strange quark density at medium values of $x$.

In Figure~\ref{FIG:latt:val}
different lattice determinations of the moment $\int_0^1 dx [x(u_v - d_v) - x(\bar{u}-\bar{d})]$ are compared
with the corresponding value obtained from PDF fits, see also \cite{Bali:2012av}.
Very similar values are obtained by the 
different fitting groups, cf.~\cite{Alekhin:2012ig}. There is still a difference between the lattice 
results and the  value from the PDF analyses, despite lower pion masses are used in the present lattice 
simulations, which has to be understood further. A determination of the gluon momentum with lattice 
methods has recently been performed in \cite{Horsley:2012pz}.
For a review on the status of the calculations
of PDF-moments with lattice methods see \cite{Renner:2010ks}. 

\noindent
%---------------------------------------------------------------------------------------------- 
\restylefloat{figure} 
\begin{figure}[H] 
\begin{center} 
\mbox{\epsfig{file=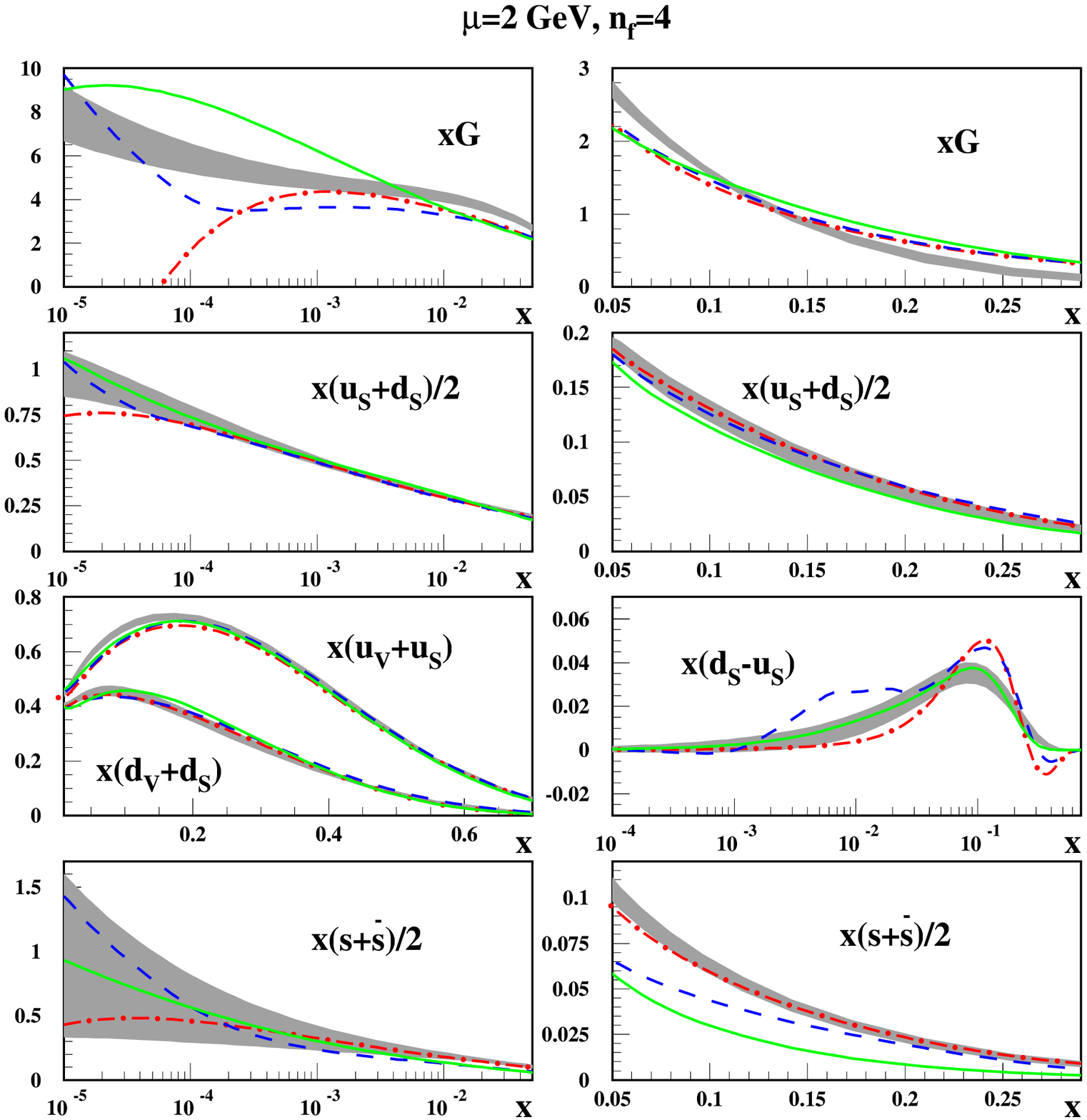,width=11cm}} 
\end{center} 
\caption[]{ 
\label{FIG:abmPDF} 
\small \sf
The 1$\sigma$ error band for the 4-flavor NNLO ABM11 PDFs at the scale of $\mu = 2 \GeV$ 
versus $x$ obtained in \cite{Alekhin:2012ig} (shaded area) compared with the ones obtained by other groups. 
Solid lines: JR09 \cite{JimenezDelgado:2008hf}, dashed dots: MSTW \cite{Martin:2009iq}, dashes: NN21 
\cite{Ball:2011mu}; from \cite{Alekhin:2012ig}.} 
\end{figure} 
%---------------------------------------------------------------------------------------------- 

\vspace*{-5mm}
%---------------------------------------------------------------------------------------------- 
\restylefloat{figure} 
\begin{figure}[H] 
\begin{center} 
\mbox{\epsfig{file=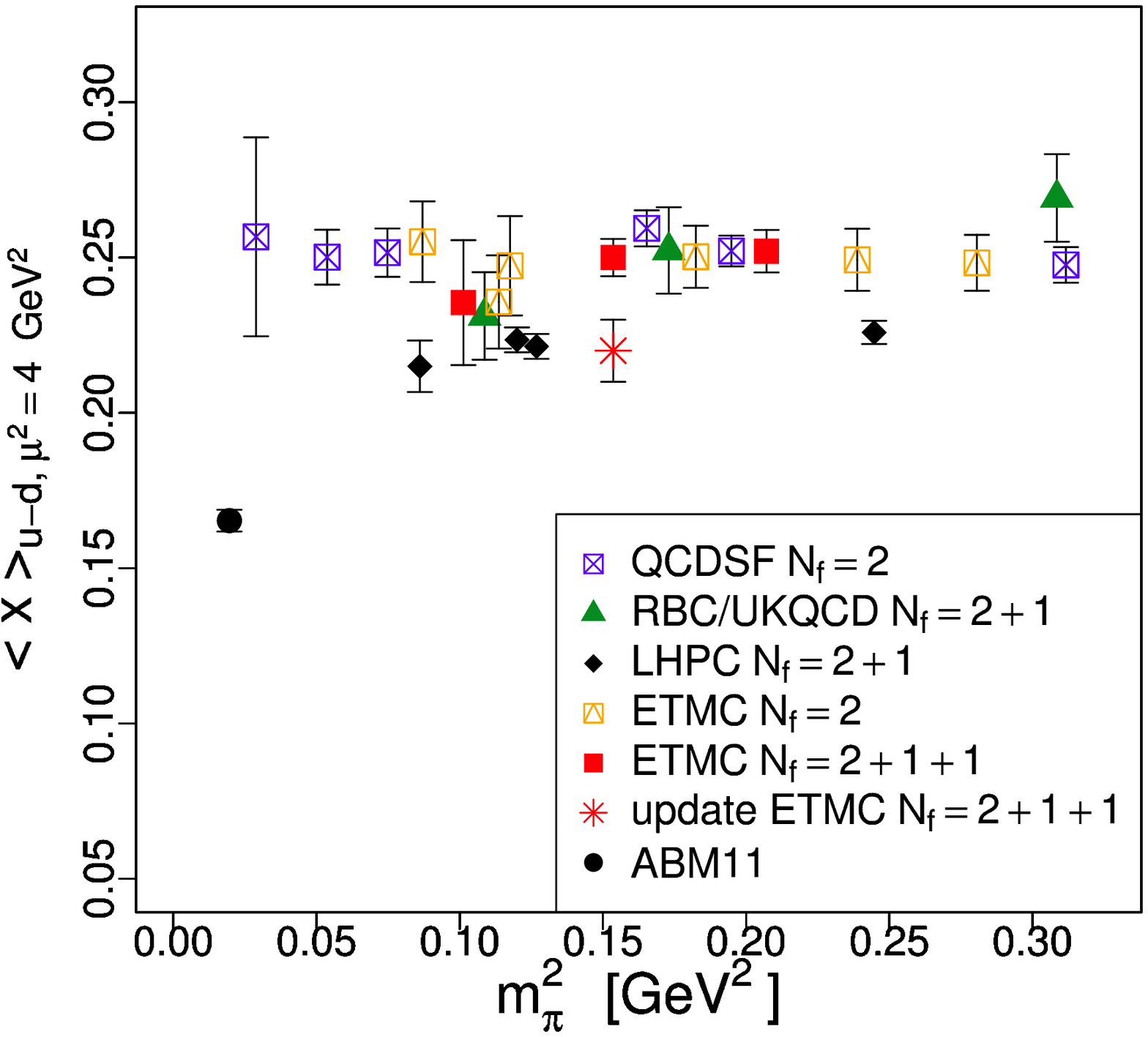,width=6cm}} 
\end{center} 
\caption[]{\label{FIG:latt:val} 
\small \sf 
Comparison of lattice computations for the second moment of the non-singlet distribution as
a function of the pion mass $m_\pi$ with the result of ABM11 \cite{Alekhin:2012ig} 
along with the uncertainties of the respective measurement; from \cite{Alekhin:2012ig}, 
by courtesy of V.~Drach.}
\end{figure} 
%---------------------------------------------------------------------------------------------- 

\noindent

%----------------------------------------------------------------------------------------------

\vspace*{-1cm}
\restylefloat{figure}
\begin{figure}[H]
\begin{center}
\mbox{\epsfig{file=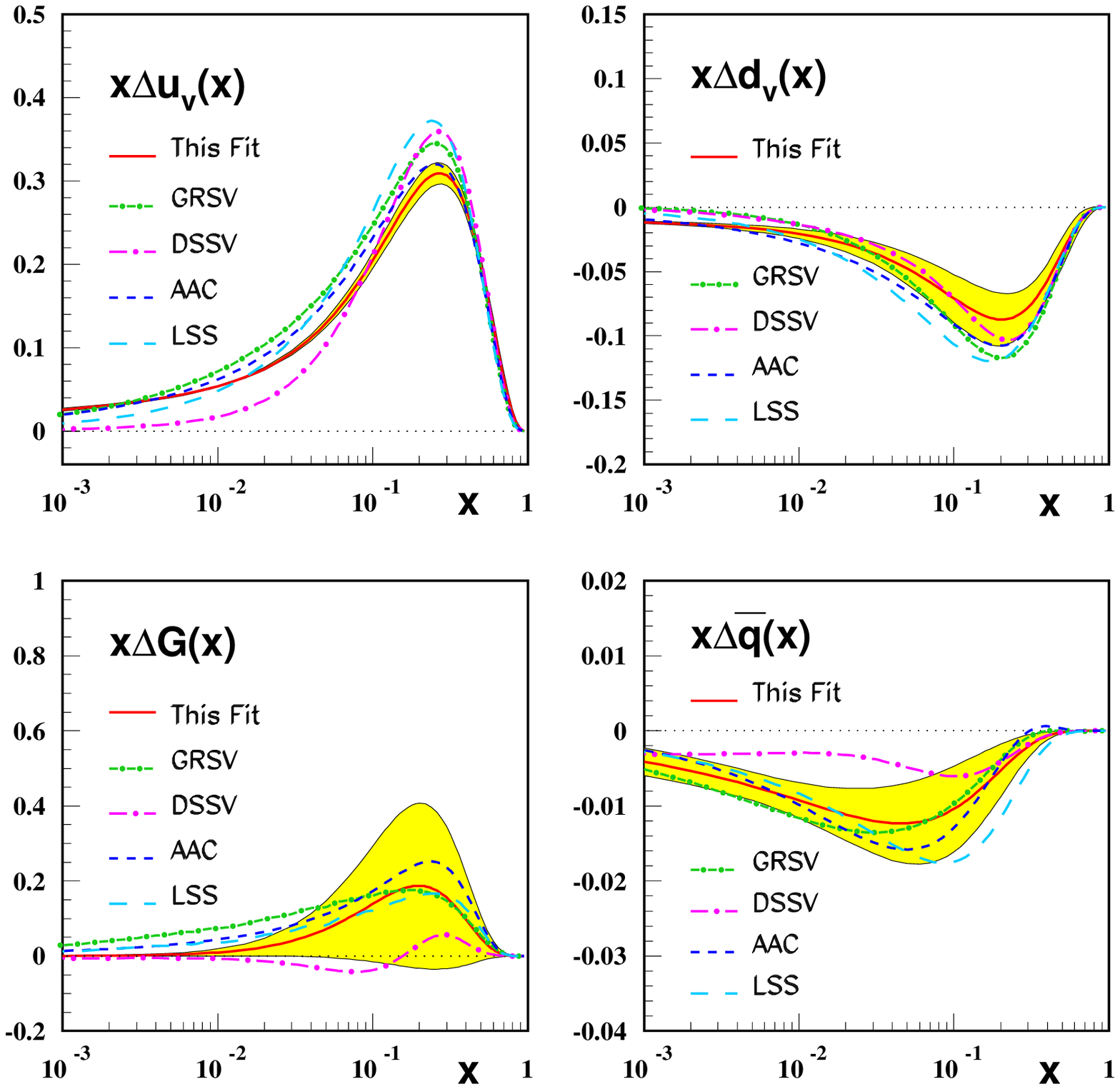,height=8cm,width=7cm}}
\mbox{\epsfig{file=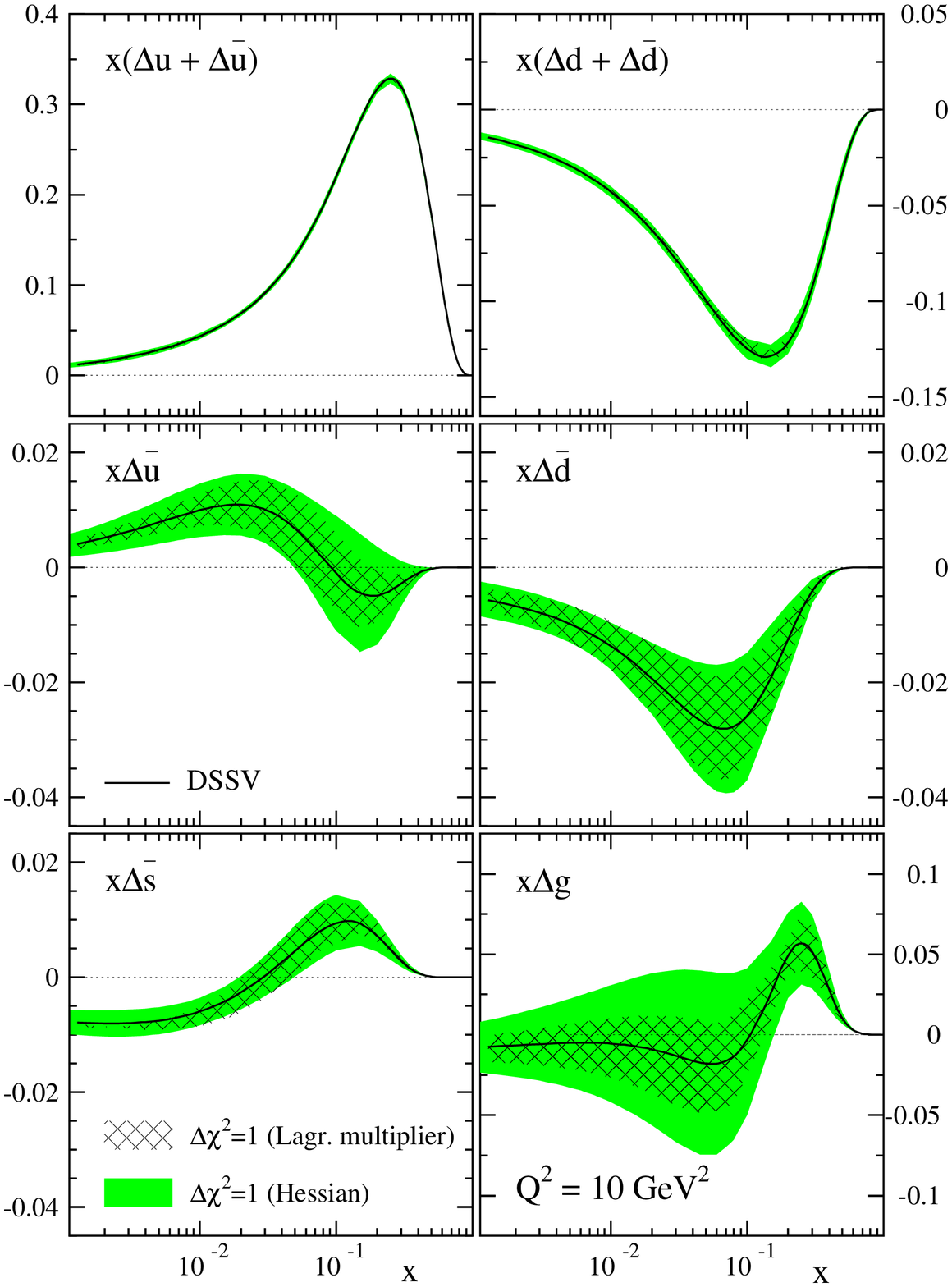,height=8.5cm,width=7cm}}
\end{center}          
\caption[]{\label{FIG:pol1} \small \sf
Left~: The NLO polarized parton BB distributions \cite{Blumlein:2010rn} at the input 
scale $Q^2_0 = 4.0 \GeV^2$ (solid line) compared to results obtained by GRSV 
\cite{Gluck:2000dy}
(dashed-dotted line), 
DSSV (long dashed-dotted line) \cite{deFlorian:2009vb}, 
AAC (dashed line) \cite{Hirai:2008aj}, and LSS (long dashed line) \cite{Leader:2001kh}. 
The shaded areas represent the fully correlated $1\sigma$ error bands calculated by Gaussian error
propagation; from \cite{Blumlein:2010rn}, \TCop (2010) by Elsevier Science.~~Right~:
The polarized DSSV PDFs \cite{deFlorian:2009vb} of the proton at $Q^2 = 10 \GeV^2$ in the
$\overline{\rm MS}$ scheme, along with their $\Delta \chi^2 = 1$ uncertainty bands computed
with Lagrange multipliers and the improved Hessian approach; from \cite{deFlorian:2009vb},
{\TCop (2009)} by the American Physical Society.}
\end{figure}
%%%%%%%%%%%%%%%%%%%%%%%%%%%%%%%%%%%%%%%%%%%%%%%%%%%%%%%%%%%%%%%%%%%%%%%
\subsection{Polarized Nucleons}
%%%%%%%%%%%%%%%%%%%%%%%%%%%%%%%%%%%%%%%%%%%%%%%%%%%%%%%%%%%%%%%%%%%%%%%

\noindent
Early determinations of the polarized parton densities were carried out at the beginning of the
1990ies at LO under a series of model assumptions, see e.g.~\cite{Bluemlein:2002be,Ramsey:1997iu}.
With the advent of the NLO polarized anomalous dimension NLO analyses were carried 
out~\cite{Gluck:1995yr,ABFR, AAC,Leader:2001kh,Bluemlein:2002be}; cf. also 
\cite{Lampe:1998eu,Kuhn:2008sy}. 
In these analyses $SU(3)_F$ symmetry was assumed for the sea quarks. In Figure~\ref{FIG:pol1} the results
of different polarized PDF-analyses are compared (left panel). The errors on these distributions
are larger than in the unpolarized case, with a basic agreement in the results for the valence distributions.
For the sea-quark and gluon distribution the errors are larger and there is a stronger variation of the current 
predictions, widely within the present errors. We note also other approaches based on statistical distributions 
\cite{STAT:SOFFER}, which allow for a very efficient modeling.

To resolve the flavor dependence of the sea-quarks data on semi-inclusive measurements, 
as e.g. \cite{Airapetian:2004zf}, are used. Furthermore, data on photo- and electro-production of hadrons and charm
and proton-proton collisions at RHIC can be used, along with the structure function data. A recent analysis based on 
these data was carried out in \cite{deFlorian:2009vb}. The semi-inclusive data were also used in the analysis 
\cite{Leader:2010rb}. The polarized PDFs of \cite{deFlorian:2009vb} are shown in Figure~\ref{FIG:pol1} (right panel) 
and allow, yet with larger errors, to derive different shapes for the polarized sea quarks. The gluon distribution is 
found to be much lower than in the case of the analysis of only the deep-inelastic structure function data.

%%%%%%%%%%%%%%%%%%%%%%%%%%%%%%%%%%%%%%%%%%%%%%%%%%%%%%%%%%%%%%%%%%%%%%%
\subsection{\boldmath{$\alpha_s(M_Z^2)$}}
%%%%%%%%%%%%%%%%%%%%%%%%%%%%%%%%%%%%%%%%%%%%%%%%%%%%%%%%%%%%%%%%%%%%%%%

\noindent
The strong coupling constant can be determined as one parameter in the QCD-analysis of the
deep inelastic world data along with the parameters of the non-perturbative parton distributions
at a given scale $Q^2_0$. In the following we will compare the results on $\alpha_s(M_Z^2)$ in different NNLO
analyses. The results are summarized in Table~\ref{tab:alphas}, cf. \cite{Alekhin:2012ig}.
%-------------------------------------------------------------------------------------------------
\restylefloat{figure}
\begin{table}[H]
\begin{center}
\renewcommand{\arraystretch}{1.2}
\begin{tabular}{|l|l|l|}
\hline
\multicolumn{1}{|c|}{ } &
\multicolumn{1}{c|}{$\alpha_s({M_Z^2})$} &
\multicolumn{1}{c|}{  } \\
\hline
BBG      & $0.1134~^{+~0.0019}_{-~0.0021}$
         & {\rm valence~analysis, NNLO}  \cite{Blumlein:2006be}           
\renewcommand{\arraystretch}{1}
\\
GRS      & $0.112 $ & {\rm valence~analysis, NNLO}  \cite{Gluck:2006yz}           
\\
ABKM           & $0.1135 \pm 0.0014$ & {\rm HQ:~FFNS~$n_f=3$} \cite{Alekhin:2009ni}             
\\
ABKM           & $0.1129 \pm 0.0014$ & {\rm HQ:~BSMN-approach} 
\cite{Alekhin:2009ni}             
\\
JR       & $0.1124 \pm 0.0020$ & {\rm
dynamical~approach} \cite{JimenezDelgado:2008hf}   
\\
JR       & $0.1158 \pm 0.0035$ & {\rm
standard~fit}  \cite{JimenezDelgado:2008hf}    
\\
ABM11            & $0.1134\pm 0.0011$ & \cite{Alekhin:2012ig} \\
MSTW & $0.1171\pm 0.0014$ &  \cite{Martin:2009bu}     \\
NN21 & $0.1173\pm 0.0007$ &  \cite{Ball:2011us}     \\
CT10 & $0.118\phantom{0} \pm 0.005$  &  \cite{CT10} \\
\hline
BBG & $0.1141~^{+~0.0020}_{-~0.0022}$
& {\rm valence~analysis, N$^3$LO$(^*)$}  \cite{Blumlein:2006be}            \\[0.5ex]
\hline
{world average} & {$0.1183 \pm 0.0010$  } & \cite{Bethke:2011tr} (2011)
\\
\hline
\end{tabular}
\end{center}
\renewcommand{\arraystretch}{1}   
\caption[]{\sf \small
\label{tab:alphas}
Summary of recent NNLO QCD analyses of the DIS world data.}
\end{table}
%--------------------------------------------------------------------------------------------------------

\noindent
There are two valence analyses \cite{Blumlein:2006be,Gluck:2006yz}, effectively limited to the region of $x \gsim 
0.3$. In \cite{Blumlein:2006be} an effective N$^3$LO fit has been performed, including the 3-loop Wilson 
coefficient, noting that the effect of the 4-loop non-singlet anomalous dimension is very small. The other analyses 
cover the whole kinematic range. In the valence analyses and 
\cite{JimenezDelgado:2008hf,Alekhin:2009ni,Alekhin:2012ig} very similar values of $\alpha_s(M_Z^2)$ are obtained, 
which 
agree within their errors. The way in which the heavy flavor corrections are treated implies a systematic error of 
$0.0006$. The N$^3$LO value is well compatible with the results obtained at NNLO. 
These values are lower than the current world average \cite{Bethke:2011tr}. 
Larger values are found in the MSTW \cite{Martin:2009bu} and NNPDF analyses \cite{Ball:2011us}.
A preliminary value has been reported by CTEQ \cite{CT10} at NNLO, yet with a rather large error.
In Ref.~\cite{Alekhin:2012ig} a detailed analysis has been performed comparing the above results also with
respect to the pulls given by the different data sets used. Despite the final $\alpha_s(M_Z^2)$ values of MSTW and
NNPDF are quite similar, there are still significant differences in the pulls. The Tevatron jet data 
do not cause the larger values. It rather seems that the response of the MSTW and NNPDF fits to the SLAC
and also the HERA data is causing this difference, which needs to be investigated further.

In different other reactions larger values of $\alpha_s(M_Z^2)$ were found. This applies to the analysis of the
3-jet rate with $\alpha_s(M_Z^2) = 0.1175 \pm 0.0025$ \cite{Dissertori:2009qa} and
inclusive $Z$-decay yielding $\alpha_s(M_Z^2) = 0.1189 \pm 0.0026$ \cite{Baikov:2008jh,Baikov:2012er}.
This also applies to the $\alpha_s(M_Z^2)$ values measured for hadronic $\tau$-decays, cf.~\cite{Bethke:2011tr}.
On the other hand, the analysis of thrust in $e^+e^-$ annihilation led to $\alpha_s(M_Z^2) = 0.1153 \pm 0.0017 \pm 
0.0023$ \cite{Gehrmann:2009eh}, resp., $\alpha_s(M_Z^2) = 0.1135 \pm 0.0011 \pm 0.0006$ \cite{Abbate:2010xh}.
In a NLO analysis of $e^+e^- \rightarrow 5~\text{jets}$ $\alpha_s(M_Z^2) = 0.1156 {\tiny \begin{array}{c} +0.0041 \\ 
-0.0034 \end{array}}$ was obtained. There are also first results on the measurements of $\alpha_s(M_Z^2)$ with
including dynamical fermions from the lattice \cite{LATalph}.
Very recently a NLO analysis of the ATLAS jet data resulted in $\alpha_s(M_Z^2) 
\sim 0.1156$ with a larger error \cite{Malaescu:2012ts}. Similar results are obtained in a multi-jet analysis 
of $e^+e^-$ data \cite{Frederix:2010ne}.
 Usually the values of $\alpha_s(M_Z^2)$ obtained at NNLO are
lower than those at NLO. In various classes of high energy reactions partly larger and lower values of 
$\alpha_s(M_Z^2)$ are obtained at present. This is due to systematics sources, which have to be understood better in 
the 
future. Further clarification can be obtained from better hard scattering data. Here the LHC jet data will be 
of importance in the near future.

%%%%%%%%%%%%%%%%%%%%%%%%%%%%%%%%%%%%%%%%%%%%%%%%%%%%%%%%%%%%%%%%%%%%%%%%
\section{Small \boldmath{$x$} Resummations}
\label{sec:smx} \renewcommand{\theequation}{\thesection.\arabic{equation}} 
\setcounter{equation}{0}
%%%%%%%%%%%%%%%%%%%%%%%%%%%%%%%%%%%%%%%%%%%%%%%%%%%%%%%%%%%%%%%%%%%%%%%%

\noindent
Approaching very small values of $x$ at a fixed virtuality $Q^2$ the criteria for the parton model, 
Section~\ref{sec:parton}, are no longer valid and new phenomena are expected to contribute. The small 
$x$ region was probed at HERA to values of $x \sim 10^{-4}$ at $Q^2 = 10 \GeV^2$ and at LHC values 
of $x \sim 10^{-6}$ at $Q^2 = 10^2 \GeV^2$ can be reached. The potentially new effects will 
influence the size and 
scaling violations of the structure functions both in the unpolarized and polarized case. To which extent 
these phenomena can be dealt with using perturbative methods and resummations 
based on them, is not finally clear at present. Two main extensions of the perturbative approach based 
on fixed order perturbative QCD were proposed:
{\it i)}~the linear BFKL small-$x$ resummation \cite{BFKL} of large logarithms 
$(\alpha_s (N_c/\pi))^k (1/x)\ln^k(1/x)/k!$, and {\it ii)}~saturation models with 
highly non-linear gluo-dynamics, cf.~\cite{GLR} related to Glauber-models \cite{GLAUBER}.\footnote{For 
early estimates of the transition line between the perturbative and non-perturbative domain and the 
onset the possible of saturation effects see \cite{Kwiecinski:1985gr,TRANS:GLR}.} 
In both approaches elements known from the parton picture
are used. Since perturbative kernels are evaluated it has to be clarified whether factorization holds. In 
case of the saturation corrections this also applies to multi-parton states. Due to the 
connection between the 
non-perturbative distributions and the perturbative kernels, usually in terms of a convolution, both 
effects at small and larger values of $x$ will contribute. In the following we will not cover the saturation
models and related theoretical developments in nucleon-nucleus and nucleus-nucleus scattering, being 
an interesting broad topic in its own right. We refer to recent detailed surveys, as e.g. given 
in~\cite{Boer:2011fh,McLerran:2011zza}, and will discuss the effects due to BFKL-type resummations.

The resummation of the leading small $x$ terms \cite{BFKL} are often confronted to those in $\ln(Q^2)$, 
stating that the first ones results from the strong ordering $x_1 \gg ... x_i \gg
x_{i+1} ...$ and the second stem from strong ordering in the transverse momentum $k_{\perp, 1}
\ll ... k_{\perp,i} \ll k_{\perp,i+1} ...$ along a ladder.\footnote{One may study this process 
under a more general point of view and consider angularly ordered emissions covering both the above 
cases~\cite{MAR}, which allows for interesting applications through Monte Carlo studies. Whereas this 
unified treatment is possible at LO, higher order corrections cannot be cast into this form in general.}
Note that this is a gauge-dependent statement and requires to use effective vertices
in case of the first approach. The second case refers to fixed-order perturbation theory using the 
renormalization group, as outlined in the previous sections. In the first approach one would calculate in 
a more systematic way the scale-invariant contributions to the evolution equation (\ref{eq:ren3}) in the 
massless case setting $\beta = 0$, i.e. considering the strong coupling as {\sf const.} One obtains
%-----------------------------------------------------------------------
\begin{equation}
E_k^n(\mu^2) = E_k^n(\mu_0^2) \left(\frac{\mu^2}{\mu_0^2}\right)^
{\frac{1}{2}\left(\gamma_{O_k} - n \gamma_{\Phi}\right)}~.
\label{eqCONF}
\end{equation}
%-----------------------------------------------------------------------
The scale invariant part of the anomalous dimension has the representation
%-----------------------------------------------------------------------
\begin{equation}
\gamma_{O_k} - n \gamma_{\Phi} = \sum_{l=1}^{\infty} 
\gamma_O^{(l)} a_s^l
\end{equation}
%-----------------------------------------------------------------------
and exponentiates to all orders. The representation (\ref{eqCONF}) 
applies also for higher order resummations under the above requirements.
To treat running coupling effects one has to systematically account for 
scale-invariance breaking effects. Furthermore, to form observables, the 
small-$x$ resummed Wilson coefficients have to be calculated as well.

Both in the flavor non-singlet and singlet cases for scattering off unpolarized 
and polarized targets the leading order small-$x$ resummations were derived in
form of evolution kernels. They seem to correspond to a resummation of the leading 
order anomalous dimensions, as the comparison in the known orders show, because
the corresponding singularity in the Wilson coefficients turns out to be one order
lower \cite{Blumlein:1995jp,Moch:2004pa,Vogt:2004mw}. 
%%%%%%%%%%%%%%%%%%%%%%%%%%%%%%%%%%%%%%%%%%%%%%%%%%%%%%%%%%%%%%%%%%%%%%%%%%%%%%%%%%%%%%%%%%%%%%%%%%%%%%%%%%%
\subsection{The Non-singlet and Polarized Singlet Contribution}
%%%%%%%%%%%%%%%%%%%%%%%%%%%%%%%%%%%%%%%%%%%%%%%%%%%%%%%%%%%%%%%%%%%%%%%%%%%%%%%%%%%%%%%%%%%%%%%%%%%%%%%%%%%

\noindent
The most singular contributions to the Mellin transforms of the 
structure--function evolution kernels at all orders in 
$a$ stem from the poles at $N = 0$ in each individual order.
In the flavor non-singlet case the kernels $K^\pm$
can be obtained from the positive and negative signature amplitudes 
studied in \cite{KL} for QCD via
%-----------------------------------------------------------------------
\begin{equation}
{\cal M} \left [ K_{x \rightarrow 0}^{\pm}(a) \right ](N) 
 \equiv \int_0^1 \! dx \, x^{N-1} K_{x \rightarrow 0}^{\pm}(x, a)
 \equiv - \frac{1}{2} \GA_{x \rightarrow 0}^{\pm}(N, a),
\end{equation}
%-----------------------------------------------------------------------
with
%-----------------------------------------------------------------------
\begin{eqnarray}
\label{eqGAA1}
 \GA_{x \rightarrow 0}^{+}(N, a) &=& - N
  \left \{ 1 - \sqrt{1 - \frac{8 a C_F}{N^2}} \right \} 
%\\
\end{eqnarray}
\begin{eqnarray}
\label{eqGAA2}
 \GA_{x \rightarrow 0}^{-}(N, a) &=& - N  \left \{ 1 - \sqrt{1 - \frac{8 a C_F}{N^2}
  \left [1 - \frac{8 a N_c}{N} \frac{d}{d N}
  \ln \left ( e^{z^2/4} D_{-1/[2N_c^2]}(z) \right ) \right ] } 
  \right \}, 
\end{eqnarray}
%-----------------------------------------------------------------------
where $z = N/\sqrt{2 N_c a}$, $D_p(z)$ denotes the function of the 
parabolic cylinder, and $N_c = 3$ in case of QCD.

The LO small $x$ evolution kernels in the case of the polarized singlet
evolution were derived in \cite{BER2}. The resummed splitting function is given by
%-----------------------------------------------------------------------
\begin{equation}
\PV(x,a_s) \equiv \sum_{l=0}^{\infty} \PV^{(l)}_{x\rightarrow 0} 
a_s^{l+1} \log^{2l} x = \frac{1}{8\pi^2} {\cal M}^{-1}[\FV_0(N,a_s)](x).
\label{eqSIPO}
\end{equation}
%-----------------------------------------------------------------------
The matrix valued function $\FV_0(N,a_s)$ is obtained as the
solution of 
%-----------------------------------------------------------------------
\begin{eqnarray}
\FV_0(N,a_s) &=& 16 \pi^2 \frac{a_s}{N} \left(\begin{array}{cc} C_F & -2T_R N_f \\ 2C_F & 
4C_A \end{array} \right)
%\MV_0 
- \frac{8a_s}{N^2} 
\FV_8(N,a_s) 
\left(\begin{array}{cc} C_F & 0 \\ 0 & C_A
\end{array} \right)
%\GV_0
+ \frac{1}{8\pi^2}\frac{1}{N} \FV_0^2(N,a_s) \\
\text{\rm with}~~~~~~~~~~&& \nonumber\\
\FV_8(N,a_s) &=& 16 \pi^2 \frac{a_s}{N} 
\MV_8 = \left(\begin{array}{cc} C_F - C_A/2 & -T_R N_f \\ C_A & 2 C_A
\end{array} \right)
%\MV_8 
+ \frac{2a_s}{N} C_A 
\frac{d}{dN} \FV_8(N,a_s) + \frac{1}{8\pi^2} \frac{1}{N} \FV_8^2(N,a_s).
\nonumber\\
\end{eqnarray}
%%%%%%%%%%%%%%%%%%%%%%%%%%%%%%%%%%%%%%%%%%%%%%%%%%%%%%%%%%%%%%%%%%%%%%%%%
%%%%%%%%%%%%%%%%%%%%%%%%%%%%%%%%%%%%%%%%%%%%%%%%%%%%%%%%%%%%%%%%%%%%%%%%%
%-----------------------------------------------------------------------
Eq.~(\ref{eqSIPO}) obeys~\cite{BV3}
%-----------------------------------------------------------------------
\begin{equation}
P_{qg}^{(l)}/(T_R N_f) = - P_{gq}^{(l)}/C_F
\end{equation}
%-----------------------------------------------------------------------
to all orders, where $T_R=1/2$ and $N_f$ denotes the number of flavors.
The leading contributions of the fixed order results
in LO and NLO ($\overline{\rm MS}$) are correctly described. In the
supersymmetric limit $C_A=C_F=N_f=1, T_R=1/2$ the relations
%-----------------------------------------------------------------------
\begin{equation}
P_{qq}^{(l)}+P_{gq}^{(l)} = P_{qg}^{(l)}+P_{gg}^{(l)}
\end{equation}
%-----------------------------------------------------------------------
are obeyed for all $l$ and Eq.~(\ref{eqSIPO}) can be given in a simple
analytic form~\cite{BV3}.
It is evident from Eqs.~(\ref{eqGAA1}--\ref{eqSIPO}) that the poles at $N = 0$ being present 
in the
individual orders are resummed into branch cuts, which usually exhibit a milder
singularity.

\begin{figure}[H]
\begin{center}
\includegraphics[width=.6\textwidth]{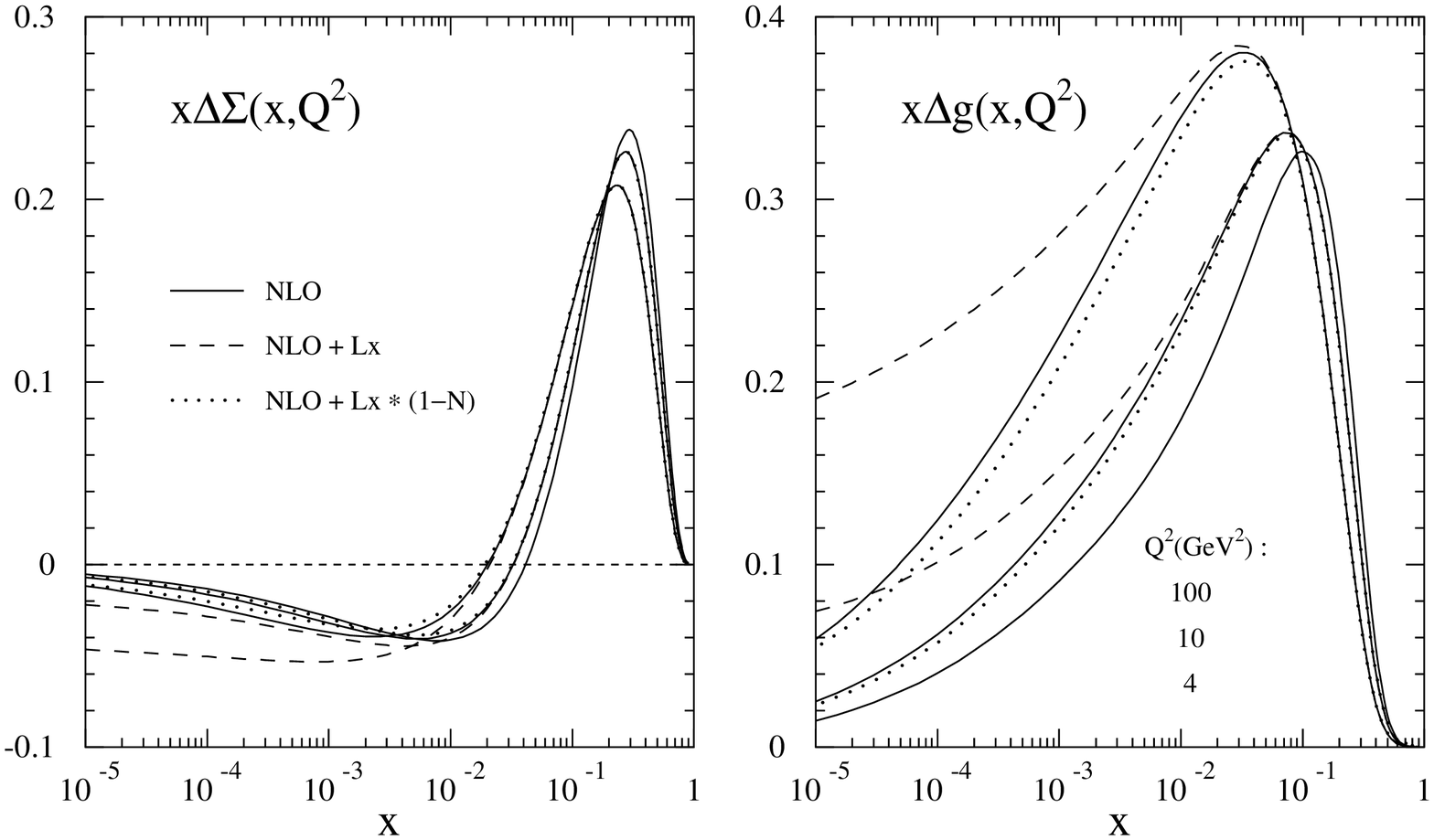}
\end{center}
\caption[]{\sf \small
The $Q^2$ evolution of the polarized quark singlet and gluon
momentum distributions evolving from $Q_0^2 = 4 \GeV^2$; from \cite{BV3}, 
\TCop (1996) by Elsevier Science.}
\label{BL-figG1}
\end{figure}

To perform numerical studies it is interesting to consider the effect of 
less singular terms to estimate the stability of the leading order resummed
terms. In case of the fixed-order anomalous dimensions they are known
and one may try Ans\"atze like 
%------------------------------------------------------------------------
\begin{alignat}{3}
\label{eq:subl}
(a)~~~\Gamma(N,a_s) &\rightarrow& \Gamma(N,a_s) - \Gamma(1,a_s) &~~~~~ 
(b)~~~\Gamma(N,a_s) &\rightarrow& \Gamma(N,a_s)(1-N)            
\nonumber\\
(c)~~~\Gamma(N,a_s) &\rightarrow& \Gamma(N,a_s)(1-N)^2          &~~~~~
(d)~~~\Gamma(N,a_s) &\rightarrow& \Gamma(N,a_s)(1-2N+N^3)~,
\end{alignat}
%------------------------------------------------------------------------
cf.~\cite{Blumlein:1997em}. It turns out that the corrections to the  
flavor non-singlet contributions are numerically very small \cite{Blumlein:1995jp}.
Larger corrections are obtained in the polarized singlet case, see Figure~\ref{BL-figG1}.
However, here sub-leading terms e.g. of the type (\ref{eq:subl} $(b)$) do nearly completely cancel 
these contributions again. Comparable results were obtained in \cite{KOD}. Large effects 
as anticipated in \cite{EMR} are not confirmed.
%------------------------------------------------------------------------
\restylefloat{figure} 
%------------------------------------------------------------------------
%---------------------------------------------------------------------------------------------- 
\restylefloat{figure} 
\begin{figure}[H] 
\begin{center} 
\mbox{\epsfig{file=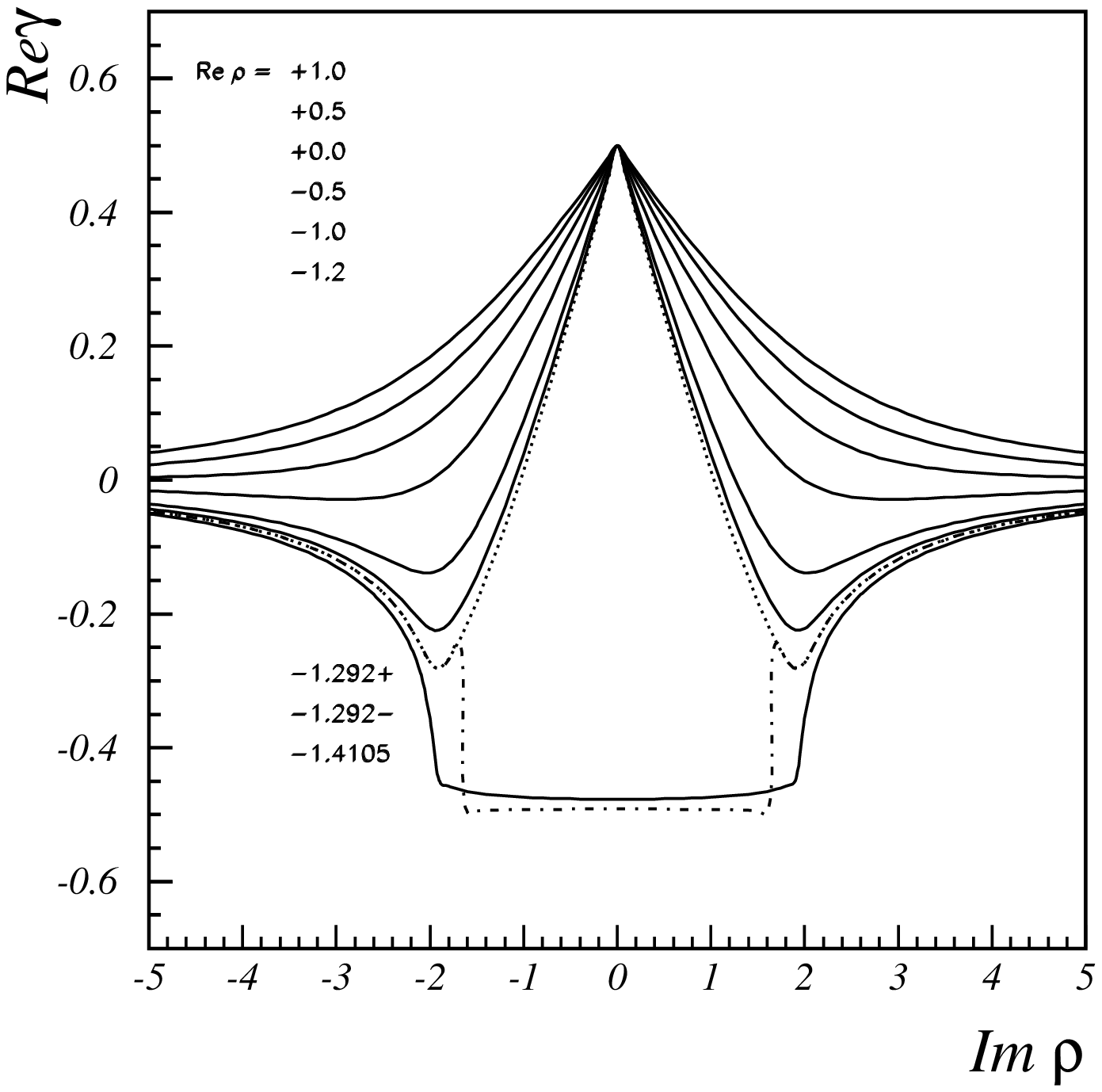,width=6cm}} 
\mbox{\epsfig{file=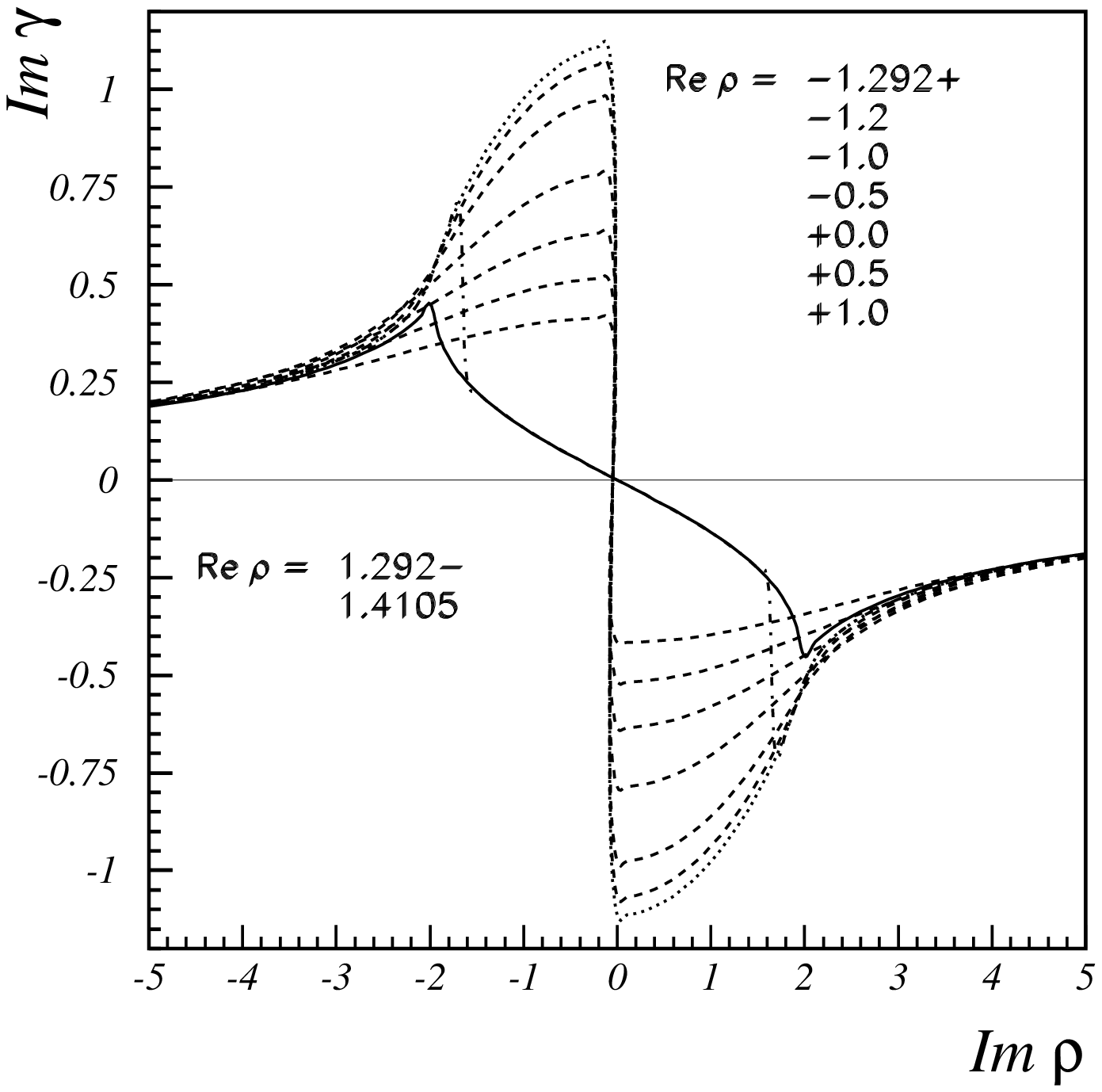,width=6cm}} 
\end{center} 
\caption[]{ 
\label{FIG:lipLO} 
\small \sf Real and imaginary part of the LO BFKL anomalous dimension; from \cite{Blumlein:1997em},
\TCop (1998) by the American Physical Society.} 
\end{figure} 
%----------------------------------------------------------------------------------------------

%%%%%%%%%%%%%%%%%%%%%%%%%%%%%%%%%%%%%%%%%%%%%%%%%%%%%%%%%%%%%%%%%%%%%%%%
\subsection{Unpolarized Singlet Distributions}
%%%%%%%%%%%%%%%%%%%%%%%%%%%%%%%%%%%%%%%%%%%%%%%%%%%%%%%%%%%%%%%%%%%%%%%%
%

\noindent
The LO resummation for the evolution kernel of the
 unpolarized singlet
distributions was derived in \cite{BFKL}. It was shown in \cite{JAR} that 
the eigenvalue 
%-----------------------------------------------------------------------
\begin{equation}
(N-1)       = \frac{\alpha_s N_c}{\pi} \chi_0(\gamma_L)
      \equiv
\frac{\alpha_s N_c}{\pi} \left[2 \psi(1) - \psi(\gamma_L) 
- \psi(1-\gamma_L)\right]
\label{eq1BF}
\end{equation}
%-----------------------------------------------------------------------
represents the LO resummed gluon-gluon anomalous dimension 
$\gamma_L = \gamma_{gg}^{(0)}(N,a_s)$. The resummed LO
gluon-quark anomalous dimension is given by $\gamma_{gq}^{(0)}(N,a_s)
= (C_F/C_A) \gamma_L$ and the quarkonic terms do not contribute in
$O[(a_s/(N-1))^l]$. Eq.~(\ref{eq1BF}) can be solved iteratively
demanding $\gamma_L(N,a_s) \rightarrow \overline{\alpha}_s/(N-1)$ as
$|N| \rightarrow \infty$ for $N~\in~\mathbb{C}$, which selects
the physical branch of the resummed anomalous dimension,
%-----------------------------------------------------------------------
\begin{equation}
\gamma_L \equiv \gamma_{gg,0}(N,\alpha_s) =
\frac{\overline{\alpha}_s}{N-1} \left\{1 + 2 \sum_{l=1}^{\infty} 
\zeta_{2l+1} \gamma_{gg,0}^{2l+1}(N,\alpha_s) \right\}.
\end{equation}
%-----------------------------------------------------------------------
\noindent
Here we rewrite
$\overline{\alpha}_s = N_c \alpha_s/\pi$. $\gamma_L$
has the serial representation
%-----------------------------------------------------------------------
\begin{equation}
\gamma_{gg,0}(N,\alpha_s) = \frac{\overline{\alpha}_s}{N-1}
+ 2 \zeta_3 \left(\frac{\overline{\alpha}_s}{N-1}\right)^4
+ 2 \zeta_5 \left(\frac{\overline{\alpha}_s}{N-1}\right)^6
+ 12 \zeta_3^2 \left(\frac{\overline{\alpha}_s}{N-1}\right)^7
+ \ldots
\label{eqPOLE}
\end{equation}
%-----------------------------------------------------------------------
Under the above conditions one may calculate $\gamma_L(N,a_s)$ in the
whole complex plane. It is a bounded function of
$\rho = (N-1)/\overline{\alpha}_s$, the singularities of which are
branch points \cite{JB95,EHW} at
%-----------------------------------------------------------------------
\begin{equation}
\rho_1 = 4 \log 2,~~~~~\rho_{2,3} = -1.41048 \pm 1.97212~i.
\end{equation}
%-----------------------------------------------------------------------
Its analytic structure is shown in Figure~\ref{FIG:lipLO}.
Note, that the resummed form of 
$\gamma_L(N,a_s)$ removes {\it all} the fixed--order
pole singularities of
Eq.~(\ref{eqPOLE}) into branch cuts. 
Since the known NLO resummed anomalous dimensions  are functions of 
$\gamma_L(N,a_s)$ which introduce no further singularities the contour
integral around the singularities of the problem has to cover the
three BFKL branch points, the singularities of the input distributions
along the real axis left of $N=1$, and the remaining singularities of the
fixed order anomalous dimensions at the non--positive 
integers~\cite{JB95,Blumlein:1997em}. 
Any finite correction to $\gamma_L$
may thus lead to essential changes of the corresponding numerical
results. Early numerical studies on the impact of the LO resummed
anomalous dimensions were performed in \cite{KWI} and  more recently
in Refs.~\cite{Blumlein:1997em,EHW,BVR1}.

The next-to-leading order resummed anomalous dimensions are given by
%-----------------------------------------------------------------------
\renewcommand{\arraystretch}{1.3}
\begin{equation}
\widehat{\gamma}_{NL}(N,\alpha_s) =  -2 \left(
\begin{array}{ll} {\ds \frac{C_F}{C_A}} \left[\gamma_{qg}^{NL} -
{\ds \frac{8}{3}} a_s T_F \right] & \gamma_{qg}^{NL} \\
\gamma_{gq}^{NL} &  \gamma_{gg}^{NL} \end{array}\right)~,
\end{equation}
\renewcommand{\arraystretch}{1.}
%-----------------------------------------------------------------------

\noindent
with $T_F = T_R N_f$.
The quarkonic contributions were calculated in Ref.~\cite{Catani:1994sq}, as well
as the resummed coefficient functions $c_2(N,a_s)$ and $c_L(N,a_s)$.
In \cite{Kirschner:2009qu} $c_L(N,a_s)$ has also been calculated, giving a result which 
differs form the one in \cite{Catani:1994sq} starting with $O(a_s^4)$.
$\gamma_{gg}^{NL}$ was computed in \cite{FL,CC2}.  
In the DIS--scheme $\gamma_{qg}^{NL}$ is found to be an
analytic, scale--independent
function of $\gamma_L(N,a_s)$ and reads
%-----------------------------------------------------------------------
\begin{equation}
\gamma_{qg}^{NL,DIS}(N,\alpha_s) = T_F \frac{\alpha_s}{6\pi}
\frac{2 + 3 \gamma_L - 3 \gamma_L^2}{3-2\gamma_L} 
\frac{\left[B(1-\gamma_l,1+\gamma_L)\right]^3}
{B(2+2\gamma_L,2-2\gamma_L)} R(\gamma_L),
\end{equation}
%-----------------------------------------------------------------------
where $R(\gamma)$ is given by
%-----------------------------------------------------------------------
\begin{equation}
R(\gamma) = \frac{1}{\gamma \sqrt{-\chi_0'(\gamma)}}
\exp\left\{\frac{1}{2}
\int_0^\gamma dz \frac{2\psi'(1)-\psi'(1-z)-\psi'(z)}{\chi_0(z)} 
+ \chi_0(z)\right\}~.
\end{equation}
%-----------------------------------------------------------------------
in \cite{Catani:1994sq}. In a re-analysis in \cite{Kirschner:2009qu}
%-----------------------------------------------------------------------
\begin{equation}
R(\gamma) = \frac{1}{-\gamma^2 \chi_0'(\gamma)}
\exp\left\{\frac{1}{2}
\int_0^\gamma dz \frac{2\psi'(1)-\psi'(1-z)-\psi'(z)}{\chi_0(z)} + \chi_0(z)\right\}~.
\end{equation}
%-----------------------------------------------------------------------
has been obtained, with a different pre-factor. A future calculation of the 4th order 
Wilson coefficient for the longitudinal structure function in the $\overline{\rm MS}$
scheme may clarify this question further.
%-----------------------------------------------------------------------
%\begin{equation}
%R(\gamma) = \left[\frac{\Gamma(1-\gamma) \chi_0(\gamma)}{- \gamma
%\Gamma(1+\gamma) \chi_0'(\gamma)}\right]^{1/2} \exp\left[\gamma \psi(1)
%+
%\int_0^\gamma dz \frac{\psi'(1)-\psi'(1-z)}{\chi_0(z)}\right]~.
%\end{equation}
%-----------------------------------------------------------------------

The NLO resummed gluon anomalous dimension $\gamma_{gg}^{NL}$ was
calculated in the $Q_0$--scheme \cite{CIA}. Here, a scale $Q_0^2 \gg \Lambda_{\rm QCD}^2$ is
introduced suppressing $k_\perp$ effects for $k^2 < Q_0^2$, see also \cite{Blumlein:1993ec}.\footnote{For a 
transformation into the DIS--scheme cf. \cite{Blumlein:1997em}.}
One has to solve the Bethe--Salpeter equation
%-----------------------------------------------------------------------
\begin{equation}
(N-1) G_N(q_1,q_2) = \delta^{D-2}(q_1-q_2) + \int d^{D-2} q_3 K(q_1,q_2)
G_N(q_3,q_2)
\end{equation}
%-----------------------------------------------------------------------
with
%-----------------------------------------------------------------------
\begin{equation}
K(q_1,q_2) = \delta^{D-2}(q_1 - q_2) 2 \omega(q_1) 
+ K_{\rm real}(q_1,q_2)
+ K_{\rm virtual}(q_1,q_2).
\end{equation}
%-----------------------------------------------------------------------
%vspace{-6mm}
%%%%%%%%%%%%%%%%%%%%%%%%%%%%%%%%%%%%%%%%%%%%%%%%%%%%%%%%%%%%%%%%%%%%%%%%%
\begin{figure}[h]
\begin{center}
\includegraphics[width=0.35\textwidth]{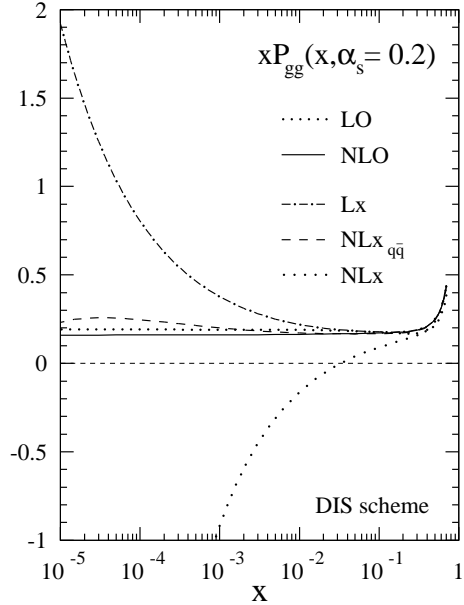}
\end{center}
\caption[]{
\label{BL-figPGG}
\small \sf
Different contributions to the resummed splitting function
$xP_{gg}(x,\alpha_s)$ in the DIS--scheme; Lx: LO BFKL; NLx$_{q\bar{q}}$: NLO BFKL quarkonic contr.; NLx:
NLO BFKL;  from \cite{BRNV}, \TCop (1998) by World Scientific.}
\end{figure}
%%%%%%%%%%%%%%%%%%%%%%%%%%%%%%%%%%%%%%%%%%%%%%%%%%%%%%%%%%%%%%%%%%%%%%%%%

\noindent
For
$q_1^2 \gg q_2^2$ one diagonalizes as in the LO case using {\it formally}
the same Ansatz~:
%-----------------------------------------------------------------------
\begin{equation}
\int d^{D-2} d q_2 K(q_1,q_2) \left(q_2^2\right)^{\gamma-1}
= \overline{\alpha}_s \left[\chi_0(\gamma) 
- \frac{\overline{\alpha}_s}{4} \delta(\gamma,q_1^2,\mu^2)\right]
\left(q_1^2\right)^{\gamma-1}~.
\end{equation}
%-----------------------------------------------------------------------
Here the scale--invariant
LO eigenvalue $\overline{\alpha}_s  \chi_0(\gamma)$ is
supplemented by the NLO correction term \newline
$(\overline{\alpha}^2_s/4)
\delta(\gamma, q_1^2,\mu^2)$,
%-----------------------------------------------------------------------
\begin{eqnarray}
\label{eqDEL}
\delta(\gamma,q_1^2,\mu^2) &=&
 - \left(\frac{67}{9} - 2 \zeta(2) -
\frac{10}{27} N_f \right) \chi_0(\gamma) 
+ 4 \Phi(\gamma) - \frac{\pi^3}
{\sin^2(\pi \gamma)} \nonumber\\ & &
+ \frac{\pi^2}{\sin^2(\pi\gamma)} \frac{\cos(\pi
\gamma)}{1- 2 \gamma} \left[(22-\beta_0)
+ \frac{\gamma(1-\gamma)}{(1+2 \gamma) (3- 2 \gamma)} \left(1 +
\frac{N_f}{3}\right)\right]
\nonumber\\
& & +
\frac{\beta_0}{3} \chi_0(\gamma) \log\left(\frac{q_1^2}{\mu^2}
\right)
+ \left[\frac{\beta_0}{6} + \frac{d}{d\gamma}\right]
\left[\chi_0^2(\gamma) + \chi_0'(\gamma)\right] - 6 \zeta_3
,
\end{eqnarray}
%-----------------------------------------------------------------------
with
%-----------------------------------------------------------------------
\begin{equation}
\Phi(\gamma) = \int_0^1 \frac{dz}{1+z} \left[z^{\gamma-1} + z^{\gamma}
\right] \left[{\rm Li}_2(1) - {\rm Li}_23(z)\right]~.
\end{equation}
%-----------------------------------------------------------------------
%----------------------------------------------------------------------------------------------
\restylefloat{figure}
\begin{figure}[H]
\begin{center}
\includegraphics[scale=0.45,angle=0]{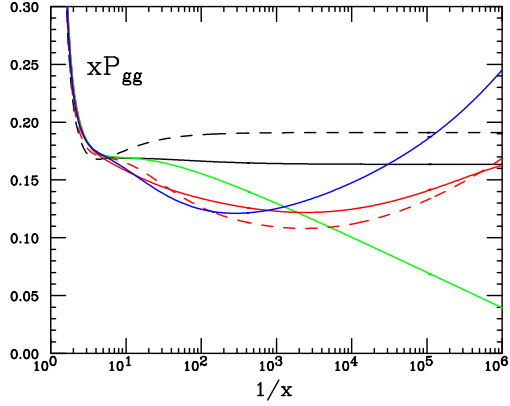}
\end{center}          
\caption[]{
\label{FIG:sx:ABP}
\small \sf
The gluon splitting functions $xP_{gg}$ plotted with $\alpha_s = 0.2$ and $N_f = 4$
The curves are (from top to bottom for $xP_{gg}$ at $x \sim 0.2$): fixed order perturbation theory
LO (black dashed), NLO (black solid), NNLO (green), resummed LO (red dashed) and NLO in $Q_0 \overline{\rm MS}$
scheme (red solid) and in the $\overline{\rm MS}$ scheme (blue); from \cite{Altarelli:2008aj}, \TCop (2008) by 
Elsevier Science.}
\end{figure}
%----------------------------------------------------------------------------------------------
Whereas the terms in the first two lines of Eq.~(\ref{eqDEL})
do contain contributions to the anomalous dimension up to $O(a_s^2)$
the third line contributes only with three--loop order. The agreement with
the perturbative results to 2-loop order has been known in the 1990ies.
The $O(a_s^3)$ from (\ref{eqDEL}), given in the DIS-scheme, agree with the perturbative
result \cite{Vogt:2004mw} after transforming it to the $\overline{\rm MS}$ scheme.
The test of even further agreement in higher orders is left for the future.
Note that the addend $-6 \zeta_3$, being numerically large, contains contributions of the gluonic 
contribution to the trajectory function $\omega(q_1^2)$. The result given in 
Ref.~\cite{FAD1} was
confirmed in a different calculation by Ref.~\cite{BNR}. A departing
value was reported in \cite{KOR}.\footnote{Despite the fact that in Ref.~\cite{KOR}
the quark contribution to $\omega(q_1^2)$ agrees with \cite{FAD1,BNR}, it may still be 
that the calculation in \cite{KOR} addresses a different quantity.}

Numerical results on the impact of the leading and next--to--leading
anomalous dimensions and coefficient functions were provided in a series
of detailed studies, see e.g.~\cite{EHW,Blumlein:1997em,BRNV,Ball:1999sh} and references 
therein. The matrix formalism for the solution of the all order evolution
equations, extending a first approach in Ref.~\cite{EKL}  to all orders, both for hadronic and photon structure 
functions, is described
in Ref.~\cite{Blumlein:1997em} in detail.
The quarkonic contributions lead to a strong enhancement of both
$F_2(x,Q^2)$ and $F_L(x,Q^2)$ at small $x$ during the evolution.
However, already simple choices for the yet unknown less singular 
contributions, cf. (\ref{eq:subl}), diminish these effects sizably so that a final conclusion
cannot be drawn at present. In the case of the resummed gluon anomalous 
dimension the NLO contributions are found to be extremely large and
negative. The large rise due to the LO BFKL term is already canceled
to the level of the fixed order contributions by the purely
quarkonic contribution to $\gamma_{gg}^{NL}$, see Figure~\ref{BL-figPGG}.
Adding also the gluonic contribution leads to negative values for
the resummed splitting function already for $\alpha_s = 0.2$ and
$x \simeq 0.01$ which has  to be regarded as unphysical. The LO and NLO
resummed contributions to the gluon anomalous dimension seem to represent
the first terms of a diverging series, which might be eventually resummed.

An interesting approach in this direction was performed in Refs.~\cite{Altarelli:2005ni,Altarelli:2008aj}, and 
related work in\cite{Ciafaloni:2006yk,Ciafaloni:2007gf}. In \cite{Altarelli:2005ni,Altarelli:2008aj} the coupled
evolution equations 
%------------------------------------------------------------------------------------
\begin{eqnarray}
\frac{d}{dt} f^+(N,t) = \gamma^+(a_s(t),N) f^+(N,t),~~~~~~~~\frac{d}{d\xi} f^+(x,M) = \chi(\hat{a_s},M) f^+(x,M),
\label{eq:DUA1}
\end{eqnarray}
%------------------------------------------------------------------------------------
with $t = \log(Q^2/Q_0^2), \xi = \ln(1/x)$, $f^+(x,M) = 
\int_{-\infty}^{+\infty} dt \exp(-Mt) f^+(x,t)$, and $N$ the Mellin variable. $\gamma^+(a_s,N)$ is 
the eigenvalue of the
singlet-evolution kernel containing poles at $N=1$ and $\chi(\hat{a}_s,M)$ denotes the BFKL-kernel, where
$\hat{a}_s$ is the operator $a_s(t \rightarrow - \partial/\partial M)$.
Eqs.~(\ref{eq:DUA1}) give rise to the so-called duality relation
%------------------------------------------------------------------------------------
\begin{eqnarray}
\chi(a_s,\gamma^+(a_s,N)) = N,~~~~~~~~\gamma^+(a_s,\chi(a_s,M)) = M
\end{eqnarray}
%------------------------------------------------------------------------------------
in the small-$x$ region. These relations are used to organize the resummation of the small-$x$
contributions, which leads to a gradual stabilization of the dominant splitting functions  at low $x$.
For the splitting function $xP_{gg}(x)$ this is illustrated in Figure~\ref{FIG:sx:ABP}, including all
the presently known information. In the $\overline{\rm MS}$ scheme significantly larger values are found
than in the 3-loop fixed order calculation at $x \sim 10^{-4}$ and below. 

Finally, we would like to mention that the effect of potential sub-leading 
contributions to the LO resummed anomalous dimension may be  studied within
$\Phi^3$ theory in $D=6$ dimensions, as a simple model with 3-boson interactions
at the perturbative level. Here the leading order resummed anomalous dimension can
be calculated for {\it all} values of $x$ solving a Bethe--Salpeter
equation~\cite{LOV}. The corresponding singularity at fixed orders is located
at $N = -1$ due to the scalar field. In \cite{BVN} the LO small-$x$ resummed terms
were compared to the complete ladder solution adding the NLO corrections. In this 
case it turns out that the pure small-$x$ resummed terms do not give the correct result. 
To further consolidate the knowledge on parton distributions in the small-$x$ region it seems
to be necessary to calculate at least one further resummed series, i.e. NNLO BFKL.

%%%%%%%%%%%%%%%%%%%%%%%%%%%%%%%%%%%%%%%%%%%%%%%%%%%%%%%%%%%%%%%%%%%%%%%
\section{\boldmath Resummation at Large Values of $x$}
\label{sec:largX}
\renewcommand{\theequation}{\thesection.\arabic{equation}}
\setcounter{equation}{0}
%%%%%%%%%%%%%%%%%%%%%%%%%%%%%%%%%%%%%%%%%%%%%%%%%%%%%%%%%%%%%%%%%%%%%%%

\noindent
The dominant contributions to the splitting functions and Wilson coefficients in the large-$x$
region result form the terms
%---------------------------------------------------------------------------------------------
\begin{eqnarray}
\left[\frac{\ln^k(1-x)}{1-x}\right]_+,~~~~~~\delta(1-x),~~~~~~\ln^l(1-x),~~~k,l \in \mathbb{N}, k \geq 0, l > 0~.
\end{eqnarray}
%---------------------------------------------------------------------------------------------
This is best seen in Mellin space, where
%---------------------------------------------------------------------------------------------
\begin{eqnarray}
\frac{1}{k!}\Mvec\left[\left(\frac{\ln^k(1-x)}{1-x}\right)_+\right](N) &=& (-1)^{k+1} 
S_{\underbrace{\mbox{\scriptsize 1, \ldots ,1}}_{\mbox{\scriptsize $k+1$}}}(N-1)
\\
\Mvec\left[\delta(1-x)\right](N) &=& 1 
\\
\frac{1}{k!}\Mvec\left[\ln^k(1-x)\right](N) &=& (-1)^k \frac{1}{N} S_{\underbrace{\mbox{\scriptsize 1, \ldots 
,1}}_{\mbox{\scriptsize $k$}}}(N)~. 
\end{eqnarray}
%---------------------------------------------------------------------------------------------
The harmonic sums with equal index form a polynomial in single harmonic sums, cf.~\cite{HSUM,Blumlein:2003gb},
%---------------------------------------------------------------------------------------------
\begin{eqnarray}
S_{\underbrace{\mbox{\scriptsize 1, \ldots ,1}}_{\mbox{\scriptsize $k$}}}(N) = 
\frac{1}{k} \sum_{l=1}^k 
S_{\underbrace{\mbox{\scriptsize 1, \ldots ,1}}_{\mbox{\scriptsize $k-l$}}}(N) S_l(N)~. 
\end{eqnarray}
%---------------------------------------------------------------------------------------------
In the limit $|N| \rightarrow \infty$ one obtains
%---------------------------------------------------------------------------------------------
\begin{eqnarray}
S_{\underbrace{\mbox{\scriptsize 1, \ldots ,1}}_{\mbox{\scriptsize $k$}}}(N) \propto \frac{1}{k!} S_1^k(N)
+ O(S_1^{k-2}(N) S_2(N)) 
\end{eqnarray}
%---------------------------------------------------------------------------------------------
and
%---------------------------------------------------------------------------------------------
\begin{eqnarray}
S_1(N) \propto \ln(\bar{N}) + \frac{1}{2N} + \frac{1}{12 N^2} - \frac{1}{120 N^4} + 
O\left(\frac{1}{N^6}\right)~.
\end{eqnarray}
%---------------------------------------------------------------------------------------------
with $\bar{N} = N + \gamma_E$.

Let us consider the splitting functions $P_{ik}$ and Wilson coefficients $C_{p,i}$
%---------------------------------------------------------------------------------------------
\begin{eqnarray}
P_{ik}(x,a_s)  &=& \sum_{j=0}^\infty a_s^{j+1} P_{ik}^{(j)}(x) \\
C_{p,i}(x,a_s) &=& \delta_{p,2(3)} \delta(1-x) + \sum_{j=1}^\infty a_s^{j} c_{p,i}^{(j)}(x)~.
\end{eqnarray}
%---------------------------------------------------------------------------------------------
The large $x$ structure for the spitting functions $P_{ik}$ and 
Wilson coefficients known up to $O(\alpha_s^3)$ \cite{Moch:2004pa,Vogt:2004mw,Vermaseren:2005qc}
is
%---------------------------------------------------------------------------------------------
\begin{eqnarray}
\label{eqRES1}
P_{kk}^{(l)}(x)       &=& A_{l+1}}{(1-x)_+ + B_{l+1} \delta(1-x) + C_{l+1} \ln(1-x) + O[(1-x)^{k \geq 1}\ln^{l+1}(1-x)]
\\ 
\label{eqRES2}
P_{i \neq k}^{(l)}(x) &=& \sum_{j=0}^{2l-1} D_{kl}^{(l,j)}\ln^{2l-j}(1-x) + O(1) 
\\
\label{eqRES3}
c_{a,l}(x) &=& \frac{(2 C_F)^l}{(l-1)!} p_{qq}(x) \ln^{2l-1}(1-x) + O[(1-x)^{k \geq -1}\ln^{2l-2}(1-x)], 
\end{eqnarray}
%---------------------------------------------------------------------------------------------
cf.~\cite{Moch:2009hr,Almasy:2010wn}. Here $c_{a,l}(x)$ denotes the non-singlet Wilson coefficient with $p_{qq}(x) = 
2/(1-x)_+ - (1+x)$. The singlet Wilson coefficients are of a similar structure.
The splitting function $P_{qq}^s(x)$ 
related to the color factor $d_{abc} d^{abc}$ (\ref{eq:3s1},\ref{eq:3s2}) behaves like $(1-x)^{k \geq 1} 
\ln(1-x)$ for large values of $x$, \cite{Moch:2009hr}.

In Mellin-space the resummed Wilson coefficients have the structure \cite{Kodaira:1981nh,CEXP1}
%---------------------------------------------------------------------------------------------
\begin{eqnarray}
C(N) = g_0(a_s) \exp\left\{\ln(N) g_1(\lambda) + g_2(\lambda) + a_s g_3(\lambda) + O(a_s^2 f(\lambda))\right\},
\end{eqnarray}
%---------------------------------------------------------------------------------------------
with $g_0(a_s)$ the normalization and $\lambda = a_s \beta_0 \ln(N)$.  In the flavor non-singlet case
the functions $g_{1,2}(a_s)$ are given by
%---------------------------------------------------------------------------------------------
\begin{eqnarray}
g_1(a_s) &=& \frac{A_1}{\beta_0 \lambda} \left[\lambda + (1-\lambda) \ln(1-\lambda) \right] \\
g_2(a_s) &=& - \frac{\gamma_E A_1 -B_1}{\beta_0} \ln(1-\lambda) - \frac{A_2}{\beta_0^2} \left[\lambda + 
\ln(1-\lambda)\right] + \frac{A_1 \beta_1}{\beta_0^3} \left[\lambda \ln(1-\lambda) + \frac{1}{2} 
\ln^2(1-\lambda)\right]~, \nonumber\\
\end{eqnarray}
%---------------------------------------------------------------------------------------------
with
%---------------------------------------------------------------------------------------------
\begin{eqnarray}
A_1 = 4 C_F,~~~A_2 = 8 C_F \left[\left(\frac{67}{18} - \zeta_2\right) C_A - \frac{5}{9} N_f\right], 
\text{\cite{Kodaira:1981nh}}~~~B_1 = -3 C_F \text{\cite{CEXP1}}~.
\end{eqnarray}
%---------------------------------------------------------------------------------------------
The universal part of $g_3 \propto a_s \ln(N)$ was derived in \cite{Vogt:1999xa}.  

The leading order resummation corrections beyond $O(a_s^3)$ valid for large values of $N$ 
in the non-singlet case have been calculated in \cite{LXRESUM1}. Next-to-leading-log contributions 
were accounted for in \cite{LXRESUM2,Moch:2009hr}.

One may investigate the large-$x$ structure of the 4-loop anomalous dimensions studying scheme-invariant
evolution equations, cf.~\cite{Furmanski:1981cw,Grunberg:1982fw,Catani:1996sc,Blumlein:2000wh}, for the $F_2, 
F_\phi$-system, with $\phi$  a scalar particle coupling to the gluon as the photon couples to the quarks.
The specific structure of the scheme-invariant kernels to the level of $a_s^3$ is assumed to hold at  $a_s^4$.
Knowing the 3-loop Wilson coefficients, one may determine under this assumption the large $x$ behaviour of 
the singlet splitting functions $P_{ij}^{(3)}(x)$ in the highest two powers in $\ln(1-x)$, cf. \cite{Soar:2009yh}.
A resummation of the contributions to $P_{qg}, P_{gq}, C_{2,g}$ and $C_{L,g}$ has been performed in 
\cite{Almasy:2010wn}. A systematic approach to obtain even higher order terms in case of the large-$x$ resummation 
has been proposed in Ref.~\cite{EXPSYST}. 

%%%%%%%%%%%%%%%%%%%%%%%%%%%%%%%%%%%%%%%%%%%%%%%%%%%%%%%%%%%%%%%%%%%%%%%
\section{Sum Rules and Integral Relations}
\label{sec:polt3}
\renewcommand{\theequation}{\thesection.\arabic{equation}}
\setcounter{equation}{0}
%%%%%%%%%%%%%%%%%%%%%%%%%%%%%%%%%%%%%%%%%%%%%%%%%%%%%%%%%%%%%%%%%%%%%%%

\noindent
Deep-inelastic structure functions obey a series of sum rules for special 
moments or even integral relations, which are of interest for experimental tests.
These relations are of different rigor. In most of the cases they receive radiative 
corrections, mass corrections, and in some cases even non-perturbative corrections a priori. 
Using current algebra techniques many sum rules have been investigated in  \cite{Ravindran:2001dk}.
A larger series of sum rules having been proposed for polarized scattering have 
been analyzed in \cite{Blumlein:1996vs}. Quark mass and QCD corrections have been given in 
\cite{Blumlein:1998sh,Ravindran:2001dk}. In the following we give a brief description of the
main sum rules and comment on the status of available QCD corrections.

\vspace*{2.5mm}
\noindent
{\underline{\sf Adler sum rule} \cite{Adler:1965ty}:}\\
This sum rule is rigorous and neither obtains QCD nor mass corrections.
%---------------------------------------------------------------------------
\begin{eqnarray}
\int_0^1 \frac{dx}{x} \left[F_2^{\bar{\nu}p}(x,Q^2) - F_2^{{\nu}p}(x,Q^2) \right] = K(N_f),
\end{eqnarray}
%---------------------------------------------------------------------------
with $K(3) = 2 + 2 \sin^2\theta_c,~(SU_F(3))$ and $K(4) = 2,~(SU_F(4))$, and $\theta_c$
the Cabibbo angle, cf~\cite{Blumlein:1998sh,Ravindran:2001dk}.

\noindent
{\underline{\sf Unpolarized Bjorken sum rule} \cite{Bjorken:1967px}:}\\
The sum rule refers to the charged current structure functions $F_1$
%---------------------------------------------------------------------------
\begin{eqnarray}
\label{eq:UBJ}
\int_0^1 {dx} \left[F_1^{\bar{\nu}p}(x,Q^2) - F_1^{\nu p}(x,Q^2) \right] = 
K(N_f) A^{\rm F_1}(N_f,Q^2),
\end{eqnarray}
%---------------------------------------------------------------------------
with $K(3(4)) = 1 +  \sin^2\theta_c,~(1)$ and $A^{\rm F}(N_f,Q^2) = 1 + O(a_s)$.
The 3-loop QCD corrections to  $A^{F_1}$ were given in \cite{Larin:1990zw}.

\noindent
{\underline{\sf Gross-Llewellyn Smith sum rule} \cite{Gross:1969jf}:}\\
Likewise, the combination for the charged current structure functions $F_3$ yields
%---------------------------------------------------------------------------
\begin{eqnarray}
\label{eq:GLL}
\int_0^1 {dx} \left[F_3^{\bar{\nu}p}(x,Q^2) + F_3^{{\nu}p}(x,Q^2) \right] = K(N_f) A^{\rm F_3}(N_f,Q^2),
\end{eqnarray}
%---------------------------------------------------------------------------
with $K(3(4)) = 6 - 2 \sin^2\theta_c,~~(6)$. The 3-loop QCD corrections
to $A^{\rm F_3}(N_f,Q^2)$  were given in \cite{Larin:1991tj} and the 4-loop corrections in~\cite{CHET:GLL}. 

\noindent
{\underline{\sf Polarized Bjorken sum rule} \cite{Bjorken:1969mm}:}\\
The sum rule refers to the flavor non-singlet combination
%---------------------------------------------------------------------------
\begin{eqnarray}
\label{eq:PBJ}
\int_0^1 {dx} \left[g_1^{ep}(x,Q^2) - g_1^{en}(x,Q^2) \right] = \frac{1}{6}\left|\frac{g_A}{g_V}\right|
A^{\rm g_1}(N_f,Q^2),
\end{eqnarray}
%---------------------------------------------------------------------------
with $g_{A,V}$ the neutron decay constants, $g_A/g_V \approx -1.26$. The 3-loop QCD corrections to $A^{\rm 
g_1}(N_f,Q^2)$  were given in \cite{Larin:1991tj} and the 4-loop corrections in\cite{Baikov:2010je}. 

\noindent
{\underline{\sf Gerasimov-Drell-Hearn sum rule} \cite{Gerasimov:1965et,Drell:1966jv}:}\\
%---------------------------------------------------------------------------
This sum rule is given by the first moment of the polarized structure function $g_1^{p,n}(x,Q^2)$
in the form, cf. e.g. \cite{Drechsel:2000ct,Lampe:1998eu,Drechsel:2004ki},
%---------------------------------------------------------------------------
\begin{eqnarray}
I_{p,n}(Q^2) = 2 \frac{M^2}{Q^2} \int_0^{x_0} dx~g_1^{p,n}(x,Q^2) = \left\{  
\begin{array}{ll} 
-\frac{1}{4} \mu_{p,n}^2, & Q^2 \rightarrow 0 \\
\frac{2 M^2}{Q^2} \Gamma_1^{p,n}, & Q^2 \rightarrow \infty \\
\end{array}
\right.
\end{eqnarray}
%---------------------------------------------------------------------------
at proton and neutron targets, with $x_0 =Q^2/(2 M m_{\pi} + m^2_{\pi} + Q^2)$, $\mu_{p,n}$ 
the 
anomalous magnetic moment of the proton or nucleon (\ref{eq:anmom}), and $\Gamma_1$ the first moment
of the structure function $g_1$ at infinite space-like momentum transfer. The sum-rule has 
a very strong $Q^2$-evolution for low values of the virtuality. In case of proton targets
it changes sign between $Q^2 = 0$ and $Q^2 \approx 1 \GeV^2$, \cite{Lampe:1998eu}.

\noindent
{\underline{\sf Burkhardt-Cottingham sum rule} \cite{Burkhardt:1970ti}:}\\
%---------------------------------------------------------------------------
\begin{eqnarray}
\int_0^1 dx g_2(x,Q^2) = 0~.
\end{eqnarray}
%---------------------------------------------------------------------------
In the derivation of the sum rule it was assumed that the large $\nu$-behaviour of the associated
amplitude $A_2$, cf. \cite{Blumlein:1996vs}, is governed by Regge-theory and it behaves as $A_2(Q^2,\nu) 
\sim \nu^{-1-\epsilon}, \epsilon < 0$. However, this may be questioned, see \cite{Jaffe:1989xx}.
It was argued in \cite{IKL} that Regge cuts spoil the sum rule, which would vanish for large $Q^2$, 
however. The sum-rule may be invalidated in case of a short-distance singularity \cite{Jaffe:1989xx}, 
see also \cite{Heimann:1973hq}. The Burkhardt-Cottingham sum rule cannot be expressed in terms of
expectation values of (axial-)vector current operators and not be derived using the light-cone expansion, 
see e.g. \cite{Blumlein:1996vs}, although a {\it formal} analytic continuation for general values of the 
Mellin variable $N$ would suggest it. It is well-known, however, that specific moments may have
a particular behaviour. The $O(\alpha_s)$ corrections \cite{Ravindran:2001dk}, target mass corrections 
\cite{Blumlein:1998nv} and massive quark corrections \cite{Kodaira:1994gh} do not alter this relation. 
A brief review on this sum rule has been given in Ref.~\cite{Jaffe:1989xx}.

\noindent
{\underline{\sf Efremov-Teryaev-Leader sum rule} \cite{Efremov:1996hd}:}\\
This sum rule refers to the valence (V) contributions of the polarized structure functions
$g_{1,2}$ related to the twist-3 operator matrix element $d_1^{\gamma q,V}$ 
%---------------------------------------------------------------------------
\begin{eqnarray}
\int_0^1 dx~x \left[g_1^{\gamma q, \rm V}(x,Q^2) + 2 g_1^{\gamma q, \rm V}(x,Q^2)\right] 
= \frac{e_q^2}{8} d_1^{\gamma q,V} = 0~.
\end{eqnarray}
%---------------------------------------------------------------------------
It receives quark mass corrections and is thus only valid in the limit $m_q \rightarrow 0$, cf. 
\cite{Blumlein:1996vs}.
It holds under certain conditions also in the presence of target mass corrections \cite{Blumlein:1998nv}.

\noindent
{\underline{\sf Ellis-Jaffe sum rule} \cite{Ellis:1973kp}:}\\
This sum rule is given by the integral
%---------------------------------------------------------------------------
\begin{eqnarray}
\label{eq:EJ}
\int_0^1 {dx} g_1^{p(n)}(x,Q^2) = C^{\rm NS}(Q^2) A^{\rm NS}(Q^2) + C^{\rm S}(Q^2) A^{\rm S}(Q^2). 
\end{eqnarray}
%---------------------------------------------------------------------------
It was originally devised to describe the nucleon spin composition. 
However,
the nucleon spin receives also gluonic and angular momentum contributions. The latter
can in principle be measured using non-forward scattering \cite{Ji:1996ek}.
Here $C^{\rm NS,S}$ denote
Wilson coefficients and  $A^{\rm NS,S}$ the corresponding OMEs in the flavor non-singlet and singlet case.
The non-singlet current is conserved for 
massless quarks, while the singlet current is not conserved, which causes a non-vanishing 
singlet anomalous dimension at $O(a_s^2)$. The 3- and 4-loop QCD corrections 
to this sum rule have been calculated in \cite{Larin:1997qq,CHET:EJ}. 

\noindent
{\underline{\sf Gottfried sum rule} \cite{Gottfried:1967kk}:}\\
%---------------------------------------------------------------------------
\begin{eqnarray}
\int_0^1 \frac{dx}{x} \left[F_2^{\gamma p}(x,Q^2) - F_2^{\gamma n}(x,Q^2) \right] = 
\frac{1}{3} \int_0^1 dx \left[
u_v(x,Q^2) - d_v(x,Q^2)
-\left(\bar{d}(x,Q^2) - \bar{u}(x,Q^2)\right)\right)]~.
\end{eqnarray}
%---------------------------------------------------------------------------
This sum rule is non-rigorous since the distributions $\bar{d}(x,Q^2)$ and $\bar{u}(x,Q^2)$ 
are found to be different and are non-perturbative quantities. It is currently studied in lattice 
simulations, cf.~Figure~\ref{FIG:latt:val}. The 3-loop corrections to the sum-rule follow from 
the Wilson coefficients \cite{Vermaseren:2005qc}. The large-$N_c$ limit was studied in 
\cite{Broadhurst:2004jx}.

\noindent
{\underline{\sf Integral Relations at Twist 2 and 3} 
\cite{Wandzura:1977qf,Blumlein:1998nv}:}\\
The twist-2 and twist-3 contributions to the structure functions $g_{1,2}(x,Q^2)$
obey the relations
%---------------------------------------------------------------------------
\begin{eqnarray}
\label{eq:WW}
g_2^{\tau = 2}(x,Q^2) &=& - g_1^{\tau = 2}(x,Q^2) + \int_x^1 \frac{dy}{y} g_2^{\tau = 2}(y,Q^2) \\
\label{eq:BT}
g_1^{\tau = 3}(x,Q^2) &=& \frac{4M^2}{Q^2} \left[
g_2^{\tau = 3}(x,Q^2) - 2 \int_x^1 \frac{dy}{y} g_2^{\tau = 3}(y,Q^2)\right]~.
\end{eqnarray}
%---------------------------------------------------------------------------
The Wandzura-Wilczek relation (\ref{eq:WW}) \cite{Wandzura:1977qf}, originally derived in the massless case,
remains valid in case of quark- and target mass corrections \cite{Blumlein:1998nv}, for the gluonic 
contributions \cite{Blumlein:2003wk}, for non-forward scattering \cite{Blumlein:1999sc,Geyer:2004bx}, and 
diffractive scattering \cite{Blumlein:2002fw,Blumlein:2008di}. It can be derived using the covariant parton 
model, as well as relations for other polarized structure functions containing twist-3 contributions, 
cf.~\cite{Jackson:1989ph,Blumlein:1996tp}.
In the limit of vanishing target masses $g_1$ does not 
receive a twist-3 contribution. Eqs.~(\ref{eq:WW},\ref{eq:BT}) allow to disentangle
the fermionic twist-2 and 3 contributions to $g_1$ and $g_2$. 
%%%%%%%%%%%%%%%%%%%%%%%%%%%%%%%%%%%%%%%%%%%%%%%%%%%%%%%%%%%%%%%%%%%%%%%%% 
\section{Higher Twist Contributions} 
\label{sec:HTC}
\renewcommand{\theequation}{\thesection.\arabic{equation}}
\setcounter{equation}{0}
%%%%%%%%%%%%%%%%%%%%%%%%%%%%%%%%%%%%%%%%%%%%%%%%%%%%%%%%%%%%%%%%%%%%%%%%%

\noindent
Beyond the leading twist contributions, i.e. twist $\tau = 2$ in Quantum Chromodynamics, a quite
different picture emerges for the structure functions both in the polarized and unpolarized 
case. With the twist-3 contributions to polarized deep-inelastic scattering the problem of 
the operator mixing becomes more involved since the number of operators contributing at total 
spin $N$ is growing with $N$. Both the anomalous dimensions and Wilson coefficients have been
calculated during the last three decades at the level of the $O(a_s)$ corrections. The twist-3 
anomalous dimensions have been derived in Refs.~\cite{Shuryak:1981pi,TW3ANOMD}. The evolution
in the non-singlet case in the large $N_c$ limit has been studied in \cite{Ali:1991em} 
and terms beyond this limit were derived in \cite{Braun:2000av}. The twist-3 non-singlet and singlet
anomalous dimensions have been calculated using light-ray operators in \cite{Geyer:1996vj,Mueller:1997yk}. 
The leading order Wilson coefficients were computed in \cite{WILSPNS} in the non-singlet and in 
\cite{WILSPSING} in the singlet case. The renormalization of gauge invariant operators contributing 
to the structure function $g_2(x,Q^2)$ has been studied in \cite{Kodaira:1997ig}. In Ref.~\cite{Braun:2000yi} 
the gluonic contributions to the structure function $g_2(x,Q^2)$ were calculated. The complete non-singlet 
and singlet evolution of the twist-3 moments was derived in \cite{Braun:2001qx}.

The data on $g_2(x,Q^2)$  
\cite{EXP:G2,E143pd} have still large errors, which can be substantially improved at a facility like the 
EIC \cite{EIC,Boer:2011fh}. For first estimates on the behaviour of the twist-3 part of $g_2(x,Q^2)$ see e.g. in 
\cite{Braun:2011aw}. Yet the experimental errors on the twist-3 contribution of the structure function $g_2(x,Q^2)$
are large \cite{Blumlein:2012se}. First lattice measurements of the lowest twist-3 moment have been performed in
\cite{Gockeler:2000ja,Dolgov:2002zm} at larger pion masses and have to be refined for realistic values of $m_\pi$ 
in the future.

In the unpolarized case the contributions of higher (dynamical) twist are of $\tau =4, 6, ...$. 
They are largely suppressed in the limit of large virtualities $Q^2$. However, the experimental 
data often exhibit a correlation between $x$ and $Q^2$ due to similar values of $S$. Furthermore, in 
neutral current deep-inelastic scattering the largest statistics is located in the region of lower 
values of $Q^2$. It is difficult to decide from which scale $Q^2$ onwards a data sample is widely 
free of higher twist contributions, for which sometimes a cut of $W^2 = 12.5 - 15 \GeV^2$ is proposed
\cite{Martin:2002dr,Blumlein:2006be}. This cut has been verified in the non-singlet analysis
\cite{Blumlein:2006be}. 

The local operators of higher twist can be constructed systematically near the light-cone.
They are formed by more external quark and gluon fields than the twist-2 operators and
potential contributions of lower twist operators in case mass scales are present. A 
twist-4 operator is given e.g. by
%---------------------------------------------------------------------------------------------
\begin{eqnarray}
:\bar{\psi}(x) \gamma_{\mu_1} \partial_{\mu_2} ... \partial_{\mu_m} \psi(x)
 \bar{\psi}(y) \gamma_{\nu_1} \partial_{\nu_2} ... \partial_{\nu_n} \psi(y):
\end{eqnarray}
%---------------------------------------------------------------------------------------------
A systematic twist-decomposition has to be performed, cf.~\cite{GEYLA}.
One forms operator matrix elements with these operators between nucleon states. A `partonic'
interpretation assumes, that all external lines can be factorized. The corresponding 
contributions to deep-inelastic structure functions are then similar as in the case of 
twist-2, with the generalization that both the Wilson coefficients and OMEs depend 
on $n-1$ dimensionless 
parameters $\sum_{i=1}^{n-1} x_i = x$, where $n$ denotes the number of external fields.
These parameters, except $x$, cannot be measured in the deep-inelastic process. 

%----------------------------------------------------------------------------------------------
\restylefloat{figure}
\begin{figure}[H]
\begin{center}
\includegraphics[scale=0.35,angle=0]{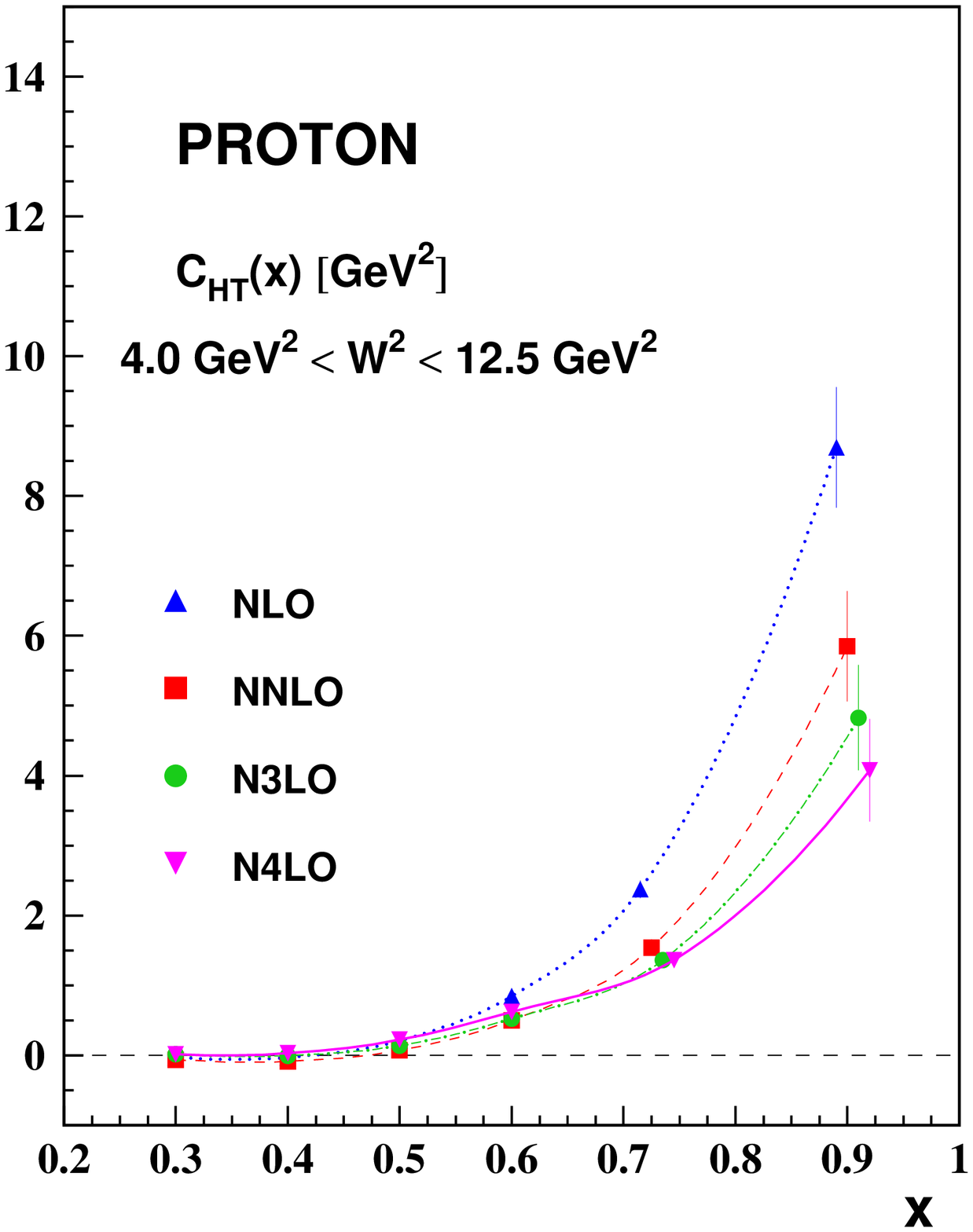}
\includegraphics[scale=0.35,angle=0]{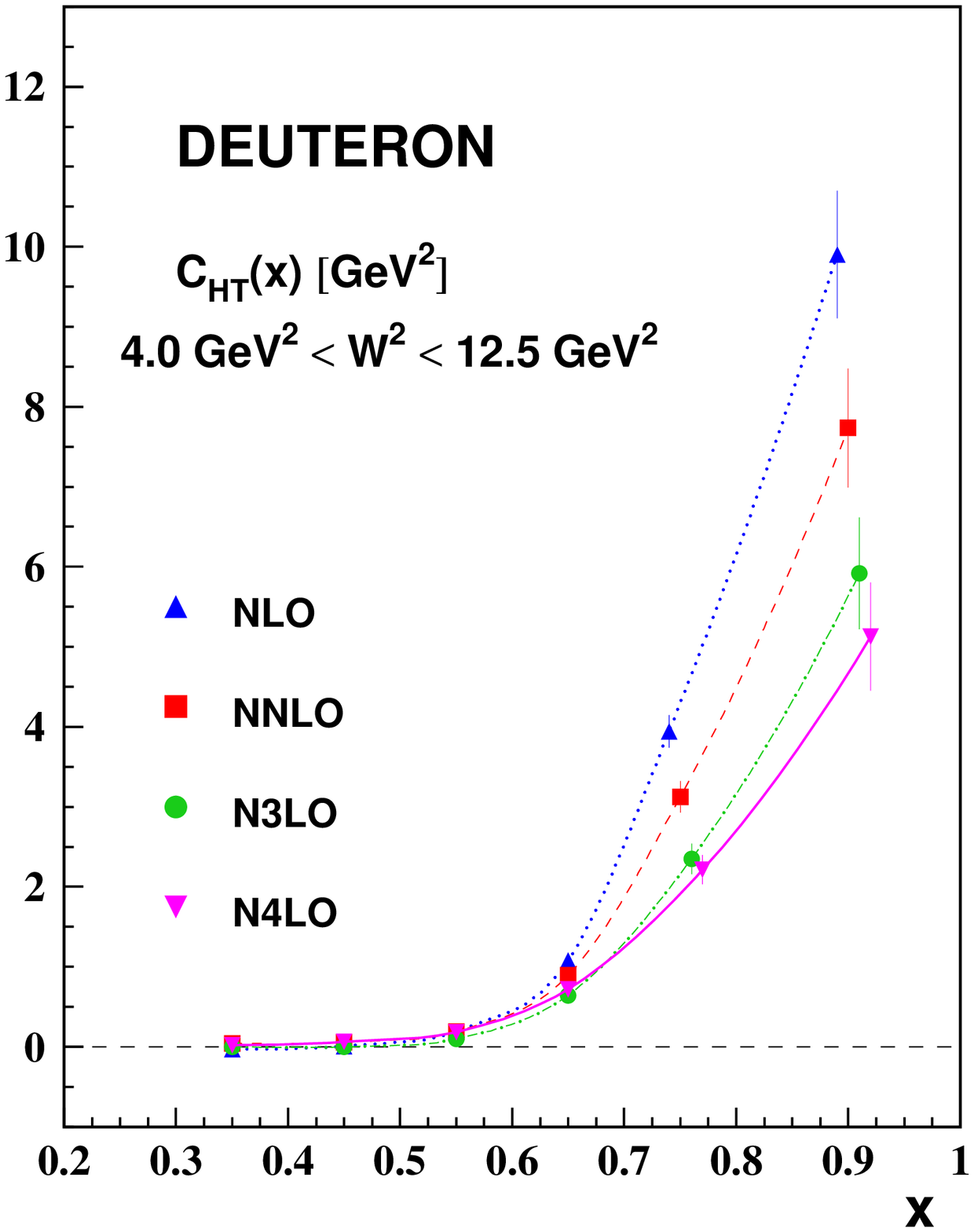}
\end{center}          
\caption[]{
\label{FIG:HT2}
\small \sf
Comparison of the higher twist coefficient $C_{HT}(x) \equiv C_2^{\rm \tau = 4}(x)$ in the large x region for 
the proton
data as function of x in a NLO (dotted lines), NNLO (dashed lines), N3LO analysis (dash-dotted lines) for the 
non-singlet QCD Wilson coefficient (full
lines). Some bin centers are slightly shifted for better visibility; from \cite{Blumlein:2008kz},
\TCop (2010) by Elsevier Science.
}
\end{figure}
%----------------------------------------------------------------------------------------------

Early theoretical investigations of the structure of higher twist operators and their
anomalous dimensions in $D = 6~\Phi^3$-theory \cite{HTphi3} and QCD 
\cite{HTQCD,Shuryak:1981pi,JAFSOL,EFP,Bukhvostov:1985rn,Bukhvostov:1987pr,Qiu:1988dn,Bartels:1999km,Braun:2000kw}
revealed 
the principal structure of these contributions. The lowest order anomalous 
dimensions have been calculated in \cite{Bukhvostov:1985rn,Bukhvostov:1987pr,Braun:2000kw}
and the Wilson coefficients, in different operator bases, in  \cite{JAFSOL,EFP,Qiu:1988dn}. More 
recently also gluonic operators were considered  \cite{Bartels:1999km}. A systematic study 
of higher twist light-cone distribution amplitudes was given in \cite{Braun:2000kw}. The 
renormalization of these operators has been worked out systematically in \cite{Braun:2009vc}.
The evolution of the lowest twist-4 moments at leading order has been studied in 
\cite{Glatzmaier:2012iu} recently.

Due to the fact that the OMEs of higher twist operators cannot be measured experimentally one may 
envisage their determination with lattice simulations. At present, investigations of this kind are 
still at the beginning. In Mellin--space the basis of higher twist operators grows considerably with 
the value of the Mellin variable $N$ and an according number of OMEs has to be measured on the
lattice, which further complicates quantifications of higher twist effects ab initio. 

Estimates of higher twist effects have been obtained studying renormalon corrections to sum-rules
and deep-inelastic structure functions, see \cite{BRAUN1} and for reviews 
Refs.~\cite{Beneke:1998ui,Beneke:2000kc}.

Because of the complications mentioned the extraction of higher twist contributions to deep-inelastic
structure functions is usually being performed applying a suitable Ansatz for these contributions:
%-------------------------------------------------------------------------------------------------
\begin{eqnarray}
F_i(x,Q^2) &=& F_i^{\rm \tau = 2}(x,Q^2) \left[ 1 
+ \frac{C_i^{\rm \tau = 4}(x,Q^2)}{Q^2}  
+ \frac{C_i^{\rm \tau = 6}(x,Q^4)}{Q^2} + ... \right]~.
\label{eq:HTPHAE}
\end{eqnarray}
%-------------------------------------------------------------------------------------------------
Sometimes one considers the higher twist terms in (\ref{eq:HTPHAE}) in additive form.
In many cases these contributions are fitted together with the PDFs,
cf.~\cite{HT:PHAE,Blumlein:2008kz}. Since the logarithmic 
scaling violations of the higher twist coefficients are in general different from those of the
lowest twist it was suggested to proceed in a different way, cf.~\cite{Blumlein:2006be,Blumlein:2008kz}.
In the flavor non-singlet case the lowest twist contributions to the structure functions may be determined in 
the region which is free of higher twist terms. One then extrapolates the twist--2 distributions
into the region, where higher twist contributions are present and measures the difference.
The size of higher twist contributions thus obtained stabilizes including higher order 
QCD corrections to the leading twist terms. The inclusion of the leading twist 3-loop Wilson coefficients
yields the dominant contribution and effectively describes the N$^3$LO terms, while the N$^4$LO term
is approximate only as it refers to the soft exponentiations \cite{LXRESUM1}.
In Figure~\ref{FIG:HT2} the corresponding results are shown
for a world analysis in the valence region for proton and deuteron data.

%----------------------------------------------------------------------------------------------
\restylefloat{figure}
\begin{figure}[H]
\begin{center}
\includegraphics[scale=0.60,angle=0]{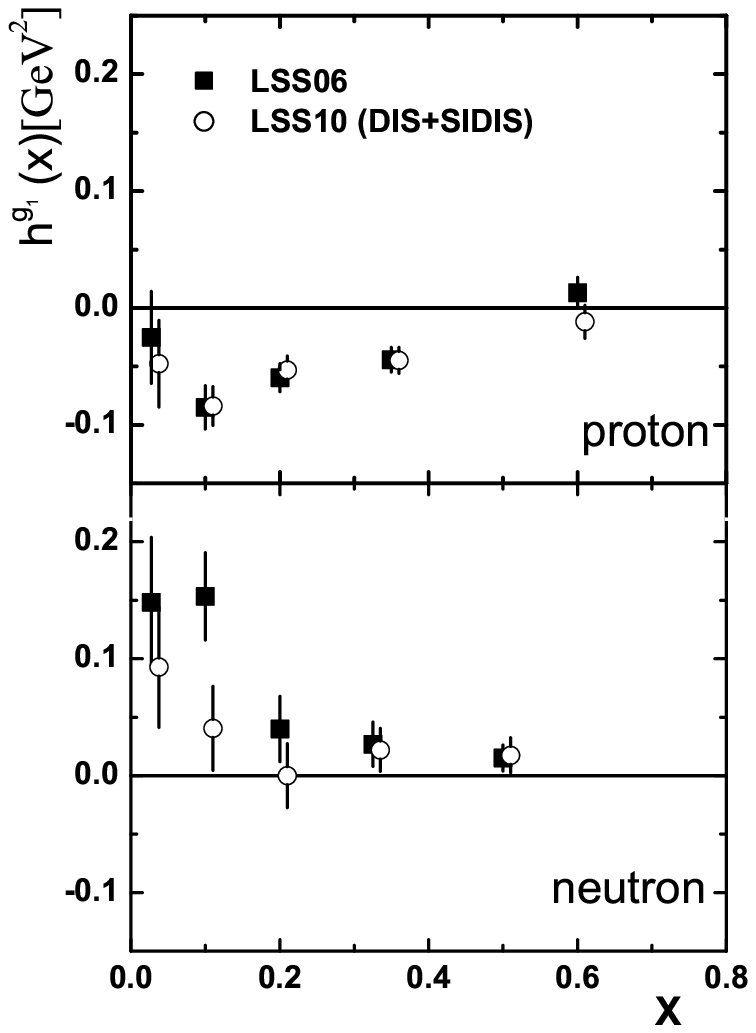}
\includegraphics[scale=0.38,angle=0]{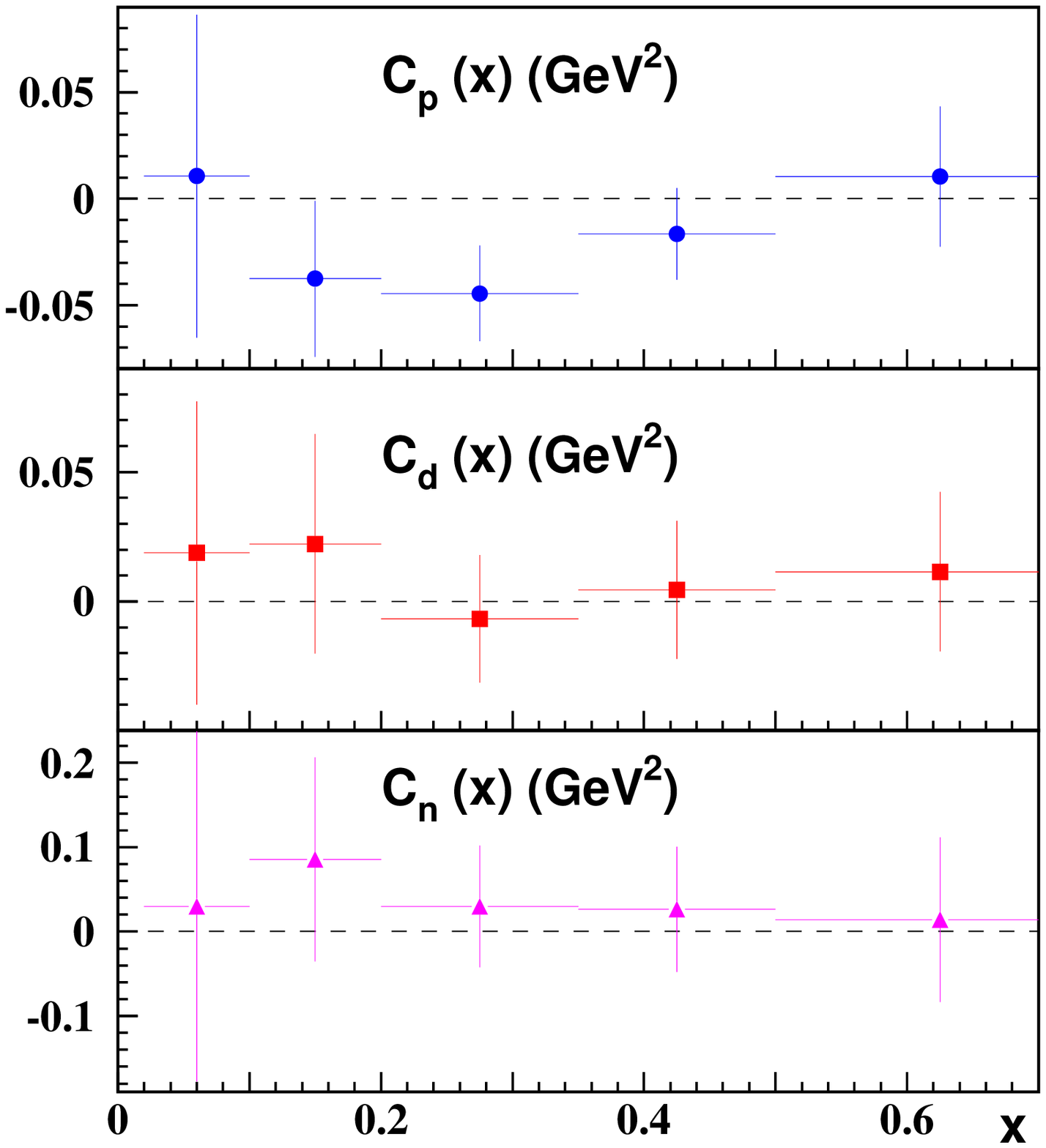}
\end{center}          
\caption[]{
\label{FIG:HT2P}
\small \sf
Left: comparison between the higher twist values corresponding to the
fits of inclusive DIS LSS06 \cite{Leader:2006xc} and combined inclusive and SIDIS
data set LSS10 \cite{Leader:2010rb} \TCop (2010) by the American Physical Society.
Right: The additive higher twist coefficients $C_p(x), C_d(x)$ and 
$C_n(x)$ \cite{Blumlein:2010rn}, \TCop (2010) by Elsevier Science.}
\end{figure}
%----------------------------------------------------------------------------------------------

A similar analysis is not possible in the polarized case at present, because of the smaller kinematic
range in $Q^2$ and the larger errors of the data for the structure function $g_1(x,Q^2)$.
Here the higher twist contributions are fitted as additive terms in an NLO analysis for individual bins, cf. 
e.g. \cite{Leader:2006xc,Leader:2010rb,Blumlein:2010rn}. In Refs.~\cite{Leader:2006xc,Leader:2010rb}
smaller errors have been obtained with an excess at lower values of $x$ in case of neutron targets and
proton targets, see Figure~\ref{FIG:HT2P}. In \cite{Blumlein:2010rn} higher twist contributions
which are widely compatible with zero have been obtained both for the proton and neutron targets,
with a 1--1.5 $\sigma$ effect in two bins for the proton data. To reveal twist--4 contributions 
to the structure function $g_1(x,Q^2)$ more precise data from future experiments are needed.

%%%%%%%%%%%%%%%%%%%%%%%%%%%%%%%%%%%%%%%%%%%%%%%%%%%%%%%%%%%%%%%%%%%%%%%%%
\section{Nuclear Parton Distribution Functions}
%%%%%%%%%%%%%%%%%%%%%%%%%%%%%%%%%%%%%%%%%%%%%%%%%%%%%%%%%%%%%%%%%%%%%%%%%

\noindent
The parton densities for free nucleons and nuclei are different. This has been
first demonstrated in detail by the EMC-experiment \cite{Aubert:1983xm}. 
One well-known effect is due to the Fermi-motion of nucleons within nuclei
\cite{BODEKRI}. Beyond these contributions other deviations have been observed,
in the region of lower values of $x$. Various phenomenological explanations have been 
proposed, like binding models of nucleons, admixture models, change of nucleon mass,
pion enhancement,  multi-quark clusters, dynamical rescaling (change of confinement size),
shadowing, and others; for reviews see 
\cite{NUCL:REV}.
%%Rith:REV,Geesaman:1995yd,Norton:2003cb,Armesto:2006ph}.
Nuclear effects also modify the Drell-Yan process, $W^\pm/Z$--production, high 
$p_\perp$ hadron production, $J/\psi$-- and  direct photon production.

\begin{figure}[H]
\begin{center}
\includegraphics[scale=0.50,angle=0]{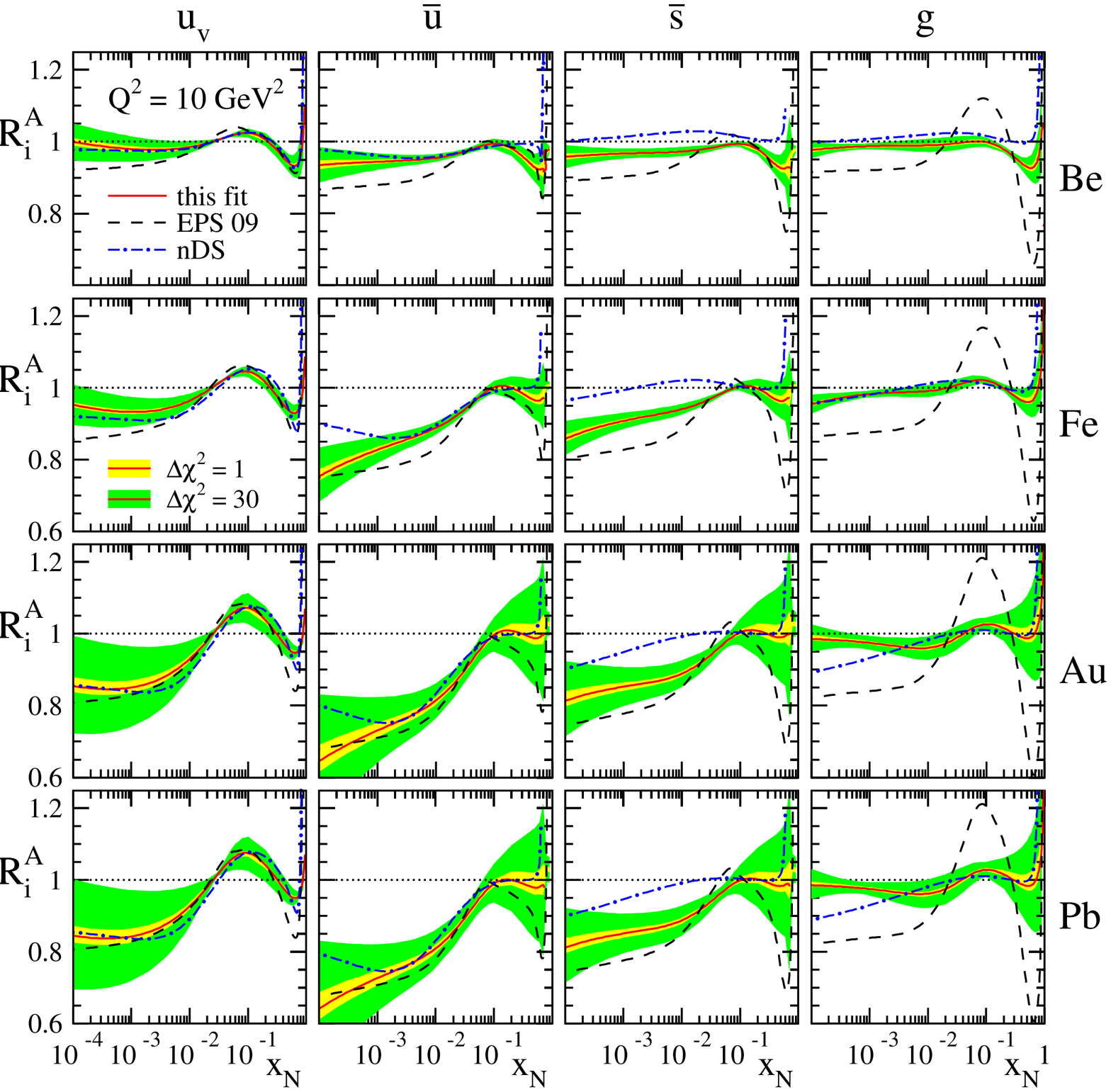}
\end{center}
\caption[]{
\label{fig:NUCL}
\small \sf
The NLO nuclear modification factors $R_i^A$ for different parton densities at $Q^2 = 10 \GeV^2$
for {\sf Be, Fe, Au} and {\sf Pb} targets as a function of $x_N$. The inner (outer) error bands
correspond to  $\Delta \chi^2 = 1 (30)$, respectively. Red lines: \cite{deFlorian:2011fp};
dash-dotted lines \cite{deFlorian:2003qf}; dashed lines \cite{Eskola:2009uj}; from
\cite{deFlorian:2011fp} \TCop (2012) by the American Physical Society.}
\end{figure}

Recent global data analyses to extract the nuclear PDFs have been carried out by different 
groups \cite{
NUC:RAT,Eskola:2009uj,Kulagin:2004ie,
Kulagin:2010gd,
deFlorian:2003qf,
deFlorian:2011fp}
extracting nuclear correction factors $R_i^A(x_N,Q^2)$ for the parton densities $f_i(x,Q^2)$ in 
free nucleons. In general the Bjorken variable $x_N$ for nuclear PDFs has support $x_N \in 
[0,A]$. However, measured distributions for values of $x_N > 1$ do rapidly fall. Therefore, one 
often parameterizes the nuclear PDFs only for the range $x_N \in [0,1]$. A recent 
parameterization was given in Ref.~\cite{deFlorian:2011fp}. The correction factors 
$R_i^A(x_N,Q^2)$ are illustrated in Figure~\ref{fig:NUCL} for a series of nuclei for the 
quark distributions $u_v, \bar{u}, s$ and the gluon distribution $g$. The $d_v$ distribution is 
assigned the same correction factor as the $u_v$ distribution. 

The nuclear corrections at very small and very large $x_N$ become stronger with the rising value 
of $A$. At large $x_N$ Fermi-motion effects dominate. Anti-shadowing effects are observed 
in the region $x_N \in [0.01, 0.1]$ and the correction factors become smaller than one for even
smaller values of $x_N$. The comparison of different fits shows that there are still quite some 
uncertainties. On the theoretical side a complete understanding of the factors $R_i^A$ is not
yet obtained, despite a large number of proposed models. 

%----------------------------------------------------------------------------------------------
\restylefloat{figure}
%----------------------------------------------------------------------------------------------   

Deep-inelastic data analyses include deuteron data to obtain an improved representation of  
the down-quark distributions due to scattering off the neutron. The deuteron data require nuclear 
corrections \cite{Arrington:2011qt}, which  mainly concern Fermi-motion and the off-shell effect, 
cf.~\cite{Alekhin:2012ig}. For the off-shell corrections the Bonn \cite{Machleidt:1987hj}
and Paris \cite{Lacombe:1980dr} potentials are used, including the induced differences into the 
systematic errors. One assumes that the nuclear model suggested in \cite{Kulagin:2004ie}
can be applied to the case of light nuclei, like deuterium, which has recently been 
confirmed for  $^3$He and $^9$Be targets \cite{Kulagin:2010gd}. Di-muon production data, 
which are 
important to measure the strange-quark distributions, are currently measured at nuclear targets, for 
which the corrections \cite{Kulagin:2007ju} were applied in \cite{Alekhin:2012ig}. Drell-Yan data of 
nuclear targets which are used in present PDF-fits are measured in a region with lower 
nuclear effects, cf.~\cite{Alde:1990im}.

%%%%%%%%%%%%%%%%%%%%%%%%%%%%%%%%%%%%%%%%%%%%%%%%%%%%%%%%%%%%%%%%%%%%%%%%%
\section{Conclusions}
%%%%%%%%%%%%%%%%%%%%%%%%%%%%%%%%%%%%%%%%%%%%%%%%%%%%%%%%%%%%%%%%%%%%%%%%%

\noindent
During the last 40 years the understanding of the sub-structure of nucleons took a vast
development. The theory of strong interactions, QCD, can now be probed at the per-cent 
level in an increasing number of reactions. High-luminosity measurements performed at 
SLAC, CERN, DESY, FERMILAB, and JLAB more and more revealed the dynamics of quarks and 
gluons at 
short distances. With the advent of the LHC this process will continue in an intimate 
partnership between precision measurements and precision calculations. During the 
coming years the data analysis of the HERA-experiments will be finalized. Here a remaining 
issue is the theoretical understanding of the $ep \rightarrow \text{\rm jet}$ 
cross sections at NNLO. Jet physics, the production of $W^\pm$ and $Z$-bosons, the 
Drell-Yan cross section, and inclusive $t\bar{t}$- as well as single-top production
will be important scattering cross sections measured at the LHC to limit the present 
experimental errors on $\alpha_s(M_Z^2)$ and to further refine the knowledge on the 
parton distributions. The precision of the data requests NNLO QCD corrections.

In the more distant future high luminosity facilities like EIC may allow to answer many 
open questions related to lower energies, like the partonic structure of polarized 
nucleons, twist-3 effects, and higher twist effects in general, and unravel transverse 
degrees of freedom, including associated spin effects in detail. This machine 
will also have the potential to measure $\alpha_s(M_Z^2)$ very precisely.
Proposed facilities like LHeC, on the other hand, allow to further progress into the 
small $x$ and high $Q^2$ domain by $ep$ scattering. Many proposed theoretical concepts 
can be tested in this way at high precision, giving a boost to even more refined
theoretical calculations than carried out at present. Probing ever shorter distances
may finally answer the more fundamental question for a possible substructures of  
quarks, which will request precision, both at the experimental and theoretical side.

\vspace*{2mm}\noindent
{\bf Acknowledgment.}~For discussions I would like to thank 
J.~Ablinger,
S.~Alekhin,
W.~Beenakker,
Chr.~Berger,
I.~Bierenbaum,
J.D.~Bjorken,
H.~B\"ottcher,
S.~Catani,
V.~Drach,
A.~De Freitas,
H.~Fritzsch,
J.~Gracey,
A.~Hasselhuhn,
P.~Jimenez-Delgado,
M.~Klein,
S.~Klein,
S.~Moch,
A.~Petermann,
H.~Quinn,
V.~Ravindran,
E.~Reya,
T.~Riemann,
K.~Sasaki,
J.~Smith,
C.~Schneider,
P.~S\"oding,
T.~Uematsu,
W.L.~van Neerven,
M.~Veltman,
J.~Vermaseren,
A.~Vogt, and
F.~Wi{\ss{}}brock. This work has been supported in part by DFG Sonderforschungsbereich Transregio 9,
Computergest\"utzte Theoretische Teilchenphysik, and EU Network {\sf LHCPHENOnet}
PITN-GA-2010-264564. Figure~4 has been drawn using {\tt Axodraw} \cite{Vermaseren:1994je}.
\nopagebreak[4]
{\small
%-------------------------------------------------------------------------------------------------

%%%%%%%%%%%%%%%%%%%%%%%%%%%%%%%%%%%%%%%%%%%%%%%%%%%%%%%%%%%%%%%%%%%%%%%%%
}

\begin{thebibliography}{99}
%-------------------------------------------------------------------------------------------------
%----------------------------------------------------------------------
%%             >> section 1 <<
%----------------------------------------------------------------------
%
%[1]
\bibitem{DALTON}
J. Dalton, {\sf A New System of Chemical Philosophy}, (A. Bickerstaff, Strand, London, 1808).
%---------------------------------------------------------------------------------------------
%
%[2]
\bibitem{Becquerel}
H. Becquerel, {\em Comptes Rendus} {\bf 122} (1896) 420; 501; 559; 689; 762.
%---------------------------------------------------------------------------------------------
%
%[3]
\bibitem{RUTHERF}
E. Rutherford,
%The Scattering of $\alpha$ and $\beta$ Particles by Matter and the Structure of the Atom,
{\em Phil. Mag. Ser.} 6 {\bf 21} (1911)  669.  %-688
%---------------------------------------------------------------------------------------------
%
%[4]
\bibitem{CHADWICK}
J. Chadwick,
%Possible Existence of a Neutron. In:
{\em Nature} {\bf 129} (1932)
%Nr. 3252, 27. Februar 1932, S.
312;
%James Chadwick:
%The Existence of a Neutron. In:
{\em Proc. Royal Soc. Lond.} Ser. {\bf A136} (1932)
%Containing Papers of a Mathematical and Physical Character. 136, Nr. 830, 1. Mai 1932, S.
692.  %–708
%---------------------------------------------------------------------------------------------
%
%[5]  
\bibitem{YUKAWA}
H. Yukawa, {\em Proc. Phys.-Math. Soc. Japan}  {\bf 17} (1935) 48;
%H. Yukawa, Proc. Phys.-Math. Soc. Japan,
{\bf 19}  (1937) 712.
%---------------------------------------------------------------------------------------------
%
%[6]  
\bibitem{ANOMp}
R. Frisch and O. Stern, {\em Z. Phys.} {\bf 85} (1933) 4.
%---------------------------------------------------------------------------------------------
%
%[7]  
\bibitem{ANOMn}
R. Bacher, {\em Phys. Rev.} {\bf 43} (1933) 1001;\\ %% -1002 (1933)
L. Alvarez and F. Bloch, {\em Phys. Rev.} {\bf 57} (1940) 111.
%---------------------------------------------------------------------------------------------
%
%[8]  
%\cite{Nakamura:2010zzi}
\bibitem{Nakamura:2010zzi}
  K.~Nakamura {\it et al.},  [Particle Data Group Collaboration],
  %``Review of particle physics,''
  {\em J.\ Phys.}\  G {\bf 37} (2010) 075021.
  %%CITATION = JPHGB,G37,075021;%%
%---------------------------------------------------------------------------------------------
%
%[9]  
\bibitem{Hofstadter:1963}
R.~Hofstadter, {\sf Electron scattering and nuclear and nucleon structure. A
  collection of reprints with an introduction}, (New York, Benjamin, 1963),
  690~p. and references therein.
%---------------------------------------------------------------------------------------------
%
%[10]  
%\cite{Olson:1961zz}
\bibitem{Olson:1961zz}
  D.~N.~Olson, H.~F.~Schopper and R.~R.~Wilson,
  %``Electromagnetic Properties of the Proton and Neutron,''
  {\em Phys.\ Rev.\ Lett.}\  {\bf 6} (1961) 286.
  %%CITATION = PRLTA,6,286;%%
%---------------------------------------------------------------------------------------------
%
%[11]
%\cite{GellMann:1964nj}
\bibitem{GellMann:1964nj}
  M.~Gell-Mann,
  %``A Schematic Model of Baryons and Mesons,''
  {\em Phys.\ Lett.}\  {\bf 8} (1964) 214.
  %%CITATION = PHLTA,8,214;%%
%--------------------------------------------------------------------------------
%
%[12]
%\cite{Zweig:1964jf}
\bibitem{Zweig:1964jf}
  G.~Zweig,
  %``An Su(3) Model For Strong Interaction Symmetry And Its Breaking. 2.,''
  CERN-TH-401, CERN-TH-412, (1964).
  %%CITATION = CERN-TH-412;%%
%--------------------------------------------------------------------------------
%
%[13]  
\bibitem{Panofsky:1968pb}
W.~K.~H. Panofsky, 
%{\sf Low $q^2$ electrodynamics, elastic and inelastic
%  electron (and muon) scattering}, 
Proc. 14th International Conference on
  High-Energy Physics, Vienna, 1968, J.~Prentki and J.~Steinberger, eds.,
  (CERN, Geneva, 1968), pp.~23.
%----------------------------------------------------------------------------------------------
%
%[14]  
\bibitem{Taylor:1969xi}
R.~E. Taylor, %{\sf Inelastic electron - proton scattering in the deep continuum
%  region}, 
Proc. 4th International Symposium on Electron and Photon
  Interactions at High Energies, Liverpool, 1969, (Daresbury Laboratory, 1969),
  eds. D.W.~Braben and R.E.~Rand, pp. 251.
%----------------------------------------------------------------------------------------------
%
%[15]
%\cite{Bloom:1969kc}
\bibitem{Bloom:1969kc}
  E.~D.~Bloom
  %,D.~H.~Coward, H.~C.~DeStaebler, J.~Drees, G.~Miller, L.~W.~Mo, R.~E.~Taylor and M.~Breidenbach 
  {\it et al.},
  %``High-Energy Inelastic e p Scattering at 6-Degrees and 10-Degrees,''
  {\em Phys.\ Rev.\ Lett.}\  {\bf 23} (1969) 930.
  %%CITATION = PRLTA,23,930;%%
%--------------------------------------------------------------------------------
%
%[16]
%\cite{Breidenbach:1969kd}
\bibitem{Breidenbach:1969kd}
  M.~Breidenbach 
  %,J.~I.~Friedman, H.~W.~Kendall, E.~D.~Bloom, D.~H.~Coward, H.~C.~DeStaebler, J.~Drees and L.~W.~Mo 
  {\it et al.},
  %``Observed Behavior of Highly Inelastic electron-Proton Scattering,''
  {\em Phys.\ Rev.\ Lett.}\  {\bf 23} (1969) 935.
  %%CITATION = PRLTA,23,935;%%
%--------------------------------------------------------------------------------
%
%[17]  
\bibitem{Taylor:1991ew}
R.~E. Taylor, {\em Rev. Mod. Phys.} {\bf 63}, 573 (1991).
%----------------------------------------------------------------------------------------------
%
%[18]  
\bibitem{Kendall:1991np}
H.~W. Kendall, {\em Rev. Mod. Phys.} {\bf 63}, 597 (1991).
%----------------------------------------------------------------------------------------------
%
%[19]  
\bibitem{Friedman:1991nq}
J.~I. Friedman, {\em Rev. Mod. Phys.} {\bf 63}, 615 (1991).
%----------------------------------------------------------------------------------------------
%
%[20]
%\cite{Callan:1969uq}
\bibitem{Callan:1969uq}
  C.~G.~Callan, Jr. and D.~J.~Gross,
  %``High-energy electroproduction and the constitution of the electric current,''
  {\em Phys.\ Rev.\ Lett.}\  {\bf 22} (1969) 156.
  %%CITATION = PRLTA,22,156;%%
%--------------------------------------------------------------------------------
%
%[21]
%\cite{Bjorken:1968dy}
\bibitem{Bjorken:1968dy}
  J.~D.~Bjorken,
  %``Asymptotic Sum Rules at Infinite Momentum,''
  {\em Phys.\ Rev.}\  {\bf 179} (1969) 1547.
  %%CITATION = PHRVA,179,1547;%%
%--------------------------------------------------------------------------------
%
%[22]
\bibitem{FEYNMAN}
%\cite{Feynman:1969wa}
%\bibitem{Feynman:1969wa}
  R.~P.~Feynman,
  %``The Behavior Of Hadron Collisions At Extreme Energies,''
 {In: Proceedings of 3rd International Conference on High Energy Collisions, Stony Brook, N.Y., 5-6 
 Sep 1969, pp 237.};
  %%CITATION = CONFP,C690905,237;%%
%\cite{Feynman:1969ej}
%\bibitem{Feynman:1969ej}
%  R.~P.~Feynman,
  %``Very high-energy collisions of hadrons,''
  {\em Phys.\ Rev.\ Lett.}\  {\bf 23} (1969) 1415.
  %%CITATION = PRLTA,23,1415;%%
%--------------------------------------------------------------------------------
%
%[23]
%\cite{Feynman:1973xc}
\bibitem{Feynman:1973xc}
  R.~P.~Feynman,
  {\sf Photon-hadron interactions}, 
  (Benjamin, Reading, MA, 1972), 282~p.
%--------------------------------------------------------------------------------
%
%[24]
\bibitem{BJ-HQ}
%\cite{Bjorken:1967ur}
%\bibitem{Bjorken:1967ur}
  J.~D.~Bjorken,
  %{\sf Theoretical ideas on inelastic electron and muon scattering}, 
  Proc. of the 1967 Int. Symposium on Electron and Photon Interactions at High Energies,
  CONF.-670923, UC-34-Physics, Stanford,CA, Sept. 1967, ed. S.M.~Berman, pp.~109;\\
  %--127
%\bibitem{Quinn:1967} 
H.~R.~Quinn, {\sf 
Radiative Corrections To Beta Decay And Some Sum Rules For Neutrino Interactions}, PhD Thesis,
Stanford University, Sept.~1967, RX-327, 93pp.
%--------------------------------------------------------------------------------
%
%[25]
\bibitem{Hand:1967}
  L. Hand,
  %{\sf Experiments on Inelastic Electron Scattering},
  Proc. of the 1967 Int. Symposium on Electron and Photon Interactions at High Energies,
  CONF.-670923, UC-34-Physics, Stanford,CA, Sept. 1967, ed. S.M.~Berman, pp.~128.%--155
%--------------------------------------------------------------------------------
%
%[26]
%\cite{Brasse:1972wk}
\bibitem{Brasse:1972wk}
  F.~W.~Brasse, {\it et al.},
%E.~Chazelas, W.~Fehrenbach, K.~H.~Frank, E.~Ganssauge, J.~Gayler, V.~Korbel and J.~May 
%  {\it et al.},
  %``Analysis of photo and electroproduction data against omega(w),''
  {\em Nucl.\ Phys.}\ B {\bf 39} (1972) 421.
  %%CITATION = NUPHA,B39,421;%%
%--------------------------------------------------------------------------------
%
%[27]
\bibitem{Nambu:1966}
Y.~Nambu, %{\sf A Systematics Of Hadrons In Subnuclear Physics}, 
  in: {\sf Preludes in Theoretical Physics}, eds. A. De-Shalit, H.~Fehsbach and L. van Hove
  (North-Holland, Amsterdam, 1966), pp. 133.
%--------------------------------------------------------------------------------
%
%[28]
%\cite{Yang:1954ek}
\bibitem{Yang:1954ek}
  C.~-N.~Yang and R.~L.~Mills,
  %``Conservation of Isotopic Spin and Isotopic Gauge Invariance,''
  {\em Phys.\ Rev.}\  {\bf 96} (1954) 191.
  %%CITATION = PHRVA,96,191;%%
%--------------------------------------------------------------------------------
%
%[29]
%\cite{Han:1965pf}
\bibitem{Han:1965pf}
  M.~Y.~Han and Y.~Nambu,
  %``Three-triplet model with double SU(3) symmetry,''
  {\em Phys.\ Rev.}\  {\bf 139} (1965) B1006.
  %%CITATION = PHRVA,139,B1006;%%
%--------------------------------------------------------------------------------
%
%[30]
%\cite{Greenberg:1964pe}
\bibitem{Greenberg:1964pe}
  O.~W.~Greenberg,
  %``Spin And Unitary Spin Independence In A Paraquark Model Of Baryons And
  %Mesons,''
  {\em Phys.\ Rev.\ Lett.}\  {\bf 13} (1964) 598.
  %%CITATION = PRLTA,13,598;%%
%--------------------------------------------------------------------------------
%
%[31]
%\cite{Faddeev:1967fc}
\bibitem{Faddeev:1967fc}
  L.~D.~Faddeev and V.~N.~Popov,
  %``Feynman Diagrams for the Yang-Mills Field,''
  {\em Phys.\ Lett.}\ B {\bf 25} (1967) 29.
  %%CITATION = PHLTA,B25,29;%%
%--------------------------------------------------------------------------------
%
%[32]
%\cite{'tHooft:1971fh}
\bibitem{'tHooft:1971fh}
  G.~'t Hooft,
  %``Renormalization of Massless Yang-Mills Fields,''
  {\em Nucl.\ Phys.}\ B {\bf 33} (1971) 173.
  %%CITATION = NUPHA,B33,173;%%
%--------------------------------------------------------------------------------
%
%[33]
\bibitem{Fritzsch:1972jv}
  H.~Fritzsch and M.~Gell-Mann, %{\sf Current algebra: Quarks and what else?},
  Proceedings of 16th International Conference on High-Energy Physics, Batavia,
  Illinois, 6-13 Sep Vol. {\bf 2}, J.D.~Jackson, A.~Roberts, R.~Donaldson,
  eds., pp. 135 (1972), hep-ph/0208010.
%%CITATION = HEP-PH/0208010;%%
%--------------------------------------------------------------------------------
%
%[34]
%\cite{Fritzsch:1973pi}
\bibitem{Fritzsch:1973pi}
  H.~Fritzsch, M.~Gell-Mann and H.~Leutwyler,
  %``Advantages of the Color Octet Gluon Picture,''
  {\em Phys.\ Lett.}\ B {\bf 47} (1973) 365.
  %%CITATION = PHLTA,B47,365;%%
%--------------------------------------------------------------------------------
%
%[35]
\bibitem{Gross:1973id}
D.~J. Gross and F.~Wilczek, {\em Phys. Rev. Lett.} {\bf 30} (1973) 1343. 
%--------------------------------------------------------------------------------
%
%[36]
\bibitem{Politzer:1973fx}
H.~D. Politzer, {\em Phys. Rev. Lett.} {\bf 30} (1973) 1346.
%--------------------------------------------------------------------------------
%
%[37]
\bibitem{Khriplovich:1969aa}
I.~B. Khriplovich, {\em Yad. Fiz.} {\bf 10}  (1969) 409.
%--------------------------------------------------------------------------------
%
%[38]
\bibitem{tHooft:unpub}
G.~t'Hooft (1972), {\rm unpublished}, cf.~Proc. Colloquium on
Renormalization of Yang-Mills Fields and Application to Particle Physics,
Marseille, 1972, ed.~C.P.~Korthals-Altes; %%G.~t'Hooft,
{\it Nucl. Phys.} B {\bf 254} (1985) 11.
%--------------------------------------------------------------------------------
%
%[39]
\bibitem{YND}
F.J. Yndurain, {\sf The Theory of Quark and Gluon Interactions}, (Springer, Berlin, 1999).
%--------------------------------------------------------------------------------
%
%[40]
\bibitem{Veltman:1963}
M.~J.~G. Veltman {\sf SCHOONSCHIP, version 1, Dec. 1963}.
%--------------------------------------------------------------------------------
%
%[41]
%\cite{Wilson:1969zs}
\bibitem{Wilson:1969zs}
  K.~G.~Wilson,
  %``Nonlagrangian models of current algebra,''
  {\em Phys.\ Rev.}\  {\bf 179} (1969) 1499.
  %%CITATION = PHRVA,179,1499;%%
%--------------------------------------------------------------------------------
%
%[42]
\bibitem{Zimmermann:1970}
W. Zimmermann, {\sf Lectures on Elementary Particle Physics and Quantum Field Theory}, Brandeis
Summer Institute, Vol.~{\bf 1} (MIT Press, Cambridge, 1970),~p.~395.
%--------------------------------------------------------------------------------
%
%[43]
%\cite{Brandt:1970kg}
\bibitem{Brandt:1970kg}
  R.~A.~Brandt and G.~Preparata,
  %``Operator product expansions near the light cone,''
  {\em Nucl.\ Phys.}\ B {\bf 27} (1972) 541;\\
  %%CITATION = NUPHA,B27,541;%%
%\cite{Christ:1972ms}
%\bibitem{Christ:1972ms}
  N.~H.~Christ, B.~Hasslacher and A.~H.~Mueller,
  %``Light cone behavior of perturbation theory,''
  {\it Phys.\ Rev.}\ D {\bf 6} (1972) 3543.
  %%CITATION = PHRVA,D6,3543;%%
%--------------------------------------------------------------------------------
%
%[44]
%\cite{Frishman:1971qn}
\bibitem{Frishman:1971qn}
  Y.~Frishman,
  %``Operator products at almost light like distances,''
  {\em Annals Phys.}\  {\bf 66} (1971) 373.
  %%CITATION = APNYA,66,373;%%
%--------------------------------------------------------------------------------
%
%[45]
%\cite{Gross:1971wn}
\bibitem{Gross:1971wn}
  D.~J.~Gross and S.~B.~Treiman,
  %``Light cone structure of current commutators in the gluon quark model,''
  {\em Phys.\ Rev.}\ D {\bf 4} (1971) 1059.
  %%CITATION = PHRVA,D4,1059;%%
%--------------------------------------------------------------------------------
%
%[46]
\bibitem{Stueckelberg:1953dz}
  E.~C.~G.~Stueckelberg and A.~Petermann,
  %``Normalization of constants in the quanta theory,''
  {\it Helv.\ Phys.}\ Acta {\bf 26} (1953) 499.
  %%CITATION = HPACA,26,499;%%
%--------------------------------------------------------------------------------
%
%[47]
%\cite{Symanzik:1970rt}
\bibitem{Symanzik:1970rt}
  K.~Symanzik,
  %``Small distance behavior in field theory and power counting,''
  {\em Commun.\ Math.\ Phys.}\  {\bf 18} (1970) 227.
  %%CITATION = CMPHA,18,227;%%
%--------------------------------------------------------------------------------
%
%[48]
%\cite{Callan:1970yg}
\bibitem{Callan:1970yg}
  C.~G.~Callan, Jr.,
  %``Broken scale invariance in scalar field theory,''
  {\em Phys.\ Rev.}\ D {\bf 2} (1970) 1541.
  %%CITATION = PHRVA,D2,1541;%%
%--------------------------------------------------------------------------------
%
%[49]
\bibitem{Petermann:1979}
A. Petermann, {\em Phys.\ Rept.} {\bf 53} (1979) 157.
%--------------------------------------------------------------------------------
%
%[50]
\bibitem{DRELL60}
%\cite{Drell:1969jm}
%\bibitem{Drell:1969jm}
  S.~D.~Drell, D.~J.~Levy and T.~-M.~Yan,
  %``A Theory of Deep Inelastic Lepton-Nucleon Scattering and Lepton Pair Annihilation Processes. 1.,''
  {\em Phys.\ Rev.}\  {\bf 187} (1969) 2159;
  %%CITATION = PHRVA,187,2159;%%
%\cite{Drell:1969wd}
%\bibitem{Drell:1969wd}
%  S.~D.~Drell, D.~J.~Levy and T.~-M.~Yan,
  %``A Theory of Deep Inelastic Lepton Nucleon Scattering and Lepton Pair Annihilation Processes. 2. Deep Inelastic electron Scattering,''
%  {\em Phys.\ Rev.}\ D {\bf 1} (1970) 
1035;
  %%CITATION = PHRVA,D1,1035;%%
%\cite{Drell:1969wb}
%\bibitem{Drell:1969wb}
%  S.~D.~Drell, D.~J.~Levy and T.~-M.~Yan,
  %``A Theory of Deep Inelastic Lepton-Nucleon Scattering and Lepton Pair Annihilation Processes. 3. Deep Inelastic electron-Positron Annihilation,''
%  {\em Phys.\ Rev.}\ D {\bf 1} (1970) 
1617.
  %%CITATION = PHRVA,D1,1617;%%
%-----------------------------------------------------------------------
%
%[51]
%\cite{Yan:1969gn}
\bibitem{Yan:1969gn}
  T.~-M.~Yan and S.~D.~Drell,
  %``A Theory Of Deep Inelastic Lepton - Nucleon Scattering And Lepton Pair Annihilation Processes. 4. Deep Inelastic Neutrino Scattering,''
  {\em Phys.\ Rev.}\ D {\bf 1} (1970) 2402.
  %%CITATION = PHRVA,D1,2402;%%
%-----------------------------------------------------------------------
%
%[52]
\bibitem{DISS:WVN}
  W.L. van Neerven, {\sf The infinite momentum frame method in field theory with some applications},
  PhD Thesis, Radboud University Nijmegen, 1975.
%-----------------------------------------------------------------------
%
%[53]
%\cite{vanRitbergen:1997va}
\bibitem{vanRitbergen:1997va}
  T.~van Ritbergen, J.~A.~M.~Vermaseren and S.~A.~Larin,
  %``The Four loop beta function in quantum chromodynamics,''
  {\em Phys.\ Lett.}\ B {\bf 400} (1997) 379
  [hep-ph/9701390].
  %%CITATION = HEP-PH/9701390;%%
%-----------------------------------------------------------------------
%
%[54]
%\cite{Czakon:2004bu}
\bibitem{Czakon:2004bu}
  M.~Czakon,
  %``The Four-loop QCD beta-function and anomalous dimensions,''
  {\em Nucl.\ Phys.}\ B {\bf 710} (2005) 485
  [hep-ph/0411261].
  %%CITATION = HEP-PH/0411261;%%
%-----------------------------------------------------------------------
%
%[55]
%\cite{Larin:1993vu}
\bibitem{Larin:1993vu}
  S.~A.~Larin, T.~van Ritbergen and J.~A.~M.~Vermaseren,
  %``The Next next-to-leading QCD approximation for nonsinglet moments of deep inelastic structure functions,''
  {\em Nucl.\ Phys.}\ B {\bf 427} (1994) 41.
  %%CITATION = NUPHA,B427,41;%%
%-----------------------------------------------------------------------
%
%[56]
%\cite{Larin:1996wd}
\bibitem{Larin:1996wd}
  S.~A.~Larin, P.~Nogueira, T.~van Ritbergen and J.~A.~M.~Vermaseren,
  %``The Three loop QCD calculation of the moments of deep inelastic structure functions,''
  {\em Nucl.\ Phys.}\ B {\bf 492} (1997) 338
  [hep-ph/9605317].
  %%CITATION = HEP-PH/9605317;%%
%-----------------------------------------------------------------------
%
%[57]
%\cite{Retey:2000nq}
\bibitem{Retey:2000nq}
  A.~Retey and J.~A.~M.~Vermaseren,
  %``Some higher moments of deep inelastic structure functions at next-to-next-to-leading order of perturbative QCD,''
  {\em Nucl.\ Phys.}\ B {\bf 604} (2001) 281
  [hep-ph/0007294].
  %%CITATION = HEP-PH/0007294;%%
%-----------------------------------------------------------------------
%
%[58]
%\cite{Blumlein:2004xt}
\bibitem{Blumlein:2004xt}
  J.~Bl\"umlein and J.~A.~M.~Vermaseren,
  %``The 16th moment of the non-singlet structure functions F(2)(x,Q**2) and F(L)(x,Q**2) to O(alpha**3(S)),''
  {\em Phys.\ Lett.}\ B {\bf 606} (2005) 130
  [hep-ph/0411111].
  %%CITATION = HEP-PH/0411111;%%
%-----------------------------------------------------------------------
%
%[59]
%\cite{Moch:2004pa}
\bibitem{Moch:2004pa}
  S.~Moch, J.~A.~M.~Vermaseren and A.~Vogt,
  %``The Three loop splitting functions in QCD: The Nonsinglet case,''
  {\em Nucl.\ Phys.}\ B {\bf 688} (2004) 101
  [hep-ph/0403192].
  %%CITATION = HEP-PH/0403192;%%
%-----------------------------------------------------------------------
%
%[60]
%\cite{Vogt:2004mw}
\bibitem{Vogt:2004mw}
  A.~Vogt, S.~Moch and J.~A.~M.~Vermaseren,
  %``The Three-loop splitting functions in QCD: The Singlet case,''
  {\em Nucl.\ Phys.}\ B {\bf 691} (2004) 129
  [hep-ph/0404111].
  %%CITATION = HEP-PH/0404111;%%
%-----------------------------------------------------------------------
%
%[61]
%\cite{Vermaseren:2005qc}
\bibitem{Vermaseren:2005qc}
  J.~A.~M.~Vermaseren, A.~Vogt and S.~Moch,
  %``The Third-order QCD corrections to deep-inelastic scattering by photon exchange,''
  {\em Nucl.\ Phys.}\ B {\bf 724} (2005) 3
  [hep-ph/0504242].
  %%CITATION = HEP-PH/0504242;%%
%-----------------------------------------------------------------------
%
%[62]
%\cite{Baikov:2006ai}
\bibitem{Baikov:2006ai}
  P.~A.~Baikov and K.~G.~Chetyrkin,
  %``New four loop results in QCD,''
  {\em Nucl.\ Phys.\ (Proc.\ Suppl.)}\  {\bf 160} (2006) 76.
  %%CITATION = NUPHZ,160,76;%%
%------------------------------------------------------------------------
%
%[63]
\bibitem{CHET_priv1}
K.~G.~Chetyrkin, private communication, 2010.
%------------------------------------------------------------------------
%
%[64]
%\cite{Velizhanin:2011es}
\bibitem{Velizhanin:2011es}
  V.~N.~Velizhanin,
  %``Four loop anomalous dimension of the second moment of the non-singlet twist-2 operator in QCD,''
  {\it Nucl.\ Phys.}\ B {\bf 860} (2012) 288
  [arXiv:1112.3954 [hep-ph]].
  %%CITATION = ARXIV:1112.3954;%%
%------------------------------------------------------------------------
%
%[65]
\bibitem{HEAV2}
%\cite{Laenen:1992zk}
%\bibitem{Laenen:1992zk}
  E.~Laenen, S.~Riemersma, J.~Smith and W.~L.~van Neerven,
  %``Complete O (alpha-s) corrections to heavy flavor structure functions in electroproduction,''
  {\em Nucl.\ Phys.}\ B {\bf 392} (1993) 162;
  %%CITATION = NUPHA,B392,162;%%
%\cite{Laenen:1992xs}
%\bibitem{Laenen:1992xs}
%  E.~Laenen, S.~Riemersma, J.~Smith and W.~L.~van Neerven,  %``O(alpha-s) corrections to heavy flavor inclusive distributions in electroproduction,''
%  Nucl.\ Phys.\ B {\bf 392} (1993) 
229.
  %%CITATION = NUPHA,B392,229;%%
%------------------------------------------------------------------------
%
%[66]
%\cite{Riemersma:1994hv}
\bibitem{Riemersma:1994hv}
  S.~Riemersma, J.~Smith and W.~L.~van Neerven,
  %``Rates for inclusive deep inelastic electroproduction of charm quarks at HERA,''
  {\em Phys.\ Lett.}\ B {\bf 347} (1995) 143
  [hep-ph/9411431].
  %%CITATION = HEP-PH/9411431;%%
%------------------------------------------------------------------------
%
%[67]
%\cite{Bierenbaum:2009mv}
\bibitem{Bierenbaum:2009mv}
  I.~Bierenbaum, J.~Bl\"umlein and S.~Klein,
  %``Mellin Moments of the O(alpha**3(s)) Heavy Flavor Contributions to unpolarized Deep-Inelastic Scattering at Q**2 >> m**2 and Anomalous Dimensions,''
  {\em Nucl.\ Phys.}\ B {\bf 820} (2009) 417
  [arXiv:0904.3563 [hep-ph]].
  %%CITATION = ARXIV:0904.3563;%%
%------------------------------------------------------------------------
%
%[68]
%\cite{Zijlstra:1993sh}
\bibitem{Zijlstra:1993sh}
  E.~B.~Zijlstra and W.~L.~van Neerven,
  %``Order alpha-s**2 corrections to the polarized structure function g1 (x,Q**2),''
  {\em Nucl.\ Phys.}\ B {\bf 417} (1994) 61
   [Erratum-ibid.\ B {\bf 426} (1994) 245]
   [Erratum-ibid.\ B {\bf 773} (2007) 105].
  %%CITATION = NUPHA,B417,61;%%
%------------------------------------------------------------------------
%
%[69]
%\cite{Mertig:1995ny}
\bibitem{Mertig:1995ny}
  R.~Mertig and W.~L.~van Neerven,
  %``The Calculation of the two loop spin splitting functions P(ij)(1)(x),''
  {\em Z.\ Phys.}\ C {\bf 70} (1996) 637
  [hep-ph/9506451].
  %%CITATION = HEP-PH/9506451;%%
%------------------------------------------------------------------------
%
%[70]
\bibitem{VOGELS:POL}
%\cite{Vogelsang:1995vh}
%\bibitem{Vogelsang:1995vh}
  W.~Vogelsang,
  %``A Rederivation of the spin dependent next-to-leading order splitting functions,''
  {\em Phys.\ Rev.}\ D {\bf 54} (1996) 2023
  [hep-ph/9512218];
  %%CITATION = HEP-PH/9512218;%%
%\cite{Vogelsang:1996im}
%\bibitem{Vogelsang:1996im}
%  W.~Vogelsang,
  %``The Spin dependent two loop splitting functions,''
  {\em Nucl.\ Phys.}\ B {\bf 475} (1996) 47
  [hep-ph/9603366].
  %%CITATION = HEP-PH/9603366;%%
%------------------------------------------------------------------------
%
%[71]
%\cite{Buza:1996xr}
\bibitem{Buza:1996xr}
  M.~Buza, Y.~Matiounine, J.~Smith and W.~L.~van Neerven,
  %``O (alpha-s**2) corrections to polarized heavy flavor production at Q**2 is >> m**2,''
  {\em Nucl.\ Phys.}\ B {\bf 485} (1997) 420
  [hep-ph/9608342].
  %%CITATION = HEP-PH/9608342;%%
%------------------------------------------------------------------------
%
%[72]
%\cite{Bierenbaum:2007pn}
\bibitem{Bierenbaum:2007pn}
  I.~Bierenbaum, J.~Bl\"umlein and S.~Klein,
  %``Two-loop massive operator matrix elements for polarized and unpolarized deep-inelastic scattering,''
  {\it PoS (ACAT)} 070, arXiv:0706.2738 [hep-ph].
  %%CITATION = ARXIV:0706.2738;%%
%------------------------------------------------------------------------ 
%
%[73]
%\cite{Vogt:2008yw}
\bibitem{Vogt:2008yw}
  A.~Vogt, S.~Moch, M.~Rogal and J.~A.~M.~Vermaseren,
  %``Towards the NNLO evolution of polarised parton distributions,''
  {\em Nucl.\ Phys.\ (Proc.\ Suppl.)}\  {\bf 183} (2008) 155
  [arXiv:0807.1238 [hep-ph]].
  %%CITATION = ARXIV:0807.1238;%%
%------------------------------------------------------------------------ 
%
%[74]
%\cite{Vermaseren:2000nd}
\bibitem{Vermaseren:2000nd}
  J.~A.~M.~Vermaseren,
  %``New features of FORM,''
  math-ph/0010025.
  %%CITATION = MATH-PH/0010025;%%
%------------------------------------------------------------------------ 
%
%[75]
%\cite{Steinhauser:2000ry}
\bibitem{Steinhauser:2000ry}
  M.~Steinhauser,
  %``MATAD: A Program package for the computation of MAssive TADpoles,''
  {\it Comput.\ Phys.\ Commun.}\  {\bf 134} (2001) 335
  [hep-ph/0009029].
  %%CITATION = HEP-PH/0009029;%%
%------------------------------------------------------------------------ 
%
%[76]
\bibitem{SIGMA}
C. Schneider, {\it J. Symbolic Comput.} {\bf 43} (2008) 611, \newblock [arXiv:0808.2543v1]; {\it Ann. 
Comb.} {\bf 9} 
(2005) 75;
{\it J. Differ. Equations Appl. }{\bf 11} (2005) 799; {\it Ann. Comb. } {\bf 14} (4) (2010),
[arXiv:0808.2596]; Proceedings of the Workshop ``Motives, Quantum Field Theory, and Pseudodifferential
Operators'', held at the Clay Mathematics
       Institute, Boston University, June 2--13, 2008,
       Clay Mathematics Proceedings {\bf 12} (2010) 285-308, Eds. A.~Carey,
       D.~Ellwood, S.~Paycha, S. Rosenberg;
{\it S\'em.~Lothar. Combin.} {\bf 56} (2007) 1, Article B56b,  Habilitationsschrift JKU Linz (2007)
and references therein;\\
%\cite{Ablinger:2010pb}
%\bibitem{Ablinger:2010pb}
  J.~Ablinger, J.~Bl\"umlein, S.~Klein, C.~Schneider,
  %``Modern Summation Methods and the Computation of 2- and 3-loop Feynman Diagrams,''
  {\it Nucl.\ Phys.\ (Proc.\ Suppl.)}\  {\bf 205-206 } (2010)  110%-115.
  [arXiv:1006.4797 [math-ph]].
%-----------------------------------------------------------------------------------
%
%[77]
%\cite{Harlander:1998dq}
\bibitem{Harlander:1998dq}
  R.~Harlander and M.~Steinhauser,
  %``Automatic computation of Feynman diagrams,''
  {\it Prog.\ Part.\ Nucl.\ Phys.}\  {\bf 43} (1999) 167
  [hep-ph/9812357].
  %%CITATION = HEP-PH/9812357;%%
%------------------------------------------------------------------------ 
%
%[78]
\bibitem{EMCp1} 
%\cite{Ashman:1987hv} 
%\bibitem{Ashman:1987hv} 
  J.~Ashman {\it et al.},  [European Muon Collaboration],
  %``A measurement of the spin asymmetry and determination of the structure  
  %function g(1) in deep inelastic muon proton scattering,''
  {\it Phys.\ Lett.}\  B {\bf 206} (1988) 364.
  %%CITATION = PHLTA,B206,364;%%
%----------------------------------------------------------------------
%
%[79]
\bibitem{NONF}
%\cite{Mueller:1998fv}
%\bibitem{Mueller:1998fv}
  D.~M\"uller, D.~Robaschik, B.~Geyer, F.~M.~Dittes and J.~Horejsi,
  %``Wave functions, evolution equations and evolution kernels from light ray operators of QCD,''
  {\em Fortsch.\ Phys.}\  {\bf 42} (1994) 101
  [hep-ph/9812448];\\
  %%CITATION = HEP-PH/9812448;%%
%------------------------------------------------------------------------
%\cite{Diehl:2003ny}
%\bibitem{Diehl:2003ny}
  M.~Diehl,
  %``Generalized parton distributions,''
  {\em Phys.\ Rept.}\  {\bf 388} (2003) 41
  [hep-ph/0307382];\\
  %%CITATION = HEP-PH/0307382;%%
%------------------------------------------------------------------------
%\cite{Belitsky:2005qn}
%\bibitem{Belitsky:2005qn}
  A.~V.~Belitsky and A.~V.~Radyushkin,
  %``Unraveling hadron structure with generalized parton distributions,''
  {\em Phys.\ Rept.}\  {\bf 418} (2005) 1
  [hep-ph/0504030].
  %%CITATION = HEP-PH/0504030;%%
%----------------------------------------------------------------------
%
%[80]
%\cite{Ji:1996ek}
\bibitem{Ji:1996ek}
  X.-D.~Ji,
  %``Gauge invariant decomposition of nucleon spin and its spin - off,''
  {\it Phys.\ Rev.\ Lett.}\  {\bf 78} (1997) 610
  [hep-ph/9603249].
  %%CITATION = HEP-PH/9603249;%%
%----------------------------------------------------------------------
%%             >> section 2 <<
%----------------------------------------------------------------------
%
%[81]
\bibitem{JBMK}
%\cite{Blumlein:1992we}
%\bibitem{Blumlein:1992we}
  J.~Bl\"umlein and M.~Klein,
  %``On the cross calibration of calorimeters at e p colliders,''
  {\em Nucl.\ Instrum.\ Meth.}\ A {\bf 329} (1993) 112.
  %%CITATION = NUIMA,A329,112;%%
%-------------------------------------------------------------------------------------------------
%
%[82]
%\cite{Amaldi:1979yh}
\bibitem{Amaldi:1979yh}
F. Jaquet and A. Blondel, in:~U.~Amaldi {\it al.}, %A.~Baroncelli, P.~Bloch, A.~Blondel, D.~Crennell, 
%F.~Dydak, F.~Eisele and C.~Geweniger 
{\sf Report From The Study Group On Detectors For Charged Current 
Events}, (DESY, Hamburg, 1979), pp.~391.
%-------------------------------------------------------------------------------------------------
%
%[83]
\bibitem{BEK}
S. Bentvelsen, J. Engelen and P. Kooijman, in: W. Buchm\"uller and G. Ingelman (eds.),
Proceedings of the Workshop on Physics at HERA, Oct. 29-30, 1991, Hamburg (DESY,
Hamburg, 1992), Vol. {\bf 1}, p. 23.
%-------------------------------------------------------------------------------------------------
%
%[84]
\bibitem{Blumlein:1994ii}
  J.~Bl\"umlein,
  %``O (alpha**2 L**2) radiative corrections to deep inelastic e p scattering for different kinematical variables,''
  {\em Z.\ Phys.}\ C {\bf 65} (1995) 293
  [hep-ph/9403342].
  %%CITATION = HEP-PH/9403342;%%
%-------------------------------------------------------------------------------------------------
%
%[85]
%\cite{Bassler:1994uq}
\bibitem{Bassler:1994uq}
  U.~Bassler and G.~Bernardi,
  %``On the kinematic reconstruction of deep inelastic scattering at HERA: The Sigma method,''
  {\em Nucl.\ Instrum.\ Meth.}\ A {\bf 361} (1995) 197
  [hep-ex/9412004].
  %%CITATION = HEP-EX/9412004;%%
%-------------------------------------------------------------------------------------------------
%
%[86]
\bibitem{FORMFACT}
%\cite{HydeWright:2004gh}
%\bibitem{HydeWright:2004gh}
  C.~E.~Hyde and K.~de Jager,
  %``Electromagnetic form factors of the nucleon and Compton scattering,''
  {\em Ann.\ Rev.\ Nucl.\ Part.\ Sci.}\  {\bf 54} (2004) 217
  [nucl-ex/0507001];\\
  %%CITATION = NUCL-EX/0507001;%%
%-------------------------------------------------------------------------------------------------
%\cite{Gao:2005cj}
%\bibitem{Gao:2005cj}
  H.~Gao,
  %``Nucleon electromagnetic form factors,''
  {\em Int.\ J.\ Mod.\ Phys.}\ A {\bf 20} (2005) 1595.
  %%CITATION = IMPAE,A20,1595;%%
%-------------------------------------------------------------------------------------------------
%
%[87]
\bibitem{Schmitz}
  N. Schmitz, {\sf Neutrinophysik}, (Teubner, Stuttgart, 1997).
%--------------------------------------------------------------------------------
%
%[88]
\bibitem{RESON}
%\cite{Feynman:1971wr}
%\bibitem{Feynman:1971wr}
  R.~P.~Feynman, M.~Kislinger and F.~Ravndal,
  %``Current matrix elements from a relativistic quark model,''
  {\em Phys.\ Rev.}\ D {\bf 3} (1971) 2706;\\
  %%CITATION = PHRVA,D3,2706;%%
%-------------------------------------------------------------------------------------------------
%\bibitem{RAVNDAL}
%\cite{Ravndal:1973xx}
%\bibitem{Ravndal:1973xx}
  F.~Ravndal,
  %``Weak production of nuclear resonances in a relativistic quark model,''
  {\em Nuovo Cim.}\ A {\bf 18} (1973) 385;
  %%CITATION = NUCIA,A18,385;%%
%\cite{Ravndal:1972ws}
%\bibitem{Ravndal:1972ws}
%  F.~Ravndal,
  %``Neutrino excitation of nucleon resonances in a quark model,''
  {\em Lett.\ Nuovo Cim.}\  {\bf 3S2} (1972) 631;\\
  %%CITATION = NCLTA,3S2,631;%%
%-------------------------------------------------------------------------------------------------
%\cite{Rein:1980wg}
%\bibitem{Rein:1980wg}
  D.~Rein and L.~M.~Sehgal,
  %``Neutrino Excitation of Baryon Resonances and Single Pion Production,''
  {\em Annals Phys.}\  {\bf 133} (1981) 79.
  %%CITATION = APNYA,133,79;%%
%-------------------------------------------------------------------------------------------------
%
%[89]
\bibitem{SCX}
%\cite{Drell:1963ej}
%\bibitem{Drell:1963ej}
  S.~D.~Drell and J.~D.~Walecka,
  %``Electrodynamic Processes with Nuclear Targets,''
  {\em Annals Phys.}\  {\bf 28} (1964) 18;\\
  %%CITATION = APNYA,28,18;%%
%--------------------------------------------------------------------------------
%\cite{DeForest:1966dn}
%\bibitem{DeForest:1966dn}
  T.~De Forest, Jr. and J.~D.~Walecka,
  %``Electron scattering and nuclear structure,''
  {\em Adv.\ Phys.}\  {\bf 15} (1966) 1;\\
  %%CITATION = ADPHA,15,1;%%
%--------------------------------------------------------------------------------
%\cite{Derman:1978iz}
%\bibitem{Derman:1978iz}
  E.~Derman,
  %``Parity Violation In Polarized Electron - Deuteron Scattering Without The Parton Model,''
  {\em Phys.\ Rev.}\ D {\bf 19} (1979) 133.
  %%CITATION = PHRVA,D19,133;%%
%--------------------------------------------------------------------------------
%
%[90]
%\bibitem{RC24}  
%\cite{Arbuzov:1995id}
\bibitem{Arbuzov:1995id}
  A.~Arbuzov, D.~Y.~Bardin, J.~Bl\"umlein, L.~Kalinovskaya and T.~Riemann,
  %``Hector 1.00: A Program for the calculation of QED, QCD and electroweak corrections to e p and lepton+- N deep inelastic neutral and charged current scattering,''
  {\em Comput.\ Phys.\ Commun.}\  {\bf 94} (1996) 128
  [hep-ph/9511434].
  %%CITATION = HEP-PH/9511434;%%
%--------------------------------------------------------------------------------
%
%[91]
%\cite{Geyer:1977gv}
\bibitem{Geyer:1977gv}
  B.~Geyer, D.~Robaschik and E.~Wieczorek,
  %``Theory of Deep Inelastic Lepton-Hadron Scattering. 1.,''
  {\em Fortsch.\ Phys.}\  {\bf 27} (1979) 75.
  %%CITATION = FPYKA,27,75;%%
%------------------------------------------------------------------------
%
%[92]
%\cite{Tangerman:1994eh}
\bibitem{Tangerman:1994eh}
  R.~D.~Tangerman and P.~J.~Mulders,
  %``Intrinsic transverse momentum and the polarized Drell-Yan process,''
  {\em Phys.\ Rev.}\ D {\bf 51} (1995) 3357
  [hep-ph/9403227].
  %%CITATION = HEP-PH/9403227;%%
%-------------------------------------------------------------------------------------------------
%
%[93]
\bibitem{Blumlein:1998nv}
  J.~Bl\"umlein and A.~Tkabladze,
  %``Target mass corrections for polarized structure functions and new sum rules,''
  {\em Nucl.\ Phys.}\ B {\bf 553} (1999) 427
  [hep-ph/9812478].
  %%CITATION = HEP-PH/9812478;%%
%------------------------------------------------------------------------
%
%[94]
%\cite{Blumlein:1996vs}
\bibitem{Blumlein:1996vs}
  J.~Bl\"umlein and N.~Kochelev,
  %``On the twist -2 and twist - three contributions to the spin dependent electroweak structure functions,''
  {\em Nucl.\ Phys.}\ B {\bf 498} (1997) 285
  [hep-ph/9612318].
  %%CITATION = HEP-PH/9612318;%%
%--------------------------------------------------------------------------------
%
%[95]
%\cite{Ji:1993ey}  
\bibitem{Ji:1993ey}  
  X.~-D.~Ji,
  %``The Nucleon structure functions from deep inelastic scattering with electroweak currents,''
  {\em Nucl.\ Phys.}\ B {\bf 402} (1993) 217.
  %%CITATION = NUPHA,B402,217;%%
%-----------------------------------------------------------------------
%
%[96]
%\cite{Maul:1996dx}
\bibitem{Maul:1996dx}
  M.~Maul, B.~Ehrnsperger, E.~Stein and A.~Sch\"afer,
  %``OPE analysis of the nucleon scattering tensor including weak interaction and finite mass effects,''
  {\em Z.\ Phys.}\ A {\bf 356} (1997) 443
  [hep-ph/9602377].
  %%CITATION = HEP-PH/9602377;%%
%-----------------------------------------------------------------------
%
%[97]
%\cite{Forte:2001ph}
\bibitem{Forte:2001ph}
  S.~Forte, M.~L.~Mangano and G.~Ridolfi,
  %``Polarized parton distributions from charged current deep inelastic scattering and future neutrino factori$
  {\em Nucl.\ Phys.}\ B {\bf 602} (2001) 585
  [hep-ph/0101192].
  %%CITATION = HEP-PH/0101192;%%
%-----------------------------------------------------------------------
%% SLAC %%
%----------------------------------------------------------------------------------------------
%
%[98]
\bibitem{SLAC}
%\bibitem{Atwood:1976ys}
W.~Atwood {\em et~al.},
\newblock {\em Phys.Lett.} B {\bf 64} (1976) 479;\\
%----------------------------------------------------------------------------------------------
%\cite{Bodek:1979rx} 
%\bibitem{Bodek:1979rx}
  A.~Bodek
  %, M.~Breidenbach, D.~L.~Dubin, J.~E.~Elias, J.~I.~Friedman, H.~W.~Kendall, 
  % J.~S.~Poucher and E.~M.~Riordan 
  {\it et al.},
  %``Experimental Studies of the Neutron and Proton Electromagnetic Structure Functions,''
  {\em Phys.\ Rev.}\ D {\bf 20} (1979) 1471;\\
  %%CITATION = PHRVA,D20,1471;%%
%----------------------------------------------------------------------------------------------
%\bibitem{Mestayer:1982ba}
M.~Mestayer {\em et~al.},
\newblock {\em Phys.Rev.} D {\bf 27} (1983) 285;\\
%----------------------------------------------------------------------------------------------
%\bibitem{Whitlow:1990dr}
L.~Whitlow,
%\cite{Whitlow:1990dr}
{\sf Deep inelastic structure functions from electron sctatering on hydrogen, deuterium, and
at iron at $0.6 \GeV^2 \leq Q^2 \leq 30 \GeV^2$}, PhD Thesis, SLAC-0357;\\
%%CITATION = SLAC-0357;%%
%----------------------------------------------------------------------------------------------
%\bibitem{Whitlow:1991uw}
L.~Whitlow, 
E.~Riordan, S.~Dasu, S.~Rock, and A.~Bodek,
\newblock {\em Phys.Lett.} B {\bf 282} (1992) 475;\\
%%CITATION = PHLTA,B282,475;%%
%----------------------------------------------------------------------------------------------
%\bibitem{Gomez:1993ri}
J.~Gomez {\em et~al.},
\newblock {\em Phys.Rev.} D {\bf 49} (1994) 4348.
%----------------------------------------------------------------------------------------------
%
%[99]
%\cite{Dasu:1993vk}
\bibitem{Dasu:1993vk}
  S.~Dasu %, P.~deBarbaro, A.~Bodek, H.~Harada, M.~W.~Krasny, K.~Lang, E.~M.~Riordan and L.~Andivahis
   {\it et al.},
  %``Measurement of kinematic and nuclear dependence of R = sigma-L / sigma-t in deep inelastic
  % electron scattering,''
  {\em Phys.\ Rev.}\ D {\bf 49} (1994) 5641.
  %%CITATION = PHRVA,D49,5641;%%
%----------------------------------------------------------------------------------------------
%% BCDMS %%
%----------------------------------------------------------------------------------------------
%
%[100]
%\cite{Benvenuti:1989rh}
\bibitem{Benvenuti:1989rh}
  A.~C.~Benvenuti {\it et al.},  [BCDMS Collaboration],
  %``A High Statistics Measurement of the Proton Structure Functions F(2) (x, Q**2) and R from Deep Inelastic Muon Scattering at High Q**2,''
  {\em Phys.\ Lett.}\ B {\bf 223} (1989) 485.
  %%CITATION = PHLTA,B223,485;%%
%----------------------------------------------------------------------------------------------
%
%[101]
\bibitem{Benvenuti:1989fm}
A.~Benvenuti {\em et~al.}, BCDMS Collaboration, 
\newblock {\em Phys.Lett.} B {\bf 237} (1990) 592.
%----------------------------------------------------------------------------------------------
%% NMC %%
%----------------------------------------------------------------------------------------------
%
%[102]
\bibitem{NMC}
%\bibitem{Arneodo:1993kz}
NMC Collaboration, M.~Arneodo {\em et~al.},
\newblock {\em Phys.Lett.} B {\bf 309}  (1993) 222;
%%CITATION = PHLTA,B309,222;%%
%----------------------------------------------------------------------------------------------
%\bibitem{Arneodo:1995cq}
%NMC Collaboration., M.~Arneodo {\em et~al.},
\newblock {\em Phys.Lett.} B {\bf 364}  (1995) 107, hep-ph/9509406;
%----------------------------------------------------------------------------------------------
%\bibitem{Arneodo:1996kd}
%NMC Collaboration, M.~Arneodo {\em et~al.},
\newblock {\em Nucl.Phys.} B {\bf 487} (1997) 3, hep-ex/9611022.
%%CITATION = HEP-EX/9611022;%%
%----------------------------------------------------------------------------------------------
%
%[103]
\bibitem{Arneodo:1996qe}
NMC Collaboration, M.~Arneodo {\em et~al.},
\newblock {\em Nucl.Phys.} B {\bf 483} (1997) 3, hep-ph/9610231.
%----------------------------------------------------------------------------------------------
%% HERA combined %%
%----------------------------------------------------------------------------------------------
%
%[104]
%\cite{herapdf:2009wt}
\bibitem{herapdf:2009wt}
  F.~D.~Aaron {\it et al.},  [H1 and ZEUS Collaboration],
  %``Combined Measurement and QCD Analysis of the Inclusive e+- p Scattering Cross Sections at HERA,''
  {\em JHEP} {\bf 1001} (2010) 109
  [arXiv:0911.0884 [hep-ex]].
  %%CITATION = ARXIV:0911.0884;%%
%----------------------------------------------------------------------------------
%% earlier data: Reviews %%
%----------------------------------------------------------------------------------------------
%
%[105]
\bibitem{SURV1}
%\cite{Eisele:1986uz}  
%\bibitem{Eisele:1986uz}
  F.~Eisele,
  %``High-energy Neutrino Interactions,''   
  {\em Rept.\ Prog.\ Phys.}\  {\bf 49} (1986) 233;\\
  %%CITATION = RPPHA,49,233;%%
%----------------------------------------------------------------------------------------------
%\cite{Diemoz:1986kt}  
%\bibitem{Diemoz:1986kt}
  M.~Diemoz, F.~Ferroni and E.~Longo,
  %``Nucleon Structure Functions From Neutrino Scattering,''
  {\em Phys.\ Rept.}\  {\bf 130} (1986) 293;\\
  %%CITATION = PRPLC,130,293;%%
%----------------------------------------------------------------------------------------------
%\cite{Sloan:1988qj}
%\bibitem{Sloan:1988qj}
  T.~Sloan, R.~Voss and G.~Smadja,
  %``The Quark Structure of the Nucleon from the CERN Muon Experiments,''
  {\em Phys.\ Rept.}\  {\bf 162} (1988) 45;\\
  %%CITATION = PRPLC,162,45;%%
%----------------------------------------------------------------------------------------------
%\cite{Mishra:1989jc}
%\bibitem{Mishra:1989jc}
  S.~R.~Mishra and F.~Sciulli,
  %``Deep Inelastic Lepton - Nucleon Scattering,''
  {\em Ann.\ Rev.\ Nucl.\ Part.\ Sci.}\  {\bf 39} (1989) 259;\\
  %%CITATION = ARNUA,39,259;%%
%----------------------------------------------------------------------------------------------
%\bibitem{WINTER}
K. Winter, {\sf Neutrino Physics}, (Cambridge, University Press, Cambridge, 1991), 670~p.
%----------------------------------------------------------------------------------------------
%% FL %%
%----------------------------------------------------------------------------------------------
%
%[106]
%\cite{Whitlow:1990gk}
\bibitem{Whitlow:1990gk}
  L.~W.~Whitlow, S.~Rock, A.~Bodek, E.~M.~Riordan and S.~Dasu,
  %``A Precise extraction of R = sigma-L / sigma-T from a global analysis of
  % the SLAC deep inelastic e p and e d scattering cross-sections,''
  {\em Phys.\ Lett.}\ B {\bf 250} (1990) 193.
  %%CITATION = PHLTA,B250,193;%%
%----------------------------------------------------------------------------------------------
%
%[107]
%\cite{Adloff:1996yz}
\bibitem{Adloff:1996yz}
  C.~Adloff {\it et al.},  [H1 Collaboration],
  %``Determination of the longitudinal proton structure function F(L) (x, Q**2) at low x,''
  {\em Phys.\ Lett.}\ B {\bf 393} (1997) 452
  [hep-ex/9611017].
  %%CITATION = HEP-EX/9611017;%%
%----------------------------------------------------------------------------------------------
%
%[108]
%\cite{Abe:1998ym}
\bibitem{Abe:1998ym}
  K.~Abe {\it et al.},  [E143 Collaboration],
  %``Measurements of R = sigma(L) / sigma(T) for 0.03<x<0.1 and fit to world data,''
  {\em Phys.\ Lett.}\ B {\bf 452} (1999) 194 [hep-ex/9808028].
  %%CITATION = HEP-EX/9808028;%%
%----------------------------------------------------------------------
%
%[109]
%\cite{Aaron:2010ry}
\bibitem{Aaron:2010ry}
  F.~D.~Aaron %, C.~Alexa, V.~Andreev, S.~Backovic, A.~Baghdasaryan, S.~Baghdasaryan, E.~Barrelet 
  %and W.~Bartel 
  {\it et al.},
  %``Measurement of the Inclusive e{\pm}p Scattering Cross Section at High Inelasticity y and of 
  %the Structure Function $F_L$,''
  {\em Eur.\ Phys.\ J.}\ C {\bf 71} (2011) 1579
  [arXiv:1012.4355 [hep-ex]].
  %%CITATION = ARXIV:1012.4355;%%
%----------------------------------------------------------------------------------------------
%----------------------------------------------------------------------------------------------
%% DRELL-YAN %%
%----------------------------------------------------------------------------------------------
%
%[110]
\bibitem{DY:EXP}
%\bibitem{Moreno:1990sf}
G.~Moreno {\em et~al.},
\newblock {\em Phys.Rev.} D {\bf 43}, 2815 (1991);\\
%----------------------------------------------------------------------------------------------
%\bibitem{Towell:2001nh}
FNAL E866/NuSea Collaboration, R.~Towell {\em et~al.},
\newblock {\em Phys.Rev.} D {\bf 64}, 052002 (2001), hep-ex/0103030;\\
%\cite{Webb:2003ps}
%\bibitem{Webb:2003ps}
  J.~C.~Webb {\it et al.},  [NuSea Collaboration],
  %``Absolute Drell-Yan dimuon cross-sections in 800 GeV / c pp and pd collisions,''
  %Submitted to: Phys.Rev.Lett.
  [hep-ex/0302019].
%%CITATION = HEP-EX/0302019;%%
%----------------------------------------------------------------------------------------------
%----------------------------------------------------------------------------------------------
%% Di-muon %%
%----------------------------------------------------------------------------------------------
%
%[111]
\bibitem{DIMUON}
%\bibitem{Bazarko:1994tt}
CCFR Collaboration, A.~Bazarko {\em et~al.},
\newblock {\em Z.Phys.} {\bf C65} (1995) 189, hep-ex/9406007;\\
%----------------------------------------------------------------------------------------------
%\bibitem{Goncharov:2001qe}
NuTeV Collaboration, M.~Goncharov {\em et~al.},
\newblock {\em Phys.Rev.} D {\bf 64} (2001) 112006, hep-ex/0102049;\\
%----------------------------------------------------------------------------------------------
%\bibitem{Tzanov:2005kr}
NuTeV Collaboration, M.~Tzanov {\em et~al.},
\newblock {\em Phys. Rev.} D {\bf 74} (2006) 012008, hep-ex/0509010;\\
%%CITATION = HEP-EX/0509010;%%
%----------------------------------------------------------------------------------------------
%\bibitem{Mason:2006qa}
D.~A. Mason, {\sf 	
Measurement of the strange-antistrange asymmetry at NLO in QCD from NuTeV dimuon data},
PhD Thesis, FERMILAB-THESIS-2006-01.
%%CITATION = FERMILAB-THESIS-2006-01 ETC.;%%
%----------------------------------------------------------------------------------------------
%% other data in fits %%
%----------------------------------------------------------------------------------------------
%
%[112]
%\cite{Martin:2009iq}
\bibitem{Martin:2009iq}
  A.~D.~Martin, W.~J.~Stirling, R.~S.~Thorne and G.~Watt,
  %``Parton distributions for the LHC,''
  {\em Eur.\ Phys.\ J.}\ C {\bf 63} (2009) 189
  [arXiv:0901.0002 [hep-ph]].
  %%CITATION = ARXIV:0901.0002;%%
%----------------------------------------------------------------------------------------------
% 
%[113]
%\cite{Ball:2011uy}
\bibitem{Ball:2011uy}
  R.~D.~Ball {\it et al.},
  %``Unbiased global determination of parton distributions and their uncertainties at NNLO and at LO,''
  {\em Nucl.\ Phys.}\ B {\bf 855} (2012) 153
  [arXiv:1107.2652 [hep-ph]].
  %%CITATION = ARXIV:1107.2652;%%
%------------------------------------------------------------------------------------------------------
%
%[114]
\bibitem{EXP1a}
%\cite{Adloff:2000qk}
%\bibitem{Adloff:2000qk}
  C.~Adloff {\it et al.},  [H1 Collaboration],
  %``Deep inelastic inclusive e p scattering at low x and a determination of alpha(s),''
  {\em Eur.\ Phys.\ J.}\ C {\bf 21} (2001) 33
  [hep-ex/0012053];
  %%CITATION = HEP-EX/0012053;%%
%\cite{Adloff:2003uh}
%\bibitem{Adloff:2003uh}
%  C.~Adloff {\it et al.},  [H1 Collaboration],
  %``Measurement and QCD analysis of neutral and charged current cross-sections at HERA,''
  {\em Eur.\ Phys.\ J.}\ C {\bf 30} (2003) 1
  [hep-ex/0304003];
  %%CITATION = HEP-EX/0304003;%%
%------------------------------------------------------------------------------------------------------
%\cite{Adloff:2000qj}
%\bibitem{Adloff:2000qj}
%  C.~Adloff {\it et al.},  [H1 Collaboration],
  %``Measurement of neutral and charged current cross-sections in electron - proton collisions at high $Q^{2}$,''
  {\em Eur.\ Phys.\ J.}\ C {\bf 19} (2001) 269
  [hep-ex/0012052];\\
  %%CITATION = HEP-EX/0012052;%%
%------------------------------------------------------------------------------------------------------
%\cite{Chekanov:2001qu}
%\bibitem{Chekanov:2001qu}
  S.~Chekanov {\it et al.},  [ZEUS Collaboration],
  %``Measurement of the neutral current cross-section and F(2) structure function for deep inelastic e + p scattering at HERA,''
  {\em Eur.\ Phys.\ J.}\ C {\bf 21} (2001) 443
  [hep-ex/0105090];
  %%CITATION = HEP-EX/0105090;%%
%------------------------------------------------------------------------------------------------------
%\cite{Chekanov:2008aa}
%\bibitem{Chekanov:2008aa}
%  S.~Chekanov {\it et al.},  [ZEUS Collaboration],
  %``Measurement of charged current deep inelastic scattering cross sections with a longitudinally polarised electron beam at HERA,''
  {\em Eur.\ Phys.\ J.}\ C {\bf 61} (2009) 223
  [arXiv:0812.4620 [hep-ex]];
  %%CITATION = ARXIV:0812.4620;%%
%------------------------------------------------------------------------------------------------------
%\cite{Chekanov:2009gm}
%\bibitem{Chekanov:2009gm}
%  S.~Chekanov {\it et al.},  [ZEUS Collaboration],
  %``Measurement of high-Q**2 neutral current deep inelastic e- p scattering cross sections with a longitudinally polarised electron beam at HERA,''
  {\em Eur.\ Phys.\ J.}\ C {\bf 62} (2009) 625
  [arXiv:0901.2385 [hep-ex]];
  %%CITATION = ARXIV:0901.2385;%%
%\cite{Chekanov:2002ej}
%\bibitem{Chekanov:2002ej}
%  S.~Chekanov {\it et al.},  [ZEUS Collaboration],
  %``Measurement of high Q**2 e- p neutral current cross-sections at HERA and the extraction of xF(3),''
  {\em Eur.\ Phys.\ J.}\ C {\bf 28} (2003) 175
  [hep-ex/0208040];
  %%CITATION = HEP-EX/0208040;%%
%------------------------------------------------------------------------------------------------------
%\cite{Chekanov:2003yv}
%\bibitem{Chekanov:2003yv}
%  S.~Chekanov {\it et al.}  [ZEUS Collaboration],
  %``High Q**2 neutral current cross-sections in e+ p deep inelastic scattering at s**(1/2) = 318-GeV,''
  {\em Phys.\ Rev.}\ D {\bf 70} (2004) 052001
  [hep-ex/0401003];
  %%CITATION = HEP-EX/0401003;%%
%------------------------------------------------------------------------------------------------------
%\cite{Chekanov:2003vw}
%\bibitem{Chekanov:2003vw}
%  S.~Chekanov {\it et al.}  [ZEUS Collaboration],
  %``Measurement of high Q**2 charged current cross-sections in e+ p deep inelastic scattering at HERA,''
  {\em Eur.\ Phys.\ J.}\ C {\bf 32} (2003) 1
  [hep-ex/0307043];\\
  %%CITATION = HEP-EX/0307043;%%
%------------------------------------------------------------------------------------------------------
%\cite{Breitweg:1998dz}
%\bibitem{Breitweg:1998dz}
  J.~Breitweg {\it et al.},  [ZEUS Collaboration],
  %``ZEUS results on the measurement and phenomenology of F(2) at low x and low Q**2,''
  {\em Eur.\ Phys.\ J.}\ C {\bf 7} (1999) 609
  [hep-ex/9809005];
  %%CITATION = HEP-EX/9809005;%%
%------------------------------------------------------------------------------------------------------
%------------------------------------------------------------------------------------------------------
%% >>>> CHARM and BOTTOM
%------------------------------------------------------------------------------------------------------
%\cite{Breitweg:1999ad}
%\bibitem{Breitweg:1999ad}
%  J.~Breitweg {\it et al.},  [ZEUS Collaboration],
  %``Measurement of D*+- production and the charm contribution to F(2) in deep inelastic scattering at HERA,''
  {\em Eur.\ Phys.\ J.}\ C {\bf 12} (2000) 35
  [hep-ex/9908012];\\
  %%CITATION = HEP-EX/9908012;%%
%------------------------------------------------------------------------------------------------------
%\cite{Chekanov:2003rb}
%\bibitem{Chekanov:2003rb}
  S.~Chekanov {\it et al.},  [ZEUS Collaboration],
  %``Measurement of D*+- production in deep inelastic e+- p scattering at HERA,''
  {\em Phys.\ Rev.}\ D {\bf 69} (2004) 012004
  [hep-ex/0308068];
  %%CITATION = HEP-EX/0308068;%%
%------------------------------------------------------------------------------------------------------
%\cite{Chekanov:2008yd}
%\bibitem{Chekanov:2008yd}
%  S.~Chekanov {\it et al.},  [ZEUS Collaboration],
  %``Measurement of D+- and D0 production in deep inelastic scattering using a lifetime tag at HERA,''
  {\em Eur.\ Phys.\ J.}\ C {\bf 63} (2009) 171
  [arXiv:0812.3775 [hep-ex]];
  %%CITATION = ARXIV:0812.3775;%%
%------------------------------------------------------------------------------------------------------
%\cite{Chekanov:2009kj}
%\bibitem{Chekanov:2009kj}
%  S.~Chekanov {\it et al.},  [ZEUS Collaboration],
  %``Measurement of charm and beauty production in deep inelastic ep scattering from decays into muons at HERA,''
  {\em Eur.\ Phys.\ J.}\ C {\bf 65} (2010) 65
  [arXiv:0904.3487 [hep-ex]];\\
  %%CITATION = ARXIV:0904.3487;%%
%------------------------------------------------------------------------------------------------------
%\cite{Adloff:1996xq}
%\bibitem{Adloff:1996xq}
  C.~Adloff {\it et al.},  [H1 Collaboration],
  %``Inclusive D0 and D*+- production in deep inelastic e p scattering at HERA,''
  {\em Z.\ Phys.}\ C {\bf 72} (1996) 593
  [hep-ex/9607012];
  %%CITATION = HEP-EX/9607012;%%
%------------------------------------------------------------------------------------------------------
%\cite{Adloff:2001zj}
%\bibitem{Adloff:2001zj}
%  C.~Adloff {\it et al.},  [H1 Collaboration],
  %``Measurement of D*+- meson production and F2(c) in deep inelastic scattering at HERA,''
  {\em Phys.\ Lett.}\ B {\bf 528} (2002) 199
  [hep-ex/0108039];\\
  %%CITATION = HEP-EX/0108039;%%
%------------------------------------------------------------------------------------------------------
%\cite{Aktas:2005iw}
%\bibitem{Aktas:2005iw}
  A.~Aktas {\it et al.},  [H1 Collaboration],
  %``Measurement of F(2)**c anti-c and F(2)**b anti-b at low Q*2 and x using the H1 vertex detector at HERA,''
  {\em Eur.\ Phys.\ J.}\ C {\bf 45} (2006) 23
  [hep-ex/0507081];
  %%CITATION = HEP-EX/0507081;%%
%------------------------------------------------------------------------------------------------------
%\cite{Aktas:2004az}
%\bibitem{Aktas:2004az}
%  A.~Aktas {\it et al.},  [H1 Collaboration],
  %``Measurement of F2($c \bar{c}$) and F2($b \bar{b}$) at high $Q^{2}$ using the H1 vertex detector at HERA,''
  {\em Eur.\ Phys.\ J.}\ C {\bf 40} (2005) 349
  [hep-ex/0411046];
  %%CITATION = HEP-EX/0411046;%%
%------------------------------------------------------------------------------------------------------
%% >>> JET
%------------------------------------------------------------------------------------------------------
%\cite{H1:2007pb}
%\bibitem{H1:2007pb}
%  A.~Aktas {\it et al.},  [H1 Collaboration],
  %``Measurement of inclusive jet production in deep-inelastic scattering at high Q**2 and determination of the strong coupling,''
  {\em Phys.\ Lett.}\ B {\bf 653} (2007) 134
  [arXiv:0706.3722 [hep-ex]];\\
  %%CITATION = ARXIV:0706.3722;%%
%------------------------------------------------------------------------------------------------------
%\cite{Chekanov:2002be}
%\bibitem{Chekanov:2002be}
  S.~Chekanov {\it et al.},  [ZEUS Collaboration],
  %``Inclusive jet cross-sections in the Breit frame in neutral current deep inelastic scattering at HERA and determination of alpha(s),''
  {\em Phys.\ Lett.}\ B {\bf 547} (2002) 164
  [hep-ex/0208037];\\
  %%CITATION = HEP-EX/0208037;%%
%------------------------------------------------------------------------------------------------------
%\cite{Chekanov:2006xr}
%\bibitem{Chekanov:2006xr}
%  S.~Chekanov {\it et al.},  [ZEUS Collaboration],
  %``Inclusive-jet and dijet cross-sections in deep inelastic scattering at HERA,''
  {\em Nucl.\ Phys.}\ B {\bf 765} (2007) 1
  [hep-ex/0608048];\\
  %%CITATION = HEP-EX/0608048;%%
%------------------------------------------------------------------------------------------------------
%% >>> E665
%------------------------------------------------------------------------------------------------------
%\cite{Adams:1996gu}
%\bibitem{Adams:1996gu}
  M.~R.~Adams {\it et al.},  [E665 Collaboration],
  %``Proton and deuteron structure functions in muon scattering at 470-GeV,''
  {\em Phys.\ Rev.}\ D {\bf 54} (1996) 3006;\\
  %%CITATION = PHRVA,D54,3006;%%
%------------------------------------------------------------------------------------------------------
%% >>> CHORUS
%------------------------------------------------------------------------------------------------------
%\cite{Onengut:2005kv}
%\bibitem{Onengut:2005kv}
  G.~Onengut {\it et al.},  [CHORUS Collaboration],
  %``Measurement of nucleon structure functions in neutrino scattering,''
  {\em Phys.\ Lett.}\ B {\bf 632} (2006) 65;\\
  %%CITATION = PHLTA,B632,65;%%
%------------------------------------------------------------------------------------------------------
%% >>> Weak boson prod
%------------------------------------------------------------------------------------------------------
%\cite{Abazov:2007pm}
%\bibitem{Abazov:2007pm}
  V.~M.~Abazov {\it et al.},  [D0 Collaboration],
  %``Measurement of the muon charge asymmetry from $W$ boson decays,''
  {\em Phys.\ Rev.}\ D {\bf 77} (2008) 011106
  [arXiv:0709.4254 [hep-ex]];\\
  %%CITATION = ARXIV:0709.4254;%%
%------------------------------------------------------------------------------------------------------
%\cite{Acosta:2005ud}
%\bibitem{Acosta:2005ud}
  D.~Acosta {\it et al.},  [CDF Collaboration],
  %``Measurement of the forward-backward charge asymmetry from $W \to e \nu$ production in $p\bar{p}$ collisions at $\sqrt{s} = 1.96$ TeV,''
  {\em Phys.\ Rev.}\ D {\bf 71} (2005) 051104
  [hep-ex/0501023];\\
  %%CITATION = HEP-EX/0501023;%%
%------------------------------------------------------------------------------------------------------
%\bibitem{CDF_Z}
CDF Collaboration, J. Han {\it et al.}, public note, May 2008, \newline
{\tt http:www-cdf.fnal.gov/physics/ewk/2008/dszdy/};\\
%------------------------------------------------------------------------------------------------------
%\cite{Aaltonen:2009ta}
%\bibitem{Aaltonen:2009ta}
  T.~Aaltonen {\it et al.},  [CDF Collaboration],
  %``Direct Measurement of the $W$ Production Charge Asymmetry in $p\bar{p}$ Collisions at $\sqrt{s} = 1.96$ TeV,''
  {\em Phys.\ Rev.\ Lett.}\  {\bf 102} (2009) 181801
  [arXiv:0901.2169 [hep-ex]];
  %%CITATION = ARXIV:0901.2169;%%
%------------------------------------------------------------------------------------------------------
%\cite{Aaltonen:2010zza}
%\bibitem{Aaltonen:2010zza}
%  T.~A.~Aaltonen {\it et al.},  [CDF Collaboration],
  %``Measurement of $d\sigma/dy$ of Drell-Yan $e^+e^-$ pairs in the $Z$ Mass Region from $p\bar{p}$ Collisions at $\sqrt{s}=1.96$ TeV,''
  {\em Phys.\ Lett.}\ B {\bf 692} (2010) 232
  [arXiv:0908.3914 [hep-ex], arXiv:0908.3914 [hep-ex]];\\
  %%CITATION = ARXIV:0908.3914;%%
%------------------------------------------------------------------------------------------------------
%\cite{Abazov:2007jy}
%\bibitem{Abazov:2007jy}
  V.~M.~Abazov {\it et al.},  [D0 Collaboration],
  %``Measurement of the shape of the boson rapidity distribution for $p \bar{p} \to Z/gamma^* \to e^{+} e^{-}$ + $X$ events produced at $\sqrt{s}$ of 1.96-TeV,''
  {\em Phys.\ Rev.}\ D {\bf 76} (2007) 012003
  [hep-ex/0702025].
  %%CITATION = HEP-EX/0702025;%%
%------------------------------------------------------------------------------------------------------
%% >>> TEVATRON JET
%------------------------------------------------------------------------------------------------------
%
%[115]
\bibitem{TEVJET}
%\cite{Abulencia:2007ez}
%\bibitem{Abulencia:2007ez}
  A.~Abulencia {\it et al.},  [CDF - Run II Collaboration],
  %``Measurement of the Inclusive Jet Cross Section using the {\boldmath $k_{\rm T}$} algorithmin
  %{\boldmath $p\overline{p}$} Collisions at{\boldmath $\sqrt{s}$} = 1.96 TeV with the CDF II 
  %Detector,''
  {\em Phys.\ Rev.}\ D {\bf 75} (2007) 092006
  [Erratum-ibid.\ D {\bf 75} (2007) 119901]
  [hep-ex/0701051];\\
  %%CITATION = HEP-EX/0701051;%%
%------------------------------------------------------------------------------------------------------
%\cite{Aaltonen:2008eq}
%\bibitem{Aaltonen:2008eq}
  T.~Aaltonen {\it et al.},  [CDF Collaboration],
  %``Measurement of the Inclusive Jet Cross Section at the Fermilab Tevatron p anti-p Collider 
  % Using a Cone-Based Jet Algorithm,''
  {\em Phys.\ Rev.}\ D {\bf 78} (2008) 052006
   [Erratum-ibid.\ D {\bf 79} (2009) 119902]
  [arXiv:0807.2204 [hep-ex]];\\
  %%CITATION = ARXIV:0807.2204;%%
%------------------------------------------------------------------------------------------------------
%\cite{D0:2008hua}
%\bibitem{D0:2008hua}
  V.~M.~Abazov {\it et al.},  [D0 Collaboration],
  %``Measurement of the inclusive jet cross-section in $p \bar{p}$ collisions at $s^{91/2)}$ =1.96-TeV,''
  {\em Phys.\ Rev.\ Lett.}\  {\bf 101} (2008) 062001
  [arXiv:0802.2400 [hep-ex]];
  %%CITATION = ARXIV:0802.2400;%%
%------------------------------------------------------------------------------------------------------
%\cite{Abazov:2010fr}
%\bibitem{Abazov:2010fr}
%  V.~M.~Abazov {\it et al.},  [D0 Collaboration],
  %``Measurement of the dijet invariant mass cross section in proton anti-proton collisions at sqrt{s} = 1.96 TeV,''
  {\em Phys.\ Lett.}\ B {\bf 693} (2010) 531
  [arXiv:1002.4594 [hep-ex]];\\
  %%CITATION = ARXIV:1002.4594;%%
%------------------------------------------------------------------------------------------------------
%\cite{Affolder:2001hn}
%\bibitem{Affolder:2001hn}
  T.~Affolder {\it et al.},  [CDF Collaboration],
  %``Measurement of the strong coupling constant from inclusive jet production at the Tevatron $\bar{p}p$ collider,''
  {\em Phys.\ Rev.\ Lett.}\  {\bf 88} (2002) 042001
  [hep-ex/0108034].
  %%CITATION = HEP-EX/0108034;%%
%------------------------------------------------------------------------------------------------------
%
%[116]
\bibitem{HERAPDF1.0}
HERAPDF1.0, {\tt https://www.desy.de/h1zeus/combined\_results/benchmark/herapdf1.0.html}
%------------------------------------------------------------------------------------------------------
%
%[117]
%\cite{Alekhin:2012ig}
\bibitem{Alekhin:2012ig}
  S.~Alekhin, J.~Bl\"umlein and S.~Moch,
  %``Parton distribution functions and benchmark cross sections at NNLO,''
  arXiv:1202.2281 [hep-ph].
  %%CITATION = ARXIV:1202.2281;%%
%----------------------------------------------------------------------------------------------
%
%[118]
%\cite{JimenezDelgado:2008hf}
\bibitem{JimenezDelgado:2008hf}
  P.~Jimenez-Delgado and E.~Reya,
  %``Dynamical NNLO parton distributions,''
  {\em Phys.\ Rev.}\ D {\bf 79} (2009) 074023
  [arXiv:0810.4274 [hep-ph]].
  %%CITATION = ARXIV:0810.4274;%%
%----------------------------------------------------------------------------------
%
%[119]
%\cite{Ball:2011mu}
\bibitem{Ball:2011mu}
  R.~D.~Ball%, V.~Bertone, F.~Cerutti, L.~Del Debbio, S.~Forte, A.~Guffanti, J.~I.~Latorre and J.~Rojo
  ~{\it et al.},
  %``Impact of Heavy Quark Masses on Parton Distributions and LHC Phenomenology,''
  {\em Nucl.\ Phys.}\ B {\bf 849} (2011) 296
  [arXiv:1101.1300 [hep-ph]].
  %%CITATION = ARXIV:1101.1300;%%
%----------------------------------------------------------------------------------
%------------------------------------------------------------------------------------------------------
%------------ DATA: polarized ----------------------------------------------------------
%------------------------------------------------------------------------------------------------------
%
%[120]
\bibitem{E142n}
%\cite{Anthony:1996mw}
%\bibitem{Anthony:1996mw}
  P.~L.~Anthony {\it et al.},  [E142 Collaboration],
  %``Deep Inelastic Scattering of Polarized Electrons by Polarized $^3$He and
  %the Study of the Neutron Spin Structure,''
  {\em Phys.\ Rev.}\  D {\bf 54} (1996) 6620
  [arXiv:hep-ex/9610007].
  %%CITATION = PHRVA,D54,6620;%%
%----------------------------------------------------------------------
%
%[121]
\bibitem{E154n}
%\cite{Abe:1997cx}
%\bibitem{Abe:1997cx}
  K.~Abe {\it et al.},  [E154 Collaboration],
  %``Precision determination of the neutron spin structure function g1(n),''
  {\em Phys.\ Rev.\ Lett.}\  {\bf 79} (1997) 26
  [arXiv:hep-ex/9705012].
  %%CITATION = PRLTA,79,26;%%
%----------------------------------------------------------------------
%
%[122]
\bibitem{E143pd}
%\cite{Abe:1998wq}
%\bibitem{Abe:1998wq}
  K.~Abe {\it et al.},  [E143 collaboration],
  %``Measurements of the proton and deuteron spin structure functions g1  and
  %g2,''
  {\em Phys.\ Rev.}\  D {\bf 58} (1998) 112003
  [arXiv:hep-ph/9802357].
  %%CITATION = PHRVA,D58,112003;%%
%----------------------------------------------------------------------
%
%[123]
\bibitem{E155d}
%\cite{Anthony:1999rm}
%\bibitem{Anthony:1999rm}
  P.~L.~Anthony {\it et al.},  [E155 Collaboration],
  %``Measurement of the deuteron spin structure function g1(d)(x) for
  %1-(GeV/c)**2 < Q**2 < 40-(GeV/c)**2,''
  {\em Phys.\ Lett.}\  B {\bf 463} (1999) 339
  [arXiv:hep-ex/9904002].   
  %%CITATION = PHLTA,B463,339;%%
%----------------------------------------------------------------------
%
%[124]
\bibitem{E155p}
%\cite{Anthony:2000fn}
%\bibitem{Anthony:2000fn}
  P.~L.~Anthony {\it et al.},  [E155 Collaboration],
  %``Measurements of the Q**2 dependence of the proton and neutron spin
  %structure functions g1(p) and g1(n),''
  {\em Phys.\ Lett.}\  B {\bf 493} (2000) 19
  [arXiv:hep-ph/0007248].
  %%CITATION = PHLTA,B493,19;%%
%----------------------------------------------------------------------
%
%[125]
\bibitem{EMCp2} 
%\cite{Ashman:1989ig}
%\bibitem{Ashman:1989ig}
  J.~Ashman {\it et al.},  [European Muon Collaboration],
  %``An investigation of the spin structure of the proton in deep inelastic
  %scattering of polarized muons on polarized protons,''
  {\em Nucl.\ Phys.}\  B {\bf 328} (1989) 1.
  %%CITATION = NUPHA,B328,1;%%
%----------------------------------------------------------------------
%
%[126]
\bibitem{SMCpd}
%\cite{Adeva:1998vv}
%\bibitem{Adeva:1998vv}
  B.~Adeva {\it et al.},  [Spin Muon Collaboration],
  %``Spin asymmetries A(1) and structure functions g1 of the proton and  the
  %deuteron from polarized high energy muon scattering,''
  {\em Phys.\ Rev.}\  D {\bf 58} (1998) 112001;
  %%CITATION = PHRVA,D58,112001;%%
%\cite{Adeva:1999pa}
%\bibitem{Adeva:1999pa}
%  B.~Adeva {\it et al.},  [Spin Muon Collaboration],
  %``Spin asymmetries A(1) of the proton and the deuteron in the low x and low Q**2 
  %region from polarized high-energy muon scattering,''
  {\em Phys.\ Rev.}\ D {\bf 60} (1999) 072004
   [Erratum-ibid.\ D {\bf 62} (2000) 079902].
  %%CITATION = PHRVA,D60,072004;%%
%----------------------------------------------------------------------
%   
%[127]
\bibitem{COMPd}
%\cite{Alexakhin:2006vx} 
%\bibitem{Alexakhin:2006vx}
  V.~Y.~Alexakhin {\it et al.},  [COMPASS Collaboration],
  %``The Deuteron Spin-dependent Structure Function g1d and its First Moment,''
  {\em Phys.\ Lett.}\  B {\bf 647} (2007) 8 
  [arXiv:hep-ex/0609038].
  %%CITATION = PHLTA,B647,8;%% 
%----------------------------------------------------------------------
%
%[128]
\bibitem{COMP1}
%\cite{:2010hc}
%\bibitem{:2010hc}
  M.~G.~Alekseev  {\it et al.}, [COMPASS Collaboration],
  %``The Spin-dependent Structure Function of the Proton g_1^p and a Test of the
  %Bjorken Sum Rule,''
  arXiv:1001.4654 [hep-ex].
  %%CITATION = ARXIV:1001.4654;%%
%----------------------------------------------------------------------
%
%[129]
%\cite{Ageev:2005gh}  
\bibitem{Ageev:2005gh}
  E.~S.~Ageev {\it et al.},  [COMPASS Collaboration],
  %``Measurement of the spin structure of the deuteron in the DIS region,''
  {\em Phys.\ Lett.}\ B {\bf 612} (2005) 154
  [hep-ex/0501073].
  %%CITATION = HEP-EX/0501073;%%
%----------------------------------------------------------------------------------
%
%[130]
\bibitem{JLABn}
X.~Zheng {\it et al.}, The JLAB Hall A collaboration, {\it Phys. Rev.} {\bf C70}
(2004) 065207.
%----------------------------------------------------------------------
%
%[131]
\bibitem{CLA1pd}
%\cite{Dharmawardane:2006zd}
%\bibitem{Dharmawardane:2006zd}
  K.~V.~Dharmawardane {\it et al.},  [CLAS Collaboration],
  %``Measurement of the $x$- and $Q^2$-Dependence of the Asymmetry $A_1$ on the
  %Nucleon,''
  {\em Phys.\ Lett.}\  B {\bf 641} (2006) 11
  [arXiv:nucl-ex/0605028].
  %%CITATION = PHLTA,B641,11;%%
%----------------------------------------------------------------------
%
%[132]
\bibitem{CLA2pd}
CLAS collaboration, private communication.
%----------------------------------------------------------------------
%
%[133]
\bibitem{HERMn}
%\cite{Ackerstaff:1997ws}
%\bibitem{Ackerstaff:1997ws}
  K.~Ackerstaff {\it et al.},  [HERMES Collaboration],
  %``Measurement of the neutron spin structure function g1(n) with a  polarized
  %He-3 internal target,''
  {\em Phys.\ Lett.}\  B {\bf 404} (1997) 383
  [arXiv:hep-ex/9703005].
  %%CITATION = PHLTA,B404,383;%%
%----------------------------------------------------------------------
%
%[134]
\bibitem{HERMpd}
%\cite{Airapetian:2007mh}
%\bibitem{Airapetian:2007mh}
  A.~Airapetian {\it et al.},  [HERMES Collaboration],
  %``Precise determination of the spin structure function g(1) of the  proton,
  %deuteron and neutron,''
  {\em Phys.\ Rev.}\  D {\bf 75} (2007) 012007
  [arXiv:hep-ex/0609039].
  %%CITATION = PHRVA,D75,012007;%%
%----------------------------------------------------------------------
%------------   g_2    ------------------------------------------------
%----------------------------------------------------------------------
%
%[135]
\bibitem{EXP:G2}
%\cite{Adams:1997tq}
%\bibitem{Adams:1997tq}
  D.~Adams {\it et al.},  [Spin Muon (SMC) Collaboration],
  %``Spin structure of the proton from polarized inclusive deep inelastic muon - proton scattering,''
  {\em Phys.\ Rev.}\ D {\bf 56} (1997) 5330
  [hep-ex/9702005];\\
  %%CITATION = HEP-EX/9702005;%%
%----------------------------------------------------------------------
%\cite{Anthony:2002hy}
%\bibitem{Anthony:2002hy}
  P.~L.~Anthony {\it et al.},  [E155 Collaboration],
  %``Precision measurement of the proton and deuteron spin structure functions g(2) and asymmetries A(2),''
  {\em Phys.\ Lett.}\ B {\bf 553} (2003) 18
  [hep-ex/0204028];\\
  %%CITATION = HEP-EX/0204028;%%
%----------------------------------------------------------------------
%\cite{Airapetian:2011wu}
%\bibitem{Airapetian:2011wu}
  A.~Airapetian 
  %,N.~Akopov, Z.~Akopov, E.~C.~Aschenauer, W.~Augustyniak, R.~Avakian, 
  %A.~Avetissian and E.~Avetisyan 
  {\it et al.}, [HERMES Collaboration],
  %``Measurement of the virtual-photon asymmetry $A_2$ and the spin-structure function 
  % $g_2$ of the proton,''
  arXiv:1112.5584 [hep-ex].
  %%CITATION = ARXIV:1112.5584;%%
%----------------------------------------------------------------------
%
%[136]
\bibitem{LHEC}
%\cite{AbelleiraFernandez:2012cc}
%\bibitem{AbelleiraFernandez:2012cc}
  J.~L.~Abelleira Fernandez {\it et al.}  [LHeC Study Group Collaboration],
  {\sf A Large Hadron Electron Collider at CERN: Report on the Physics and Design Concepts for Machine and 
  Detector}
  J.\ Phys.\ G  {\bf 39} (2012) 075001
  [arXiv:1206.2913 [physics.acc-ph]].
  %%CITATION = ARXIV:1206.2913;%%
%----------------------------------------------------------------------
%
%[137]
\bibitem{EIC}
E. Aschenauer {\it et al.}, The EIC Collaboration,
{\sf A High Luminosity, High Energy Electron-Ion-Collider}, \newline
A White Paper Prepared for the NSAC LRP 2007.
%----------------------------------------------------------------------
%
%[138]
%\cite{Boer:2011fh}
\bibitem{Boer:2011fh}
  D.~Boer, M.~Diehl, R.~Milner, R.~Venugopalan, W.~Vogelsang, D.~Kaplan, H.~Montgomery and 
  S.~Vigdor {\it et al.},
  %``Gluons and the quark sea at high energies: Distributions, polarization, tomography,''
  arXiv:1108.1713 [nucl-th].
  %%CITATION = ARXIV:1108.1713;%%
%----------------------------------------------------------------------------------
%----------------------------------------------------------------------
%%             >> section 3 <<
%----------------------------------------------------------------------
%-------------------------------------------------------------------------------------------------
%
%[139]
\bibitem{RL}
{\sf Bernhard Riemann's Gesammelte mathematische Werke und Wissenschaftlicher Nachlass}, Eds.
H. Weber and R. Dedekind, (Teubner, Leipzig, 1876), p.~241;\\
H. Lebesgue, {\it Ann. scient. de l' {\'{E}.N.S.}} {$3^{e}$} {s\'{e}rie}, {\bf 20} (1903) 
453.%--485
%-------------------------------------------------------------------------------------------------
%
%[140]
\bibitem{GEYLA}
%\cite{Geyer:1999uq}
%\bibitem{Geyer:1999uq}
  B.~Geyer, M.~Lazar and D.~Robaschik,
  %``Decomposition of nonlocal light cone operators into harmonic operators of definite twist,''
  {\em Nucl.\ Phys.}\ B {\bf 559} (1999) 339
  [hep-th/9901090];\\
  %%CITATION = HEP-TH/9901090;%%
%----------------------------------------------------------------------------------------
%\cite{Geyer:2000ig}
%\bibitem{Geyer:2000ig}
  B.~Geyer and M.~Lazar,
  %``Twist decomposition of nonlocal light cone operators. 2. General tensors of 2nd rank,''
  {\em Nucl.\ Phys.}\ B {\bf 581} (2000) 341
  [hep-th/0003080].
  %%CITATION = HEP-TH/0003080;%%
%----------------------------------------------------------------------------------------
%
%[141]
\bibitem{FACT}
H.D.~Politzer, {\it Nucl. Phys.} B {\bf 129} (1977) 301;\\
D.~Amati, R.~Petronzio, and G.~Veneziano,
{\it Nucl. Phys.} B {\bf 140} (1978) 54;
%\cite{Amati:1978by}
%\bibitem{Amati:1978by}
%  D.~Amati, R.~Petronzio and G.~Veneziano,
  %``Relating Hard QCD Processes Through Universality of Mass Singularities. 2.,''
%  Nucl.\ Phys.\ B 
{\bf 146} (1978) 29;\\
  %%CITATION = NUPHA,B146,29;%%
S.B.~Libby and G.~Sterman,
{\it Phys. Rev.} D {\bf 18} (1978) 3252, 4737;\\
A.H.~Mueller,
{\it Phys. Rev.} D {\bf 18} (1978) 3705;\\
J.C.~Collins and G.~Sterman,
{\it Nucl. Phys.} B {\bf 185} (1981) 172;\\
J.C.~Collins, D.~Soper, and G.~Sterman,
{\it Nucl. Phys.} B {\bf 261} (1985) 104;\\
G.T.~Bodwin,
{\it Phys. Rev.} D {\bf 31} (1985) 2616.
%----------------------------------------------------------------------------------------
%
%[142]
\bibitem{MELLIN}
H. Mellin, 
%"Ueber die fundamentelle Wichtigkeit des Satzes von Cauchy für die Theorie der Gamma- und hypergeometrischen Funktionen" 
{\it Acta Soc. Sci. Fennica}, {\bf 21} (1896) 1;~ %–115
%H. Mellin, 
%"Ueber den Zusammenhang zwischen linearen Differential- und Differenzengleichungen" 
{\it Acta Math.} {\bf 25} (1902) 139.
%-----------------------------------------------------------------------
%----------------------------------------------------------------------
%%             >> section 4 <<
%----------------------------------------------------------------------
%-----------------------------------------------------------------------
%
%[143]
\bibitem{MUENST}
A. M\"unster, {\sf Statistische Thermodynamik}, (Springer, Berlin, 1956), pp.~112.
%--------------------------------------------------------------------------------------
%--------------------------------------------------------------------------------
%
%[144]
\bibitem{Stratmann:1995fn}
  M.~Stratmann, A.~Weber and W.~Vogelsang,
  %``Spin dependent nonsinglet structure functions in next-to-leading order,''
  Phys.\ Rev.\ D {\bf 53} (1996) 138
  [hep-ph/9509236].
  %%CITATION = HEP-PH/9509236;%%
%--------------------------------------------------------------------------------
%
%[145]
%\cite{Blumlein:1996tp}
\bibitem{Blumlein:1996tp}
  J.~Bl\"umlein and N.~Kochelev,
  %``On the twist-2 contributions to polarized structure functions and new sum rules,''
  {\em Phys.\ Lett.}\ B {\bf 381} (1996) 296
  [hep-ph/9603397].
  %%CITATION = HEP-PH/9603397;%%
%--------------------------------------------------------------------------------
%
%[146]
%\cite{Drell:1970yt}
\bibitem{Drell:1970yt}
  S.~D.~Drell and T.~-M.~Yan,
  %``Partons and their Applications at High-Energies,''
  {\em Annals Phys.}\  {\bf 66} (1971) 578
   [{\em Annals Phys.}\  {\bf 281} (2000) 450].
  %%CITATION = APNYA,66,578;%%
%--------------------------------------------------------------------------------------
%
%[147]
\bibitem{Weinberg:1966jm}
  S.~Weinberg,
  %``Dynamics at infinite momentum,''
  {\em Phys.\ Rev.}\  {\bf 150} (1966) 1313.
  %%CITATION = PHRVA,150,1313;%%
%-------------------------------------------------------------------------------------------------
%
%[148]
\bibitem{ZKIN}
%\cite{Barnett:1976kh}
%\bibitem{Barnett:1976kh}
  R.~M.~Barnett,
  %``Evidence in Neutrino Scattering for Righthanded Currents Associated with Heavy Quarks,''
  {\em Phys.\ Rev.}\ D {\bf 14} (1976) 70;\\
  %%CITATION = PHRVA,D14,70;%%
%-------------->
%\cite{Ellis:1976ig}
%\bibitem{Ellis:1976ig}
  R.~K.~Ellis, R.~Petronzio and G.~Parisi,
  %``Mass Dependent Corrections to the Bjorken Scaling Law,''
  {\em Phys.\ Lett.}\ B {\bf 64} (1976) 97;\\
  %%CITATION = PHLTA,B64,97;%%
%-------------->
%\cite{Barbieri:1976bj}
%\bibitem{Barbieri:1976bj}
  R.~Barbieri {\it et al.}, %, J.~R.~Ellis, M.~K.~Gaillard and G.~G.~Ross,
  %``A Quest for a Wholly Scaling Variable,''
  {\em Phys.\ Lett.}\ B {\bf 64} (1976) 171;
  %%CITATION = PHLTA,B64,171;%%
%-------------->
%\cite{Barbieri:1976rd}
%\bibitem{Barbieri:1976rd}
%  R.~Barbieri, J.~R.~Ellis, M.~K.~Gaillard and G.~G.~Ross,
  %``Mass Corrections to Scaling in Deep Inelastic Processes,''
  {\em Nucl.\ Phys.}\ B {\bf 117} (1976) 50;\\
  %%CITATION = NUPHA,B117,50;%%
%-------------->
%\cite{Brock:1979zi}
%\bibitem{Brock:1979zi}
  R.~Brock,
  %``Dimuon Measurements And The Strange Quark Sea,''
  {\em Phys.\ Rev.\ Lett.}\  {\bf 44} (1980) 1027.
  %%CITATION = PRLTA,44,1027;%%
%--------------------------------------------------------------------------------------
%
%[149]
\bibitem{COVP}
%\cite{Nash:1971aw}
%\bibitem{Nash:1971aw}
  C.~Nash,
  %``Polarization effects in the nonperturbative parton model of deep inelastic scattering,''
  {\em Nucl.\ Phys.}\ B {\bf 31} (1971) 419;\\
  %%CITATION = NUPHA,B31,419;%%
%-------------->
%\cite{Landshoff:1971xb}
%\bibitem{Landshoff:1971xb}
  P.~V.~Landshoff and J.~C.~Polkinghorne,
  %``Models for hadronic and leptonic processes at high-energy,''
  {\em Phys.\ Rept.}\  {\bf 5} (1972) 1.
  %%CITATION = PRPLC,5,1;%%
%--------------------------------------------------------------------------------------
%
%[150]
%\cite{Jackson:1989ph}
\bibitem{Jackson:1989ph}
  J.~D.~Jackson, G.~G.~Ross and R.~G.~Roberts,
  %``Polarized Structure Functions In The Parton Model,''
  {\em Phys.\ Lett.}\ B {\bf 226} (1989) 159;\\
  %%CITATION = PHLTA,B226,159;%%
%\cite{Roberts:1996ub}
%\bibitem{Roberts:1996ub}
  R.~G.~Roberts and G.~G.~Ross,
  %``Quark model description of polarized deep inelastic scattering and the prediction of g(2),''
  Phys.\ Lett.\ B {\bf 373} (1996) 235
  [hep-ph/9601235].
  %%CITATION = HEP-PH/9601235;%%
%--------------------------------------------------------------------------------------
%----------------------------------------------------------------------
%%             >> section 5 <<
%----------------------------------------------------------------------
%--------------------------------------------------------------------------------------
%
%[151]
\bibitem{CURNC}
%\cite{Gracey:2000am}    
%\bibitem{Gracey:2000am} 
  J.~A.~Gracey, 
  %``Three loop MS-bar tensor current anomalous dimension in QCD,''
  {\it Phys.\ Lett.}\ B {\bf 488} (2000) 175  
  [hep-ph/0007171];\\
  %%CITATION = HEP-PH/0007171;%%
%\cite{Blumlein:2001ca}
%\bibitem{Blumlein:2001ca}
  J.~Bl\"umlein,
  %``On the anomalous dimension of the transversity distribution h(1)(x,Q**2),''
  {\it Eur.\ Phys.\ J}.\ C {\bf 20} (2001) 683
  [hep-ph/0104099].
  %%CITATION = HEP-PH/0104099;%%
%--------------------------------------------------------------------------------------
%
%[152]
%\cite{Furmanski:1981cw}
\bibitem{Furmanski:1981cw}
  W.~Furmanski and R.~Petronzio,
  %``Lepton - Hadron Processes Beyond Leading Order in Quantum Chromodynamics,''
  {\it Z.\ Phys.}\ C {\bf 11} (1982) 293 and references therein.
  %%CITATION = ZEPYA,C11,293;%%
%--------------------------------------------------------------------------------
%
%[153]
%\cite{Grunberg:1982fw}
\bibitem{Grunberg:1982fw}
  G.~Grunberg,
  %``Renormalization Scheme Independent QCD and QED: The Method of Effective Charges,''
  {\it Phys.\ Rev.}\ D {\bf 29} (1984) 2315.
  %%CITATION = PHRVA,D29,2315;%%
%--------------------------------------------------------------------------------------
%
%[154]
%\cite{Catani:1996sc}
\bibitem{Catani:1996sc}
  S.~Catani,
  %``Physical anomalous dimensions at small x,''
  {\it Z.\ Phys.}\ C {\bf 75} (1997) 665
  [hep-ph/9609263].
  %%CITATION = HEP-PH/9609263;%%
%--------------------------------------------------------------------------------------
%
%[155]
%\cite{Blumlein:2000wh}
\bibitem{Blumlein:2000wh}
  J.~Bl\"umlein, V.~Ravindran and W.~L.~van Neerven,
  %``On the Drell-Levy-Yan relation to O(alpha(s)**2),''
  {\it Nucl.\ Phys.}\ B {\bf 586} (2000) 349
  [hep-ph/0004172].
  %%CITATION = HEP-PH/0004172;%%
%--------------------------------------------------------------------------------------
%
%[156]
\bibitem{MASSREN}
%\cite{Tarrach:1980up}
%\bibitem{Tarrach:1980up}
  R.~Tarrach,
  %``The Pole Mass In Perturbative QCD,''
  {\it Nucl.\ Phys.}\  B {\bf 183} (1981) 384; \\
  %%CITATION = NUPHA,B183,384;%%
%\cite{Nachtmann:1981zg}
%\bibitem{Nachtmann:1981zg}   
  O.~Nachtmann and W.~Wetzel,
  %``The Beta Function For Effective Quark Masses To Two Loops In QCD,''
  {\it Nucl.\ Phys.}\  B {\bf 187} (1981) 333; \\
  %%CITATION = NUPHA,B187,333;%%
%\cite{Gray:1990yh}
%\bibitem{Gray:1990yh}
  N.~Gray, D.~J.~Broadhurst, W.~Grafe and K.~Schilcher,
  %``Three Loop Relation Of Quark (Modified) Ms And Pole Masses,''
  {\it Z.\ Phys.}\  C {\bf 48},  (1990) 673;\\
  %%CITATION = ZEPYA,C48,673;%%
%\cite{Broadhurst:1991fy}    
%\bibitem{Broadhurst:1991fy}  
  D.~J.~Broadhurst, N.~Gray and K.~Schilcher,
  %``Gauge invariant on-shell Z(2) in QED, QCD and the effective field theory of
  %a static quark,''
  {\it Z.\ Phys.}\  C {\bf 52} (1991) 111;\\
  %%CITATION = ZEPYA,C52,111;%%
%\cite{Chetyrkin:1999qi}
%\bibitem{Chetyrkin:1999qi}
  K.~G.~Chetyrkin and M.~Steinhauser,
  %``The relation between the MS-bar and the on-shell quark mass at order
  %alpha(s)**3,''
  {\it Nucl.\ Phys.}\  B {\bf 573}, 617 (2000)
  [arXiv:hep-ph/9911434]; \\
  %%CITATION = NUPHA,B573,617;%%
%\cite{Chetyrkin:1999ys}
%\bibitem{Chetyrkin:1999ys}
%  K.~G.~Chetyrkin and M.~Steinhauser,
  %``Short distance mass of a heavy quark at order alpha(s)**3,''
  {\it Phys.\ Rev.\ Lett.}\  {\bf 83}, 4001 (1999)
  [arXiv:hep-ph/9907509].
  %%CITATION = PRLTA,83,4001;%%
%------------------------------------------------------------------------
%
%[157]
\bibitem{BGF}
%\cite{Abbott:1980hw}
%\bibitem{Abbott:1980hw}
  L.~F.~Abbott,
  %``The Background Field Method Beyond One Loop,''
  {\it Nucl.\ Phys.}\  B {\bf 185} (1981) 189;\\
  %%CITATION = NUPHA,B185,189;%%
%\cite{Rebhan:1985yf}
%\bibitem{Rebhan:1985yf}
  A.~Rebhan,
  %``MOMENTUM SUBTRACTION SCHEME AND THE BACKGROUND FIELD METHOD IN QCD,''
  {\it Z.\ Phys.}\  C {\bf 30} (1986) 309;\\
  %%CITATION = ZEPYA,C30,309;%%
%\cite{Jegerlehner:1998zg}
%\bibitem{Jegerlehner:1998zg}  
  F.~Jegerlehner and O.~V.~Tarasov,
  %``Exact mass dependent two-loop alpha(s)-bar(Q**2) in the background MOM
  %renormalization scheme,''
  {\it Nucl.\ Phys.}\  B {\bf 549} (1999) 481
  [arXiv:hep-ph/9809485].
  %%CITATION = NUPHA,B549,481;%%
%--------------------------------------------------------------------------------
%----------------------------------------------------------------------
%%             >> section 6 <<
%----------------------------------------------------------------------
%--------------------------------------------------------------------------------
%
%[158]
%\cite{Duke:1984ge}
\bibitem{Duke:1984ge}
  D.~W.~Duke and R.~G.~Roberts,
  %``Determinations of the QCD Strong Coupling alpha-s and the Scale Lambda (QCD),''
  {\it Phys.\ Rept.}\  {\bf 120} (1985) 275.
  %%CITATION = PRPLC,120,275;%%
%--------------------------------------------------------------------------------
%
%[159]
%\cite{Bardeen:1978yd}
\bibitem{Bardeen:1978yd}
  W.~A.~Bardeen, A.~J.~Buras, D.~W.~Duke and T.~Muta,
  %``Deep Inelastic Scattering Beyond the Leading Order in Asymptotically Free Gauge Theories,''
  {\it Phys.\ Rev.}\ D {\bf 18} (1978) 3998.
  %%CITATION = PHRVA,D18,3998;%%
%--------------------------------------------------------------------------------
%
%[160]
%\cite{Caswell:1974gg}
\bibitem{Caswell:1974gg}
  W.~E.~Caswell,
  %``Asymptotic Behavior of Nonabelian Gauge Theories to Two Loop Order,''
  {\it Phys.\ Rev.\ Lett.}\  {\bf 33} (1974) 244.
  %%CITATION = PRLTA,33,244;%%
%--------------------------------------------------------------------------------
%
%[161]
%\cite{Jones:1974mm}
\bibitem{Jones:1974mm}
  D.~R.~T.~Jones,
  %``Two Loop Diagrams in Yang-Mills Theory,''
  {\it Nucl.\ Phys.}\ B {\bf 75} (1974) 531.
  %%CITATION = NUPHA,B75,531;%%
%--------------------------------------------------------------------------------
%
%[162]
%\cite{Tarasov:1980au}
\bibitem{Tarasov:1980au}
  O.~V.~Tarasov, A.~A.~Vladimirov and A.~Y.~Zharkov,
  %``The Gell-Mann-Low Function of QCD in the Three Loop Approximation,''
  {\it Phys.\ Lett.}\ B {\bf 93} (1980) 429.
  %%CITATION = PHLTA,B93,429;%%
%--------------------------------------------------------------------------------
%
%[163]
%\cite{Larin:1993tp}
\bibitem{Larin:1993tp}
  S.~A.~Larin and J.~A.~M.~Vermaseren,
  %``The Three loop QCD Beta function and anomalous dimensions,''
  {\it Phys.\ Lett.}\ B {\bf 303} (1993) 334
  [hep-ph/9302208].
  %%CITATION = HEP-PH/9302208;%%
%--------------------------------------------------------------------------------
%
%[164]
\bibitem{CHET11}
K. Chetyrkin {\it et al.},
{\tt http://indico.cern.ch/conferenceOtherViews.py?view=standard\&confId=154661},
contribution to RADCOR 2011;\\
%\cite{Baikov:2012rr}
%\bibitem{Baikov:2012rr}
  P.~A.~Baikov, K.~G.~Chetyrkin, J.~H.~K\"uhn and C.~Sturm,
  %``The relation between the QED charge renormalized in MSbar and on-shell schemes at four loops, the QED on-shell beta-function at five loops and asymptotic contributions to the muon anomaly at five and six loops,''
  arXiv:1207.2199 [hep-ph].
  %%CITATION = ARXIV:1207.2199;%%
%--------------------------------------------------------------------------------
%
%[165]
%\cite{Blumlein:1994kw}
\bibitem{Blumlein:1994kw}
  J.~Bl\"umlein and J.~Botts,
  %``Do deep inelastic scattering data favor a light gluino?,''
  {\it Phys.\ Lett.}\ B {\bf 325} (1994) 190
   [Erratum-ibid.\ B {\bf 331} (1994) 450]
  [hep-ph/9401291].
  %%CITATION = HEP-PH/9401291;%%
%--------------------------------------------------------------------------------
%
%[166]
%\cite{Chetyrkin:1997sg}
\bibitem{Chetyrkin:1997sg}
  K.~G.~Chetyrkin, B.~A.~Kniehl and M.~Steinhauser,
  %``Strong coupling constant with flavor thresholds at four loops in the MS scheme,''
  {\it Phys.\ Rev.\ Lett.}\  {\bf 79} (1997) 2184
  [hep-ph/9706430].
  %%CITATION = HEP-PH/9706430;%%
%--------------------------------------------------------------------------------
%
%[167]
%\cite{Bernreuther:1981sg}
\bibitem{Bernreuther:1981sg}
  W.~Bernreuther and W.~Wetzel,
  %``Decoupling of Heavy Quarks in the Minimal Subtraction Scheme,''
  {\it Nucl.\ Phys.}\ B {\bf 197} (1982) 228
   [Erratum-ibid.\ B {\bf 513} (1998) 758].
  %%CITATION = NUPHA,B197,228;%%
%--------------------------------------------------------------------------------
%
%[168]
%\cite{Marciano:1983pj}
\bibitem{Marciano:1983pj}
  W.~J.~Marciano,
  %``Flavor Thresholds and Lambda in the Modified Minimal Subtraction Prescription,''
  {\it Phys.\ Rev.}\ D {\bf 29} (1984) 580.
  %%CITATION = PHRVA,D29,580;%%
%--------------------------------------------------------------------------------
%---------- splitting functions -------------------------------------------------
%--------------------------------------------------------------------------------
%
%[169]
\bibitem{CARLS}
E. Carlson, Thesis, Univ. Uppsala, 1914;\\
E.C. Titchmarsh, Theory of Functions, (Oxford University Press, Oxford, 1939), Chapt. 9.5.
%-----------------------------------------------------------------------
%
%[170]
\bibitem{MEL:UN}
%\cite{Parisi:1973nx}
%\bibitem{Parisi:1973nx}
  G.~Parisi,
  %``Experimental limits on the values of anomalous dimensions,''
  {\it Phys.\ Lett.}\ B {\bf 43} (1973) 207;
%-----------------------------------------------------------------------
%\cite{Parisi:1974sq}
%\bibitem{Parisi:1974sq}
%  G.~Parisi,
  %``Detailed Predictions for the p n Structure Functions in Theories with Computable Large Momenta Behavior,''
  {\it Phys.\ Lett.}\ B {\bf 50} (1974) 367;\\
  %%CITATION = PHLTA,B50,367;%%
%-----------------------------------------------------------------------
%\bibitem{Gross:1974fm}
  D.~J.~Gross,
  %``How to Test Scaling in Asymptotically Free Theories,''
  {\it Phys.\ Rev.\ Lett.}\  {\bf 32} (1974) 1071;\\
  %%CITATION = PRLTA,32,1071;%%
%-----------------------------------------------------------------------
%\cite{Zee:1974du}
%\bibitem{Zee:1974du}
  A.~Zee, F.~Wilczek and S.~B.~Treiman,
  %``Scaling Deviations for Neutrino Reactions in Asymptotically Free Field Theories,''
  {\it Phys.\ Rev.}\ D {\bf 10} (1974) 2881.
  %%CITATION = PHRVA,D10,2881;%%
%-----------------------------------------------------------------------
%
%[171]
%\bibitem{SP1}
%\cite{Gross:1973ju}
\bibitem{Gross:1973ju}
  D.~J.~Gross and F.~Wilczek,
  %``Asymptotically Free Gauge Theories. 1,''
  {\it Phys.\ Rev.}\ D {\bf 8} (1973) 3633.
  %%CITATION = PHRVA,D8,3633;%%
%-----------------------------------------------------------------------
%
%[172]
%\cite{Gross:1974cs}
\bibitem{Gross:1974cs}
  D.~J.~Gross and F.~Wilczek,
  %``Asymptotically Free Gauge Theories. 2.,''
  {\it Phys.\ Rev.}\ D {\bf 9} (1974) 980.
  %%CITATION = PHRVA,D9,980;%%%
%-----------------------------------------------------------------------
%
%[173]
%\bibitem{SP2}
%\cite{Georgi:1951sr}
\bibitem{Georgi:1951sr}
  H.~Georgi and H.~D.~Politzer,
  %``Electroproduction scaling in an asymptotically free theory of strong interactions,''
  {\it Phys.\ Rev.}\ D {\bf 9} (1974) 416.
  %%CITATION = PHRVA,D9,416;%%
%-----------------------------------------------------------------------
%
%[174]
\bibitem{HSUM}
%\cite{Vermaseren:1998uu}
%\bibitem{Vermaseren:1998uu}
  J.~A.~M.~Vermaseren,
  %``Harmonic sums, Mellin transforms and integrals,''
  {\it Int.\ J.\ Mod.\ Phys.}\ A {\bf 14} (1999) 2037
  [hep-ph/9806280];\\
  %%CITATION = HEP-PH/9806280;%%
%-----------------------------------------------------------------------
%\cite{Blumlein:1998if}
%\bibitem{Blumlein:1998if}
  J.~Bl\"umlein and S.~Kurth,
  %``Harmonic sums and Mellin transforms up to two loop order,''
  {\it Phys.\ Rev.}\ D {\bf 60} (1999) 014018
  [hep-ph/9810241].
  %%CITATION = HEP-PH/9810241;%%
%-----------------------------------------------------------------------
%
%[175]
\bibitem{QED1}
%\cite{Fermi:1924tc}
%\bibitem{Fermi:1924tc}
  E.~Fermi,
  %``On the Theory of the impact between atoms and electrically charged particles,''
  {\em Z.\ Phys.}\  {\bf 29} (1924) 315;\\
  %%CITATION = ZEPYA,29,315;%%
%--------------------------------------------------------------------------------
%\bibitem{Williams}
E.J. Williams, {\em Proc. Roy. Soc. London} (A) {\bf 139} (1933) 163; 
%\cite{Williams:1934ad}
%\bibitem{Williams:1934ad}
%  E.~J.~Williams,
  %``Nature of the high-energy particles of penetrating radiation and status of ionization and radiation formulae,''
  {\em Phys.\ Rev.}\  {\bf 45} (1934) 729;
Mat. Fys. Medd. {\bf 13} (1935) 4;\\
  %%CITATION = PHRVA,45,729;%%
%--------------------------------------------------------------------------------
%\cite{vonWeizsacker:1934sx}
%\bibitem{vonWeizsacker:1934sx}
  C.~F.~von Weizs\"acker,
  %``Radiation emitted in collisions of very fast electrons,''
  {\em Z.\ Phys.}\  {\bf 88} (1934) 612;\\
  %%CITATION = ZEPYA,88,612;%%
%--------------------------------------------------------------------------------
%\bibitem{KESSLER}
P. Kessler, {\it Nuovo Cim.} {\bf 17} (1960) 809;\\
%--------------------------------------------------------------------------------
%\bibitem{LANDAU4}
{{\cyr V.B. Berestetski\u{i}, E.M. Lifshits, L.P. Pitaevski\u{i}, {\sf Relyativist{skaya} Kvantovaya 
Teoriya, Chastp1}} {\rm I}, \S 96, {{\cyr (Nauka, Moskva, 1967);}} German translation, 
(Akademie Verlag, Berlin, 1970). 
%--------------------------------------------------------------------------------
%
%[176]
\bibitem{QED2}
%\cite{Kim:1973he}
%\bibitem{Kim:1973he}
  K.~J.~Kim and Y.-S.~Tsai,
  %``Improved Weizsacker-williams Method And Its Application To Lepton And W Boson Pair Production,''
  {\it Phys.\ Rev.}\ D {\bf 8} (1973) 3109;\\
  %%CITATION = PHRVA,D8,3109;%%
%\cite{Baier:1973ms}
%\bibitem{Baier:1973ms}
  V.~N.~Baier, V.~S.~Fadin and V.~A.~Khoze,
  %``Quasireal electron method in high-energy quantum electrodynamics,''
  {\it Nucl.\ Phys.}\ B {\bf 65} (1973) 381;\\
  %%CITATION = NUPHA,B65,381;%%
%--------------------------------------------------------------------------------
%\cite{Chen:1975sh}
%\bibitem{Chen:1975sh}
  M.~-S.~Chen and P.~M.~Zerwas,
  %``Equivalent-Particle Approximations in electron and Photon Processes of Higher Order QED,''
  {\it Phys.\ Rev.}\ D {\bf 12} (1975) 187.
  %%CITATION = PHRVA,D12,187;%%
%--------------------------------------------------------------------------------
%
%[177]
%\bibitem{RC28}
%\cite{Gribov:1972rt}
\bibitem{Gribov:1972rt}
  V.~N.~Gribov and L.~N.~Lipatov,
  %``e+ e- pair annihilation and deep inelastic e p scattering in perturbation theory,''
  {\em Sov.\ J.\ Nucl.\ Phys.}\  {\bf 15} (1972) 675
   [{\em Yad.\ Fiz.}\  {\bf 15} (1972) 1218].
  %%CITATION = SJNCA,15,675;%%
%--------------------------------------------------------------------------------
%
%[178]
%\bibitem{RC29}
%\cite{Gribov:1972ri}
\bibitem{Gribov:1972ri}
  V.~N.~Gribov and L.~N.~Lipatov,
  %``Deep inelastic e p scattering in perturbation theory,''
  {\em Sov.\ J.\ Nucl.\ Phys.}\  {\bf 15} (1972) 438
   [{\em Yad.\ Fiz.}\  {\bf 15} (1972) 781].
  %%CITATION = SJNCA,15,438;%%
%--------------------------------------------------------------------------------
%
%[179]
\bibitem{CHRIST72}
N. Christ, {\sf Perturbation theory on the lightcone}, paper \# 485 in:
Proceedings of the XVI International Conference on High Energy Physics,
Batavia, USA, Sept. 6-13, 1972, eds. J.D. Jackson, A. Roberts, R.~Donaldson,
(Fermilab, Batavia, 1972), Vol.~2, pp.~122.
%-----------------------------------------------------------------------
%
%[180]
%\bibitem{SP3}
%\cite{Lipatov:1974qm}
\bibitem{Lipatov:1974qm}
  L.~N.~Lipatov,
  %``The parton model and perturbation theory,''
  {\em Sov.\ J.\ Nucl.\ Phys.}\  {\bf 20} (1975) 94
   [{\em Yad.\ Fiz.}\  {\bf 20} (1974) 181].
  %%CITATION = SJNCA,20,94;%%
%-----------------------------------------------------------------------
%
%[181]
%\cite{Bukhvostov:1974uu}
\bibitem{Bukhvostov:1974uu}
  A.~P.~Bukhvostov, L.~N.~Lipatov and N.~P.~Popov,
  %``Parton distribution functions in perturbation theory,''
  {\em Yad.\ Fiz.}\  {\bf 20} (1974) 532.
  %%CITATION = YAFIA,20,532;%%
%-----------------------------------------------------------------------
%
%[182]
%\cite{Parisi:1976qj}
\bibitem{Parisi:1976qj}
  G.~Parisi, {\sf An Introduction to Scaling Violations},
  LNF-76-25-P, Int. Meeting on Neutrino Physics, Flaine, March 6--12, 1976. 
  %%CITATION = LNF-76-25-P;%%
%-----------------------------------------------------------------------
%
%[183]
%\bibitem{SP4}
%\cite{Altarelli:1977zs}
\bibitem{Altarelli:1977zs}
  G.~Altarelli and G.~Parisi,
  %``Asymptotic Freedom in Parton Language,''
  {\it Nucl.\ Phys.}\ B {\bf 126} (1977) 298.
  %%CITATION = NUPHA,B126,298;%%
%-----------------------------------------------------------------------
%
%[184]
%\bibitem{SP5}
%\cite{Kim:1977hp}
\bibitem{Kim:1977hp}
  K.~J.~Kim and K.~Schilcher,
  %``Scaling Violation in the Infinite Momentum Frame,''
  {\it Phys.\ Rev.}\ D {\bf 17} (1978) 2800.
  %%CITATION = PHRVA,D17,2800;%%
%-----------------------------------------------------------------------
%
%[185]
%\bibitem{SP6}
%\cite{Dokshitzer:1977sg}
\bibitem{Dokshitzer:1977sg}
  Y.~L.~Dokshitzer,
  %``Calculation of the Structure Functions for Deep Inelastic Scattering and e+ e- Annihilation by Perturbation Theory in Quantum Chromodynamics.,''
  {\it Sov.\ Phys.\ JETP} {\bf 46} (1977) 641
   [{\it Zh.\ Eksp.\ Teor.\ Fiz.}\  {\bf 73} (1977) 1216].
  %%CITATION = SPHJA,46,641;%%
%-----------------------------------------------------------------------
%
%[186]
\bibitem{Bukhvostov:1985rn}
  A.~P.~Bukhvostov, G.~V.~Frolov, L.~N.~Lipatov and E.~A.~Kuraev,
  %``Evolution Equations for Quasi-Partonic Operators,''
  {\it Nucl.\ Phys.}\ B {\bf 258} (1985) 601.
  %%CITATION = NUPHA,B258,601;%%
%-----------------------------------------------------------------------
%
%[187]
\bibitem{DKMT}
Y.L. Dokshitzer, V.A. Khoze, A.H. Mueller, and S.I. Troyan, {\sf Basics of perturbative QCD},
(Editions Frontieres, Paris, 1991).
%-----------------------------------------------------------------------
%
%[188]
%\bibitem{PO1}
%\cite{Sasaki:1975hk}
\bibitem{Sasaki:1975hk}
  K.~Sasaki,
  %``Polarized Electroproduction in Asymptotically Free Gauge Theories,''
  {\it Prog.\ Theor.\ Phys.}\  {\bf 54} (1975) 1816.
  %%CITATION = PTPKA,54,1816;%%
%-----------------------------------------------------------------------
%
%[189]
%\cite{Ahmed:1975tj}
\bibitem{Ahmed:1975tj}
  M.~A.~Ahmed and G.~G.~Ross,
  %``Spin-Dependent Deep Inelastic electron Scattering in an Asymptotically Free Gauge Theory,''
  {\it Phys.\ Lett.}\ B {\bf 56} (1975) 385.
  %%CITATION = PHLTA,B56,385;%%
%-----------------------------------------------------------------------
%
%[190]
%\cite{Ito:1975pf}
\bibitem{Ito:1975pf}
  H.~Ito,
  %``Polarized Electroproduction in an Asymptotically Free Gauge Theory,''
  {\it Prog.\ Theor.\ Phys.}\  {\bf 54} (1975) 555.
  %%CITATION = PTPKA,54,555;%%
%-----------------------------------------------------------------------
%
%[191]
\bibitem{SPLIT:NLO}
%\cite{Floratos:1977au}
%\bibitem{Floratos:1977au}
  E.~G.~Floratos, D.~A.~Ross and C.~T.~Sachrajda,
  %``Higher Order Effects in Asymptotically Free Gauge Theories: The Anomalous Dimensions of Wilson Operators,''
  {\it Nucl.\ Phys.}\ B {\bf 129} (1977) 66
   [Erratum-ibid.\ B {\bf 139} (1978) 545];
  %%CITATION = NUPHA,B129,66;%%
%-----------------------------------------------------------------------
%\cite{Floratos:1978ny}
%\bibitem{Floratos:1978ny}
%  E.~G.~Floratos, D.~A.~Ross and C.~T.~Sachrajda,
  %``Higher Order Effects in Asymptotically Free Gauge Theories. 2. Flavor Singlet Wilson Operators and Coefficient Functions,''
  {\it Nucl.\ Phys.}\ B {\bf 152} (1979) 493;\\
  %%CITATION = NUPHA,B152,493;%%
%-----------------------------------------------------------------------
%\bibitem{SP8}
%\cite{GonzalezArroyo:1979df}
%\bibitem{GonzalezArroyo:1979df}
  A.~Gonzalez-Arroyo, C.~Lopez and F.~J.~Yndurain,
  %``Second Order Contributions to the Structure Functions in Deep Inelastic Scattering. 1. Theoretical Calculations,''
  {\it Nucl.\ Phys.}\ B {\bf 153} (1979) 161;
  %%CITATION = NUPHA,B153,161;%% 
%----------------------------------------------------------------------- % 
%\cite{GonzalezArroyo:1979ng} 
%\bibitem{GonzalezArroyo:1979ng} 
% A.~Gonzalez-Arroyo, C.~Lopez and F.~J.~Yndurain,
  %``Second Order Contributions To The Structure Functions In Deep Inelastic Scattering. Ii. Comparison With Experiment For The Nonsinglet Contributions To E, Mu Nucleon Scattering,''
  {\it Nucl.\ Phys.}\ B {\bf 159} (1979) 512;\\
  %%CITATION = NUPHA,B159,512;%%
%-----------------------------------------------------------------------
%\bibitem{SP11}
%\cite{Curci:1980uw}
%\bibitem{Curci:1980uw}
  G.~Curci, W.~Furmanski and R.~Petronzio,
  %``Evolution of Parton Densities Beyond Leading Order: The Nonsinglet Case,''
  {\it Nucl.\ Phys.}\ B {\bf 175} (1980) 27;\\
  %%CITATION = NUPHA,B175,27;%%
%-----------------------------------------------------------------------
%\cite{Furmanski:1980cm}
%\bibitem{Furmanski:1980cm}
  W.~Furmanski and R.~Petronzio,
  %``Singlet Parton Densities Beyond Leading Order,''
  {\it Phys.\ Lett.}\ B {\bf 97} (1980) 437;\\
  %%CITATION = PHLTA,B97,437;%%
%-----------------------------------------------------------------------
%\bibitem{SP10}
%\cite{Floratos:1980hk}
%\bibitem{Floratos:1980hk}
  E.~G.~Floratos, R.~Lacaze and C.~Kounnas,
  %``Space and Timelike Cut Vertices in QCD Beyond the Leading Order. 1. Nonsinglet Sector,''
  {\it Phys.\ Lett.}\ B {\bf 98} (1981) 89;
  %%CITATION = PHLTA,B98,89;%%
%-----------------------------------------------------------------------
%\bibitem{SP10A}
%\cite{Floratos:1981hs}
%\bibitem{Floratos:1981hs}
%  E.~G.~Floratos, C.~Kounnas and R.~Lacaze,
  %``Higher Order QCD Effects in Inclusive Annihilation and Deep Inelastic Scattering,''
  {\it Nucl.\ Phys.}\ B {\bf 192} (1981) 417;
  %%CITATION = NUPHA,B192,417;%%
%-----------------------------------------------------------------------
%\bibitem{SP12}
%\cite{Floratos:1980hm}
%\bibitem{Floratos:1980hm}
%  E.~G.~Floratos, R.~Lacaze and C.~Kounnas,
  %``Space And Timelike Cut Vertices In Qcd Beyond The Leading Order. 2. The Singlet Sector,''
  {\it Phys.\ Lett.}\ B {\bf 98} (1981) 285;\\
  %%CITATION = PHLTA,B98,285;%%
%-----------------------------------------------------------------------
%\bibitem{SP9}
%\cite{GonzalezArroyo:1979he}
%\bibitem{GonzalezArroyo:1979he}
  A.~Gonzalez-Arroyo and C.~Lopez,
  %``Second Order Contributions to the Structure Functions in Deep Inelastic Scattering. 3. The Singlet Case,''
  {\it Nucl.\ Phys.}\ B {\bf 166} (1980) 429;\\
  %%CITATION = NUPHA,B166,429;%%
%-----------------------------------------------------------------------
%\cite{Hamberg:1991qt}
%\bibitem{Hamberg:1991qt}
  R.~Hamberg and W.~L.~van Neerven,
  %``The Correct renormalization of the gluon operator in a covariant gauge,''
  {\it Nucl.\ Phys.}\ B {\bf 379} (1992) 143;\\
  %%CITATION = NUPHA,B379,143;%%
%-----------------------------------------------------------------------
%\cite{Ellis:1996nn}
%\bibitem{Ellis:1996nn}
  R.~K.~Ellis and W.~Vogelsang,
  %``The Evolution of parton distributions beyond leading order: The Singlet case,''
  hep-ph/9602356.
  %%CITATION = HEP-PH/9602356;%%
%-----------------------------------------------------------------------
%
%[192]
%\cite{Moch:1999eb}
\bibitem{Moch:1999eb}
  S.~Moch and J.~A.~M.~Vermaseren,
  %``Deep inelastic structure functions at two loops,''
  {\it Nucl.\ Phys.}\ B {\bf 573} (2000) 853
  [hep-ph/9912355].
  %%CITATION = HEP-PH/9912355;%%
%-----------------------------------------------------------------------
%
%[193]
%\cite{Gracey:1994nn}
\bibitem{Gracey:1994nn}
  J.~A.~Gracey,
  %``Anomalous dimension of nonsinglet Wilson operators at O (1 / N(f)) in deep inelastic scattering,''
  {\it Phys.\ Lett.}\ B {\bf 322} (1994) 141
  [hep-ph/9401214].
  %%CITATION = HEP-PH/9401214;%%
%------------------------------------------------------------------------------------------
%
%[194]
%\cite{Ablinger:2010ty}
\bibitem{Ablinger:2010ty}
  J.~Ablinger, J.~Bl\"umlein, S.~Klein, C.~Schneider and F.~Wissbrock,
  %``The O(\alpha_s^3) Massive Operator Matrix Elements of O(n_f) for the Structure Function F_2(x,Q^2) and Transversity,''
  {\it Nucl.\ Phys.}\ B {\bf 844} (2011) 26
  [arXiv:1008.3347 [hep-ph]].
  %%CITATION = ARXIV:1008.3347;%%
%------------------------------------------------------------------------------------------
%
%[195]
\bibitem{BHS12a}
J.~Bl\"umlein, A. Hasselhuhn, S. Klein, and C. Schneider, 
%\cite{Blumlein:2012vq}
%\bibitem{Blumlein:2012vq}
%  J.~Blumlein, A.~Hasselhuhn, S.~Klein and C.~Schneider,
  %``The $O(\alpha_s^3 n_f T_F^2 C_{A,F})$} Contributions to the Gluonic Massive Operator Matrix Elements,''
  arXiv:1205.4184 [hep-ph].
  %%CITATION = ARXIV:1205.4184;%%
%------------------------------------------------------------------------------------------
%
%[196]
%\cite{Blumlein:2009ta}
\bibitem{Blumlein:2009ta}
  J.~Bl\"umlein,
  %``Structural Relations of Harmonic Sums and Mellin Transforms up to Weight w = 5,''
  {\it Comput.\ Phys.\ Commun.}\  {\bf 180} (2009) 2218
  [arXiv:0901.3106 [hep-ph]];
  %%CITATION = ARXIV:0901.3106;%%
%\cite{Blumlein:2009fz}
%\bibitem{Blumlein:2009fz}
%  J.~Bl\"umlein,
  %``Structural Relations of Harmonic Sums and Mellin Transforms at Weight w = 6,''
  %Submitted to: Clay Math.Proc.
  [arXiv:0901.0837 [math-ph]].
  %%CITATION = ARXIV:0901.0837;%%
%------------------------------------------------------------------------------------------
%
%[197]
\bibitem{ANCONT}
%\cite{Blumlein:2000hw}
%\bibitem{Blumlein:2000hw}
  J.~Bl\"umlein,
  %``Analytic continuation of Mellin transforms up to two loop order,''
  {\it Comput.\ Phys.\ Commun.}\  {\bf 133} (2000) 76
  [hep-ph/0003100];
\\
  %%CITATION = HEP-PH/0003100;%%
%\cite{Blumlein:2005jg}
%\bibitem{Blumlein:2005jg}
  J.~Bl\"umlein and S.~-O.~Moch,
  %``Analytic continuation of the harmonic sums for the 3-loop anomalous dimensions,''
  {\it Phys.\ Lett.}\ B {\bf 614} (2005) 53
  [hep-ph/0503188];\\
  %%CITATION = HEP-PH/0503188;%%
%\cite{Kotikov:2005gr}
%\bibitem{Kotikov:2005gr}
  A.~V.~Kotikov and V.~N.~Velizhanin,
  %``Analytic continuation of the Mellin moments of deep inelastic structure functions,''
  hep-ph/0501274.
  %%CITATION = HEP-PH/0501274;%%
%------------------------------------------------------------------------------------------
%
%[198]
%\cite{Remiddi:1999ew}  
\bibitem{Remiddi:1999ew}
  E.~Remiddi and J.~A.~M.~Vermaseren,
  %``Harmonic polylogarithms,''
  {\it Int.\ J.\ Mod.\ Phys.}\ A {\bf 15} (2000) 725
  [hep-ph/9905237].
  %%CITATION = HEP-PH/9905237;%%
%------------------------------------------------------------------------------------------
%
%[199]
%\cite{Blumlein:2006be}
\bibitem{Blumlein:2006be}
  J.~Bl\"umlein, H.~B\"ottcher and A.~Guffanti,
  %``Non-singlet QCD analysis of deep inelastic world data at O(alpha(s)**3),''
  {\it Nucl.\ Phys.}\ B {\bf 774} (2007) 182
  [hep-ph/0607200].
  %%CITATION = HEP-PH/0607200;%%
%------------------------------------------------------------------------------------------
%---------------------------- Large N_f Limit             ---------------------------------
%------------------------------------------------------------------------------------------
%
%[200]
\bibitem{VPH}
A.N. Vasil'ev, Yu.M. Pis'mak, and J.R. Honkonen, {\it Theor. Math. Phys.} {\bf 46} (1981) 157;
%A.N. Vasil'ev, Yu.M. Pis'mak, and J.R. Honkonen, 
%Theor. Math. Phys. 
{\bf 47} (1981) 291.
%------------------------------------------------------------------------------------------
%
%[201]
%\cite{Gracey:1993sn}
\bibitem{Gracey:1993sn}
  J.~A.~Gracey,
  %``Electron mass anomalous dimension at O(1/(Nf(2)) in quantum electrodynamics,''
  {\it Phys.\ Lett.}\ B {\bf 317} (1993) 415
  [hep-th/9309092].
  %%CITATION = HEP-TH/9309092;%%
%------------------------------------------------------------------------------------------
%
%[202]
%\cite{Ciuchini:1999wy}
\bibitem{Ciuchini:1999wy}
  M.~Ciuchini, S.~E.~Derkachov, J.~A.~Gracey and A.~N.~Manashov,
  %``Computation of quark mass anomalous dimension at O(1 / N**2(f)) in quantum chromodynamics,''
  {\it Nucl.\ Phys.}\ B {\bf 579} (2000) 56
  [hep-ph/9912221].
  %%CITATION = HEP-PH/9912221;%%
%------------------------------------------------------------------------------------------
%
%[203]
%\cite{Ali:2001ng}
\bibitem{Ali:2001ng}
  D.~B.~Ali and J.~A.~Gracey,
  %``Anomalous dimension of nonsinglet quark currents at O(1/N(f)**2) in QCD,''
  {\it Phys.\ Lett.}\ B {\bf 518} (2001) 188
  [hep-ph/0105038].
  %%CITATION = HEP-PH/0105038;%%
%------------------------------------------------------------------------
%
%[204]
%\cite{Gracey:1996up}
\bibitem{Gracey:1996up}
  J.~A.~Gracey,
  %``The QCD Beta function at O(1/N(f)),''
  {\it Phys.\ Lett.}\ B {\bf 373} (1996) 178
  [hep-ph/9602214].
  %%CITATION = HEP-PH/9602214;%%
%------------------------------------------------------------------------------------------
%
%[205]
%\cite{Bennett:1997ch}
\bibitem{Bennett:1997ch}
  J.~F.~Bennett and J.~A.~Gracey,
  %``Determination of the anomalous dimension of gluonic operators in deep inelastic scattering at O (1/N(f)),''
  {\it Nucl.\ Phys.}\ B {\bf 517} (1998) 241
  [hep-ph/9710364].
  %%CITATION = HEP-PH/9710364;%%
%------------------------------------------------------------------------------------------
%
%[206]
%\cite{Gracey:1996tb}
\bibitem{Gracey:1996tb}
  J.~A.~Gracey,
  %``QCD and QED renormalization group functions: A Large N(f) approach,''
  {\it Nucl.\ Phys.\ (Proc.\ Suppl.)}\  {\bf 51C} (1996) 24
  [hep-ph/9604426].
  %%CITATION = HEP-PH/9604426;%%
%------------------------------------------------------------------------------------------
%
%[207]
\bibitem{GRAC2}
%\cite{Gracey:1996ad}
%\bibitem{Gracey:1996ad}
  J.~A.~Gracey,
  %``Anomalous dimensions of operators in polarized deep inelastic scattering at O(1/N(f)),''
  {\it Nucl.\ Phys.}\ B {\bf 480} (1996) 73
  [hep-ph/9609301];\\
  %%CITATION = HEP-PH/9609301;%%
%------------------------------------------------------------------------------------------
%\cite{Bennett:1998sr}
%\bibitem{Bennett:1998sr}
  J.~F.~Bennett and J.~A.~Gracey,
  %``Anomalous dimension of gluonic operator in polarized deep inelastic scattering at O(1 / N(f)),''
  {\it Phys.\ Lett.}\ B {\bf 432} (1998) 209
  [hep-ph/9803446].
  %%CITATION = HEP-PH/9803446;%%
%------------------------------------------------------------------------------------------
%
%[208]
\bibitem{GRACTR}
%\cite{Gracey:2003yr}
%\bibitem{Gracey:2003yr}
  J.~A.~Gracey,
  %``Three loop anomalous dimension of non-singlet quark currents in the RI'
  %scheme,''
  {\it Nucl.\ Phys.}\  B {\bf 662} (2003) 247   
  [arXiv:hep-ph/0304113];
  %%CITATION = NUPHA,B662,247;%%
%\cite{Gracey:2003mr}
%\bibitem{Gracey:2003mr}
%  J.~A.~Gracey,
  %``Three loop anomalous dimension of the second moment of the transversity
  %operator in the MS-bar and RI' schemes,''
  {\it Nucl.\ Phys.}\  B {\bf 667} (2003) 242
  [arXiv:hep-ph/0306163].
  %%CITATION = NUPHA,B667,242;%%
%\cite{Gracey:2006zr}
%\bibitem{Gracey:2006zr}
%  J.~A.~Gracey,
  %``Three loop anomalous dimensions of higher moments of the non-singlet
  %twist-2 Wilson and transversity operators in the MSbar and RI' schemes,''
  {\it JHEP} {\bf 0610} (2006) 040
  [arXiv:hep-ph/0609231];
  %%CITATION = JHEPA,0610,040;%%
%\cite{Gracey:2006ah}
%\bibitem{Gracey:2006ah}
%  J.~A.~Gracey,
  %``Three loop MS-bar transversity operator anomalous dimensions for fixed
  %moment n <= 8,''
  {\it Phys.\ Lett.}\  B {\bf 643} (2006) 374
  [arXiv:hep-ph/0611071].
  %%CITATION = PHLTA,B643,374;%%
%--------------------------------------------------------------------------------
%---------- Massless Wilson coefficients ----------------------------------------
%--------------------------------------------------------------------------------
%
%[209]
\bibitem{DY}
%\cite{Hamberg:1990np}
%\bibitem{Hamberg:1990np}
  R.~Hamberg, W.~L.~van Neerven and T.~Matsuura,
  %``A Complete calculation of the order $\alpha-s^{2}$ correction to the Drell-Yan $K$ factor,''
  {\it Nucl.\ Phys.}\ B {\bf 359} (1991) 343
   [Erratum-ibid.\ B {\bf 644} (2002) 403];\\
  %%CITATION = NUPHA,B359,343;%%
%\cite{Harlander:2002wh}    
%\bibitem{Harlander:2002wh} 
  R.~V.~Harlander and W.~B.~Kilgore,
  %``Next-to-next-to-leading order Higgs production at hadron colliders,''
  {\it Phys.\ Rev.\ Lett.}\  {\bf 88} (2002) 201801
  [hep-ph/0201206].
%--------------------------------------------------------------------------------
%
%[210]
\bibitem{HIGGS}
%\cite{Anastasiou:2002yz}
%\bibitem{Anastasiou:2002yz}
  C.~Anastasiou and K.~Melnikov,
  %``Higgs boson production at hadron colliders in NNLO QCD,''
  {\it Nucl.\ Phys.}\ B {\bf 646} (2002) 220
  [hep-ph/0207004];\\
  %%CITATION = HEP-PH/0207004;%%
%\cite{Ravindran:2003um}
%\bibitem{Ravindran:2003um}
  V.~Ravindran, J.~Smith and W.~L.~van Neerven,
  %``NNLO corrections to the total cross-section for Higgs boson production in hadron hadron collisions,''
  {\it Nucl.\ Phys.}\ B {\bf 665} (2003) 325
  [hep-ph/0302135];\\
  %%CITATION = HEP-PH/0302135;%%
%\cite{Catani:2007vq}
%\bibitem{Catani:2007vq}
  S.~Catani and M.~Grazzini,
  %``An NNLO subtraction formalism in hadron collisions and its application to Higgs boson production at the LHC,''
  {\it Phys.\ Rev.\ Lett.}\  {\bf 98} (2007) 222002
  [hep-ph/0703012].
  %%CITATION = HEP-PH/0703012;%%
%--------------------------------------------------------------------------------
%
%[211]
%\cite{Altarelli:1979ub}
\bibitem{Altarelli:1979ub}
  G.~Altarelli, R.~K.~Ellis and G.~Martinelli,
  %``Large Perturbative Corrections to the Drell-Yan Process in QCD,''
  {\it Nucl.\ Phys.}\ B {\bf 157} (1979) 461.
  %%CITATION = NUPHA,B157,461;%%
%--------------------------------------------------------------------------------
%
%[212]
%\cite{Humpert:1980uv}
\bibitem{Humpert:1980uv}
  B.~Humpert and W.~L.~van Neerven,
  %``Infrared And Mass Regularization In Af Field Theories 2. Qcd,''
  {\it Nucl.\ Phys.}\ B {\bf 184} (1981) 225.
  %%CITATION = NUPHA,B184,225;%%
%--------------------------------------------------------------------------------
%
%[213]
%\cite{Bodwin:1989nz}
\bibitem{Bodwin:1989nz}
  G.~T.~Bodwin and J.~-W.~Qiu,
  %``The Gluonic Contribution to G(1) and Its Relationship to the Spin Dependent Parton Distributions,''
  {\it Phys.\ Rev.}\ D {\bf 41} (1990) 2755.
  %%CITATION = PHRVA,D41,2755;%%
%--------------------------------------------------------------------------------
%
%[214]
%\cite{Vogelsang:1990ug}
\bibitem{Vogelsang:1990ug}
  W.~Vogelsang,
  %``The Gluonic contribution to g(1)-p (x,Q**2) in the parton model,''
  {\it Z.\ Phys.}\ C {\bf 50} (1991) 275.
  %%CITATION = ZEPYA,C50,275;%%
%--------------------------------------------------------------------------------
%
%[215]
%\cite{Gluck:1995yr}
\bibitem{Gluck:1995yr}
  M.~Gl\"uck, E.~Reya, M.~Stratmann and W.~Vogelsang,
  %``Next-to-leading order radiative parton model analysis of polarized deep inelastic lepton - nucleon scattering,''
  {\it Phys.\ Rev.}\ D {\bf 53} (1996) 4775
  [hep-ph/9508347].
  %%CITATION = HEP-PH/9508347;%%
%--------------------------------------------------------------------------------
%
%[216]
\bibitem{F2FL}
%\cite{Zijlstra:1991qc}
%\bibitem{Zijlstra:1991qc}
  E.~B.~Zijlstra and W.~L.~van Neerven,
  %``Contribution of the second order gluonic Wilson coefficient to the deep inelastic structure function,''
  {\it Phys.\ Lett.}\ B {\bf 273} (1991) 476;
  %%CITATION = PHLTA,B273,476;%%
%\cite{vanNeerven:1991nn}
%\bibitem{vanNeerven:1991nn}
%  W.~L.~van Neerven and E.~B.~Zijlstra,
  %``Order alpha-s**2 contributions to the deep inelastic Wilson coefficient,''
  {\it Phys.\ Lett.}\ B {\bf 272} (1991) 127;
  %%CITATION = PHLTA,B272,127;%%
%\cite{Zijlstra:1992qd}
%\bibitem{Zijlstra:1992qd}
%  E.~B.~Zijlstra and W.~L.~van Neerven,
  %``Order alpha-s**2 QCD corrections to the deep inelastic proton structure functions F2 and F(L),''
  {\it Nucl.\ Phys.}\ B {\bf 383} (1992) 525.
  %%CITATION = NUPHA,B383,525;%%
%--------------------------------------------------------------------------------
%
%[217]
\bibitem{FL_I}
%\cite{Kazakov:1987jk}
%\bibitem{Kazakov:1987jk}
  D.~I.~Kazakov and A.~V.~Kotikov,
  %``TOTAL ALPHA-s CORRECTION TO DEEP INELASTIC SCATTERING CROSS-SECTION RATIO, R = sigma-L / sigma-t IN QCD. CALCULATION OF LONGITUDINAL STRUCTURE FUNCTION,''
  {\it Nucl.\ Phys.}\ B {\bf 307} (1988) 721
   [Erratum-ibid.\ B {\bf 345} (1990) 299];\\
  %%CITATION = NUPHA,B307,721;%%
%\cite{Kazakov:1990fu}
%\bibitem{Kazakov:1990fu}
  D.~I.~Kazakov, A.~V.~Kotikov, G.~Parente, O.~A.~Sampayo and J.~Sanchez Guillen,
  %``Complete quartic (alpha(s)**2) correction to the deep inelastic longitudinal structure function F(L) in QCD,''
  {\it Phys.\ Rev.\ Lett.}\  {\bf 65} (1990) 1535
   [Erratum-ibid.\  {\bf 65} (1990) 2921];\\
  %%CITATION = PRLTA,65,1535;%%
%\cite{SanchezGuillen:1990iq}
%\bibitem{SanchezGuillen:1990iq}
  J.~Sanchez Guillen, J.~Miramontes, M.~Miramontes, G.~Parente and O.~A.~Sampayo,
  %``Next-to-leading order analysis of the deep inelastic R = sigma-L / sigma-total,''
  {\it Nucl.\ Phys.}\ B {\bf 353} (1991) 337.
  %%CITATION = NUPHA,B353,337;%%
%--------------------------------------------------------------------------------
%
%[218]
%\cite{Zijlstra:1992kj}
\bibitem{Zijlstra:1992kj}
  E.~B.~Zijlstra and W.~L.~van Neerven,
  %``Order alpha-s**2 correction to the structure function F3 (x, Q**2) in deep inelastic neutrino - hadron scattering,''
  {\it Phys.\ Lett.}\ B {\bf 297} (1992) 377.
  %%CITATION = PHLTA,B297,377;%%
%--------------------------------------------------------------------------------
%
%[219]
%\cite{Larin:1991fv}
\bibitem{Larin:1991fv}
  S.~A.~Larin and J.~A.~M.~Vermaseren,
  %``Two loop QCD corrections to the coefficient functions of the deep inelastic structure functions F-2 AND F-L,''
  {\it Z.\ Phys.}\ C {\bf 57} (1993) 93.
  %%CITATION = ZEPYA,C57,93;%%
%--------------------------------------------------------------------------------
%
%[220]
%\cite{Blumlein:2009tj}
\bibitem{Blumlein:2009tj}
  J.~Bl\"umlein, M.~Kauers, S.~Klein and C.~Schneider,
  %``Determining the closed forms of the O(a**3(s)) anomalous dimensions and Wilson coefficients from Mellin moments by means of computer algebra,''
  {\it Comput.\ Phys.\ Commun.}\  {\bf 180} (2009) 2143
  [arXiv:0902.4091 [hep-ph]].
  %%CITATION = ARXIV:0902.4091;%%
%--------------------------------------------------------------------------------
%
%[221]
\bibitem{JBVR}
%\cite{Blumlein:2005im}
%\bibitem{Blumlein:2005im}
  J.~Bl\"umlein and V.~Ravindran,
  %``Mellin moments of the next-to-next-to leading order coefficient functions for the Drell-Yan process and hadronic Higgs-boson production,''
  {\it Nucl.\ Phys.}\ B {\bf 716} (2005) 128
  [hep-ph/0501178];
  %%CITATION = HEP-PH/0501178;%%
%\cite{Blumlein:2006rr}
%\bibitem{Blumlein:2006rr}
%  J.~Bl\"umlein and V.~Ravindran,
  %``O (alpha**2(s)) Timelike Wilson Coefficients for Parton-Fragmentation Functions in Mellin Space,''
  {\it Nucl.\ Phys.}\ B {\bf 749} (2006) 1
  [hep-ph/0604019].
  %%CITATION = HEP-PH/0604019;%%
%--------------------------------------------------------------------------------
%
%[222]
\bibitem{XF3}
%\cite{Moch:2007gx}
%\bibitem{Moch:2007gx}
  S.~Moch and M.~Rogal,
  %``Charged current deep-inelastic scattering at three loops,''
  {\it Nucl.\ Phys.}\ B {\bf 782} (2007) 51
  [arXiv:0704.1740 [hep-ph]];\\
  %%CITATION = ARXIV:0704.1740;%%
%\cite{Moch:2007rq}
%\bibitem{Moch:2007rq}
  S.~Moch, M.~Rogal and A.~Vogt,
  %``Differences between charged-current coefficient functions,''
  {\it Nucl.\ Phys.}\ B {\bf 790} (2008) 317
  [arXiv:0708.3731 [hep-ph]];\\
  %%CITATION = ARXIV:0708.3731;%%
%\cite{Moch:2008fj}
%\bibitem{Moch:2008fj}
  S.~Moch, J.~A.~M.~Vermaseren and A.~Vogt,
  %``Third-order QCD corrections to the charged-current structure function F(3),''
  {\it Nucl.\ Phys.}\ B {\bf 813} (2009) 220
  [arXiv:0812.4168 [hep-ph]].
  %%CITATION = ARXIV:0812.4168;%%
%--------------------------------------------------------------------------------
%
%[223]
%\cite{Blumlein:2003gb}
\bibitem{Blumlein:2003gb}
  J.~Bl\"umlein,
  %``Algebraic relations between harmonic sums and associated quantities,''
  {\it Comput.\ Phys.\ Commun.}\  {\bf 159} (2004) 19
  [hep-ph/0311046].
  %%CITATION = HEP-PH/0311046;%%
%--------------------------------------------------------------------------------
%--------------------------------------------------------------------------------
%-----------------   QED Radiative Corrections    -------------------------------
%--------------------------------------------------------------------------------
%
%[224]
\bibitem{RC1} 
%\cite{Mo:1968cg}
%\bibitem{Mo:1968cg}
  L.~W.~Mo and Y.~-S.~Tsai,
  %``Radiative Corrections to Elastic and Inelastic e p and mu p Scattering,''
  {\em Rev.\ Mod.\ Phys.}\  {\bf 41} (1969) 205.
  %%CITATION = RMPHA,41,205;%%
%--------------------------------------------------------------------------------
%
%[225]
\bibitem{RC2} 
%\cite{Bardin:1980ii}
%\bibitem{Bardin:1980ii}
  D.~Y.~Bardin, O.~M.~Fedorenko and N.~M.~Shumeiko,
  %``RADIATIVE CORRECTIONS TO P ODD ASYMMETRIES IN DEEP INELASTIC SCATTERING OF POLARIZED MUONS ON NUCLEONS AT TeV ENERGIES,''
  {\em J.\ Phys.}\  G {\bf 7} (1981) 1331.
  %%CITATION = JPHGB,G7,1331;%%
%--------------------------------------------------------------------------------
%
%[226]
\bibitem{RC3} 
%\cite{Consoli:1980nm}
%\bibitem{Consoli:1980nm}
  M.~Consoli and M.~Greco,
  %``Radiative Corrections To Very High-energy Electron - Proton Scattering,''
  {\em Nucl.\ Phys.}\ B {\bf 186} (1981) 519.
  %%CITATION = NUPHA,B186,519;%%
%--------------------------------------------------------------------------------
%
%[227]
\bibitem{RC55} 
%\bibitem{RC5} 
%\cite{Marciano:1980pb}
%\bibitem{Marciano:1980pb}
  W.~J.~Marciano and A.~Sirlin,
  %``Radiative Corrections to Neutrino Induced Neutral Current Phenomena in the SU(2)-L x U(1) Theory,''
  {\em Phys.\ Rev.}\ D {\bf 22} (1980) 2695
   [Erratum-ibid.\ D {\bf 31} (1985) 213];
  %%CITATION = PHRVA,D22,2695;%%
%\cite{Sirlin:1981yz}
%\bibitem{Sirlin:1981yz}
%  A.~Sirlin and W.~J.~Marciano,
  %``Radiative Corrections to Muon-neutrino N ---> mu- X and their Effect on the Determination of rho**2 and sin**2-Theta(W),''
  {\em Nucl.\ Phys.}\ B {\bf 189} (1981) 442;\\
  %%CITATION = NUPHA,B189,442;%%
%--------------------------------------------------------------------------------
%\bibitem{RC6} 
%\cite{Sarantakos:1982bp}
%\bibitem{Sarantakos:1982bp}
  S.~Sarantakos, A.~Sirlin and W.~J.~Marciano,
  %``Radiative Corrections to Neutrino-Lepton Scattering in the SU(2)-L x U(1) Theory,''
  {\em Nucl.\ Phys.}\ B {\bf 217} (1983) 84;\\
  %%CITATION = NUPHA,B217,84;%%
%--------------------------------------------------------------------------------
%\bibitem{RC7} 
%\cite{Llewellyn Smith:1981yk}
%\bibitem{Llewellyn Smith:1981yk}
  C.~H.~Llewellyn Smith and J.~F.~Wheater,
  %``Electroweak Radiative Corrections and the Value of sin**2-Theta-W,''
  {\em Phys.\ Lett.}\ B {\bf 105} (1981) 486;
  %%CITATION = PHLTA,B105,486;%%
%---------------->>
%\cite{Wheater:1982yk}
%\bibitem{Wheater:1982yk}
%  J.~F.~Wheater and C.~H.~Llewellyn Smith,
  %``Electroweak Radiative Corrections To Neutrino And Electron Scattering And The Value Of Sin**2-theta-w,''
  {\em Nucl.\ Phys.}\ B {\bf 208} (1982) 27
   [Erratum-ibid.\ B {\bf 226} (1983) 547];\\
  %%CITATION = NUPHA,B208,27;%%
%--------------------------------------------------------------------------------
%\bibitem{RC8} 
%\cite{Paschos:1981zx}
%\bibitem{Paschos:1981zx}
  E.~A.~Paschos and M.~Wirbel,
  %``CORRECTIONS TO SIN**2-theta-w IN NEUTRINO EXPERIMENTS,''
  {\em Nucl.\ Phys.}\ B {\bf 194} (1982) 189;\\
  %%CITATION = NUPHA,B194,189;%%
%--------------------------------------------------------------------------------
%\bibitem{RC9}  
%\cite{Wirbel:1982cw}
%\bibitem{Wirbel:1982cw}
  M.~Wirbel,
  %``WEAK AND ELECTROMAGNETIC CORRECTIONS TO sin**2-theta-w IN DEEP INELASTIC NEUTRINO - NUCLEON SCATTERING,''
  {\em Z.\ Phys.}\ C {\bf 14} (1982) 293;\\
  %%CITATION = ZEPYA,C14,293;%%
%--------------------------------------------------------------------------------
%\bibitem{RC10} 
%\cite{Liede:1983hw}
%\bibitem{Liede:1983hw}
  I.~Liede,
  %``On The Electromagnetic Corrections In Deep Inelastic Neutrino N Scattering,''
  {\em Nucl.\ Phys.}\ B {\bf 229} (1983) 499;\\
  %%CITATION = NUPHA,B229,499;%%
%--------------------------------------------------------------------------------
%\bibitem{RC11} 
%\cite{Bardin:1986bc}
%\bibitem{Bardin:1986bc}
  D.~Y.~Bardin and V.~A.~Dokuchaeva,
  %``On The Radiative Corrections To The Neutrino Deep Inelastic Scattering,''
  JINR-E2-86-260.
  %%CITATION = JINR-E2-86-260;%%
%--------------------------------------------------------------------------------
%
%[228]
\bibitem{RC66}
%\cite{Diener:2003ss}
%\bibitem{Diener:2003ss}
  K.-P.~O.~Diener, S.~Dittmaier and W.~Hollik,
  %``Electroweak radiative corrections to deep inelastic neutrino scattering: Implications for NuTeV?,''
  {\em Phys.\ Rev.}\ D {\bf 69} (2004) 073005
  [hep-ph/0310364];\\
  %%CITATION = HEP-PH/0310364;%%
%--------------------------------------------------------------------------------
%\cite{Arbuzov:2004zr}
%\bibitem{Arbuzov:2004zr}
  A.~B.~Arbuzov, D.~Y.~Bardin and L.~V.~Kalinovskaya,
  %``Radiative corrections to neutrino deep inelastic scattering revisited,''
  {\em JHEP} {\bf 0506} (2005) 078
  [arXiv:hep-ph/0407203];\\
%--------------------------------------------------------------------------------
%\cite{Diener:2005me}
%\bibitem{Diener:2005me}
  K.-P.~O.~Diener, S.~Dittmaier and W.~Hollik,
  %``Electroweak higher-order effects and theoretical uncertainties in deep-inelastic neutrino scattering,''
  {\em Phys.\ Rev.}\ D {\bf 72} (2005) 093002
  [hep-ph/0509084];\\
  %%CITATION = HEP-PH/0509084;%%
%--------------------------------------------------------------------------------
%\cite{Park:2009ft}
%\bibitem{Park:2009ft}
  K.~Park, U.~Baur and D.~Wackeroth,
  %``Electroweak radiative corrections to neutrino--nucleon scattering at NuTeV,''
  arXiv:0910.5013 [hep-ph].
  %%CITATION = ARXIV:0910.5013;%%
%--------------------------------------------------------------------------------
%
%[229]
%\cite{Zeller:2001hh}
\bibitem{Zeller:2001hh}
  G.~P.~Zeller {\it et al.},  [NuTeV Collaboration],
  %``A Precise determination of electroweak parameters in neutrino nucleon scattering,''
  {\it Phys.\ Rev.\ Lett.}\  {\bf 88} (2002) 091802
   [Erratum-ibid.\  {\bf 90} (2003) 239902]
  [hep-ex/0110059].
  %%CITATION = HEP-EX/0110059;%%
%--------------------------------------------------------------------------------
%
%[230]
\bibitem{KLN}
%\cite{Kinoshita:1962ur}
%\bibitem{Kinoshita:1962ur}
  T.~Kinoshita,
  %``Mass singularities of Feynman amplitudes,''
  {\em J.\ Math.\ Phys.}\  {\bf 3} (1962) 650;\\
  %%CITATION = JMAPA,3,650;%%
%\cite{Lee:1964is}
%\bibitem{Lee:1964is}
  T.~D.~Lee and M.~Nauenberg,
  %``Degenerate Systems and Mass Singularities,''
  {\em Phys.\ Rev.}\  {\bf 133} (1964) B1549.
  %%CITATION = PHRVA,133,B1549;%%
%--------------------------------------------------------------------------------
%
%[231]
\bibitem{RC4} 
%\cite{De Rujula:1979jj}
%\bibitem{De Rujula:1979jj}
  A.~De Rujula, R.~Petronzio and A.~Savoy-Navarro,
  %``Radiative Corrections to High-Energy Neutrino Scattering,''
  {\em Nucl.\ Phys.}\ B {\bf 154} (1979) 394.
  %%CITATION = NUPHA,B154,394;%%
%--------------------------------------------------------------------------------
%
%[232]
%\bibitem{RC25a} 
%\cite{Bohm:1986na}
\bibitem{Bohm:1986na}
  M.~B\"ohm and H.~Spiesberger,
  %``Radiative Corrections To Neutral Current Deep Inelastic Lepton Nucleon Scattering At Hera Energies,''
  {\em Nucl.\ Phys.}\ B {\bf 294} (1987) 1081.
  %%CITATION = NUPHA,B294,1081;%% 
%-------------------------------------------------------------------------------- 
%
%[233]
%\bibitem{RC25b} 
%\cite{Bohm:1987cg} 
\bibitem{Bohm:1987cg}
  M.~B\"ohm and H.~Spiesberger,
  %``Radiative Corrections To Charged Current Deep Inelastic Electron - Proton Scattering At Hera,''
  {\em  Nucl.\ Phys.}\ B {\bf 304} (1988) 749.
  %%CITATION = NUPHA,B304,749;%%
%--------------------------------------------------------------------------------
%
%[234]
%\cite{Bardin:1987rz}
\bibitem{Bardin:1987rz}
  D.~Y.~Bardin, C.~Burdik, P.~K.~Christova and T.~Riemann,
  %``Study Of Electroweak Radiative Corrections To Deep Inelastic Scattering At Hera,''
  JINR-E2-87-595.
  %%CITATION = JINR-E2-87-595;%%
%--------------------------------------------------------------------------------
%
%[235]
%\bibitem{RC12a} 
%\cite{Bardin:1988by}
\bibitem{Bardin:1988by}
  D.~Y.~Bardin, C.~Burdik, P.~C.~Christova and T.~Riemann,
  %``Electroweak Radiative Corrections To Deep Inelastic Scattering At Hera. Neutral Current Scattering,''
  {\em Z.\ Phys.}\ C {\bf 42} (1989) 679.
  %%CITATION = ZEPYA,C42,679;%%
%--------------------------------------------------------------------------------
%
%[236]
%\bibitem{RC12b} 
%\cite{Bardin:1989vz}
\bibitem{Bardin:1989vz}
  D.~Y.~Bardin, K.~C.~Burdik, P.~K.~Christova and T.~Riemann,
  %``Electroweak Radiative Corrections To Deep Inelastic Scattering At Hera! Charged Current Scattering,''
  {\em Z.\ Phys.}\ C {\bf 44} (1989) 149.
  %%CITATION = ZEPYA,C44,149;%%
%--------------------------------------------------------------------------------
%
%[237]
%\bibitem{RC13} 
%\cite{Kuraev:1988xn}
\bibitem{Kuraev:1988xn}
  E.~A.~Kuraev, N.~P.~Merenkov and V.~S.~Fadin,
  %``Calculation Of Radiative Corrections To Electron Nucleus Scattering Cross-section By The Structure Functions Method. (in Russian),''
  {\em Sov.\ J.\ Nucl.\ Phys.}\  {\bf 47} (1988) 1009
   [{\em Yad.\ Fiz.}\  {\bf 47} (1988) 1593].
  %%CITATION = SJNCA,47,1009;%%
%--------------------------------------------------------------------------------
%
%[238]
\bibitem{BEBEN} 
W.~Beenakker, F.~Berends, and W.~van Neerven, in: Proc.
of the Workshop on Electroweak Radiative Corrections, Ringberg, Germany,
ed.~J.~K\"uhn, (Springer, Berlin, 1989), p.~3.
%--------------------------------------------------------------------------------
%
%[239]
%\cite{Blumlein:1989gk}
\bibitem{Blumlein:1989gk}
  J.~Bl\"umlein,
  %``Leading Log Radiative Corrections To Deep Inelastic Neutral And Charged Current Scattering At Hera,''
  {\em Z.\ Phys.}\ C {\bf 47} (1990) 89.
  %%CITATION = ZEPYA,C47,89;%%
%--------------------------------------------------------------------------------
%
%[240]
%\cite{Blumlein:1990wz}
\bibitem{Blumlein:1990wz}
  J.~Bl\"umlein,
  %``Leading log radiative corrections to e p scattering including jet measurement,''
  {\em Phys.\ Lett.}\ B {\bf 271} (1991) 267.
  %%CITATION = PHLTA,B271,267;%%
%--------------------------------------------------------------------------------
%
%[241]
%\cite{Blumlein:1991ag}
\bibitem{Blumlein:1991ag}
  J.~Bl\"umlein, {\tt HELIOS 1.00}, PHE 91--16, 
       Proc. of the 1992 HERA Workshop, Vol. {\bf 3},
       eds.~W.~Buchm\"uller and G.~Ingelman, (DESY, Hamburg, 1992),
       p.~1269.%--1284. 
%--------------------------------------------------------------------------------
%
%[242]
%\cite{Montagna:1991ku}
\bibitem{Montagna:1991ku}
  G.~Montagna, O.~Nicrosini and L.~Trentadue,
  %``QED radiative corrections to lepton scattering in the structure function formalism,''
  {\em Nucl.\ Phys.}\ B {\bf 357} (1991) 390.
  %%CITATION = NUPHA,B357,390;%%
%--------------------------------------------------------------------------------
%
%[243]
%\cite{Kripfganz:1987yu}
\bibitem{Kripfganz:1987yu}
  J.~Kripfganz and H.~J.~M\"ohring,
  %``Electromagnetic Corrections To Deep Inelastic Scattering At Hera,''
  {\em Z.\ Phys.}\ C {\bf 38} (1988) 653.
  %%CITATION = ZEPYA,C38,653;%%
%--------------------------------------------------------------------------------
%
%[244]
%\bibitem{RC18}  
%\cite{Kripfganz:1990vm}
\bibitem{Kripfganz:1990vm}
  J.~Kripfganz, H.~J.~M\"ohring and H.~Spiesberger,
  %``Higher order leading logarithmic QED corrections to deep inelastic e p scattering at very high-energies,''
  {\em Z.\ Phys.}\ C {\bf 49} (1991) 501.
  %%CITATION = ZEPYA,C49,501;%%
%--------------------------------------------------------------------------------
%
%[245]
%\bibitem{RC20}  
%\cite{Akhundov:1995na}
\bibitem{Akhundov:1995na}
  A.~A.~Akhundov, D.~Y.~Bardin, L.~Kalinovskaya and T.~Riemann,
  %``Leptonic QED corrections to the process e p ---> e X in Jaquet-Blondel variables,''
  {\em Phys.\ Lett.}\ B {\bf 301} (1993) 447
  [hep-ph/9507278].
  %%CITATION = HEP-PH/9507278;%%
%--------------------------------------------------------------------------------
%
%[246]
%\bibitem{RC22}  
%\cite{Akhundov:1994my}
\bibitem{Akhundov:1994my}
  A.~A.~Akhundov, D.~Y.~Bardin, L.~Kalinovskaya and T.~Riemann,
  %``Model independent QED corrections to the process e p ---> e X,''
  {\em Fortsch.\ Phys.}\  {\bf 44} (1996) 373
  [hep-ph/9407266].
  %%CITATION = HEP-PH/9407266;%%
%--------------------------------------------------------------------------------
%
%[247]
%\bibitem{RC23}  
%\cite{Bardin:1995mf}
\bibitem{Bardin:1995mf}
  D.~Y.~Bardin, P.~Christova, L.~Kalinovskaya and T.~Riemann,
  %``Complete O(alpha) QED corrections to the process e p ---> e X in mixed variables,''
  {\em Phys.\ Lett.}\ B {\bf 357} (1995) 456
  [hep-ph/9504423].
  %%CITATION = HEP-PH/9504423;%%
%--------------------------------------------------------------------------------
%
%[248]
%\bibitem{RC26}  
%\cite{Spiesberger:1990fa}
\bibitem{Spiesberger:1990fa}
  H.~Spiesberger,
  %``Radiative Corrections To Charged Current Deep Inelastic Scattering At Hera,''
  {\em Nucl.\ Phys.}\ B {\bf 349} (1991) 109.
  %%CITATION = NUPHA,B349,109;%%
%--------------------------------------------------------------------------------
%
%[249]
%\bibitem{RC27}
%\cite{Kwiatkowski:1990es}
\bibitem{Kwiatkowski:1990es}
  A.~Kwiatkowski, H.~Spiesberger and H.~J.~M\"ohring,
  %``HERACLES: AN EVENT GENERATOR FOR e p INTERACTIONS AT HERA ENERGIES INCLUDING RADIATIVE PROCESSES: VERSION 1.0,''
  {\em Comput.\ Phys.\ Commun.}\  {\bf 69} (1992) 155.
  %%CITATION = CPHCB,69,155;%%
%--------------------------------------------------------------------------------
%
%[250]
%\cite{Blumlein:1993ef}
\bibitem{Blumlein:1993ef}
  J.~Bl\"umlein, G.~Levman and H.~Spiesberger,
  %``On the measurement of the proton structure at small Q**2,''
  {\em J.\ Phys.}\  G {\bf 19} (1993) 1695.
  %%CITATION = JPHGB,G19,1695;%%
%--------------------------------------------------------------------------------
%
%[251]
%\cite{Blumlein:2002fy}
\bibitem{Blumlein:2002fy}
  J.~Bl\"umlein and H.~Kawamura,
  %``O(alpha**2 L) radiative corrections to deep inelastic ep scattering,''
  {\em Phys.\ Lett.}\ B {\bf 553} (2003) 242
  [hep-ph/0211191].
  %%CITATION = HEP-PH/0211191;%%
%--------------------------------------------------------------------------------
%
%[252]
%\cite{Kripfganz:1988bd}
\bibitem{Kripfganz:1988bd}
  J.~Kripfganz and H.~Perlt,
  %``Electroweak Radiative Corrections And Quark Mass Singularities,''
  {\em Z.\ Phys.}\ C {\bf 41} (1988) 319.
  %%CITATION = ZEPYA,C41,319;%%
%--------------------------------------------------------------------------------
%
%[253]
\bibitem{QUARK:lat}
%\cite{Spiesberger:1994dm}
%\bibitem{Spiesberger:1994dm}
  H.~Spiesberger,
  %``QED radiative corrections for parton distributions,''
  {\em Phys.\ Rev.}\ D {\bf 52} (1995) 4936
  [hep-ph/9412286];\\
  %%CITATION = HEP-PH/9412286;%%
%--------------------------------------------------------------------------------
%\cite{Roth:2004ti}
%\bibitem{Roth:2004ti}
  M.~Roth and S.~Weinzierl,
  %``QED corrections to the evolution of parton distributions,''
  {\em Phys.\ Lett.}\ B {\bf 590} (2004) 190
  [hep-ph/0403200].
  %%CITATION = HEP-PH/0403200;%%
%--------------------------------------------------------------------------------
%
%[254]
%\cite{Bardin:1996ch}
\bibitem{Bardin:1996ch}
  D.~Y.~Bardin, J.~Bl\"umlein, P.~Christova and L.~Kalinovskaya,
  %``O (alpha) QED corrections to neutral current polarized deep inelastic lepton - nucleon scattering,''
  {\em Nucl.\ Phys.}\ B {\bf 506} (1997) 295
  [hep-ph/9612435].
  %%CITATION = HEP-PH/9612435;%%
%--------------------------------------------------------------------------------
%
%[255]
%\cite{Fadin:1989dw}
\bibitem{Fadin:1989dw}
  V.~S.~Fadin,
  %``Applications of structure function method to polarized particles,''
  In: Conference Proceedings No. 201, {\sf QED Structure Functions}, Ann Arbor, 1989, 
  Ed. G.~Bonvicini, (APS, New York, 1989), pp.~118. 
%--------------------------------------------------------------------------------
%
%[256]
\bibitem{MC:POLQED}
%\cite{Akushevich:1998dz}
%\bibitem{Akushevich:1998dz}
  I.~V.~Akushevich, A.~N.~Ilyichev and N.~M.~Shumeiko,
  %``Radiative electroweak effects in deep inelastic scattering of polarized leptons by polarized nucleons,''
  {\em J.\ Phys.}\ G  {\bf 24} (1998) 1995
  [hep-ph/9804361];\\
  %%CITATION = HEP-PH/9804361;%%
%--------------------------------------------------------------------------------
%\cite{Akushevich:2011zy}
%\bibitem{Akushevich:2011zy}
  I.~V.~Akushevich, O.~F.~Filoti, A.~Ilyichev and N.~Shumeiko,
  %``Monte Carlo generator ELRADGEN 2.0 for simulation of radiative events in elastic ep-scattering of polarized particles,''
  arXiv:1104.0039 [hep-ph].
  %%CITATION = ARXIV:1104.0039;%%
%--------------------------------------------------------------------------------
%
%[257]
%\cite{Skrzypek:1992vk}
\bibitem{Skrzypek:1992vk}
  M.~Skrzypek,
  %``Leading logarithmic calculations of QED corrections at LEP,''
  {\em Acta Phys.\ Polon.}\ B {\bf 23} (1992) 135.
  %%CITATION = APPOA,B23,135;%%
%--------------------------------------------------------------------------------
%
%[258]
%\cite{Jezabek:1991bx}
\bibitem{Jezabek:1991bx}
  M.~Jezabek,
  %``A Perturbative solution to Gribov-Lipatov equation,''
  {\em Z.\ Phys.}\ C {\bf 56} (1992) 285.
  %%CITATION = ZEPYA,C56,285;%%
%--------------------------------------------------------------------------------
%
%[259]
\bibitem{QED5th}
%\cite{Przybycien:1992qe}
%\bibitem{Przybycien:1992qe}
  M.~Przybycien,
  %``A Fifth order perturbative solution to the Gribov-Lipatov equation,''
  {\em Acta Phys.\ Polon.}\ B {\bf 24} (1993) 1105
  [hep-th/9511029];\\
  %%CITATION = HEP-TH/9511029;%%
%--------------------------------------------------------------------------------
%\cite{Arbuzov:1999cq}
%\bibitem{Arbuzov:1999cq}
  A.~B.~Arbuzov,
  %``Nonsinglet splitting functions in QED,''
  {\em Phys.\ Lett.}\ B {\bf 470} (1999) 252
  [hep-ph/9908361].
  %%CITATION = HEP-PH/9908361;%%
%--------------------------------------------------------------------------------
%
%[260]
%\cite{Blumlein:2002bg}
\bibitem{Blumlein:2002bg}
  J.~Bl\"umlein and H.~Kawamura,
  %``Universal QED corrections to polarized electron scattering in higher orders,''
  {\em Acta Phys.\ Polon.}\ B {\bf 33} (2002) 3719
  [hep-ph/0207259];
  %%CITATION = HEP-PH/0207259;%%
%--------------------------------------------------------------------------------
%\cite{Blumlein:2004bs}
%\bibitem{Blumlein:2004bs}
%  J.~Bl\"umlein and H.~Kawamura,
  %``Universal higher order QED corrections to polarized lepton scattering,''
  {\em Nucl.\ Phys.}\ B {\bf 708} (2005) 467
  [hep-ph/0409289].
  %%CITATION = HEP-PH/0409289;%%
%--------------------------------------------------------------------------------
%
%[261]
%\cite{Blumlein:2007kx}
\bibitem{Blumlein:2007kx}
  J.~Bl\"umlein and H.~Kawamura,
  %``Universal higher order singlet QED corrections to unpolarized lepton scattering,''
  {\em Eur.\ Phys.}\ J.\ C {\bf 51} (2007) 317
  [hep-ph/0701019].
  %%CITATION = HEP-PH/0701019;%%
%--------------------------------------------------------------------------------
%
%[262]
%\cite{Blumlein:1996yz}
\bibitem{Blumlein:1996yz}
  J.~Bl\"umlein, S.~Riemersma and A.~Vogt,
  %``On the resummation of the alpha ln**2 z terms for QED corrections to deep inelastic e p scattering and e+ e- annihilation,''
  {\em Eur.\ Phys.\ J.}\ C {\bf 1} (1998) 255
  [hep-ph/9611214].
  %%CITATION = HEP-PH/9611214;%%
%--------------------------------------------------------------------------------
%----------------------------------------------------------------------
%%             >> Heavy Flavor Wilson Coefficients: Section 10 <<
%----------------------------------------------------------------------
%--------------------------------------------------------------------------------
%
%[263]
\bibitem{HQ:LO}
%\cite{Witten:1975bh}
%\bibitem{Witten:1975bh}
  E.~Witten,
  %``Heavy Quark Contributions to Deep Inelastic Scattering,''
  {\it Nucl.\ Phys.}\ B {\bf 104} (1976) 445;\\
  %%CITATION = NUPHA,B104,445;%%
%--
%\cite{Babcock:1977fi}
%\bibitem{Babcock:1977fi}
  J.~Babcock, D.~W.~Sivers and S.~Wolfram,
  %``QCD Estimates for Heavy Particle Production,''
  {\it Phys.\ Rev.}\ D {\bf 18} (1978) 162;\\
  %%CITATION = PHRVA,D18,162;%%
%--
%\cite{Shifman:1977yb}
%\bibitem{Shifman:1977yb}
  M.~A.~Shifman, A.~I.~Vainshtein and V.~I.~Zakharov,
  %``Remarks on Charm Electroproduction in QCD,''
  {\it Nucl.\ Phys.}\ B {\bf 136} (1978) 157
   [{\it Yad.\ Fiz.}\  {\bf 27} (1978) 455];\\
  %%CITATION = NUPHA,B136,157;%%
%--
%\cite{Leveille:1978px}
%\bibitem{Leveille:1978px}
  J.~P.~Leveille and T.~J.~Weiler,
  %``Characteristics of Heavy Quark Leptoproduction in QCD,''
  {\it Nucl.\ Phys.}\ B {\bf 147} (1979) 147.
  %%CITATION = NUPHA,B147,147;%%
%--------------------------------------------------------------------------------
%
%[264]
%\cite{Watson:1981ce}
\bibitem{Watson:1981ce}
  A.~D.~Watson,
  %``Spin Spin Asymmetries In Inclusive Muon Proton Charm Production,''
  {\it Z.\ Phys.}\ C {\bf 12} (1982) 123.
  %%CITATION = ZEPYA,C12,123;%%
%--------------------------------------------------------------------------------
%
%[265]
%\cite{Blumlein:2003wk}
\bibitem{Blumlein:2003wk}
  J.~Bl\"umlein, V.~Ravindran and W.~L.~van Neerven,
  %``Twist-2 heavy flavor contributions to the structure function g(2)(x,Q**2),''
  {\it Phys.\ Rev.}\ D {\bf 68} (2003) 114004
  [hep-ph/0304292].
  %%CITATION = HEP-PH/0304292;%%
%--------------------------------------------------------------------------------
%
%[266]
%\cite{Wandzura:1977qf}
\bibitem{Wandzura:1977qf}
  S.~Wandzura and F.~Wilczek,
  %``Sum Rules for Spin Dependent Electroproduction: Test of Relativistic Constituent Quarks,''
  {\it Phys.\ Lett.}\ B {\bf 72} (1977) 195.
  %%CITATION = PHLTA,B72,195;%%
%--------------------------------------------------------------------------------
%
%[267]
%\cite{Alekhin:2003ev}
\bibitem{Alekhin:2003ev}
  S.~I.~Alekhin and J.~Bl\"umlein,
  %``Mellin representation for the heavy flavor contributions to deep inelastic structure functions,''
  {\it Phys.\ Lett.}\ B {\bf 594} (2004) 299
  [hep-ph/0404034].
  %%CITATION = HEP-PH/0404034;%%
%--------------------------------------------------------------------------------
%
%[268]
%\cite{Collins:1998rz}
\bibitem{Collins:1998rz}
  J.~C.~Collins,
  %``Hard scattering factorization with heavy quarks: A General treatment,''
  {\it Phys.\ Rev.}\ D {\bf 58} (1998) 094002
  [hep-ph/9806259].
  %%CITATION = HEP-PH/9806259;%%
%-----------------------------------------------------------------------------------
%
%[269]
%\cite{Blumlein:2011mi}
\bibitem{Blumlein:2011mi}
  J.~Bl\"umlein, A.~De Freitas and W.~van Neerven,
  %``Two-loop QED Operator Matrix Elements with Massive External Fermion
  %Lines,''
  {\it Nucl.\ Phys.}\  B {\bf 855} (2012) 508
  [arXiv:1107.4638 [hep-ph]].
  %%CITATION = NUPHA,B855,508;%%
%--------------------------------------------------------------------------------
%
%[270]
%\cite{Bierenbaum:2007qe}
\bibitem{Bierenbaum:2007qe}
  I.~Bierenbaum, J.~Bl\"umlein and S.~Klein,
  %``Two-Loop Massive Operator Matrix Elements and Unpolarized Heavy Flavor Production at Asymptotic Values Q**2 >> m**2,''
  {\it Nucl.\ Phys.}\ B {\bf 780} (2007) 40
  [hep-ph/0703285].
  %%CITATION = HEP-PH/0703285;%%
%--------------------------------------------------------------------------------
%
%[271]
%\cite{Bierenbaum:2008yu}
\bibitem{Bierenbaum:2008yu}
  I.~Bierenbaum, J.~Bl\"umlein, S.~Klein and C.~Schneider,
  %``Two-Loop Massive Operator Matrix Elements for Unpolarized Heavy Flavor Production to O(epsilon),''
  {\it Nucl.\ Phys.}\ B {\bf 803} (2008) 1
  [arXiv:0803.0273 [hep-ph]].
  %%CITATION = ARXIV:0803.0273;%%
%--------------------------------------------------------------------------------
%
%[272]
\bibitem{CCHQ}
%\cite{Gottschalk:1980rv}
%\bibitem{Gottschalk:1980rv}
  T.~Gottschalk,
  %``Chromodynamic Corrections To Neutrino Production Of Heavy Quarks,''
  {\it Phys.\ Rev.}\ D {\bf 23} (1981) 56;\\
  %%CITATION = PHRVA,D23,56;%%
%\cite{Gluck:1996ve}
%\bibitem{Gluck:1996ve}
  M.~Gl\"uck, S.~Kretzer and E.~Reya,
  %``The Strange sea density and charm production in deep inelastic charged current processes,''
  {\it Phys.\ Lett.}\ B {\bf 380} (1996) 171
   [Erratum-ibid.\ B {\bf 405} (1997) 391]
  [hep-ph/9603304];\\
  %%CITATION = HEP-PH/9603304;%%
%\cite{Blumlein:2011zu}
%\bibitem{Blumlein:2011zu}
  J.~Bl\"umlein, A.~Hasselhuhn, P.~Kovacikova and S.~Moch,
  %``$O(\alpha_s)$ Heavy Flavor Corrections to Charged Current Deep-Inelastic Scattering in Mellin Space,''
  {\it Phys.\ Lett.}\ B {\bf 700} (2011) 294
  [arXiv:1104.3449 [hep-ph]];\\
  %%CITATION = ARXIV:1104.3449;%%
%\cite{Buza:1997mg}
%\bibitem{Buza:1997mg}
  M.~Buza and W.~L.~van Neerven,
  %``O (alpha-s**2) contributions to charm production in charged current deep inelastic lepton - hadron scattering,''
  {\it Nucl.\ Phys.}\ B {\bf 500} (1997) 301
  [hep-ph/9702242].
  %%CITATION = HEP-PH/9702242;%%
%-----------------------------------------------------------------------------------
%
%[273]
%\cite{Bierenbaum:2010jp}
\bibitem{Bierenbaum:2010jp}
  I.~Bierenbaum, J.~Bl\"umlein and S.~Klein,
  %``Logarithmic $O(\alpha_s^3)$ contributions to the DIS Heavy Flavor Wilson Coefficients at $Q^2 \gg m^2$,''
  {\it PoS} {\bf DIS2010} (2010) 148
  [arXiv:1008.0792 [hep-ph]].
  %%CITATION = ARXIV:1008.0792;%%
%--------------------------------------------------------------------------------
%
%[274]
\bibitem{ABHKSW12}
J.~Ablinger, J.~Bl\"umlein, A.~Hasselhuhn, S.~Klein, C.~Schneider, F.~Wi\ss{}brock, 
%\cite{Ablinger:2012qm}
%\bibitem{Ablinger:2012qm}
%  J.~Ablinger, J.~Blumlein, A.~Hasselhuhn, S.~Klein, C.~Schneider and F.~Wisbrock,
  %``Massive 3-loop Ladder Diagrams for Quarkonic Local Operator Matrix Elements,''
  {\it Nucl.\ Phys.}\ B {\bf 864} (2012) 52
  [arXiv:1206.2252 [hep-ph]].
  %%CITATION = ARXIV:1206.2252;%%
%--------------------------------------------------------------------------------
%
%[275]
\bibitem{ABS12}
J.~Ablinger, J.~Bl\"umlein, and C. Schneider, in preparation.
%--------------------------------------------------------------------------------
%
%[276]
%\cite{Blumlein:2006mh}
\bibitem{Blumlein:2006mh}
  J.~Bl\"umlein, A.~De Freitas, W.~L.~van Neerven and S.~Klein,
  %``The longitudinal heavy quark structure function F(L)(Q anti-Q) in the
  %region Q**2 >> m**2 at O(alpha(s)**3),''
  {\it Nucl.\ Phys.}\  B {\bf 755} (2006) 272
  [arXiv:hep-ph/0608024].
  %%CITATION = NUPHA,B755,272;%%
%--------------------------------------------------------------------------------
%
%[277]
%\cite{Blumlein:2009rg}
\bibitem{Blumlein:2009rg}
  J.~Bl\"umlein, S.~Klein and B.~T\"odtli,
  %``O(alpha(s)**2) and O(alpha(s)**3) Heavy Flavor Contributions to Transversity at Q**2 >>m**2,''
  {\it Phys.\ Rev.}\ D {\bf 80} (2009) 094010
  [arXiv:0909.1547 [hep-ph]].
  %%CITATION = ARXIV:0909.1547;%%
%--------------------------------------------------------------------------------
%
%[278]
%\cite{Alekhin:2010sv}
\bibitem{Alekhin:2010sv}
  S.~Alekhin and S.~Moch,
  %``Heavy-quark deep-inelastic scattering with a running mass,''
  {\it Phys.\ Lett.}\ B {\bf 699} (2011) 345
  [arXiv:1011.5790 [hep-ph]].
  %%CITATION = ARXIV:1011.5790;%%
%-----------------------------------------------------------------------------------
%
%[279]
\bibitem{THR:UNP}
%\cite{Laenen:1998kp}
%\bibitem{Laenen:1998kp}
  E.~Laenen and S.-O.~Moch,
  %``Soft gluon resummation for heavy quark electroproduction,''
  {\it Phys.\ Rev.}\ D {\bf 59} (1999) 034027
  [hep-ph/9809550];\\
  %%CITATION = HEP-PH/9809550;%%
%\cite{Presti:2010pd}
%\bibitem{Presti:2010pd}
  N.~A.~Lo Presti, H.~Kawamura, S.~Moch and A.~Vogt,
  %``Threshold-improved predictions for charm production in deep-inelastic scattering,''
  {\it PoS} {\bf DIS2010} (2010) 163
  [arXiv:1008.0951 [hep-ph]].
  %%CITATION = ARXIV:1008.0951;%%
%--------------------------------------------------------------------------------
%
%[280]
%\cite{Eynck:2000gz}
\bibitem{Eynck:2000gz}
  T.~O.~Eynck and S~.Moch,
  %``Soft gluon resummation for polarized deep inelastic production of heavy quarks,''
  {\it Phys.\ Lett.}\ B {\bf 495} (2000) 87
  [hep-ph/0008108].
  %%CITATION = HEP-PH/0008108;%%
%--------------------------------------------------------------------------------
%
%[281]
%\cite{Gluck:1993dpa}
\bibitem{Gluck:1993dpa}
  M.~Gl\"uck, E.~Reya and M.~Stratmann,
  %``Heavy quarks at high-energy colliders,''
  {\it Nucl.\ Phys.}\ B {\bf 422} (1994) 37.
  %%CITATION = NUPHA,B422,37;%%
%--------------------------------------------------------------------------------
%
%[282]
%\cite{Alekhin:2009ni}
\bibitem{Alekhin:2009ni}
  S.~Alekhin, J.~Bl\"umlein, S.~Klein and S.~Moch,
  %``The 3, 4, and 5-flavor NNLO Parton from Deep-Inelastic-Scattering Data and at Hadron Colliders,''
  {\it Phys.\ Rev.}\ D {\bf 81} (2010) 014032
  [arXiv:0908.2766 [hep-ph]].
  %%CITATION = ARXIV:0908.2766;%%
%--------------------------------------------------------------------------------
%
%[283]
%\cite{Buza:1996wv}
\bibitem{Buza:1996wv}
  M.~Buza, Y.~Matiounine, J.~Smith and W.~L.~van Neerven,
  %``Charm electroproduction viewed in the variable flavor number scheme versus fixed order 
  % perturbation theory,''
  {\it Eur.\ Phys.\ J.}\ C {\bf 1} (1998) 301
  [hep-ph/9612398].
  %%CITATION = HEP-PH/9612398;%%
%-----------------------------------------------------------------------------------
%
%[284]
%\cite{Bierenbaum:2009zt}
\bibitem{Bierenbaum:2009zt}
  I.~Bierenbaum, J.~Bl\"umlein and S.~Klein,
  %``The Gluonic Operator Matrix Elements at O(alpha(s)**2) for DIS Heavy Flavor Production,''
  {\it Phys.\ Lett.}\ B {\bf 672} (2009) 401
  [arXiv:0901.0669 [hep-ph]].
  %%CITATION = ARXIV:0901.0669;%%
%-----------------------------------------------------------------------------------
%
%[285]
\bibitem{ACOT}
%\cite{Aivazis:1993kh}
%\bibitem{Aivazis:1993kh}
  M.~A.~G.~Aivazis, F.~I.~Olness and W.~-K.~Tung,
  %``Leptoproduction of heavy quarks. 1. General formalism and kinematics of charged current and neutral current production processes,''
  {\it Phys.\ Rev.}\ D {\bf 50} (1994) 3085
  [hep-ph/9312318];\\
  %%CITATION = HEP-PH/9312318;%%
%\cite{Aivazis:1993pi}
%\bibitem{Aivazis:1993pi}
  M.~A.~G.~Aivazis, J.~C.~Collins, F.~I.~Olness and W.~-K.~Tung,
  %``Leptoproduction of heavy quarks. 2. A Unified QCD formulation of charged and neutral current processes from fixed target to collider energies,''
  {\it Phys.\ Rev.}\ D {\bf 50} (1994) 3102;
  [hep-ph/9312319];\\
  %%CITATION = HEP-PH/9312319;%%
%\cite{Tung:2001mv}
%\bibitem{Tung:2001mv}
  W.~-K.~Tung, S.~Kretzer and C.~Schmidt,
  %``Open heavy flavor production in QCD: Conceptual framework and implementation issues,''
  {\it J.\ Phys.}\  G {\bf 28} (2002) 983
  [hep-ph/0110247].
  %%CITATION = HEP-PH/0110247;%%
%-----------------------------------------------------------------------------------
%
%[286]
\bibitem{TR}
%\cite{Thorne:1997ga}
%\bibitem{Thorne:1997ga}
  R.~S.~Thorne and R.~G.~Roberts,
  %``An Ordered analysis of heavy flavor production in deep inelastic scattering,''
  {\it Phys.\ Rev.}\ D {\bf 57} (1998) 6871
  [hep-ph/9709442];\\
  %%CITATION = HEP-PH/9709442;%%
%\cite{Thorne:2006qt}
%\bibitem{Thorne:2006qt}
  R.~S.~Thorne,
  %``A Variable-flavor number scheme for NNLO,''
  {\it Phys.\ Rev.}\ D {\bf 73} (2006) 054019
  [hep-ph/0601245].
  %%CITATION = HEP-PH/0601245;%%
%-----------------------------------------------------------------------------------
%
%[287]
%\cite{Forte:2010ta}
\bibitem{Forte:2010ta}
  S.~Forte, E.~Laenen, P.~Nason and J.~Rojo,
  %``Heavy quarks in deep-inelastic scattering,''
  {\it Nucl.\ Phys.}\  B {\bf 834} (2010) 116
  [arXiv:1001.2312 [hep-ph]].
  %%CITATION = NUPHA,B834,116;%%
%-----------------------------------------------------------------------------------
%
%[288]
%\cite{Blumlein:1998sh}
\bibitem{Blumlein:1998sh}
  J.~Bl\"umlein and W.~L.~van Neerven,
  %``Heavy flavor contributions to the deep inelastic scattering sum rules,''
  {\it Phys.\ Lett.}\ B {\bf 450} (1999) 417
  [hep-ph/9811351].
  %%CITATION = HEP-PH/9811351;%%
%-----------------------------------------------------------------------------------
%
%[289]
\bibitem{MBMC}
%\cite{Ablinger:2011pb}
%\bibitem{Ablinger:2011pb}
  J.~Ablinger, J.~Bl\"umlein, S.~Klein, C.~Schneider and F.~Wissbrock,
  %``3-Loop Heavy Flavor Corrections to DIS with two Massive Fermion Lines,''
  arXiv:1106.5937 [hep-ph];\\
  %%CITATION = ARXIV:1106.5937;%%
%\cite{Ablinger:2012qj}
%\bibitem{Ablinger:2012qj}
  J.~Ablinger, {\it et al.}, %J.~Bl\"umlein, A.~Hasselhuhn, S.~Klein, C.~Schneider and F.~Wissbrock,
  %``New Heavy Flavor Contributions to the DIS Structure Function $F_2(x,Q^2)$ at $O(\alpha_s^3),''
  arXiv:1202.2700 [hep-ph].
  %%CITATION = ARXIV:1202.2700;%%
%--------------------------------------------------------------------------------
%
%[290]
\bibitem{QEXP}
%\cite{Harlander:1997zb}
%\bibitem{Harlander:1997zb}
  R.~Harlander, T.~Seidensticker, M.~Steinhauser,
  %``Complete corrections of Order alpha alpha-s to the decay of the Z boson into bottom quarks,''
  {\it Phys.\ Lett.}\  {\bf B426 } (1998)  125, %-132.
  [hep-ph/9712228];\\
%\cite{Seidensticker:1999bb}
%\bibitem{Seidensticker:1999bb}
  T.~Seidensticker,
  %``Automatic application of successive asymptotic expansions of Feynman diagrams,''
  [hep-ph/9905298].
%-----------------------------------------------------------------------------------
%----------------------------------------------------------------------
%%             >> Target-Mass Corrections: Section 11 <<
%----------------------------------------------------------------------
%--------------------------------------------------------------------------------
%
%[291]
\bibitem{Nachtmann:1973mr}
  O.~Nachtmann,
  %``Positivity constraints for anomalous dimensions,''
  {\it Nucl.\ Phys.}\ B {\bf 63} (1973) 237.
  %%CITATION = NUPHA,B63,237;%%
%------------------------------------------------------------------------
%
%[292]
%\cite{Wandzura:1977ce}
\bibitem{Wandzura:1977ce}
  S.~Wandzura,
  %``Projection of Wilson Coefficients from Deep Inelastic Leptoproduction Data,''
  {\it Nucl.\ Phys.}\ B {\bf 122} (1977) 412.
  %%CITATION = NUPHA,B122,412;%%
%------------------------------------------------------------------------
%
%[293]
%\cite{Georgi:1976ve}
\bibitem{Georgi:1976ve}
  H.~Georgi and H.~D.~Politzer,
  %``Freedom at Moderate Energies: Masses in Color Dynamics,''
  {\it Phys.\ Rev.}\ D {\bf 14} (1976) 1829.
  %%CITATION = PHRVA,D14,1829;%%
%------------------------------------------------------------------------
%
%[294]
%\cite{Schienbein:2007gr}
\bibitem{Schienbein:2007gr}
  I.~Schienbein {\it et al.}, %V.~A.~Radescu, G.~P.~Zeller, M.~E.~Christy, C.~E.~Keppel, K.~S.~McFarland, 
  %W.~Melnitchouk and F.~I.~Olness {\it et al.},
  %``A Review of Target Mass Corrections,''
  {\it J.\ Phys.}\ G  {\bf 35} (2008) 053101
  [arXiv:0709.1775 [hep-ph]].
  %%CITATION = ARXIV:0709.1775;%%
%------------------------------------------------------------------------
%
%[295]
%\cite{Blumlein:2012se}
\bibitem{Blumlein:2012se}
  J.~Bl\"umlein and H.~B\"ottcher,
  %``Higher Twist contributions to the Structure Functions F_2(x,Q^2) and g_2(x,Q^2),''
  arXiv:1207.3170 [hep-ph].
  %%CITATION = ARXIV:1207.3170;%%
%------------------------------------------------------------------------
%
%[296]
%\cite{Bitar:1978cj}
\bibitem{Bitar:1978cj}
  K.~Bitar, P.~W.~Johnson and W.~K.~Tung,
  %``QCD Asymptotics And Kinematic Thresholds In Deep Inelastic Scattering,''
  {\it Phys.\ Lett.}\  B {\bf 83} (1979) 114.
  %%CITATION = PHLTA,B83,114;%%
%------------------------------------------------------------------------
%
%[297]
%\cite{Miramontes:1988fz}
\bibitem{Miramontes:1988fz}
  J.~L.~Miramontes and J.~Sanchez Guillen,
  %``UNDERSTANDING HIGHER TWIST: OPERATOR APPROACH TO POWER CORRECTIONS,''
  {\it Z.\ Phys.}\  C {\bf 41} (1988) 247.
  %%CITATION = ZEPYA,C41,247;%%
%------------------------------------------------------------------------
%
%[298]
%\cite{Accardi:2008ne}
\bibitem{Accardi:2008ne}
  A.~Accardi and J.~-W.~Qiu,
  %``Collinear factorization for deep inelastic scattering structure functions at large Bjorken x(B),''
  {\it JHEP} {\bf 0807} (2008) 090
  [arXiv:0805.1496 [hep-ph]].
  %%CITATION = ARXIV:0805.1496;%%
%------------------------------------------------------------------------
%
%[299]
%\cite{Piccione:1997zh}
\bibitem{Piccione:1997zh}
  A.~Piccione and G.~Ridolfi,
  %``Target mass effects in polarized deep inelastic scattering,''
  {\it Nucl.\ Phys.}\ B {\bf 513} (1998) 301
  [hep-ph/9707478].
  %%CITATION = HEP-PH/9707478;%%
%------------------------------------------------------------------------
%
%[300]
%\cite{Belitsky:2001hz}
\bibitem{Belitsky:2001hz}
  A.~V.~Belitsky and D.~M\"uller,
  %``Resummation of target mass corrections in two photon processes: Twist two sector,''
  {\it Phys.\ Lett.}\ B {\bf 507} (2001) 173
  [hep-ph/0102224].
  %%CITATION = HEP-PH/0102224;%%
%------------------------------------------------------------------------
%
%[301]
%\cite{Geyer:2004bx}
\bibitem{Geyer:2004bx}
  B.~Geyer, D.~Robaschik and J.~Eilers,
  %``Target mass corrections for virtual Compton scattering at twist-2 and generalized, non-forward Wandzura-Wilczek and Callan-Gross relations,''
  {\it Nucl.\ Phys.}\ B {\bf 704} (2005) 279
  [hep-ph/0407300].
  %%CITATION = HEP-PH/0407300;%%
%------------------------------------------------------------------------
%
%[302]
%\cite{Blumlein:2008di}
\bibitem{Blumlein:2008di}
  J.~Bl\"umlein, D.~Robaschik and B.~Geyer,
  %``Target mass and finite t corrections to diffractive deeply inelastic scattering,''
  {\it Eur.\ Phys.\ J.}\ C {\bf 61} (2009) 279
  [arXiv:0812.1899 [hep-ph]].
  %%CITATION = ARXIV:0812.1899;%%
%--------------------------------------------------------------------------------
%----------   PARTON PHENOMENOLOGY   --------------------------------------------
%--------------------------------------------------------------------------------
%
%[303]
%\cite{Blumlein:1997em}
\bibitem{Blumlein:1997em}
  J.~Bl\"umlein and A.~Vogt,
  %``The Evolution of unpolarized singlet structure functions at small x,''
  {\em Phys.\ Rev.}\ D {\bf 58} (1998) 014020
  [hep-ph/9712546].
  %%CITATION = HEP-PH/9712546;%%
%------------------------------------------------------------------------
%
%[304]
\bibitem{EKL}
%\cite{Ellis:1993rb}
%\bibitem{Ellis:1993rb}
  R.~K.~Ellis, Z.~Kunszt and E.~M.~Levin,
  %``The Evolution of parton distributions at small x,''
  {\em Nucl.\ Phys.}\ B {\bf 420} (1994) 517
   [Erratum-ibid.\ B {\bf 433} (1995) 498].
  %%CITATION = NUPHA,B420,517;%%
%------------------------------------------------------------------------
%
%[305]
%\cite{Gluck:1989ze}
\bibitem{Gluck:1989ze}
  M.~Gl\"uck, E.~Reya and A.~Vogt,
  %``Radiatively generated parton distributions for high-energy collisions,''
  {\it Z.\ Phys.}\ C {\bf 48} (1990) 471.
  %%CITATION = ZEPYA,C48,471;%%
%------------------------------------------------------------------------
%
%[306]
%\cite{Vogt:2004ns}
\bibitem{Vogt:2004ns}
  A.~Vogt,
  %``Efficient evolution of unpolarized and polarized parton distributions with QCD-PEGASUS,''
  {\it Comput.\ Phys.\ Commun.}\  {\bf 170} (2005) 65
  [hep-ph/0408244].
  %%CITATION = HEP-PH/0408244;%%
%------------------------------------------------------------------------
%
%[307]
%\cite{Salam:2008qg}
\bibitem{Salam:2008qg}
  G.~P.~Salam and J.~Rojo,
  %``A Higher Order Perturbative Parton Evolution Toolkit (HOPPET),''
  {\it Comput.\ Phys.\ Commun.}\  {\bf 180} (2009) 120
  [arXiv:0804.3755 [hep-ph]].
  %%CITATION = ARXIV:0804.3755;%%
%------------------------------------------------------------------------
%
%[308]
%\cite{Botje:2010ay}
\bibitem{Botje:2010ay}
  M.~Botje,
  %``QCDNUM: Fast QCD Evolution and Convolution,''
  {\it Comput.\ Phys.\ Commun.}\  {\bf 182} (2011) 490
  [arXiv:1005.1481 [hep-ph]], \newline {\tt http://www.nikhef.nl/~h24/qcdnum/}.
  %%CITATION = ARXIV:1005.1481;%%
%------------------------------------------------------------------------
%
%[309]
\bibitem{OPENQCDRAD}
S. Alekhin {\it et al.},
{\tt http://www-zeuthen.desy.de/~alekhin/OPENQCDRAD/}
%------------------------------------------------------------------------
%
%[310]
\bibitem{ORTHPOL}
%% BERNSTEIN
%\cite{Yndurain:1977wz}
%\bibitem{Yndurain:1977wz}
  F.~J.~Yndurain,
  %``Reconstruction of the Deep Inelastic Structure Functions from their Moments,''
  {\it Phys.\ Lett.}\  B {\bf 74 } (1978)  68;\\
%-------
%% JACOBI
%\cite{Parisi:1978jv}
%\bibitem{Parisi:1978jv}
  G.~Parisi and  N.~Sourlas,
  %``A Simple Parametrization Of The Q**2 Dependence Of The Quark Distributions In Qcd,''
  {\it Nucl.\ Phys.}\  B {\bf 151 } (1979)  421;\\
%\cite{Kobayashi:1994hy}
%\bibitem{Kobayashi:1994hy}
  R.~Kobayashi, M.~Konuma, S.~Kumano,
  %``FORTRAN program for a numerical solution of the nonsinglet Altarelli-Parisi equation,''
  {\it Comput.\ Phys.\ Commun.}\  {\bf 86 } (1995)  264,%-278.
  [hep-ph/9409289]; \newline
  {\it Z.\ Phys}.\  {\bf C31 } (1986)  151;\\%-161.
%\cite{Chyla:1986eb}
%\bibitem{Chyla:1986eb}
  J.~Chyla and J.~Rames,
  %``On Methods Of Analyzing Scaling Violation In Deep Inelastic Scattering,''
  {\it Z.\ Phys.}\  C {\bf 31} (1986) 151;\\
  %%CITATION = ZEPYA,C31,151;%%
%\cite{Krivokhizhin:1987rz}
%\bibitem{Krivokhizhin:1987rz}
  V.~G.~Krivokhizhin {\it et al.}. 
% S.~P.~Kurlovich, V.~V.~Sanadze, I.~A.~Savin, A.~V.~Sidorov and N.~B.~Skachkov,
  %``Qcd Analysis Of Singlet Structure Functions Using Jacobi Polynomials: The Description Of The Method,''
  {\it Z.\ Phys.}\ C {\bf 36} (1987) 51;
  %%CITATION = ZEPYA,C36,51;%%
%\cite{Krivokhizhin:1990ct}
%\bibitem{Krivokhizhin:1990ct}
%  V.~G.~Krivokhizhin {\it et al.}
% S.~P.~Kurlovich, R.~Lednicky, S.~Nemecek, V.~V.~Sanadze, I.~A.~Savin, 
%A.~V.~Sidorov and N.~B.~Skachkov,
  %``Next-to-leading order QCD analysis of structure functions with the help of Jacobi polynomials,''
  {\it Z.\ Phys.}\ C {\bf 48} (1990) 347;\\
  %%CITATION = ZEPYA,C48,347;%%
%\cite{Blumlein:1989pd}
%\bibitem{Blumlein:1989pd}
  J.~Bl\"umlein, M.~Klein, G.~Ingelman and R.~R\"uckl,
  %``Testing QCD Scaling Violations In The Hera Energy Range,''
  {\it Z.\ Phys.}\  C {\bf 45} (1990) 501.
  %%CITATION = ZEPYA,C45,501;%%
%----------------------------------------------------------------------------------------
%
%[311]
\bibitem{FP}
%\cite{Furmanski:1981ja}
%\bibitem{Furmanski:1981ja}
  W.~Furmanski, R.~Petronzio,
  %``A Method Of Analyzing The Scaling Violation Of Inclusive Spectra In Hard Processes,''  
  {\it Nucl.\ Phys.}\  B {\bf 195 } (1982)  237.
%----------------------------------------------------------------------------------------
%
%[312]
%\cite{Forte:2002fg}
\bibitem{Forte:2002fg}
  S.~Forte, L.~Garrido, J.~I.~Latorre and A.~Piccione,
  %``Neural network parametrization of deep inelastic structure functions,''
  {\it JHEP} {\bf 0205} (2002) 062
  [hep-ph/0204232].
  %%CITATION = HEP-PH/0204232;%%
%----------------------------------------------------------------------------------------
%
%[313]
\bibitem{DeRoeck:2011na}
  A.~De Roeck and R.~S.~Thorne,
  %``Structure Functions,''
  {\it Prog.\ Part.\ Nucl.\ Phys.}\  {\bf 66} (2011) 727
  [arXiv:1103.0555 [hep-ph]].
  %%CITATION = ARXIV:1103.0555;%%
%------------------------------------------------------------------------
%
%[314]
\bibitem{Perez:2012um}
  E.~Perez and E.~Rizvi,
  %``The Quark and Gluon Structure of the Proton,''
  arXiv:1208.1178 [hep-ex].
  %%CITATION = ARXIV:1208.1178;%%
%------------------------------------------------------------------------
%
%[315]
%\cite{Buras:1977yj}
\bibitem{Buras:1977yj}
  A.~J.~Buras and K.~J.~F.~Gaemers,
  %``Simple Parametrizations of Parton Distributions with q**2 Dependence Given by Asymptotic Freedom,''
  {\it Nucl.\ Phys.}\ B {\bf 132} (1978) 249.
  %%CITATION = NUPHA,B132,249;%%
%-----------------------------------------------------------------------------------------------------
%
%[316]
%\cite{Gluck:1980cp}
\bibitem{Gluck:1980cp}
  M.~Gl\"uck, E.~Hoffmann and E.~Reya,
  %``Scaling Violations and the Gluon Distribution of the Nucleon,''
  {\it Z.\ Phys.}\ C {\bf 13} (1982) 119.
  %%CITATION = ZEPYA,C13,119;%%
%-----------------------------------------------------------------------------------------------------
%
%[317]
%\cite{Duke:1983gd}
\bibitem{Duke:1983gd}
  D.~W.~Duke and J.~F.~Owens,
  %``Q**2 Dependent Parametrizations of Parton Distribution Functions,''
  {\it Phys.\ Rev.}\ D {\bf 30} (1984) 49.
  %%CITATION = PHRVA,D30,49;%%
%-----------------------------------------------------------------------------------------------------
%
%[318]
%\cite{Eichten:1984eu}
\bibitem{Eichten:1984eu}
  E.~Eichten, I.~Hinchliffe, K.~D.~Lane and C.~Quigg,
  %``Super Collider Physics,''
  {\it Rev.\ Mod.\ Phys.}\  {\bf 56} (1984) 579
   [Addendum-ibid.\  {\bf 58} (1986) 1065].
  %%CITATION = RMPHA,56,579;%%
%-----------------------------------------------------------------------------------------------------
%
%[319]
\bibitem{GRV}
%\cite{Gluck:1991ng}
%\bibitem{Gluck:1991ng}
  M.~Gl\"uck, E.~Reya and A.~Vogt,
  %``Parton distributions for high-energy collisions,''
  {\it Z.\ Phys.}\ C 
  {\bf 53} (1992) 127;
  %%CITATION = ZEPYA,C53,127;%%
%\cite{Gluck:1994uf}
%\bibitem{Gluck:1994uf}
%  M.~Gl\"uck, E.~Reya and A.~Vogt,
  %``Dynamical parton distributions of the proton and small x physics,''
%  {\it Z.\ Phys.}\ C 
{\bf 67} (1995) 433;
  %%CITATION = ZEPYA,C67,433;%%
%\cite{Gluck:1998xa}
%\bibitem{Gluck:1998xa}
%  M.~Gl\"uck, E.~Reya and A.~Vogt,
  %``Dynamical parton distributions revisited,''
  {\it Eur.\ Phys.\ J.}\ C {\bf 5} (1998) 461
  [hep-ph/9806404].
  %%CITATION = HEP-PH/9806404;%%
%-----------------------------------------------------------------------------------------------------
%
%[320]
%\cite{Martin:1987vw}
\bibitem{Martin:1987vw}
  A.~D.~Martin, R.~G.~Roberts and W.~J.~Stirling,
  %``Structure Function Analysis and psi, Jet, W, Z Production: Pinning Down the Gluon,''
  {\it Phys.\ Rev.}\ D {\bf 37} (1988) 1161.
  %%CITATION = PHRVA,D37,1161;%%
%-----------------------------------------------------------------------------------------------------
%
%[321]
\bibitem{MT}
%\cite{Tung:1988sn}
%\bibitem{Tung:1988sn}
  W.~-K.~Tung,
  %``Small x Behavior of Parton Distribution Functions in the Next-To-Leading Order QCD Parton Model,''
  {\it Nucl.\ Phys.}\ B {\bf 315} (1989) 378;\\
  %%CITATION = NUPHA,B315,378;%%
%-----------------------------------------------------------------------------------------------------
%\cite{Morfin:1990ck}
%\bibitem{Morfin:1990ck}
  J.~G.~Morfin and W.~-K.~Tung,
  %``Parton distributions from a global QCD analysis of deep inelastic scattering and lepton pair production,''
  {\it Z.\ Phys.}\ C {\bf 52} (1991) 13.
  %%CITATION = ZEPYA,C52,13;%%
%-----------------------------------------------------------------------------------------------------
%
%[322]
%\cite{Lai:2010vv}
\bibitem{Lai:2010vv}
  H.~-L.~Lai, {\it et al.}, %M.~Guzzi, J.~Huston, Z.~Li, P.~M.~Nadolsky, J.~Pumplin and C.~-P.~Yuan,
  %``New parton distributions for collider physics,''
  {\it Phys.\ Rev.}\ D {\bf 82} (2010) 074024
  [arXiv:1007.2241 [hep-ph]] and NNLO analysis in preparation.
  %%CITATION = ARXIV:1007.2241;%%
%-----------------------------------------------------------------------------------------------------
%
%[323]
\bibitem{HERAPDF1.5}
HERAPDF collab. {\footnotesize \tt
https://www.desy.de/h1zeus/combined results/proton structure/Fits/HERAPDF1.0 NNLO 1145.LHgrid.gz}
%-----------------------------------------------------------------------------------------------------
%
%[324]
%\cite{Gluck:2007ck}
\bibitem{Gluck:2007ck}
  M.~Gl\"uck, P.~Jimenez-Delgado and E.~Reya,
  %``Dynamical parton distributions of the nucleon and very small-x physics,''
  {\it Eur.\ Phys.\ J.}\  C {\bf 53} (2008) 355
  [arXiv:0709.0614 [hep-ph]].
  %%CITATION = EPHJA,C53,355;%%
%-----------------------------------------------------------------------------------------------------
%
%[325]
%\cite{Alekhin:2006zm}
\bibitem{Alekhin:2006zm}
  S.~Alekhin, K.~Melnikov and F.~Petriello,
  %``Fixed target Drell-Yan data and NNLO QCD fits of parton distribution
  %functions,''
  {\it Phys.\ Rev.}\  D {\bf 74} (2006) 054033
  [arXiv:hep-ph/0606237].
  %%CITATION = PHRVA,D74,054033;%%
%-----------------------------------------------------------------------------------------------------
%
%[326]
%\cite{Bali:2012av}
\bibitem{Bali:2012av}
  G.~S.~Bali, S.~Collins, M.~Deka, B.~Glaessle, M.~G\"ockeler, J.~Najjar, A.~Nobile and D.~Pleiter {\it et al.},
  %``<x>_{u-d} from lattice QCD at nearly physical quark masses,''
  arXiv:1207.1110 [hep-lat].
  %%CITATION = ARXIV:1207.1110;%%
%-----------------------------------------------------------------------------------------------------
%
%[327]
%\cite{Horsley:2012pz}
\bibitem{Horsley:2012pz}
  R.~Horsley {\it et al.}  [QCDSF and UKQCD Collaborations],
  %``A Lattice Study of the Glue in the Nucleon,''
  {\it Phys.\ Lett.}\ B {\bf 714} (2012) 312
  [arXiv:1205.6410 [hep-lat]].
  %%CITATION = ARXIV:1205.6410;%%
%-----------------------------------------------------------------------------------------------------
%
%[328]
%\cite{Renner:2010ks}
\bibitem{Renner:2010ks}
  D.~B.~Renner,
  %``Status and prospects for the calculation of hadron structure from lattice
  %QCD,''
  arXiv:1002.0925 [hep-lat].
  %%CITATION = ARXIV:1002.0925;%%
%----------------------------------------------------------------------
%
%[329]
%\cite{Blumlein:2010rn}
\bibitem{Blumlein:2010rn}
  J.~Bl\"umlein and H.~B\"ottcher,
  %``QCD Analysis of Polarized Deep Inelastic Scattering Data,''
  {\it Nucl.\ Phys.}\ B {\bf 841} (2010) 205
  [arXiv:1005.3113 [hep-ph]].
  %%CITATION = ARXIV:1005.3113;%%
%----------------------------------------------------------------------
%
%[330]
%\cite{Gluck:2000dy}
\bibitem{Gluck:2000dy}
  M.~Gl\"uck, E.~Reya, M.~Stratmann and W.~Vogelsang,
  %``Models for the polarized parton distributions of the nucleon,''
  {\it Phys.\ Rev.}\ D {\bf 63} (2001) 094005
  [hep-ph/0011215].
  %%CITATION = HEP-PH/0011215;%%
%----------------------------------------------------------------------
%
%[331]
%\cite{deFlorian:2009vb}
\bibitem{deFlorian:2009vb}
  D.~de Florian, R.~Sassot, M.~Stratmann and W.~Vogelsang,
  %``Extraction of Spin-Dependent Parton Densities and Their Uncertainties,''
  {\it Phys.\ Rev.}\ D {\bf 80} (2009) 034030
  [arXiv:0904.3821 [hep-ph]].
  %%CITATION = ARXIV:0904.3821;%%
%----------------------------------------------------------------------
%
%[332]
%\cite{Hirai:2008aj}
\bibitem{Hirai:2008aj}
  M.~Hirai {\it et al.}  [Asymmetry Analysis Collaboration],
  %``Determination of gluon polarization from deep inelastic scattering and collider data,''
  {\it Nucl.\ Phys.}\ B {\bf 813} (2009) 106
  [arXiv:0808.0413 [hep-ph]].
  %%CITATION = ARXIV:0808.0413;%%
%----------------------------------------------------------------------
%
%[333]
%\cite{Leader:2001kh}
\bibitem{Leader:2001kh}
  E.~Leader, A.~V.~Sidorov and D.~B.~Stamenov,
  %``A New evaluation of polarized parton densities in the nucleon,''
  {\it Eur.\ Phys.}\ J.\ C {\bf 23} (2002) 479
  [hep-ph/0111267]. 
  %%CITATION = HEP-PH/0111267;%%
%----------------------------------------------------------------------
%
%[334]
%\cite{Bluemlein:2002be}
\bibitem{Bluemlein:2002be}
  J.~Bl\"umlein and H.~B\"ottcher,
  %``QCD analysis of polarized deep inelastic data and parton distributions,''
  {\it Nucl.\ Phys.}\ B {\bf 636} (2002) 225
  [hep-ph/0203155].
  %%CITATION = HEP-PH/0203155;%%
%----------------------------------------------------------------------
%
%[335]
%\cite{Ramsey:1997iu}
\bibitem{Ramsey:1997iu}
  G.~P.~Ramsey,
  %``Probing nucleon spin structure,''
  {\it Prog.\ Part.\ Nucl.\ Phys.}\  {\bf 39} (1997) 599
  [hep-ph/9702227].
  %%CITATION = HEP-PH/9702227;%%
%----------------------------------------------------------------------
%
%[336]
\bibitem{ABFR}
%\cite{Altarelli:1996nm}
%\bibitem{Altarelli:1996nm}
  G.~Altarelli, R.~D.~Ball, S.~Forte and G.~Ridolfi,
  %``Determination of the Bjorken sum and strong coupling from polarized structure functions,''
  {\it Nucl.\ Phys.}\ B {\bf 496} (1997) 337
  [hep-ph/9701289];\\
  %%CITATION = HEP-PH/9701289;%%
%\cite{Altarelli:1998gn}
%\bibitem{Altarelli:1998gn}
  G.~Altarelli, S.~Forte and G.~Ridolfi,
  %``On positivity of parton distributions,''
  {\it Nucl.\ Phys.}\ B {\bf 534} (1998) 277
  [hep-ph/9806345].
  %%CITATION = HEP-PH/9806345;%%
%----------------------------------------------------------------------
%
%[337]
\bibitem{AAC}
%\cite{Goto:1999by}
%\bibitem{Goto:1999by}
  Y.~Goto {\it et al.}  [Asymmetry Analysis Collaboration],
  %``Polarized parton distribution functions in the nucleon,''
  {\it Phys.\ Rev.}\ D {\bf 62} (2000) 034017
  [hep-ph/0001046].
  %%CITATION = HEP-PH/0001046;%%
%----------------------------------------------------------------------
%
%[338]
%\bibitem{SURV:POL}
%\cite{Lampe:1998eu}
\bibitem{Lampe:1998eu}
  B.~Lampe and E.~Reya,
  %``Spin physics and polarized structure functions,''
  {\it Phys.\ Rept.}\  {\bf 332} (2000) 1
  [hep-ph/9810270].
  %%CITATION = HEP-PH/9810270;%%
%------------------------------------------------------------------------
%
%[339]
%\cite{Kuhn:2008sy}
\bibitem{Kuhn:2008sy}
  S.~E.~Kuhn, J.~-P.~Chen and E.~Leader,
  %``Spin Structure of the Nucleon - Status and Recent Results,''
  {\it Prog.\ Part.\ Nucl.\ Phys.}\  {\bf 63} (2009) 1
  [arXiv:0812.3535 [hep-ph]].
  %%CITATION = ARXIV:0812.3535;%%
%----------------------------------------------------------------------
%
%[340]
\bibitem{STAT:SOFFER}
%\cite{Bourrely:2005kw}
%\bibitem{Bourrely:2005kw}
  C.~R.~V.~Bourrely, J.~Soffer and F.~Buccella,
  %``The Statistical parton distributions: Status and prospects,''
  {\it Eur.\ Phys.}\ J.\ C {\bf 41} (2005) 327  
  [hep-ph/0502180];
  %%CITATION = HEP-PH/0502180;%% 
%\cite{Bourrely:2007if}
%\bibitem{Bourrely:2007if}   
%  C.~Bourrely, J.~Soffer and F.~Buccella,
  %``Strangeness asymmetry of the nucleon in the statistical parton model,''
  {\it Phys.\ Lett.}\ B {\bf 648} (2007) 39
  [hep-ph/0702221];\\
  %%CITATION = HEP-PH/0702221;%%
%\cite{Soffer:2010ud}
%\bibitem{Soffer:2010ud}
  J.~Soffer,
  %``Recent Progress in the Statistical Approach of Parton Distributions,''
  AIP Conf.\ Proc.\  {\bf 1350} (2011) 305
  [arXiv:1011.3209 [hep-ph]].    
  %%CITATION = ARXIV:1011.3209;%%
%----------------------------------------------------------------------
%
%[341]
%\cite{Airapetian:2004zf}
\bibitem{Airapetian:2004zf}
  A.~Airapetian {\it et al.}  [HERMES Collaboration],
  %``Quark helicity distributions in the nucleon for up, down, and strange quarks from semi-inclusive deep-inelast$
  {\it Phys.\ Rev.}\ D {\bf 71} (2005) 012003
  [hep-ex/0407032].
  %%CITATION = HEP-EX/0407032;%%
%----------------------------------------------------------------------
%
%[342]
%\cite{Leader:2010rb}
\bibitem{Leader:2010rb}
  E.~Leader, A.~V.~Sidorov and D.~B.~Stamenov,
  %``Determination of Polarized PDFs from a QCD Analysis of Inclusive and Semi-inclusive
  % Deep Inelastic Scattering Data,''
  {\em Phys.\ Rev.}\ D {\bf 82} (2010) 114018
  [arXiv:1010.0574 [hep-ph]].
  %%CITATION = ARXIV:1010.0574;%%
%----------------------------------------------------------------------------------------
%
%[343]
%\cite{Gluck:2006yz}
\bibitem{Gluck:2006yz}
  M.~Gl\"uck, E.~Reya and C.~Schuck,
  %``Non-singlet QCD analysis of F(2)(x,Q**2) up to NNLO,''
  {\it Nucl.\ Phys.}\ B {\bf 754} (2006) 178
  [hep-ph/0604116].
  %%CITATION = HEP-PH/0604116;%%
%----------------------------------------------------------------------------------------
%
%[344]
%\cite{Martin:2009bu}
\bibitem{Martin:2009bu}
  A.~D.~Martin, W.~J.~Stirling, R.~S.~Thorne and G.~Watt,
  %``Uncertainties on alpha(S) in global PDF analyses and implications for predicted hadronic cross sections,''
  {\it Eur.\ Phys.\ J.}\ C {\bf 64} (2009) 653
  [arXiv:0905.3531 [hep-ph]].
  %%CITATION = ARXIV:0905.3531;%%
%----------------------------------------------------------------------------------------
%
%[345]
%\cite{Ball:2011us}
\bibitem{Ball:2011us}
  R.~D.~Ball, {\it et al.} %V.~Bertone, L.~Del Debbio, S.~Forte, A.~Guffanti,
  %J.~I.~Latorre, S.~Lionetti and J.~Rojo {\it et al.},
  %``Precision NNLO determination of alpha_s(M_Z) using an unbiased global parton set,''
  {\it Phys.\ Lett.}\ B {\bf 707} (2012) 66
  [arXiv:1110.2483 [hep-ph]].
  %%CITATION = ARXIV:1110.2483;%%
%----------------------------------------------------------------------------------------
%
%[346]
\bibitem{CT10}
P. Nadolsky,
{\tt http://indico.desy.de/conferenceDisplay.py?confId=4211} (unpublished).
%----------------------------------------------------------------------
%
%[347]
%\cite{Bethke:2011tr}
\bibitem{Bethke:2011tr}
  S.~Bethke, {\it et al.},
  %A.~H.~Hoang, S.~Kluth, J.~Schieck, I.~W.~Stewart, S.~Aoki,
  %M.~Beneke and S.~Bethke {\it et al.},
  {\sf Workshop on Precision Measurements of $\alpha_s$},
  arXiv:1110.0016 [hep-ph].
  %%CITATION = ARXIV:1110.0016;%%
%----------------------------------------------------------------------
%
%[348]
%\cite{Dissertori:2009qa}
\bibitem{Dissertori:2009qa}
  G.~Dissertori, A.~Gehrmann-De Ridder, T.~Gehrmann, E.~W.~N.~Glover, G.~Heinrich and H.~Stenzel,
  %``Precise determination of the strong coupling constant at NNLO in QCD from the three-jet rate in electron--pos$
  {\it Phys.\ Rev.\ Lett.}\  {\bf 104} (2010) 072002
  [arXiv:0910.4283 [hep-ph]].
  %%CITATION = ARXIV:0910.4283;%%
%----------------------------------------------------------------------
%
%[349]
%\cite{Baikov:2008jh}   
\bibitem{Baikov:2008jh}
  P.~A.~Baikov, K.~G.~Chetyrkin and J.~H.~K\"uhn,
  %``Order alpha**4(s) QCD Corrections to Z and tau Decays,''
  {\it Phys.\ Rev.\ Lett.}\  {\bf 101} (2008) 012002
  [arXiv:0801.1821 [hep-ph]].
%%CITATION = ARXIV:0801.1821;%%
%----------------------------------------------------------------------
%
%[350]
%\cite{Baikov:2012er}
\bibitem{Baikov:2012er}
  P.~A.~Baikov, K.~G.~Chetyrkin, J.~H.~K\"uhn and J.~Rittinger,
  %``Complete ${\cal O}(\alpha_s^4)$ QCD Corrections to Hadronic $Z$-Decays,''
  {\it Phys.\ Rev.\ Lett.}\  {\bf 108} (2012) 222003
  [arXiv:1201.5804 [hep-ph]].
  %%CITATION = ARXIV:1201.5804;%%
%----------------------------------------------------------------------
%
%[351]
%\cite{Gehrmann:2009eh}
\bibitem{Gehrmann:2009eh}
  T.~Gehrmann, M.~Jaquier and G.~Luisoni,
  %``Hadronization effects in event shape moments,''
  {\it Eur.\ Phys.\ J.}\ C {\bf 67} (2010) 57
  [arXiv:0911.2422 [hep-ph]].
  %%CITATION = ARXIV:0911.2422;%%
%----------------------------------------------------------------------
%
%[352]
\bibitem{Abbate:2010xh}
  R.~Abbate, M.~Fickinger, A.~H.~Hoang, V.~Mateu and I.~W.~Stewart,
  %``Thrust at N^3LL with Power Corrections and a Precision Global Fit for alphas(mZ),''
  {\it Phys.\ Rev.}\ D {\bf 83} (2011) 074021
  [arXiv:1006.3080 [hep-ph]].
  %%CITATION = ARXIV:1006.3080;%%
%----------------------------------------------------------------------
%
%[353]
\bibitem{LATalph}  
%\cite{Aoki:2009tf}
%\bibitem{Aoki:2009tf}
  S.~Aoki {\it et al.}  [PACS-CS Collaboration],
  %``Precise determination of the strong coupling 
  %constant in N(f) = 2+1 lattice QCD with the Schrodinger functional scheme,''
  {\it JHEP} {\bf 0910} (2009) 053
  [arXiv:0906.3906 [hep-lat]];\\  
%%CITATION = ARXIV:0906.3906;%%
%\cite{McNeile:2010ji}
%\bibitem{McNeile:2010ji}
  C.~McNeile, C.~T.~H.~Davies, E.~Follana, K.~Hornbostel and G.~P.~Lepage,
  %``High-Precision c and b Masses, and QCD Coupling from Current-Current Correlators in Lattice and Continuum QCD$
  {\it Phys.\ Rev.}\ D {\bf 82} (2010) 034512
  [arXiv:1004.4285 [hep-lat]];\\
  %%CITATION = ARXIV:1004.4285;%%
%\cite{Blossier:2012ef}
%\bibitem{Blossier:2012ef}
  B.~Blossier, P.~Boucaud, M.~Brinet, F.~De Soto, X.~Du, V.~Morenas, O.~Pene and K.~Petrov {\it et al.},
  %``The Strong running coupling at $\tau$ and $Z_0$ mass scales from lattice QCD,''
  arXiv:1201.5770 [hep-ph];\\
  %%CITATION = ARXIV:1201.5770;%%
%\cite{Fritzsch:2012wq}
%\bibitem{Fritzsch:2012wq}
  P.~Fritzsch, F.~Knechtli, B.~Leder, M.~Marinkovic, S.~Schaefer, R.~Sommer and F.~Virotta,
  %``The strange quark mass and Lambda parameter of two flavor QCD,''
  arXiv:1205.5380 [hep-lat];\\
  %%CITATION = ARXIV:1205.5380;%%
%\cite{Bazavov:2012ka}
%\bibitem{Bazavov:2012ka}
  A.~Bazavov, N.~Brambilla, X.~Garcia i Tormo, P.~Petreczky, J.~Soto and A.~Vairo,
  %``Determination of $\alpha_s$ from the QCD static energy,''
  arXiv:1205.6155 [hep-ph]; X.~Garcia i Tormo,  arXiv:1208.4850.
%----------------------------------------------------------------------
%
%[354]
%\cite{Malaescu:2012ts}
\bibitem{Malaescu:2012ts}
  B.~Malaescu and P.~Starovoitov,
  %``Evaluation of the Strong Coupling Constant alpha_s Using the ATLAS Inclusive Jet Cross-Section Data,''
  {\it Eur.\ Phys.\ J.}\ C {\bf 72} (2012) 2041
  [arXiv:1203.5416 [hep-ph]].
  %%CITATION = ARXIV:1203.5416;%%
%----------------------------------------------------------------------
%
%[355]
%\cite{Frederix:2010ne}
\bibitem{Frederix:2010ne}
  R.~Frederix, S.~Frixione, K.~Melnikov and G.~Zanderighi,
  %``NLO QCD corrections to five-jet production at LEP and the extraction of $\alpha_s(M_Z)$,''
  {\it JHEP} {\bf 1011} (2010) 050
  [arXiv:1008.5313 [hep-ph]].
  %%CITATION = ARXIV:1008.5313;%%
%----------------------------------------------------------------------
%--------------------------------------------------------------------------------
%----------    SMALL X RESUMMATION   --------------------------------------------
%--------------------------------------------------------------------------------
%
%[356]
\bibitem
{BFKL}
%\cite{Fadin:1975cb}
%\bibitem{Fadin:1975cb}
  V.~S.~Fadin, E.~A.~Kuraev and L.~N.~Lipatov,
  %``On the Pomeranchuk Singularity in Asymptotically Free Theories,''
  {\em Phys.\ Lett.}\ B {\bf 60} (1975) 50;\\
  %%CITATION = PHLTA,B60,50;%%
%>>---------
%\cite{Lipatov:1976zz}
%\bibitem{Lipatov:1976zz}
  L.~N.~Lipatov,
  %``Reggeization of the Vector Meson and the Vacuum Singularity in Nonabelian Gauge Theories,''
  {\em Sov.\ J.\ Nucl.\ Phys.}\  {\bf 23} (1976) 338
   [{\em Yad.\ Fiz.}\  {\bf 23} (1976) 642];\\
  %%CITATION = SJNCA,23,338;%%
%>>---------
%\cite{Kuraev:1977fs}
%\bibitem{Kuraev:1977fs}
  E.~A.~Kuraev, L.~N.~Lipatov and V.~S.~Fadin,
  %``The Pomeranchuk Singularity in Nonabelian Gauge Theories,''
  {\em Sov.\ Phys.\ JETP}\ {\bf 45} (1977) 199
   [{\em Zh.\ Eksp.\ Teor.\ Fiz.}\  {\bf 72} (1977) 377];\\
  %%CITATION = SPHJA,45,199;%%
%>>---------
%\cite{Balitsky:1978ic}
%\bibitem{Balitsky:1978ic}
  I.~I.~Balitsky and L.~N.~Lipatov,
  %``The Pomeranchuk Singularity in Quantum Chromodynamics,''
  {\em Sov.\ J.\ Nucl.\ Phys.}\  {\bf 28} (1978) 822
   [Yad.\ Fiz.\  {\bf 28} (1978) 1597];\\
  %%CITATION = SJNCA,28,822;%%
%>>---------
%\cite{Lipatov:1985uk}
%\bibitem{Lipatov:1985uk}
  L.~N.~Lipatov,
  %``The Bare Pomeron in Quantum Chromodynamics,''
  {\em Sov.\ Phys.\ JETP} {\bf 63} (1986) 904
   [{\em Zh.\ Eksp.\ Teor.\ Fiz.}\  {\bf 90} (1986) 1536];\\
  %%CITATION = SPHJA,63,904;%%
%>>---------
%\cite{Ciafaloni:1987ur}
%\bibitem{Ciafaloni:1987ur}
  M.~Ciafaloni,
  %``Coherence Effects in Initial Jets at Small q**2 / s,''
  {\em Nucl.\ Phys.}\ B {\bf 296} (1988) 49.
  %%CITATION = NUPHA,B296,49;%%
%------------------------------------------------------------------------
%
%[357]
\bibitem{GLR}
%\cite{Gribov:1981ac}
%\bibitem{Gribov:1981ac}
  L.~V.~Gribov, E.~M.~Levin and M.~G.~Ryskin,
  %``Singlet Structure Function at Small x: Unitarization of Gluon Ladders,''
  {\em Nucl.\ Phys.}\ B {\bf 188} (1981) 555;\\
  %%CITATION = NUPHA,B188,555;%%
%>>---------
%\cite{Mueller:1985wy}
%\bibitem{Mueller:1985wy}
  A.~H.~Mueller and J.~W.~Qiu,
  %``Gluon Recombination and Shadowing at Small Values of x,''
  {\em Nucl.\ Phys.}\ B {\bf 268} (1986) 427.
  %%CITATION = NUPHA,B268,427;%%
%------------------------------------------------------------------------
%
%[358]
\bibitem{GLAUBER}
R.J. Glauber, {\sf Lectures in Theoretical Physics}, (Interscience Publishers, 1959).
%------------------------------------------------------------------------
%
%[359]
%\cite{Kwiecinski:1985gr}
\bibitem{Kwiecinski:1985gr}
  J.~Kwiecinski,
  %``Estimate Of The Screening Corrections To Gluon Distributions In The Small X Region,''
  {\em Z.\ Phys.}\ C {\bf 29} (1985) 147.
  %%CITATION = ZEPYA,C29,147;%%
%------------------------------------------------------------------------
%
%[360]
\bibitem{TRANS:GLR}
%\cite{Collins:1990cw}
%\bibitem{Collins:1990cw}
  J.~C.~Collins and J.~Kwiecinski,
  %``Shadowing In Gluon Distributions In The Small X Region,''
  {\it Nucl.\ Phys.}\ B {\bf 335} (1990) 89;\\
  %%CITATION = NUPHA,B335,89;%%
%\cite{Bartels:1990zk}
%\bibitem{Bartels:1990zk}
  J.~Bartels, G.~A.~Schuler and J.~Bl\"umlein,
  %``A Numerical study of the small x behavior of deep inelastic structure functions in QCD,''
  {\it Z.\ Phys.}\ C {\bf 50} (1991) 91
   {\it [Nucl.\ Phys.\ Proc.\ Suppl.}\  {\bf 18C} (1991) 147];\\
  %%CITATION = ZEPYA,C50,91;%%
%\cite{Altmann:1992vm}
%\bibitem{Altmann:1992vm}
  M.~Altmann, M.~Gl\"uck and E.~Reya,
  %``Possible signatures for low x shadowing effects at high-energy e p colliders,''
  {\it Phys.\ Lett.}\ B {\bf 285} (1992) 359.
  %%CITATION = PHLTA,B285,359;%%
%------------------------------------------------------------------------
%
%[361]
%\cite{McLerran:2011zza}
\bibitem{McLerran:2011zza}
  L.~McLerran, (ed.), J.~Dunlop, (ed.), D.~Morrison, (ed.) and R.~Venugopalan, (ed.),
  {\sf Saturation the color glass condensate and the glasma: What have we learned from RHIC?}, 
  Proceedings, Workshop, Upton, Brookhaven, USA, May 10-12, 2010, 256 pp. 
  %%CITATION = INSPIRE-901469;%%
%------------------------------------------------------------------------
%
%[362]
\bibitem{MAR}
G. Marchesini, in: {\sf QCD at 200 TeV}, ed. by L. Ciffarelli and Yu.L. Dokshitser, 
(Plenum Press, New York, 1992) pp.~183 and references therein.
%------------------------------------------------------------------------
%
%[363]
%\cite{Blumlein:1995jp}
\bibitem{Blumlein:1995jp}
  J.~Bl\"umlein and A.~Vogt,
  %``On the behavior of nonsinglet structure functions at small x,''
  {\it Phys.\ Lett.}\ B {\bf 370} (1996) 149
  [hep-ph/9510410];
  %%CITATION = HEP-PH/9510410;%%
%\cite{Blumlein:1996dd}
%\bibitem{Blumlein:1996dd}
%  J.~Bl\"umlein and A.~Vogt,
  %``On the resummation of alpha ln**2 x terms for nonsinglet structure functions in QED and QCD,''
  {\em Acta Phys.\ Polon.}\ B {\bf 27} (1996) 1309
  [hep-ph/9603450].
  %%CITATION = HEP-PH/9603450;%%
%------------------------------------------------------------------------
%
%[364]
\bibitem{KL}
%\cite{Kirschner:1983di}
%\bibitem{Kirschner:1983di}
  R.~Kirschner and L.~N.~Lipatov,
  %``Double Logarithmic Asymptotics and Regge Singularities of Quark Amplitudes with Flavor Exchange,''
  {\em Nucl.\ Phys.}\ B {\bf 213} (1983) 122.
  %%CITATION = NUPHA,B213,122;%%
%------------------------------------------------------------------------
%
%[365]
\bibitem{BER2}
%\cite{Bartels:1996wc}
%\bibitem{Bartels:1996wc}
  J.~Bartels, B.~I.~Ermolaev and M.~G.~Ryskin,
  %``Flavor singlet contribution to the structure function G(1) at small x,''
  {\it Z.\ Phys.}\ C {\bf 72} (1996) 627
  [hep-ph/9603204].
  %%CITATION = HEP-PH/9603204;%%
%------------------------------------------------------------------------
%
%[366]
\bibitem{BV3}
%\cite{Blumlein:1996hb}
%\bibitem{Blumlein:1996hb}
  J.~Bl\"umlein and A.~Vogt,
  %``The Singlet contribution to the structure function g1 (x, Q**2) at small x,''
  {\em Phys.\ Lett.}\ B {\bf 386} (1996) 350
  [hep-ph/9606254].
  %%CITATION = HEP-PH/9606254;%%
%------------------------------------------------------------------------
%
%[367]
\bibitem{KOD}
%\cite{Kiyo:1996si}
%\bibitem{Kiyo:1996si}
  Y.~Kiyo, J.~Kodaira and H.~Tochimura,
  %``Does leading ln x resummation predict the rise of g(1) at small x?,''
  {\em Z.\ Phys.}\ C {\bf 74} (1997) 631
  [hep-ph/9701365].
  %%CITATION = HEP-PH/9701365;%%
%------------------------------------------------------------------------
%
%[368]
\bibitem{EMR}
%\cite{Ermolaev:1995fx}
%\bibitem{Ermolaev:1995fx}
  B.~I.~Ermolaev, S.~I.~Manaenkov and M.~G.~Ryskin,
  %``Nonsinglet structure functions at small x,''
  {\em Z.\ Phys.}\ C {\bf 69} (1996) 259
  [hep-ph/9502262];\\
  %%CITATION = HEP-PH/9502262;%%
%\bibitem{BER1}
%\cite{Bartels:1995iu}
%\bibitem{Bartels:1995iu}
  J.~Bartels, B.~I.~Ermolaev and M.~G.~Ryskin,
  %``Nonsinglet contributions to the structure function g1 at small x,''
  {\em Z.\ Phys.}\ C {\bf 70} (1996) 273
  [hep-ph/9507271].
  %%CITATION = HEP-PH/9507271;%%
%------------------------------------------------------------------------
%
%[369]
\bibitem{JAR}
%\cite{Jaroszewicz:1980mq}
%\bibitem{Jaroszewicz:1980mq}
  T.~Jaroszewicz,
  %``Infrared Divergences and Regge Behavior in QCD,''
  {\it Acta Phys.\ Polon.}\ B {\bf 11} (1980) 965;
  %%CITATION = APPOA,B11,965;%%
%>>---------
%\cite{Jaroszewicz:1982gr}
%\bibitem{Jaroszewicz:1982gr}  
%T.~Jaroszewicz,
  %``Gluonic Regge Singularities and Anomalous Dimensions in QCD,''
  {\em Phys.\ Lett.}\ B {\bf 116} (1982) 291.
  %%CITATION = PHLTA,B116,291;%%
%------------------------------------------------------------------------
%
%[370]
\bibitem{EHW}
%\cite{Ellis:1995gv}
%\bibitem{Ellis:1995gv}
  R.~K.~Ellis, F.~Hautmann and B.~R.~Webber,
  %``QCD scaling violation at small x,''
  {\em Phys.\ Lett.}\ B {\bf 348} (1995) 582
  [hep-ph/9501307].
  %%CITATION = HEP-PH/9501307;%%
%------------------------------------------------------------------------
%
%[371]
\bibitem{JB95}
%\cite{Blumlein:1995mi}
%\bibitem{Blumlein:1995mi}
  J.~Bl\"umlein,
  %{\sf On the k(T) dependent gluon density in hadrons and in the photon},
  Proc. {\sf QCD and high energy hadronic interactions}, Les Arcs, ed. Tran Than Van, pp.~191,
  [hep-ph/9506446].
  %%CITATION = HEP-PH/9506446;%%
%------------------------------------------------------------------------
%
%[372]
\bibitem{KWI} 
%\cite{Kwiecinski:1985cq}
%\bibitem{Kwiecinski:1985cq}
  J.~Kwiecinski,
  %``The Gluon Distributions in the Small x Region Beyond the Leading Order,''
  {\em Z.\ Phys.}\ C {\bf 29} (1985) 561.
  %%CITATION = ZEPYA,C29,561;%%
%------------------------------------------------------------------------
%
%[373]
\bibitem{BVR1}
%\cite{Blumlein:1996aw}
%\bibitem{Blumlein:1996aw}
  J.~Bl\"umlein, S.~Riemersma and A.~Vogt,
  %``The Evolution of unpolarized and polarized structure functions at small x,''
  {\em Nucl.\ Phys.\ (Proc.\ Suppl.)}\  {\bf 51C} (1996) 30
  [hep-ph/9608470].
  %%CITATION = HEP-PH/9608470;%%
%>>---------
%\cite{Blumlein:1996sy}
%\bibitem{Blumlein:1996sy}
%  J.~Bl\"umlein, S.~Riemersma and A.~Vogt,
  %``The Small x evolution of unpolarized and polarized structure functions,''
  {\em Acta Phys.\ Polon.}\ B {\bf 28} (1997) 577
  [hep-ph/9610427].
  %%CITATION = HEP-PH/9610427;%%
%------------------------------------------------------------------------
%
%[374]
%\cite{Catani:1994sq}
\bibitem{Catani:1994sq}
   S.~Catani and F.~Hautmann,
   %``High-energy factorization and small x deep inelastic scattering beyond leading order,''
   {\em Nucl.\ Phys.}\ B {\bf 427} (1994) 475
   [hep-ph/9405388].
   %%CITATION = HEP-PH/9405388;%%
%--------------------------------------------------------------------------------------
%
%[375]
%\cite{Kirschner:2009qu}
\bibitem{Kirschner:2009qu}
  R.~Kirschner and M.~Segond,
  %``Small x resummation in collinear factorisation,''
  {\em Eur.\ Phys.\ J.}\ C {\bf 68} (2010) 425
  [arXiv:0910.5443 [hep-ph]].
  %%CITATION = ARXIV:0910.5443;%%
%------------------------------------------------------------------------
%
%[376]
\bibitem{FL}
%\cite{Fadin:1998py}
%\bibitem{Fadin:1998py}
  V.~S.~Fadin and L.~N.~Lipatov,
  %``BFKL pomeron in the next-to-leading approximation,''
  {\em Phys.\ Lett.}\ B {\bf 429} (1998) 127
  [hep-ph/9802290];\\
  %%CITATION = HEP-PH/9802290;%%
%>>---------
%\cite{Fadin:1998sh}
%\bibitem{Fadin:1998sh}
  V.~S.~Fadin,
  %``BFKL news,''
  hep-ph/9807528.
  %%CITATION = HEP-PH/9807528;%%
%------------------------------------------------------------------------
%
%[377]
\bibitem{CC2}
%\cite{Camici:1997ta}
%\bibitem{Camici:1997ta}
  G.~Camici and M.~Ciafaloni,
  %``k factorization and small x anomalous dimensions,''
  {\em Nucl.\ Phys.}\ B {\bf 496} (1997) 305
   [Erratum-ibid.\ B {\bf 607} (2001) 431]
  [hep-ph/9701303].
  %%CITATION = HEP-PH/9701303;%%
%>>---------
%\cite{Ciafaloni:1998gs}
%\bibitem{Ciafaloni:1998gs}
%  M.~Ciafaloni and G.~Camici,
  %``Energy scale(s) and next-to-leading BFKL equation,''
  {\em Phys.\ Lett.}\ B {\bf 430} (1998) 349
  [hep-ph/9803389].
  %%CITATION = HEP-PH/9803389;%%
%------------------------------------------------------------------------
%
%[378]
\bibitem{CIA}
%\cite{Ciafaloni:1995bn}
%\bibitem{Ciafaloni:1995bn}
  M.~Ciafaloni,
  %``k(T) factorization versus renormalization group: A Small x consistency argument,''
  {\em Phys.\ Lett.}\ B {\bf 356} (1995) 74
  [hep-ph/9507307].
  %%CITATION = HEP-PH/9507307;%%
%------------------------------------------------------------------------
%
%[379]
%\cite{Blumlein:1993ec}
\bibitem{Blumlein:1993ec}
  J.~Blumlein,
  %``The Longitudinal structure function F-L (x,Q**2) at small x,''
  {\it J.\ Phys.}\  G {\bf 19} (1993) 1623.
  %%CITATION = JPHGB,G19,1623;%%
%------------------------------------------------------------------------
%
%[380]
\bibitem{BRNV}
%\cite{Blumlein:1998pp}
%\bibitem{Blumlein:1998pp}
  J.~Bl\"umlein, V.~Ravindran, W.~L.~van Neerven and A.~Vogt,
  {\sf The Unpolarized gluon anomalous dimension at small $x$},
  \emph{DIS98, 6th International Workshop on Deep Inelastic Scattering
  and QCD, Brussels, Belgium, April, 1998}, ed. by Gh. Coremans and
  R. Rosen (World Scientific, Singapore, 1998),  
  211-216
  [hep-ph/9806368].
%------------------------------------------------------------------------
%
%[381]
%\cite{Altarelli:2008aj}
\bibitem{Altarelli:2008aj}
  G.~Altarelli, R.~D.~Ball and S.~Forte,
  %``Small x Resummation with Quarks: Deep-Inelastic Scattering,''
  {\it Nucl.\ Phys.}\ B {\bf 799} (2008) 199
  [arXiv:0802.0032 [hep-ph]].
  %%CITATION = ARXIV:0802.0032;%%
%------------------------------------------------------------------------
%
%[382]
\bibitem{FAD1}
%\cite{Fadin:1996tb}
%\bibitem{Fadin:1996tb}
  V.~S.~Fadin, R.~Fiore and M.~I.~Kotsky,
  %``Gluon Regge trajectory in the two loop approximation,''
  {\em Phys.\ Lett.}\ B {\bf 387} (1996) 593
  [hep-ph/9605357].
  %%CITATION = HEP-PH/9605357;%%
%------------------------------------------------------------------------
%
%[383]
\bibitem{BNR}
%\cite{Blumlein:1998ib}
%\bibitem{Blumlein:1998ib}
  J.~Bl\"umlein, V.~Ravindran and W.~L.~van Neerven,
  %``On the gluon Regge trajectory in O alpha-s**2,''
  {\em Phys.\ Rev.}\ D {\bf 58} (1998) 091502
  [hep-ph/9806357].
  %%CITATION = HEP-PH/9806357;%%
%------------------------------------------------------------------------
%
%[384]
\bibitem{KOR}
%\cite{Korchemskaya:1996je}
%\bibitem{Korchemskaya:1996je}
  I.~A.~Korchemskaya and G.~P.~Korchemsky,
  %``Evolution equation for gluon Regge trajectory,''
  {\em Phys.\ Lett.}\ B {\bf 387} (1996) 346
  [hep-ph/9607229].
  %%CITATION = HEP-PH/9607229;%%
%------------------------------------------------------------------------
%
%[385]
\cite{Ball:1999sh}
\bibitem{Ball:1999sh}
  R.~D.~Ball and S.~Forte,
  %``The Small x behavior of Altarelli-Parisi splitting functions,''
  Phys.\ Lett.\ B {\bf 465} (1999) 271
  [hep-ph/9906222].
  %%CITATION = HEP-PH/9906222;%%
%------------------------------------------------------------------------
%
%[386]
%\cite{Altarelli:2005ni}
\bibitem{Altarelli:2005ni}
  G.~Altarelli, R.~D.~Ball and S.~Forte,
  %``Perturbatively stable resummed small x evolution kernels,''
  {\em Nucl.\ Phys.}\ B {\bf 742} (2006) 1
  [hep-ph/0512237] and references quoted therein.
  %%CITATION = HEP-PH/0512237;%%
%------------------------------------------------------------------------
%
%[387]
%\cite{Ciafaloni:2006yk}
\bibitem{Ciafaloni:2006yk}
  M.~Ciafaloni, D.~Colferai, G.~P.~Salam and A.~M.~Stasto,
  %``Minimal subtraction vs. physical factorisation schemes in small-x QCD,''
  {\it Phys.\ Lett.}\ B {\bf 635} (2006) 320
  [hep-ph/0601200].
  %%CITATION = HEP-PH/0601200;%%
%------------------------------------------------------------------------
%
%[388]
%\cite{Ciafaloni:2007gf}
\bibitem{Ciafaloni:2007gf}
  M.~Ciafaloni, D.~Colferai, G.~P.~Salam and A.~M.~Stasto,
  %``A Matrix formulation for small-x singlet evolution,''
  {\it JHEP} {\bf 0708} (2007) 046
  [arXiv:0707.1453 [hep-ph]] and references quoted therein.
  %%CITATION = ARXIV:0707.1453;%%
%------------------------------------------------------------------------
%
%[389]
\bibitem{LOV}
%\cite{Lovelace:1974mi}
%\bibitem{Lovelace:1974mi}
  C.~Lovelace,
  %``Regge Behavior Under Asymptotic Freedom,''
  {\em Phys.\ Lett.}\ B {\bf 55} (1975) 187;
  %%CITATION = PHLTA,B55,187;%%
%>>---------
%\cite{Lovelace:1975iq}
%\bibitem{Lovelace:1975iq}
%  C.~Lovelace,
  %``Regge Behavior of a Soluble Model with Asymptotic Freedom,''
  {\em Nucl.\ Phys.}\ B {\bf 95} (1975) 12.
  %%CITATION = NUPHA,B95,12;%%
%------------------------------------------------------------------------
%
%[390]
\bibitem{BVN}
%\cite{Blumlein:1998mg}
%\bibitem{Blumlein:1998mg}
  J.~Bl\"umlein and W.~L.~van Neerven,
  %``Less singular terms and small x evolution in a soluble model,''
  {\em Phys.\ Lett.}\ B {\bf 450} (1999) 412
  [hep-ph/9811519].
  %%CITATION = HEP-PH/9811519;%%
%------------------------------------------------------------------------
%-------------------------------------------------------------------------------------------------
%-------->> Large X  <<-------------------------------------------------------------------------
%-------------------------------------------------------------------------------------------------
%
%[391]
%\cite{Moch:2009hr}
\bibitem{Moch:2009hr}
  S.~Moch and A.~Vogt,
  %``On non-singlet physical evolution kernels and large-x coefficient functions in perturbative QCD,''
  {\it JHEP} {\bf 0911} (2009) 099
  [arXiv:0909.2124 [hep-ph]].
  %%CITATION = ARXIV:0909.2124;%%
%--------------------------------------------------------------------------------
%
%[392]
%\cite{Almasy:2010wn}
\bibitem{Almasy:2010wn}
  A.~A.~Almasy, G.~Soar and A.~Vogt,
  %``Generalized double-logarithmic large-x resummation in inclusive deep-inelastic scattering,''
  {\it JHEP} {\bf 1103} (2011) 030
  [arXiv:1012.3352 [hep-ph]].
  %%CITATION = ARXIV:1012.3352;%%
%--------------------------------------------------------------------------------
%
%[393]
%\cite{Kodaira:1981nh}
\bibitem{Kodaira:1981nh}
  J.~Kodaira and L.~Trentadue,
  %``Summing Soft Emission in QCD,''
  {\it Phys.\ Lett.}\ B {\bf 112} (1982) 66.
  %%CITATION = PHLTA,B112,66;%%
%-------------------------------------------------------------------------------------------------
%
%[394]
\bibitem{CEXP1}
%\cite{Sterman:1986aj}
%\bibitem{Sterman:1986aj}
  G.~F.~Sterman,
  %``Summation of Large Corrections to Short Distance Hadronic Cross-Sections,''
  {\it Nucl.\ Phys.}\ B {\bf 281} (1987) 310;\\
  %%CITATION = NUPHA,B281,310;%%
%--------------------------------------------
%\cite{Catani:1989ne}
%\bibitem{Catani:1989ne}
  S.~Catani and L.~Trentadue,
  %``Resummation of the QCD Perturbative Series for Hard Processes,''
  {\it Nucl.\ Phys.}\ B {\bf 327} (1989) 323;
  %%CITATION = NUPHA,B327,323;%%
%--------------------------------------------
%\cite{Catani:1990rp}
%\bibitem{Catani:1990rp}
%  S.~Catani and L.~Trentadue,
  %``Comment on QCD exponentiation at large x,''
  {\it Nucl.\ Phys.}\ B {\bf 353} (1991) 183;
  %%CITATION = NUPHA,B353,183;%%
%--------------------------------------------
%\cite{Catani:1996yz}
%\bibitem{Catani:1996yz}
  S.~Catani, M.~L.~Mangano, P.~Nason and L.~Trentadue,
  %``The Resummation of soft gluons in hadronic collisions,''
  {\it Nucl.\ Phys.}\ B {\bf 478} (1996) 273
  [hep-ph/9604351];\\
  %%CITATION = HEP-PH/9604351;%%
%--------------------------------------------
%\cite{Contopanagos:1996nh}
%\bibitem{Contopanagos:1996nh}
  H.~Contopanagos, E.~Laenen and G.~F.~Sterman,
  %``Sudakov factorization and resummation,''
  {\it Nucl.\ Phys.}\ B {\bf 484} (1997) 303
  [hep-ph/9604313].
  %%CITATION = HEP-PH/9604313;%%
%-------------------------------------------------------------------------------------------------
%
%[395]
%\cite{Vogt:1999xa}
\bibitem{Vogt:1999xa}
  A.~Vogt,
  %``On soft gluon effects in deep inelastic structure functions,''
  Phys.\ Lett.\ B {\bf 471} (1999) 97
  [hep-ph/9910545].
  %%CITATION = HEP-PH/9910545;%%
%-------------------------------------------------------------------------------------------------
%
%[396]
\bibitem{LXRESUM1}
%\cite{Moch:2005ba}
%\bibitem{Moch:2005ba}
  S.~Moch, J.~A.~M.~Vermaseren and A.~Vogt,
  %``Higher-order corrections in threshold resummation,''
  {\it Nucl.\ Phys.}\ B {\bf 726} (2005) 317
  [hep-ph/0506288] and Erratum;\\
  %%CITATION = HEP-PH/0506288;%%
%--------------------------------------------------------------------------------
%\cite{Ravindran:2006cg}
%\bibitem{Ravindran:2006cg}
  V.~Ravindran,
  %``Higher-order threshold effects to inclusive processes in QCD,''
  {\it Nucl.\ Phys.}\ B {\bf 752} (2006) 173
  [hep-ph/0603041].
  %%CITATION = HEP-PH/0603041;%%
%--------------------------------------------------------------------------------
%
%[397]
\bibitem{LXRESUM2}
%\cite{Laenen:2008ux}
%\bibitem{Laenen:2008ux}
  E.~Laenen, L.~Magnea and G.~Stavenga,
  %``On next-to-eikonal corrections to threshold resummation for the Drell-Yan and DIS cross sections,''
  {\it Phys.\ Lett.}\ B {\bf 669} (2008) 173
  [arXiv:0807.4412 [hep-ph]].
  %%CITATION = ARXIV:0807.4412;%%
%--------------------------------------------------------------------------------
%
%[398]
%\cite{Soar:2009yh}
\bibitem{Soar:2009yh}
  G.~Soar, S.~Moch, J.~A.~M.~Vermaseren and A.~Vogt,
  %``On Higgs-exchange DIS, physical evolution kernels and fourth-order splitting functions at large x,''
  {\it Nucl.\ Phys.}\ B {\bf 832} (2010) 152
  [arXiv:0912.0369 [hep-ph]].
  %%CITATION = ARXIV:0912.0369;%%
%------------------------------------------------------------------------
%
%[399]
\bibitem{EXPSYST}
%\cite{Laenen:2008gt}
%\bibitem{Laenen:2008gt}
  E.~Laenen, G.~Stavenga and C.~D.~White,
  %``Path integral approach to eikonal and next-to-eikonal exponentiation,''
  {\it JHEP} {\bf 0903} (2009) 054
  [arXiv:0811.2067 [hep-ph]];\\
  %%CITATION = ARXIV:0811.2067;%%
%\cite{Laenen:2010uz}
%\bibitem{Laenen:2010uz}
  E.~Laenen, L.~Magnea, G.~Stavenga and C.~D.~White,
  %``Next-to-eikonal corrections to soft gluon radiation: a diagrammatic approach,''
  {\it JHEP} {\bf 1101} (2011) 141
  [arXiv:1010.1860 [hep-ph]].
  %%CITATION = ARXIV:1010.1860;%%
%--------------------------------------------------------------------------------
%-------------------------------------------------------------------------------------------------
%-------->> sum rules  <<-------------------------------------------------------------------------
%-------------------------------------------------------------------------------------------------
%
%[400]
%\cite{Ravindran:2001dk}
\bibitem{Ravindran:2001dk}
  V.~Ravindran and W.~L.~van Neerven,
  %``QCD and power corrections to sum rules in deep inelastic lepton nucleon scattering,''
  {\it Nucl.\ Phys.}\ B {\bf 605} (2001) 517
  [hep-ph/0102280].
  %%CITATION = HEP-PH/0102280;%%
%--------------------------------------------------------------------------------
%
%[401]
%\cite{Adler:1965ty}
\bibitem{Adler:1965ty}
  S.~L.~Adler,
  %``Sum rules giving tests of local current commutation relations in high-energy neutrino reactions,''
  {\it Phys.\ Rev.}\  {\bf 143} (1966) 1144.
  %%CITATION = PHRVA,143,1144;%%
%--------------------------------------------------------------------------------
%
%[402]
%\cite{Bjorken:1967px}
\bibitem{Bjorken:1967px}
  J.~D.~Bjorken,
  %``Inequality for Backward electron-Nucleon and Muon-Nucleon Scattering at High Momentum Transfer,''
  {\it Phys.\ Rev.}\  {\bf 163} (1967) 1767.
  %%CITATION = PHRVA,163,1767;%%
%--------------------------------------------------------------------------------
%
%[403]
%\cite{Larin:1990zw}
\bibitem{Larin:1990zw}
  S.~A.~Larin, F.~V.~Tkachov and J.~A.~M.~Vermaseren,
  %``The alpha(s**3) correction to the Bjorken sum rule,''
  {\it Phys.\ Rev.\ Lett.}\  {\bf 66} (1991) 862.
  %%CITATION = PRLTA,66,862;%%
%--------------------------------------------------------------------------------
%
%[404]
%\cite{Gross:1969jf}
\bibitem{Gross:1969jf}
  D.~J.~Gross and C.~H.~Llewellyn Smith,
  %``High-energy neutrino - nucleon scattering, current algebra and partons,''
  {\it  Nucl.\ Phys.}\ B {\bf 14} (1969) 337.
  %%CITATION = NUPHA,B14,337;%%
%--------------------------------------------------------------------------------
%
%[405]
%\cite{Larin:1991tj}
\bibitem{Larin:1991tj}
  S.~A.~Larin and J.~A.~M.~Vermaseren,
  %``The alpha-s**3 corrections to the Bjorken sum rule for polarized electroproduction and to the Gross-Llewellyn Smith sum rule,''
  {\it Phys.\ Lett.}\ B {\bf 259} (1991) 345.
  %%CITATION = PHLTA,B259,345;%%
%--------------------------------------------------------------------------------
%
%[406]
\bibitem{CHET:GLL}
%\cite{Baikov:2012zn}
          %\bibitem{Baikov:2012zn}
          P.~A.~Baikov, K.~G.~Chetyrkin, J.~H.~Kuhn and J.~Rittinger,
          %``Adler Function, Sum Rules and Crewther Relation of OrderO(alpha_s^4): the Singlet Case,''
          Phys.\ Lett.\ B {\bf 714} (2012) 62
          [arXiv:1206.1288 [hep-ph]].
%--------------------------------------------------------------------------------
%
%[407]
%\cite{Bjorken:1969mm}
\bibitem{Bjorken:1969mm}
  J.~D.~Bjorken,
  %``Inelastic Scattering of Polarized Leptons from Polarized Nucleons,''
  {\it Phys.\ Rev.}\ D {\bf 1} (1970) 1376.
  %%CITATION = PHRVA,D1,1376;%%
%--------------------------------------------------------------------------------
%
%[408]
%\cite{Baikov:2010je}
\bibitem{Baikov:2010je}
  P.~A.~Baikov, K.~G.~Chetyrkin and J.~H.~K\"uhn,
  %``Adler Function, Bjorken Sum Rule, and the Crewther Relation to Order alpha_s^4 in a General Gauge Theory,''
  {\it Phys.\ Rev.\ Lett.}\  {\bf 104} (2010) 132004
  [arXiv:1001.3606 [hep-ph]].
  %%CITATION = ARXIV:1001.3606;%%
%--------------------------------------------------------------------------------
%
%[409]
%\cite{Gerasimov:1965et}
\bibitem{Gerasimov:1965et}
  S.~B.~Gerasimov,
  %``A Sum rule for magnetic moments and the damping of the nucleon magnetic moment in nuclei,''
  {\it Sov.\ J.\ Nucl.\ Phys.}\  {\bf 2} (1966) 430
   [{\it Yad.\ Fiz.}\  {\bf 2} (1965) 598].
  %%CITATION = SJNCA,2,430;%%
%--------------------------------------------------------------------------------
%
%[410]
%\cite{Drell:1966jv}
\bibitem{Drell:1966jv}
  S.~D.~Drell and A.~C.~Hearn,
  %``Exact Sum Rule for Nucleon Magnetic Moments,''
  {\it Phys.\ Rev.\ Lett.}\  {\bf 16} (1966) 908.
  %%CITATION = PRLTA,16,908;%%
%-----------------------------------------------------------------------------------------
%
%[411]
%\cite{Drechsel:2000ct}
\bibitem{Drechsel:2000ct}
  D.~Drechsel, S.~S.~Kamalov and L.~Tiator,
  %``The GDH sum rule and related integrals,''
  {\it Phys.\ Rev.}\ D {\bf 63} (2001) 114010
  [hep-ph/0008306].
  %%CITATION = HEP-PH/0008306;%%
%-----------------------------------------------------------------------------------------
%
%[412]
%\cite{Drechsel:2004ki}
\bibitem{Drechsel:2004ki}
  D.~Drechsel and L.~Tiator,
  %``The Gerasimov-Drell-Hearn sum rule and the spin structure of the nucleon,''
  {\it Ann.\ Rev.\ Nucl.\ Part.\ Sci.}\  {\bf 54} (2004) 69
  [nucl-th/0406059].
  %%CITATION = NUCL-TH/0406059;%%
%-----------------------------------------------------------------------------------------
%
%[413]
%\cite{Burkhardt:1970ti}
\bibitem{Burkhardt:1970ti}
  H.~Burkhardt and W.~N.~Cottingham,
  %``Sum rules for forward virtual Compton scattering,''
  {\it Annals Phys.}\  {\bf 56} (1970) 453.
  %%CITATION = APNYA,56,453;%%
%--------------------------------------------------------------------------------
%
%[414]
%\cite{Jaffe:1989xx}
\bibitem{Jaffe:1989xx}
  R.~L.~Jaffe,
  %``G(2): The Nucleon's Other Spin Dependent Structure Function,''
  {\it Comments Nucl.\ Part.\ Phys.}\  {\bf 19} (1990) 239.
  %%CITATION = CNPPA,19,239;%%
%--------------------------------------------------------------------------------
%
%[415]
\bibitem{IKL}
B.L. Ioffe, V.A. Khoze, and L.N. Lipatov, {\sf Hard Processes}, Vol.~1 (North-Holland, Amsterdam, 1984).
%--------------------------------------------------------------------------------
%
%[416]
%\cite{Heimann:1973hq}
\bibitem{Heimann:1973hq}
  R.~L.~Heimann,   
  %``Spin dependent high frequency inelastic electron scattering and helicity flip couplings,''
  {\it Nucl.\ Phys.}\ B {\bf 64} (1973) 429.
  %%CITATION = NUPHA,B64,429;%%
%--------------------------------------------------------------------------------
%
%[417]
%\cite{Kodaira:1994gh}
\bibitem{Kodaira:1994gh}
  J.~Kodaira, S.~Matsuda, T.~Uematsu and K.~Sasaki,
  %``More on the Burkhardt-Cottingham sum rule in QCD,''
  {\it Phys.\ Lett.}\ B {\bf 345} (1995) 527
  [hep-ph/9408353].
  %%CITATION = HEP-PH/9408353;%%
%--------------------------------------------------------------------------------
%
%[418]
%\cite{Efremov:1996hd}
\bibitem{Efremov:1996hd}
  A.~V.~Efremov, O.~V.~Teryaev and E.~Leader,
  %``An Exact sum rule for transversely polarized DIS,''
  {\it Phys.\ Rev.}\ D {\bf 55} (1997) 4307
  [hep-ph/9607217].
  %%CITATION = HEP-PH/9607217;%%
%------------------------------------------------------------------------
%
%[419]
%\cite{Ellis:1973kp}
\bibitem{Ellis:1973kp}
  J.~R.~Ellis and R.~L.~Jaffe,
  %``A Sum Rule for Deep Inelastic Electroproduction from Polarized Protons,''
  {\it Phys.\ Rev.}\ D {\bf 9} (1974) 1444
   [Erratum-ibid.\ D {\bf 10} (1974) 1669].
  %%CITATION = PHRVA,D9,1444;%%
%--------------------------------------------------------------------------------
%
%[420]
%\cite{Larin:1997qq}
\bibitem{Larin:1997qq}
  S.~A.~Larin, T.~van Ritbergen and J.~A.~M.~Vermaseren,
  %``The Alpha-s**3 approximation of quantum chromodynamics to the Ellis-Jaffe sum rule,''
  {\it Phys.\ Lett.}\ B {\bf 404} (1997) 153
  [hep-ph/9702435].
  %%CITATION = HEP-PH/9702435;%%
%--------------------------------------------------------------------------------
%
%[421]
\bibitem{CHET:EJ}
K. Chetyrkin {\it et al.}, to appear.
%--------------------------------------------------------------------------------
%
%[422]
%\cite{Gottfried:1967kk}
\bibitem{Gottfried:1967kk}
  K.~Gottfried,
  %``Sum rule for high-energy electron - proton scattering,''
  {\it Phys.\ Rev.\ Lett.}\  {\bf 18} (1967) 1174.
  %%CITATION = PRLTA,18,1174;%%
%--------------------------------------------------------------------------------
%
%[422]
%\cite{Broadhurst:2004jx}
\bibitem{Broadhurst:2004jx} 
  D.~J.~Broadhurst, A.~L.~Kataev and C.~J.~Maxwell,
  %``Comparison of the Gottfried and Adler sum rules within the large N(c) expansion,''
  {\it Phys.\ Lett.}\ B {\bf 590}, 76 (2004)
  [hep-ph/0403037].
  %%CITATION = HEP-PH/0403037;%%
%--------------------------------------------------------------------------------
%
%[423]
%\cite{Blumlein:1999sc}
\bibitem{Blumlein:1999sc}
  J.~Bl\"umlein, B.~Geyer and D.~Robaschik,
  %``The Virtual Compton amplitude in the generalized Bjorken region: twist-2 contributions,''
  {\it Nucl.\ Phys.}\ B {\bf 560} (1999) 283
  [hep-ph/9903520].
  %%CITATION = HEP-PH/9903520;%%
%--------------------------------------------------------------------------------
%
%[424]
%\cite{Blumlein:2002fw}
\bibitem{Blumlein:2002fw}
  J.~Bl\"umlein and D.~Robaschik,
  %``Polarized deep inelastic diffractive e p scattering: Operator approach,''
  {\it Phys.\ Rev.}\ D {\bf 65} (2002) 096002
  [hep-ph/0202077].
  %%CITATION = HEP-PH/0202077;%%
%--------------------------------------------------------------------------------
%-------------------------------------------------------------------------------------------------
%-------->> Higher Twist  <<----------------------------------------------------------------------
%-------------------------------------------------------------------------------------------------
%--------------------------------------------------------------------------------
%
%[425]
%\cite{Shuryak:1981pi}
\bibitem{Shuryak:1981pi}
  E.~V.~Shuryak and A.~I.~Vainshtein,
  %``Theory of Power Corrections to Deep Inelastic Scattering in Quantum Chromodynamics. 2. Q**4 Effects: Polarized Target,''
  {\it Nucl.\ Phys.}\ B {\bf 201} (1982) 141.
  %%CITATION = NUPHA,B201,141;%%
%--------------------------------------------------------------------------------
%
%[426]
\bibitem{TW3ANOMD}
%\cite{Bukhvostov:1983te}
%\bibitem{Bukhvostov:1983te}
  A.~P.~Bukhvostov, E.~A.~Kuraev and L.~N.~Lipatov,
  %``Evolution equations for higher twist operators,''
  {\it Sov.\ J.\ Nucl.\ Phys.}\  {\bf 38} (1983) 263
  {\it  [Yad.\ Fiz.}\  {\bf 38} (1983) 439];
  %%CITATION = SJNCA,38,263;%%
%--------------------------------------------------------------------------------
%\cite{Bukhvostov:1984as}
%\bibitem{Bukhvostov:1984as}
%  A.~P.~Bukhvostov, E.~A.~Kuraev and L.~N.~Lipatov,
  %``Deep Inelastic electron Scattering by a Polarized Target in Quantum Chromodynamics,''
  {\it JETP Lett.}\  {\bf 37} (1983) 482
   {\it [Pisma Zh.\ Eksp.\ Teor.\ Fiz.}\  {\bf 37} (1983) 406];
   {\it [Sov.\ Phys.\ JETP} {\bf 60} (1984) 22]
   {\it [Zh.\ Eksp.\ Teor.\ Fiz.}\  {\bf 87} (1984) 37];
  %%CITATION = JTPLA,37,482;%%
%--------------------------------------------------------------------------------
%\cite{Buchvostov:1984xf}
%\bibitem{Buchvostov:1984xf}
%  A.~P.~Bukhvostov, E.~A.~Kuraev and L.~N.~Lipatov,
  %``Deep inelastic scattering on a polarized target in Abelian gauge theory,''
  {\it Sov.\ J.\ Nucl.\ Phys.}\  {\bf 39} (1984) 121
  {\it  [Yad.\ Fiz.}\  {\bf 39} (1984) 194];\\
  %%CITATION = SJNCA,39,121;%%
%--------------------------------------------------------------------------------
%\cite{Ratcliffe:1985mp}
%\bibitem{Ratcliffe:1985mp}
  P.~G.~Ratcliffe,
  %``Transverse Spin And Higher Twist In Qcd,''
  {\it Nucl.\ Phys.}\ B {\bf 264} (1986) 493;\\
  %%CITATION = NUPHA,B264,493;%%
%--------------------------------------------------------------------------------
%\cite{Balitsky:1987bk}
%\bibitem{Balitsky:1987bk}
  I.~I.~Balitsky and V.~M.~Braun,
  %``Evolution Equations for QCD String Operators,''
  {\it Nucl.\ Phys.}\ B {\bf 311} (1989) 541;\\
  %%CITATION = NUPHA,B311,541;%%
%--------------------------------------------------------------------------------
%\cite{Ji:1990br}
%\bibitem{Ji:1990br}
  X.~D.~Ji and C.-H.~Chou,
  %``QCD radiative corrections to the transverse spin structure function g2 (x, Q**2): 1. Nonsinglet operators,''
  {\it Phys.\ Rev.}\ D {\bf 42} (1990) 3637;\\
  %%CITATION = PHRVA,D42,3637;%%
%--------------------------------------------------------------------------------
%\cite{Kodaira:1994ge}
%\bibitem{Kodaira:1994ge}
  J.~Kodaira, Y.~Yasui and T.~Uematsu,
  %``Spin structure function g2 (x, Q**2) and twist - three operators in QCD,''
  {\it Phys.\ Lett.}\ B {\bf 344} (1995) 348
  [hep-ph/9408354];\\
  %%CITATION = HEP-PH/9408354;%%
%--------------------------------------------------------------------------------
%\cite{Kodaira:1996md}
%\bibitem{Kodaira:1996md}
  J.~Kodaira, Y.~Yasui, K.~Tanaka and T.~Uematsu,
  %``QCD corrections to the nucleon's spin structure function g2 (x, Q**2),''
  {\it Phys.\ Lett.}\ B {\bf 387} (1996) 855
  [hep-ph/9603377].
%--------------------------------------------------------------------------------
%
%[427]
%\cite{Ali:1991em}
\bibitem{Ali:1991em}
  A.~Ali, V.~M.~Braun and G.~Hiller,
  %``Asymptotic solutions of the evolution equation for the polarized nucleon structure function g-2 (x, Q**2),''
  {\it Phys.\ Lett.}\ B {\bf 266} (1991) 117.
  %%CITATION = PHLTA,B266,117;%%
%--------------------------------------------------------------------------------
%
%[428]
%\cite{Braun:2000av}
\bibitem{Braun:2000av}
  V.~M.~Braun, G.~P.~Korchemsky and A.~N.~Manashov,
  %``Evolution of twist - three parton distributions in QCD beyond the large N(c) limit,''
  {\it Phys.\ Lett.}\ B {\bf 476} (2000) 455
  [hep-ph/0001130].
  %%CITATION = HEP-PH/0001130;%%
%--------------------------------------------------------------------------------
%
%[429]
%\cite{Geyer:1996vj}
\bibitem{Geyer:1996vj}
  B.~Geyer, D.~M\"uller and D.~Robaschik,
  %``Evolution kernels of twist - three light ray operators in polarized deep inelastic scattering,''
  {\it Nucl.\ Phys.\ (Proc.\ Suppl.)}\  {\bf 51C} (1996) 106
  [hep-ph/9606320].
  %%CITATION = HEP-PH/9606320;%%
%--------------------------------------------------------------------------------
%
%[430]
%\cite{Mueller:1997yk}
\bibitem{Mueller:1997yk}
  D.~M\"uller,
  %``Calculation of higher twist evolution kernels for polarized deep inelastic scattering,''
  {\it Phys.\ Lett.}\ B {\bf 407} (1997) 314
  [hep-ph/9701338].
  %%CITATION = HEP-PH/9701338;%%
%--------------------------------------------------------------------------------
%
%[431]
\bibitem{WILSPNS}
%\cite{Kodaira:1978sh}
%\bibitem{Kodaira:1978sh}
  J.~Kodaira, S.~Matsuda, T.~Muta, K.~Sasaki and T.~Uematsu,
  %``QCD Effects in Polarized Electroproduction,''
  {\it Phys.\ Rev.}\ D {\bf 20} (1979) 627.
  %%CITATION = PHRVA,D20,627;%%
%\cite{Kodaira:1979ib}
%\bibitem{Kodaira:1979ib}
  J.~Kodaira, S.~Matsuda, K.~Sasaki and T.~Uematsu,
  %``QCD Higher Order Effects in Spin Dependent Deep Inelastic Electroproduction,''
  {\it Nucl.\ Phys.}\ B {\bf 159} (1979) 99;\\
  %%CITATION = NUPHA,B159,99;%%
%\cite{Ji:2000ny}
%\bibitem{Ji:2000ny}
  X.~-D.~Ji, W.~Lu, J.~Osborne and X.~-T.~Song,
  %``One loop factorization of the nucleon g(2) structure function in the nonsinglet case,''
  {\it Phys.\ Rev.}\ D {\bf 62} (2000) 094016
  [hep-ph/0006121].
  %%CITATION = HEP-PH/0006121;%%
%--------------------------------------------------------------------------------
%
%[432]
\bibitem{WILSPSING}
%\cite{Kodaira:1979pa}
%\bibitem{Kodaira:1979pa}
  J.~Kodaira,
  %``QCD Higher Order Effects in Polarized Electroproduction: Flavor Singlet Coefficient Functions,''
  {\it Nucl.\ Phys.}\ B {\bf 165} (1980) 129;\\
  %%CITATION = NUPHA,B165,129;%%
%\cite{Belitsky:2000pb}
%\bibitem{Belitsky:2000pb}
  A.~V.~Belitsky, X.~-D.~Ji, W.~Lu and J.~Osborne,
  %``The Singlet g(2) structure function in the next-to-leading order,''
  {\it Phys.\ Rev.}\ D {\bf 63} (2001) 094012
  [hep-ph/0007305].
  %%CITATION = HEP-PH/0007305;%%
%--------------------------------------------------------------------------------
%
%[433]
%\cite{Kodaira:1997ig}
\bibitem{Kodaira:1997ig}
  J.~Kodaira, T.~Nasuno, H.~Tochimura, K.~Tanaka and Y.~Yasui,
  %``Renormalization of gauge invariant operators for the structure function g2(x, Q**2),''
  {\it Prog.\ Theor.\ Phys.}\  {\bf 99} (1998) 315
  [hep-ph/9712395].
  %%CITATION = HEP-PH/9712395;%%
%--------------------------------------------------------------------------------
%
%[434]
%\cite{Braun:2000yi}
\bibitem{Braun:2000yi}
 V.~M.~Braun, G.~P.~Korchemsky and A.~N.~Manashov,
  %``Gluon contribution to the structure function g(2)(x, Q**2),''
  {\it Nucl.\ Phys.}\ B {\bf 597} (2001) 370
  [hep-ph/0010128].
  %%CITATION = HEP-PH/0010128;%%
%--------------------------------------------------------------------------------
%
%[435]
%\cite{Braun:2001qx}
\bibitem{Braun:2001qx}
  V.~M.~Braun, G.~P.~Korchemsky and A.~N.~Manashov,
  %``Evolution equation for the structure function g(2) (x, Q**2),''
  {\it Nucl.\ Phys.}\ B {\bf 603} (2001) 69
  [hep-ph/0102313].
  %%CITATION = HEP-PH/0102313;%%
%--------------------------------------------------------------------------------
%
%[436]
%\cite{Braun:2011aw}
\bibitem{Braun:2011aw}
  V.~M.~Braun, T.~Lautenschlager, A.~N.~Manashov and B.~Pirnay,
  %``Higher twist parton distributions from light-cone wave functions,''
  {\it Phys.\ Rev.}\ D {\bf 83} (2011) 094023
  [arXiv:1103.1269 [hep-ph]].
  %%CITATION = ARXIV:1103.1269;%%
%--------------------------------------------------------------------------------
%
%[437]
%\cite{Gockeler:2000ja}
\bibitem{Gockeler:2000ja}
  M.~G\"ockeler, R.~Horsley, W.~Kurzinger, H.~Oelrich, D.~Pleiter, P.~E.~L.~Rakow, A.~Sch\"afer and G.~Schierholz,
  %``A Lattice calculation of the nucleon's spin dependent structure function g(2) revisited,''
  {\it Phys.\ Rev.}\ D {\bf 63} (2001) 074506
  [hep-lat/0011091].
  %%CITATION = HEP-LAT/0011091;%%
%--------------------------------------------------------------------------------
%
%[438]
%\cite{Dolgov:2002zm}
\bibitem{Dolgov:2002zm}
  D.~Dolgov {\it et al.}  [LHPC and TXL Collaborations],
  %``Moments of nucleon light cone quark distributions calculated in full lattice QCD,''
  {\it Phys.\ Rev.}\ D {\bf 66} (2002) 034506
  [hep-lat/0201021].
  %%CITATION = HEP-LAT/0201021;%%
%--------------------------------------------------------------------------------
%
%[439]
%\cite{Martin:2002dr}  
\bibitem{Martin:2002dr}
  A.~D.~Martin, R.~G.~Roberts, W.~J.~Stirling and R.~S.~Thorne,
  %``NNLO global parton analysis,''   
  {\em Phys.\ Lett.}\ B {\bf 531} (2002) 216
  [hep-ph/0201127].
  %%CITATION = HEP-PH/0201127;%%
%----------------------------------------------------------------------------------------
%
%[440]
\bibitem{HTphi3}
%\bibitem{HT1}
S. Gottlieb, Nucl. Phys. B {\bf 139} (1978) 125; PhD thesis
{\sf A New Twist in Deep Inelastic Scattering}, Princeton University,
1978;
%\cite{Gottlieb:1978zj}
%\bibitem{Gottlieb:1978zj}
%  S.~A.~Gottlieb,
  %``Contribution Of Twist Four Operators To Deep Inelastic Scattering,''
  {\em Nucl.\ Phys.}\ B {\bf 139} (1978) 125;\\
  %%CITATION = NUPHA,B139,125;%%   
%\cite{Lam:1984qr}
%\bibitem{Lam:1984qr}
  C.~S.~Lam and M.~A.~Walton,
  %``PERMUTATION SYMMETRY AND ANOMALOUS-DIMENSIONS OF FOUR FIELD OPERATORS I
  % N phi**3 in six-dimensions THEORY,''
  {\em Can.\ J.\ Phys.}\  {\bf 63} (1985) 1042.
  %%CITATION = CJPHA,63,1042;%%
%----------------------------------------------------------------------------------------
%
%[441]
\bibitem{HTQCD}
%\cite{Politzer:1980me}
%\bibitem{Politzer:1980me}
  H.~D.~Politzer,
  %``Power Corrections at Short Distances,''
  {\em Nucl.\ Phys.}\ B {\bf 172} (1980) 349;\\
  %%CITATION = NUPHA,B172,349;%%
%\bibitem{OKAWA}
%\cite{Okawa:1980ei}
%\bibitem{Okawa:1980ei}
  M.~Okawa,
  %``Higher Twist Effects In Asymptotically Free Gauge Theories: The Anomalous Dimensions
  % Of Four Quark Operators,''
  {\em Nucl.\ Phys.}\ B {\bf 172} (1980) 481;
  %%CITATION = NUPHA,B172,481;%%
%\cite{Okawa:1981uw}   
%\bibitem{Okawa:1981uw}  
%  M.~Okawa,
  %``Higher Twist Effects In Asymptotically Free Gauge Theories,''
%  Nucl.\ Phys.\ B
  {\bf 187} (1981) 71;\\
  %%CITATION = NUPHA,B187,71;%%
%----------------------------------------------------------------------------------------
%\bibitem{WADA}
%\cite{Wada:1981hx}
%\bibitem{Wada:1981hx}
  S.~Wada,
  %``Analysis Of Twist Four Two Quark Process By The Renormalization Mixing,''
  {\em Nucl.\ Phys.}\ B {\bf 202}, 201 (1982);
  %%CITATION = NUPHA,B202,201;%%
%\cite{Wada:1982my}    
%\bibitem{Wada:1982my}   
%  S.~Wada,
  %``Twist - Four Four Quark Process In The Parton Picture,''
  {\em Phys.\ Lett.}\ B {\bf 119} (1982) 427;\\
  %%CITATION = PHLTA,B119,427;%%
%----------------------------------------------------------------------------------------
%\cite{Luttrell:1981ke}
%\bibitem{Luttrell:1981ke}
  S.~P.~Luttrell, S.~Wada and B.~R.~Webber,
  %``The Wilson Coefficient Functions Of Four Quark Operators And The Four
  % Quark Process In Deep Inelastic Scattering,''
  {\em Nucl.\ Phys.}\ B {\bf 188} (1981) 219;\\
  %%CITATION = NUPHA,B188,219;%%
%----------------------------------------------------------------------------------------
%\cite{Luttrell:1982zu}
%\bibitem{Luttrell:1982zu}
  S.~P.~Luttrell and S.~Wada,
  %``The Current Product Expansion Of The Two Quark Process At The Twist Four Level,''
  {\em Nucl.\ Phys.}\ B {\bf 197} (1982) 290
   [Erratum-ibid.\ B {\bf 206} (1982) 497];\\
  %%CITATION = NUPHA,B197,290;%%
%----------------------------------------------------------------------------------------
%\cite{Shuryak:1981dg}
%\bibitem{Shuryak:1981dg}
  E.~V.~Shuryak and A.~I.~Vainshtein,
  %``Qcd Power Corrections To Deep Inelastic Scattering,''
  {\em Phys.\ Lett.}\ B {\bf 105} (1981) 65;
  %%CITATION = PHLTA,B105,65;%%
%\cite{Shuryak:1981kj}
%\bibitem{Shuryak:1981kj}
%  E.~V.~Shuryak and A.~I.~Vainshtein,
  %``Theory of Power Corrections to Deep Inelastic Scattering in Quantum Chromodynamics.
  % 1. Q**2 Effects,''
  {\em Nucl.\ Phys.}\ B {\bf 199} (1982) 451.
  %%CITATION = NUPHA,B199,451;%%
%----------------------------------------------------------------------------------------
%
%[442]
\bibitem{JAFSOL}
%\cite{Jaffe:1981td}
%\bibitem{Jaffe:1981td}
  R.~L.~Jaffe and M.~Soldate,
  %``Twist Four in the QCD Analysis of Leptoproduction,''
  {\em Phys.\ Lett.}\ B {\bf 105} (1981) 467;
  %%CITATION = PHLTA,B105,467;%%
%\cite{Jaffe:1982pm}
%\bibitem{Jaffe:1982pm}
%  R.~L.~Jaffe and M.~Soldate,
  %``Twist Four in Electroproduction: Canonical Operators and Coefficient Functions,''
  {\em Phys.\ Rev.}\ D {\bf 26} (1982) 49;\\
  %%CITATION = PHRVA,D26,49;%%
%\cite{Jaffe:1983hp}
%\bibitem{Jaffe:1983hp}
  R.~L.~Jaffe,
  %``Parton Distribution Functions for Twist Four,''
  {\em Nucl.\ Phys.}\ B {\bf 229} (1983) 205.
  %%CITATION = NUPHA,B229,205;%%
%----------------------------------------------------------------------------------------
%
%[443]
\bibitem{EFP}
%\cite{Ellis:1982wd}
%\bibitem{Ellis:1982wd}
  R.~K.~Ellis, W.~Furmanski and R.~Petronzio,
  %``Power Corrections to the Parton Model in QCD,''
  {\em Nucl.\ Phys.}\ B {\bf 207} (1982) 1;
  %%CITATION = NUPHA,B207,1;%%
%\cite{Ellis:1982cd}
%\bibitem{Ellis:1982cd}
%  R.~K.~Ellis, W.~Furmanski and R.~Petronzio,
  %``Unraveling Higher Twists,''
  {\em Nucl.\ Phys.}\ B {\bf 212} (1983) 29.
  %%CITATION = NUPHA,B212,29;%%
%----------------------------------------------------------------------------------------
%
%[444]
%\cite{Bukhvostov:1987pr}
\bibitem{Bukhvostov:1987pr}
  A.~P.~Bukhvostov and G.~V.~Frolov,
  %``Anomalous Dimensionalities Of Quasiparton Twist = Four Operators. (in Russian),''
  {\em Yad.\ Fiz.}\  {\bf 45} (1987) 1136, {\em Sov. J. Nucl. Phys.} {\bf 45} (1987) 704.
  %%CITATION = YAFIA,45,1136;%%
%----------------------------------------------------------------------------------------
%
%[445]
%\cite{Qiu:1988dn}
\bibitem{Qiu:1988dn}
  J.~-W.~Qiu,
  %``Twist Four Contributions To The Parton Structure Functions,''
  {\em Phys.\ Rev.}\ D {\bf 42} (1990) 30.
  %%CITATION = PHRVA,D42,30;%%
%----------------------------------------------------------------------------------------
%
%[446]
%\cite{Bartels:1999km}
\bibitem{Bartels:1999km}
  J.~Bartels, C.~Bontus and H.~Spiesberger,
  %``Factorization of twist four gluon operator contributions,''
  hep-ph/9908411.
  %%CITATION = HEP-PH/9908411;%%
%----------------------------------------------------------------------------------------
%
%[451]
%\cite{Braun:2000kw}
\bibitem{Braun:2000kw}
  V.~Braun, R.~J.~Fries, N.~Mahnke and E.~Stein,
  %``Higher twist distribution amplitudes of the nucleon in QCD,''
  {\em Nucl.\ Phys.}\ B {\bf 589} (2000) 381
   [Erratum-ibid.\ B {\bf 607} (2001) 433]
  [hep-ph/0007279].
  %%CITATION = HEP-PH/0007279;%%
%----------------------------------------------------------------------------------------
%
%[452]
%\cite{Braun:2009vc}
\bibitem{Braun:2009vc}
%\cite{Braun:2008ia}
%\bibitem{Braun:2008ia}
  V.~M.~Braun, A.~N.~Manashov and J.~Rohrwild,
  %``Baryon Operators of Higher Twist in QCD and Nucleon Distribution Amplitudes,''
  {\it Nucl.\ Phys.}\ B {\bf 807} (2009) 89
  [arXiv:0806.2531 [hep-ph]];
  %%CITATION = ARXIV:0806.2531;%%
%  V.~M.~Braun, A.~N.~Manashov and J.~Rohrwild,
  %``Renormalization of Twist-Four Operators in QCD,''
  {\em Nucl.\ Phys.}\ B {\bf 826} (2010) 235
  [arXiv:0908.1684 [hep-ph]].
  %%CITATION = ARXIV:0908.1684;%%
%----------------------------------------------------------------------------------------
%
%[448]
%\cite{Glatzmaier:2012iu}
\bibitem{Glatzmaier:2012iu}
  M.~J.~Glatzmaier, S.~Mantry and M.~J.~Ramsey-Musolf,
  %``Higher Twist in Electroproduction: Flavor Non-Singlet QCD Evolution,''
  arXiv:1208.2998 [hep-ph].
  %%CITATION = ARXIV:1208.2998;%%
%----------------------------------------------------------------------------------------
%
%[447]
\bibitem{BRAUN1}
%\cite{Braun:1986ty}
%\bibitem{Braun:1986ty}
  V.~M.~Braun and A.~V.~Kolesnichenko,
  %``Power Corrections To Bjorken And Gross-llewellyn Smith Sum Rules In Qcd,''
  {\it Nucl.\ Phys.}\ B {\bf 283} (1987) 723;\\
  %%CITATION = NUPHA,B283,723;%%
%\cite{Balitsky:1989uj}
%\bibitem{Balitsky:1989uj}
  I.~I.~Balitsky, V.~M.~Braun and A.~V.~Kolesnichenko,
  %``Power law corrections to parton sum rules for deep inelastic scattering by polarized target,''
  {\it JETP Lett.}\  {\bf 50} (1989) 61
  {\it [Pisma Zh.\ Eksp.\ Teor.\ Fiz.}\  {\bf 50} (1989) 54];
  %%CITATION = JTPLA,50,61;%%
%\cite{Balitsky:1989jb}
%\bibitem{Balitsky:1989jb}
%  I.~I.~Balitsky, V.~M.~Braun and A.~V.~Kolesnichenko,
  %``Power corrections 1 / Q**2 to parton sum rules for deep inelastic scattering from polarized targets,''
  {\it Phys.\ Lett.}\ B {\bf 242} (1990) 245
   [Erratum-ibid.\ B {\bf 318} (1993) 648]
  [hep-ph/9310316].
  %%CITATION = HEP-PH/9310316;%%
%----------------------------------------------------------------------------------------
%
%[449]
%\cite{Beneke:1998ui}
\bibitem{Beneke:1998ui}
  M.~Beneke,
  %``Renormalons,''
  {\it Phys.\ Rept.}\  {\bf 317} (1999) 1
  [hep-ph/9807443]. 
  %%CITATION = HEP-PH/9807443;%%
%----------------------------------------------------------------------------------------
%
%[450]
%\cite{Beneke:2000kc}
\bibitem{Beneke:2000kc}
  M.~Beneke and V.~M.~Braun,
  %``Renormalons and power corrections,''
  In: M.~Shifman, M. (ed.): {\sf At the frontier of particle physics}, {\bf 3}, 1719, (World Scientific,
  Singapore, 2001), 
  [hep-ph/0010208]. 
  %%CITATION = HEP-PH/0010208;%%
%----------------------------------------------------------------------------------------
%
%[453]
\bibitem{HT:PHAE}
%\cite{Eisele:1981nk}
%\bibitem{Eisele:1981nk}
  F.~Eisele, M.~Gl\"uck, E.~Hoffmann and E.~Reya,
  %``Limits On Higher Twist Contributions To Deep Inelastic Scattering,''
  {\it Phys.\ Rev.} D {\bf 26} (1982) 41;\\
  %%CITATION = PHRVA,D26,41;%%
%----------------------------------------------------------------------------------------
%\cite{Varvell:1987qu}
%\bibitem{Varvell:1987qu}
  K.~Varvell {\it et al.},  [BEBC WA59 Collaboration],
  %``Measurement Of The Structure Functions F2 And Xf3 And Comparison With Qcd predictions
  % Including Kinematical And Dynamical Higher Twist Effects,''
  {\em Z.\ Phys.}\ C {\bf 36} (1987) 1;\\
  %%CITATION = ZEPYA,C36,1;%%
%----------------------------------------------------------------------------------------
%\cite{Virchaux:1991jc}
%\bibitem{Virchaux:1991jc}
  M.~Virchaux and A.~Milsztajn,
  %``A Measurement of alpha-s and higher twists from a QCD analysis of high statistics
  % F-2 data on hydrogen and deuterium targets,''
  {\em Phys.\ Lett.}\ B {\bf 274} (1992) 221;\\
  %%CITATION = PHLTA,B274,221;%%
%---------------------------------------------------------------------------------------- 
%\cite{Kataev:1997nc}
%\bibitem{Kataev:1997nc}
  A.~L.~Kataev, A.~V.~Kotikov, G.~Parente and A.~V.~Sidorov,
  %``Next to next-to-leading order QCD analysis of the revised CCFR data for
  % xF3 structure function and the higher twist contributions,''
  {\em Phys.\ Lett.}\ B {\bf 417} (1998) 374
  [hep-ph/9706534];\\
  %%CITATION = HEP-PH/9706534;%%
%----------------------------------------------------------------------------------------
%\cite{Alekhin:1998df}
%\bibitem{Alekhin:1998df}
  S.~I.~Alekhin and A.~L.~Kataev,
  %``The nlo DGLAP extraction of alpha(s) and higher twist terms from ccfr xf(3) and f(2)
  % structure functions data for neutrino n dis,''
  {\em Phys.\ Lett.}\ B {\bf 452} (1999) 402
  [hep-ph/9812348];\\
  %%CITATION = HEP-PH/9812348;%%
%----------------------------------------------------------------------------------------
%\cite{Botje:1999dj}
%\bibitem{Botje:1999dj}
  M.~Botje,
  %``A QCD analysis of HERA and fixed target structure function data,''
  {\em Eur.\ Phys.\ J.}\ C {\bf 14} (2000) 285
  [hep-ph/9912439];\\
  %%CITATION = HEP-PH/9912439;%%
%----------------------------------------------------------------------------------------
%\cite{Alekhin:2000ch}
%\bibitem{Alekhin:2000ch}
  S.~I.~Alekhin,
  %``Global fit to the charged leptons DIS data: alpha(s) parton distributions, and high twists,''
  {\em Phys.\ Rev.}\ D {\bf 63} (2001) 094022
  [hep-ph/0011002];\\
  %%CITATION = HEP-PH/0011002;%%
%----------------------------------------------------------------------------------------
%\cite{Alekhin:2003qq}
%\bibitem{Alekhin:2003qq}
  S.~I.~Alekhin, S.~A.~Kulagin and S.~Liuti,
  %``Isospin dependence of power corrections in deep inelastic scattering,''
  {\em Phys.\ Rev.}\ D {\bf 69} (2004) 114009
  [hep-ph/0304210];\\   
  %%CITATION = HEP-PH/0304210;%%
%----------------------------------------------------------------------------------------
%\cite{Alekhin:2007fh}
%\bibitem{Alekhin:2007fh}
  S.~I.~Alekhin, S.~A.~Kulagin and R.~Petti,
  %``Modeling lepton-nucleon inelastic scattering from high to low momentum transfer,''
  {\em AIP Conf.\ Proc.}\  {\bf 967} (2007) 215
  [arXiv:0710.0124 [hep-ph]].
  %%CITATION = ARXIV:0710.0124;%%
%----------------------------------------------------------------------------------------
%
%[454]
%\cite{Blumlein:2008kz}
\bibitem{Blumlein:2008kz}
  J.~Bl\"umlein and H.~B\"ottcher,
  %``Higher Twist Contributions to the Structure Functions F(2)**p(x,Q**2) 
  %and F(2)**d(x,Q**2) at Large x and Higher Orders,''
  {\em Phys.\ Lett.}\ B {\bf 662} (2008) 336
  [arXiv:0802.0408 [hep-ph]].
  %%CITATION = ARXIV:0802.0408;%%
%----------------------------------------------------------------------------------------
%
%[455]
%\cite{Leader:2006xc}
\bibitem{Leader:2006xc}
  E.~Leader, A.~V.~Sidorov and D.~B.~Stamenov,
  %``Impact of CLAS and COMPASS data on Polarized Parton Densities and Higher Twist,''
  {\em Phys.\ Rev.}\ D {\bf 75} (2007) 074027
  [hep-ph/0612360].
  %%CITATION = HEP-PH/0612360;%%
%----------------------------------------------------------------------------------------
%------------------------ Nuclear PDFs ----------------------------------
%----------------------------------------------------------------------------------------
%
%[456]
%\cite{Aubert:1983xm}
\bibitem{Aubert:1983xm}
  J.~J.~Aubert {\it et al.},  [European Muon Collaboration],
  %``The ratio of the nucleon structure functions $F2_n$ for iron and deuterium,''
  {\em Phys.\ Lett.}\ B {\bf 123} (1983) 275.
  %%CITATION = PHLTA,B123,275;%%
%------------------------------------------------------------------------
%
%[457]
\bibitem{BODEKRI}
%\cite{Bodek:1980ar}
%\bibitem{Bodek:1980ar}
  A.~Bodek and J.~L.~Ritchie,
  %``Fermi Motion Effects in Deep Inelastic Lepton Scattering from Nuclear Targets,''
  {\it Phys.\ Rev.}\ D {\bf 23} (1981) 1070;
  %%CITATION = PHRVA,D23,1070;%%
%\cite{Bodek:1981wr}
%\bibitem{Bodek:1981wr}
%  A.~Bodek and J.~L.~Ritchie,
  %``Further Studies of Fermi Motion Effects in Lepton Scattering from Nuclear Targets,''
  {\it Phys.\ Rev.}\ D {\bf 24} (1981) 1400.
  %%CITATION = PHRVA,D24,1400;%%
%------------------------------------------------------------------------
%
%[458]
\bibitem{NUCL:REV}
%\bibitem{Rith:REV}
%\cite{Rith:1986ku}
%\bibitem{Rith:1986ku}
  K.~Rith,
  %``The Emc Effect: Quarks, Gluons And Nuclear Structure,''
  {\em Ann. Phys.} ( N.Y.) {\bf 176} (1987) 344;
%-358. Mit Cambridge - CTP-1433 (86,REC.JAN.87) 23p
%\cite{Rith:1988qf}
%\bibitem{Rith:1988qf}
%  K.~Rith,
  %``The Emc Effect - Status And Perspectives,''
  {\em Z.\ Phys.}\ C {\bf 38} (1988) 317;\\
  %%CITATION = ZEPYA,C38,317;%%
%------------------------------------------------------------------------
%\cite{Geesaman:1995yd}
%\bibitem{Geesaman:1995yd}
  D.~F.~Geesaman, K.~Saito and A.~W.~Thomas,
  %``The nuclear EMC effect,''
  {\em Ann.\ Rev.\ Nucl.\ Part.\ Sci.}\  {\bf 45} (1995) 337;\\
  %%CITATION = ARNUA,45,337;%%
%------------------------------------------------------------------------
%\cite{Norton:2003cb}
%\bibitem{Norton:2003cb}
  P.~R.~Norton,
  %``The EMC effect,''
  {\it Rept.\ Prog.\ Phys.}\  {\bf 66} (2003) 1253;\\
  %%CITATION = RPPHA,66,1253;%%
%--------------------------------------------------------------------------------
%\cite{Armesto:2006ph}
%\bibitem{Armesto:2006ph}
  N.~Armesto,
  %``Nuclear shadowing,''
  {\em J.\ Phys.}\ G {\bf 32} (2006) R367
  [hep-ph/0604108].
  %%CITATION = HEP-PH/0604108;%% 
%------------------------------------------------------------------------ 
%
%[459]
\bibitem{NUC:RAT}
%\cite{Kumano:1992ef}
%\bibitem{Kumano:1992ef}
  S.~Kumano,
  %``Nuclear shadowing in a parton recombination model,''
  {\em Phys.\ Rev.}\ C {\bf 48} (1993) 2016
  [hep-ph/9303306];\\
  %%CITATION = HEP-PH/9303306;%%
%------------------------------------------------------------------------
%\cite{Hirai:2007sx}
%\bibitem{Hirai:2007sx}
  M.~Hirai, S.~Kumano and T.~-H.~Nagai,
  %``Determination of nuclear parton distribution functions and their uncertainties in next-to-leading order,''
  {\em Phys.\ Rev.}\ C {\bf 76} (2007) 065207
  [arXiv:0709.3038 [hep-ph]];\\
  %%CITATION = ARXIV:0709.3038;%%
%------------------------------------------------------------------------
%\cite{Hirai:2010xs}
%\bibitem{Hirai:2010xs}
  M.~Hirai, S.~Kumano, K.~Saito and T.~Watanabe,
  %``Clustering aspects in nuclear structure functions,''
  {\em Phys.\ Rev.}\ C {\bf 83} (2011) 035202
  [arXiv:1008.1313 [hep-ph]];\\
  %%CITATION = ARXIV:1008.1313;%%
%------------------------------------------------------------------------
%\cite{Eskola:2008ca}
%\bibitem{Eskola:2008ca}
  K.~J.~Eskola, H.~Paukkunen and C.~A.~Salgado,
  %``An Improved global analysis of nuclear parton distribution functions including RHIC data,''
  {\em JHEP} {\bf 0807} (2008) 102
  [arXiv:0802.0139 [hep-ph]];
  %%CITATION = ARXIV:0802.0139;%%
%------------------------------------------------------------------------
%\cite{Eskola:2010jh}
%\bibitem{Eskola:2010jh}
%  K.~J.~Eskola, H.~Paukkunen and C.~A.~Salgado,
  %``Nuclear PDFs at NLO - status report and review of the EPS09 results,''
  {\em Nucl.\ Phys.}\ A {\bf 855} (2011) 150
  [arXiv:1011.6534 [hep-ph]];\\
  %%CITATION = ARXIV:1011.6534;%%
%------------------------------------------------------------------------
%\cite{Schienbein:2007fs}
%\bibitem{Schienbein:2007fs}
  I.~Schienbein, {\it et al.}, %J.~Y.~Yu, C.~Keppel, J.~G.~Morfin, F.~Olness and J.~F.~Owens,
  %``Nuclear parton distribution functions from neutrino deep inelastic scattering,''
  {\em Phys.\ Rev.}\ D {\bf 77} (2008) 054013
  [arXiv:0710.4897 [hep-ph]];\\
  %%CITATION = ARXIV:0710.4897;%%
%------------------------------------------------------------------------
%\cite{Schienbein:2009kk} 
%\bibitem{Schienbein:2009kk}
%  I.~Schienbein, {\it et al.}, %J.~Y.~Yu, K.~Kovarik, C.~Keppel, J.~G.~Morfin, F.~Olness and J.~F.~Owens,
  %``PDF Nuclear Corrections for Charged and Neutral Current Processes,''
  {\em Phys.\ Rev.}\ D {\bf 80} (2009) 094004
  [arXiv:0907.2357 [hep-ph]].
  %%CITATION = ARXIV:0907.2357;%%
%------------------------------------------------------------------------
%\cite{Eskola:2009uj}
\bibitem{Eskola:2009uj}
  K.~J.~Eskola, H.~Paukkunen and C.~A.~Salgado,
  %``EPS09: A New Generation of NLO and LO Nuclear Parton Distribution Functions,''
  {\it JHEP} {\bf 0904} (2009) 065
  [arXiv:0902.4154 [hep-ph]].
  %%CITATION = ARXIV:0902.4154;%%
%------------------------------------------------------------------------
%
%[460]
%\cite{Kulagin:2004ie}
\bibitem{Kulagin:2004ie}
  S.~A.~Kulagin and R.~Petti,
  %``Global study of nuclear structure functions,''
  {\em Nucl.\ Phys.}\ A {\bf 765} (2006) 126
  [hep-ph/0412425].
  %%CITATION = HEP-PH/0412425;%%
%------------------------------------------------------------------------
%
%[461]
%\cite{Kulagin:2010gd}
\bibitem{Kulagin:2010gd}
  S.~A.~Kulagin and R.~Petti,
  %``Structure functions for light nuclei,''
  {\em Phys.\ Rev.}\ C {\bf 82} (2010) 054614
  [arXiv:1004.3062 [hep-ph]].
  %%CITATION = ARXIV:1004.3062;%%
%------------------------------------------------------------------------
%
%[462]
%\cite{deFlorian:2003qf}
\bibitem{deFlorian:2003qf}
  D.~de Florian and R.~Sassot,
  %``Nuclear parton distributions at next-to-leading order,''
  {\it Phys.\ Rev.}\ D {\bf 69} (2004) 074028
  [hep-ph/0311227].
  %%CITATION = HEP-PH/0311227;%%
%------------------------------------------------------------------------
%
%[463]
%\cite{deFlorian:2011fp}
\bibitem{deFlorian:2011fp}
  D.~de Florian, R.~Sassot, P.~Zurita and M.~Stratmann,
  %``Global Analysis of Nuclear Parton Distributions,''
  {\it Phys.\ Rev.}\ D {\bf 85} (2012) 074028
  [arXiv:1112.6324 [hep-ph]].
  %%CITATION = ARXIV:1112.6324;%%
%------------------------------------------------------------------------
% 
%[464] 
%\cite{Arrington:2011qt}
\bibitem{Arrington:2011qt}
  J.~Arrington, J.~G.~Rubin and W.~Melnitchouk,
  %``How well do we know the neutron structure function?,''
  arXiv:1110.3362 [hep-ph].
  %%CITATION = ARXIV:1110.3362;%%
%------------------------------------------------------------------------
% 
%[465] 
%\cite{Machleidt:1987hj}
\bibitem{Machleidt:1987hj}
  R.~Machleidt, K.~Holinde and C.~Elster,
  %``The Bonn Meson Exchange Model for the Nucleon Nucleon Interaction,''
  {\it Phys.\ Rept.}\  {\bf 149} (1987) 1.
  %%CITATION = PRPLC,149,1;%%
%------------------------------------------------------------------------
% 
%[466] 
%\cite{Lacombe:1980dr}
\bibitem{Lacombe:1980dr}
  M.~Lacombe {\it et al.}, %B.~Loiseau, J.~M.~Richard, R.~Vinh Mau, J.~Cote, P.~Pires and R.~De Tourreil,
  %``Parametrization of the Paris n n Potential,''
  {\it Phys.\ Rev.}\ C {\bf 21} (1980) 861.
  %%CITATION = PHRVA,C21,861;%%
%------------------------------------------------------------------------
% 
%[467] 
%\cite{Kulagin:2007ju}
\bibitem{Kulagin:2007ju}
  S.~A.~Kulagin and R.~Petti,
  %``Neutrino inelastic scattering off nuclei,''
  {\it Phys.\ Rev.}\ D {\bf 76} (2007) 094023
  [hep-ph/0703033].
  %%CITATION = HEP-PH/0703033;%%
%------------------------------------------------------------------------
% 
%[468] 
%\cite{Alde:1990im}
\bibitem{Alde:1990im}
  D.~M.~Alde {\it et al.},
%H.~W.~Baer, T.~A.~Carey, G.~T.~Garvey, A.~Klein, C.~Lee, M.~J.~Leitch and
%  J.~W.~Lillberg {\it et al.},
  %``Nuclear dependence of dimuon production at 800-GeV. FNAL-772 experiment,''
  {\it Phys.\ Rev.\ Lett.}\  {\bf 64} (1990) 2479.
  %%CITATION = PRLTA,64,2479;%%
%------------------------------------------------------------------------
%
%[469]
%\cite{Vermaseren:1994je}
\bibitem{Vermaseren:1994je}
  J.~A.~M.~Vermaseren,
  %``Axodraw,''
  {\it Comput.\ Phys.\ Commun.}\  {\bf 83} (1994) 45.
  %%CITATION = CPHCB,83,45;%%
%----------------------------------------------------------------------
}
%%%%%%%%%%%%%%%%%%%%%%%%%%%%%%%%%%%%%%%%%%%%%%%%%%%%%%%%%%%%%%%%%%%%%%%%%
\end{thebibliography}
\end{document}